\documentclass[11pt,a4paper,twoside,openright]{report}
\pdfoutput=1

\usepackage{color}
\definecolor{MyDarkBlue}{rgb}{0.15,0.25,0.65}

\usepackage[british]{babel}
\usepackage{fancyhdr}
\usepackage{lastpage}
\usepackage[margin=2.5cm, includehead, includefoot]{geometry}  

\usepackage{tikz}
\usepackage{verbatim}

\usepackage{makeidx}

\let\fn\footnote
\renewcommand{\footnote}[1]{\linespread{1.1}\fn{#1}\linespread{1.29}}

\makeatletter
\renewcommand\@idxitem{\par\hangindent 40\p@}
\renewcommand\subitem{\@idxitem \hspace*{20\p@}}
\renewcommand\subsubitem{\@idxitem \hspace*{30\p@}}
\renewcommand\indexspace{\par \vskip 10\p@ \@plus5\p@ \@minus3\p@\relax}

\makeatletter\renewcommand{\section}{\@startsection {section}{1}{\z@}{-3.5ex plus -1ex minus -.2ex}{2.3ex plus .2ex}{\bf\mathversion{bold}\large }}

\makeatletter\renewcommand{\subsection}{\@startsection{subsection}{2}{\z@}{-3.25ex plus -1ex minus  -.2ex}{1.5ex plus .2ex}{\bf\mathversion{bold} }}

\makeatletter\renewcommand{\subsubsection}{\@startsection{subsubsection}{3}{\z@}{-3.25ex plus -1ex minus -.2ex}{1.5ex plus .2ex}{\it }}

\renewcommand{\thesection}{\arabic{chapter}.\arabic{section}.}
\renewcommand{\thesubsection}{\arabic{chapter}.\arabic{section}.\arabic{subsection}.}

\renewcommand{\@seccntformat}[1]{\@nameuse{the#1}~}

\renewcommand{\theequation}{\thechapter.\arabic{equation}}
\makeatletter \@addtoreset{equation}{chapter}

\makeatletter \@addtoreset{figure}{chapter}

\renewcommand*\l@chapter[2]{%
  \ifnum \c@tocdepth >\m@ne
    \addpenalty{-\@highpenalty}%
    \vskip 0.0em \@plus\p@   
    \setlength\@tempdima{1.5em}
    \begingroup
      \def\numberline##1{\hb@xt@\@tempdima{##1.\hfil}}
      \parindent \z@ \rightskip \@pnumwidth
      \parfillskip -\@pnumwidth
      \leavevmode 
      \advance\leftskip\@tempdima
      \hskip -\leftskip
      #1\nobreak\mdseries
      \leaders\hbox{$\m@th
        \mkern \@dotsep mu\hbox{.}\mkern \@dotsep
        mu$}\hfill
      \nobreak\hb@xt@\@pnumwidth{\hss #2}\par
      \penalty\@highpenalty
    \endgroup
  \fi}
  
\renewcommand*\l@section{\@dottedtocline{2}{1.5em}{2.2em}}
\renewcommand*\l@subsection{\@dottedtocline{3}{3.7em}{3em}}

\renewcommand\tableofcontents{%
    \section*{\large\contentsname
        \@mkboth{%
          \MakeUppercase\contentsname}{\MakeUppercase\contentsname}}%
       {\baselineskip=15pt plus 2pt minus 1pt
    \@starttoc{toc}}%
}

\renewenvironment{thebibliography}[1]
     {\baselineskip=16pt plus 2pt minus 1pt
     \thispagestyle{empty}
     \vspace*{4cm}
      \section*{\large\refname
        \@mkboth{\MakeUppercase\refname}{\MakeUppercase\refname}}%
       \addcontentsline{toc}{chapter}{References}
     \list{\@biblabel{\@arabic\c@enumiv}}%
           {\settowidth\labelwidth{\@biblabel{#1}}%
            \leftmargin\labelwidth
            \advance\leftmargin\labelsep
            \@openbib@code
            \usecounter{enumiv}%
            \let\p@enumiv\@empty
            \renewcommand\theenumiv{\@arabic\c@enumiv}}%
      \sloppy
      \clubpenalty4000
      \@clubpenalty \clubpenalty
      \widowpenalty4000%
      \sfcode`\.\@m
 \catcode`\^^M=10%
}

\newcommand{\appendices}{
\chapter*{Appendices}\label{appendices}
\clearpage{\pagestyle{empty}\cleardoublepage}
\setcounter{chapter}{0}
\renewcommand{\thechapter}{\Alph{chapter}}
\addcontentsline{toc}{chapter}{Appendices}
\setcounter{equation}{0}
\makeatletter
\renewcommand{\theequation}{\Alph{chapter}.\arabic{equation}}
\renewcommand{\thesection}{\Alph{chapter}.\arabic{section}.}
\renewcommand{\thesubsection}{\Alph{chapter}.\arabic{section}.\arabic{subsection}.}
\@addtoreset{equation}{chapter}
\makeatother
}

\newcommand{\clearemptydoublepage}{\newpage\phantom{}\thispagestyle{empty}\newpage}

\usepackage[nosort]{cite}  
\usepackage[bulletsep]{collref}  

\usepackage{amsmath,amssymb,amsthm}
\usepackage{amsfonts}
\usepackage{mathrsfs}
\usepackage{bbm}
\usepackage{bm}
\usepackage{epsfig,rotating}
\usepackage{enumerate}

\newtheorem{definition}{Definition}
\numberwithin{definition}{chapter}

\newtheorem{theorem}[definition]{Theorem}

\tikzstyle{mybox} = [draw=black!50, fill=gray!20,  thick,
    rectangle, rounded corners, inner sep=10pt, inner ysep=20pt]
\tikzstyle{fancytitle} =[fill=black!50, text=white]

\newcounter{exe}
\numberwithin{exe}{chapter}
\renewcommand{\theexe}{\thechapter.\roman{exe}}

\usepackage[Bjornstrup]{fncychap}
  
\usepackage[linktocpage=true]{hyperref}
\hypersetup{
colorlinks=true,
citecolor=MyDarkBlue,
linkcolor=MyDarkBlue,
urlcolor=MyDarkBlue,
pdfauthor={Andrea Fontanella}, 
pdftitle={PhD thesis}, 
breaklinks=true
}
\usepackage{slashed}
\usepackage{tikz-cd}
\usepackage{soul}

\usepackage{float}

\newtheorem{thm}{Theorem}[chapter]
\newtheorem{cnj}{Conjecture}[chapter]

\usepackage{ dsfont }

\newcommand{\be}{\begin{eqnarray}}
\newcommand{\ee}{\end{eqnarray}}
\newcommand{\bea}{\begin{eqnarray}}
\newcommand{\eea}{\end{eqnarray}}
\newcommand{\ba}{\begin{array}}
\newcommand{\ea}{\end{array}}

\newenvironment{frcseries}{\fontfamily{frc}\selectfont}{}

\usepackage{mathtools}
\usepackage{textgreek}

\usepackage{tikz-feynman}

\DeclarePairedDelimiter\abs{\lvert}{\rvert}

\makeatletter
\let\oldabs\abs
\def\abs{\@ifstar{\oldabs}{\oldabs*}}
\let\oldnorm\norm
\def\norm{\@ifstar{\oldnorm}{\oldnorm*}}
\makeatother

\usepackage{ mathrsfs }

\def\d{\delta}

\def\e{\epsilon}
\def\D{\Delta}

\def\G{\Gamma}
\def\O{\Omega}

\def\la{\lambda}

\def\cA{{\cal A}}

\def\cL{{\cal L}}

\def\hn{{\hat{\nu}}}

\def\hN{{\hat{N}}}

\def\le{\mathbf{e}}

\def\bcc#1{\buildrel \circ \over #1}

\def\rsq{{\cal{O}}(r^2)}

\def \cH {{\cal H}}
\def\cI{{\cal I}}
\def\cS{{\cal S}}

\def\w {\wedge}

\def\bcc#1{\buildrel \circ \over #1}

\def\hh{{\bcc h}}
\def\hhn{{\bcc \nabla}}

\def\hW{{\bcc W}}

\def\hD{{\bcc \Delta}}

\def\h\gamma{{\bcc \gamma}}

\def\hN{{\bcc N}}
\def\hY{{\bcc Y}}
\def\hW{{\bcc W}}
\def\hPhi{{\bcc \Phi}}
\def\e{{e^{- {\bcc \Phi}}}}

\def\hn {{\tilde{\nabla}}}

\usepackage[nosort]{cite}  
\usepackage[bulletsep]{collref}  

\begin{document}



\begin{titlepage}

\setcounter{page}{0}
\renewcommand{\thefootnote}{\fnsymbol{footnote}}

\begin{center}

    {\huge\textbf{\mathversion{bold}Black Horizons and Integrability \\ in String Theory}\par}

    \vspace{2.5cm}

    {\Large\bf Andrea Fontanella}

    \vspace{2cm}

    Thesis submitted to the University of Surrey\\ for the degree of Doctor of Philosophy

    \vspace{2cm}

    {\it\large Department of Mathematics\\ University of Surrey\\ Guildford GU2 7XH, United Kingdom}

    \vspace{2.5cm}

    \href{http://www.surrey.ac.uk/}{\includegraphics[width=3cm]{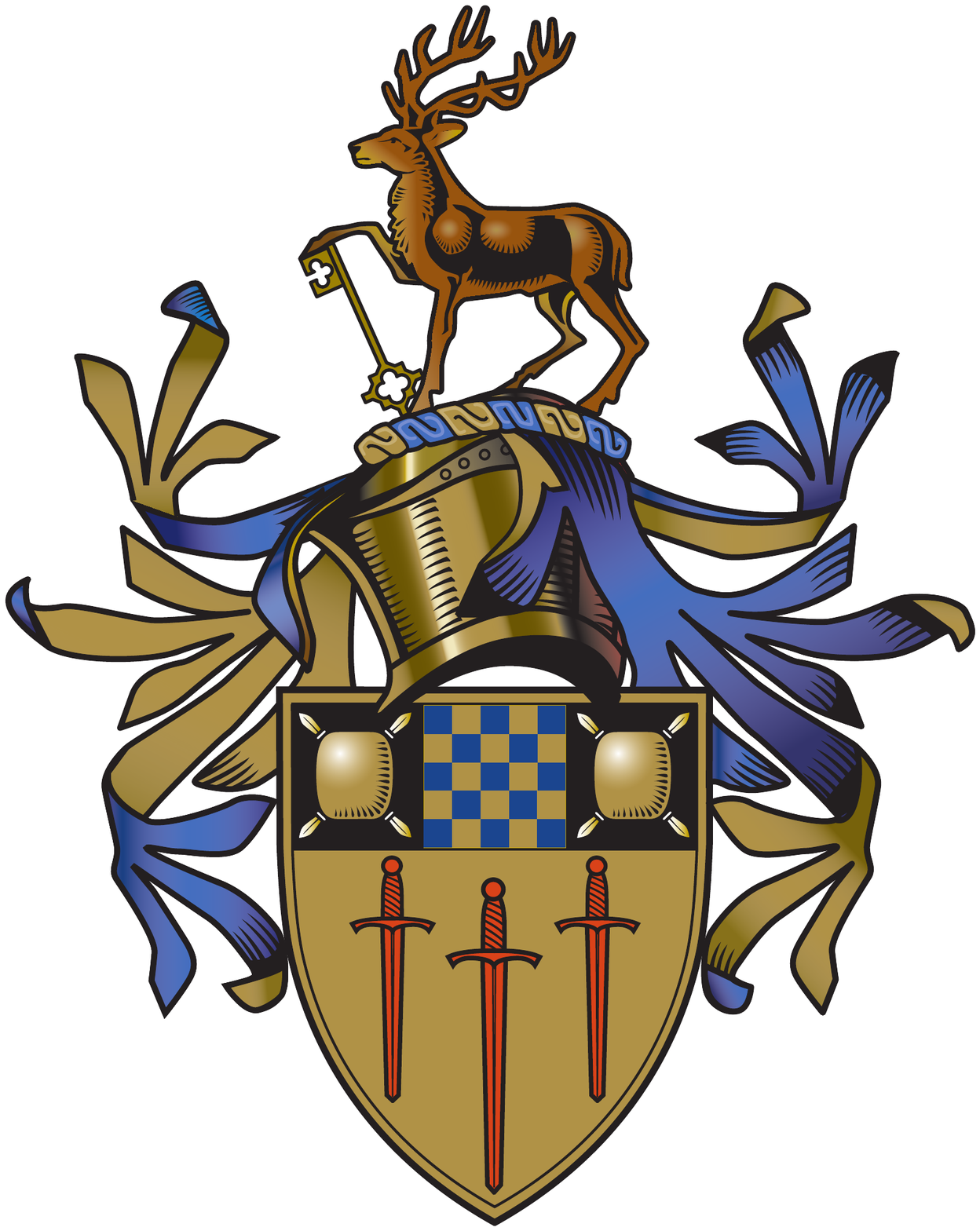}}

    \vfill
    
    Copyright \copyright\ 2018 by Andrea Fontanella. All rights reserved.\\
    E-mail address: \href{mailto:a.fontanella@surrey.ac.uk}{\ttfamily a.fontanella@surrey.ac.uk}

\end{center}

\setcounter{footnote}{0}\renewcommand{\thefootnote}{\arabic{thefootnote}}

\end{titlepage}

\clearemptydoublepage
\pagenumbering{roman}

\section*{Scientific abstract}

This thesis is devoted to the study of geometric aspects of black holes and integrable structures in string theory. 
In the first part, symmetries of the horizon and its bulk extension will be investigated. 
We investigate the horizon conjecture beyond the supergravity approximation, by considering $\alpha'$ corrections of heterotic supergravity in perturbation theory, and show that standard global techniques can no longer be applied.  
A sufficient condition to establish the horizon conjecture will be identified.
As a consequence of our analysis, we find a no-go theorem for $AdS_2$ backgrounds in heterotic theory.

The bulk extension of a prescribed near-horizon geometry will then be considered in various theories. The horizon fields will be expanded at first order in the radial coordinate.  
The moduli space of radial deformations will be proved to be finite dimensional, by showing that the moduli must satisfy elliptic PDEs.

In the second part, geometric aspects and spectral properties of integrable anti-de Sitter backgrounds will be discussed.
We formulate a Bethe ansatz in $AdS_2 \times S^2 \times T^6$ type IIB superstring, overcoming the problem of the lack of pseudo-vacuum state affecting this background. 

In $AdS_3 \times S^3 \times T^4$ type IIB superstring, we show that the S-matrix is annihilated by the boost generator of the $q$-deformed Poincar\'e superalgebra, and interpret this condition as a parallel equation for the S-matrix with respect to a connection on a fibre bundle. This hints that the algebraic problem associated with the scattering process can be geometrically rewritten. This allows us to propose a Universal S-matrix.

\vfill
\noindent
Keywords and AMS Classification Codes:

\noindent Black holes (83C57), Supergravity (83E50), Extension of spaces (54D35), Graded Lie (super)algebras (17B70), Exactly solvable models; Bethe ansatz (82B23), Fiber bundles (55R10). 
\vspace{1cm}

\clearemptydoublepage

\section*{Lay summary}

For much of the 20th century, particles in nature were mathematically described by points. 
The idea of String Theory is to replace the point particle with a string, whose characteristic length scale is assumed to be the Planck length. 
This simple substitution brings great consequences. Most notably, it is formulated in order to unify all fundamental interactions of nature, known as \emph{Einstein's dream}. 
The quantum theory of interactions between point-like particles brings short-distance divergences which one must remove, and in the case of the gravitational interaction, such removal procedure cannot be applied. Remarkably, the quantisation of String Theory does not bring short-distance divergences at all. As such, String Theory is the most promising candidate theory of Quantum Gravity.  
However, for consistency reasons, String Theory must live in a ten-dimensional spacetime, even though the Universe at human scales requires only four dimensions. This problem is solved by assuming that the extra dimensions are wrapped into tiny tubes, smaller than the Planck length.  

Part of my research aims to study Black Holes in String Theory. In four dimensions, Uniqueness Theorems state that black holes are spherical.  The presence of extra dimensions in String Theory violates such theorems, and more exotic types of black holes appear. 
I am interested in studying the black hole horizons, which describe the shape of the black holes, by investigating their symmetries, classifying their geometries, and studying how to extend them away from the region near to the horizon.

In the context of String Theory, there is a notion of a holographic principle. This states that a theory of gravity in a certain spacetime is equivalent to a theory of only light and nuclear forces, without gravity, living on the boundary of the spacetime, like a hologram. 
I am interested in understanding whether the holographic principle is a general feature of nature, by considering string theories in different spacetimes which retain a nice mathematical structure, i.e. integrability.  Integrable mathematical systems typically
do not exhibit chaotic properties, but are instead ``exactly solvable''.

\clearemptydoublepage

\section*{Acknowledgements}

I deeply acknowledge the debt of gratitude I owe to Dr. Jan Gutowski, my supervisor, and Dr. Alessandro Torrielli, my co-supervisor. There are no words to express how much I am grateful to them for all their support, encouragement, commitment, joy for every my little success, and for sharing with me many ideas. 
They are a pleasure to work with, and I hope the future will give us many opportunities to continue to work together. 

I am profoundly grateful to my parents Eleonora and Franco for believing in everything I have done in my life and for always showing support in these years away from home. 

A special thanks goes to my partner Lia, for her invaluable support during difficult occasions, and for many joyful moments together. In few words, for being as she is. 

I thank Prof. Philip Candelas and Prof. Xenia de la Ossa for having been such a wonderful hosts during my stay in Oxford in February 2018, for the memorable dinner together, and for many stimulating discussions.

I also thank various people who invited me to give talks in their institutions during the last year, namely Prof. Maciej Dunajski, Prof. Dietmar Klemm, Dr. Sara Pasquetti, Dr. Christiana Pantelidou, Dr. Daniele Dorigoni, and Prof. Douglas Smith. I thank Prof. Matthias Staudacher for kindly having been my host for my Humboldt Fellowship application.

I want to say thanks to all speakers and participants of the conference \emph{String Geometry, Supersymmetric Theories and Dualities} which I organised in Surrey, and in particular to Prof. Xenia de la Ossa, Prof. Dario Martelli and Prof. Nick Dorey for their lectures.

A general thanks goes to the other members of the \emph{Fields, Strings and Geometry} group in Surrey, in particular Dr. Martin Wolf, Dr. Jock McOrist, Dr. James Grant and Dr. Andrea Prinsloo, and to the doctoral students Dr. Fabrizio Nieri, Dr. Antonio Pittelli, Joakim Str\"omwall (a.k.a Gioacchino Strombavalle for us), Roberto Sisca and Lorenzo Raspollini. I thank all of them for providing a stimulating research environment and for several discussions together.  

I broadly thank the Department of Mathematics at the University of Surrey and its members for providing good facilities and a friendly working environment. I thank the H.o.D. Prof. Ian Roulstone for his very generous financial support of the conference I organised, and the departmental administrator Miss Niki Martin for her crucial help. 

I thank Dr. Paul Skerritt, Prof. Michele Bartuccelli, my Spanish friend Jes\'us Montero Arag\'on and the other doctoral students in Surrey for several jokes shared together, and for several funny, and usually also crazy, drinks, dinners and moments together. 

My apologize goes to the people who I am missing from this list, which I am sure is far from being complete.

\clearemptydoublepage

\section*{Declaration}

This thesis is a result of my own efforts. 
The work to which it refers is based on my PhD research projects, done in collaboration with Dr. Jan Gutowski, Dr. Alessandro Torrielli and Prof. George Papadopoulos, which are

\begin{enumerate}
\item {\bf A.~Fontanella} and A.~Torrielli,
  ``Massless $AdS_2$ scattering and Bethe ansatz,''
  JHEP {\bf 1709} (2017) 075
 \href{https://arxiv.org/abs/1706.02634}{[arXiv:1706.02634 [hep-th]]}.
  
  \item {\bf A.~Fontanella} and J.~B.~Gutowski,
  ``Moduli Spaces of Transverse Deformations of Near-Horizon Geometries,''
  J.\ Phys.\ A {\bf 50} (2017) no.21,  215202
 \href{https://arxiv.org/abs/1610.09949}{[arXiv:1610.09949 [hep-th]]}.
  
  \item {\bf A.~Fontanella} and A.~Torrielli,
  ``Massless sector of AdS$_3$ superstrings: A geometric interpretation,''
  Phys.\ Rev.\ D {\bf 94} (2016) no.6,  066008
  \href{https://arxiv.org/abs/1608.01631}{[arXiv:1608.01631 [hep-th]]}.
  
  \item {\bf A.~Fontanella}, J.~B.~Gutowski and G.~Papadopoulos,
  ``Anomaly Corrected Heterotic Horizons,''
  JHEP {\bf 1610} (2016) 121
  \href{https://arxiv.org/abs/1605.05635}{[arXiv:1605.05635 [hep-th]]}.
\end{enumerate}

Any ideas, data, images or text resulting from the work of others are clearly identified as such within the work and attributed to the authors in the text or bibliography. This thesis has not been submitted for any other academic degree or professional qualification. 
The University of Surrey reserves the right to require an electronic version of the final document as submitted for assessment as above.

\vspace{2cm}

\begin{flushright}
Andrea Fontanella
\end{flushright}
\clearemptydoublepage

\tableofcontents 
\bigskip
\bigskip
\hrule
\bigskip
\bigskip



\pagestyle{fancy}
\renewcommand{\sectionmark}[1]{\markboth{\thesection~#1}{}} 

\fancyhead[RO,LE]{\thepage}
\fancyhead[CO,CE]{}
\fancyhead[LO,RE]{\leftmark}
\fancyfoot{}

%


\newpage

\begin{center}
\Large {\bf Notation} 
\end{center}
\normalsize
\vspace{5mm}
\begin{center}
\begin{tabular}{cc}
$\cH$    &   Killing horizon \\
$\mu, \nu, \rho, ... $ & spacetime indices \\
$\cS$   &  spatial cross section of $\cH$ \\
$I, J, K ...$  & indices on $\cS$ \\
$i, j, k, ...$ & flat indices on $\cS$ \\
$\nabla$ & Levi-Civita connection of the spacetime \\
$\tilde{\nabla}$ & Levi-Civita connection of $\cS$  \\
$\{\D, h_I, \gamma_{IJ}\}$ & near-horizon data \\
$\mathbf{e}^{\pm}$ & light-cone directions \\
$\nabla^{(\pm)}$ & spacetime connection with torsion\\
$\tilde{\nabla}^{(\pm)}$ & connection with torsion of $\cS$ \\
$R, \tilde{R}, \tilde{R}^{(\pm)}$ & curvature of $\nabla, \tilde{\nabla}, \tilde{\nabla}^{(\pm)}$ \\
$\epsilon$ & spacetime spinor \\
$\eta_{\pm}$ & spinors on $\cS$ \\
${\buildrel \circ \over \xi}$ & generic field $\xi \big|_{r=0}$\\
$\d \xi$ & first order radial deformation of $\xi$
\end{tabular}
\end{center}

\vspace{2cm}

\begin{center}
\begin{tabular}{cc}
$\mathcal{Q}, \mathcal{S}$    &   fermionic generators \\
$\mathcal{P}, \mathcal{K}, \mathcal{H}$  &  bosonic generators \\
$\mathcal{J}$ & boost generator \\
$h$   &    string modes coupling constant \\ 
$p$   &  particle momentum \\
$\theta$ & particle rapidity  \\
$\mathcal{T}_0$ & monodromy matrix \\
$\mathcal{T}$ & transfer matrix \\
$\mathsf{E}_{ij}$ & basis of $2\times 2$ matrices \\
$\Delta ( \cdot )$ & coproduct of $(\cdot)$ \\
$\mathfrak{a}^{(0)}_i, \mathfrak{a}^{(1)}_{\alpha}$ & generic graded generators \\
$\Pi$ & permutation operator \\
 
\end{tabular}
\end{center}


\clearemptydoublepage
\pagenumbering{arabic}

\chapter{\textbf{Introduction}}

The 20\textsuperscript{th} century was particularly fruitful in terms of discoveries in theoretical physics, which completely changed the way we describe nature. The two most important and fascinating discoveries of this period are \emph{General Relativity} and \emph{Quantum Mechanics}. 
In the 18\textsuperscript{th} century, there was a belief that the gravitational and the electric forces share the same behaviour. This is motivated by the Newton's and Coulomb's laws, which classically describe the force of the two interactions as  
\begin{equation}
\notag
F^{g} = G \frac{M_1 M_2}{r^2} \ , \qquad\qquad
F^{e} = k \frac{Q_1 Q_2}{r^2} \ . 
\end{equation}
The 20\textsuperscript{th} century completely changed this picture. It was  discovered that gravity is complicated to describe, and in particular its analogy with the electric force is no longer satisfactory. 

The first theory of gravity to be obtained from an action principle was developed by A. Einstein and nowadays it is known as General Relativity. Einstein introduced the scientific community to two extremely important concepts: the \emph{spacetime} and gravity as the \emph{curvature} of spacetime. 
The first one is related to the fact that space and time are the same concept, in contrast to the standard picture of Newton's mechanics where time is just a parameter governing the evolution of a system which moves in space. 
A key motivation for developing this context is the inconsistency of Maxwell's equation with the Galilean transformations. This led Einstein to formulate the so-called theory of Special Relativity, where the Galilean transformations are substituted by the Lorentz transformations. In Special Relativity, time and space are treated at the same footing, and both become coordinates of the four-dimensional spacetime. 

After the discovery of Special Relativity, Einstein came up with the idea that an object with a certain mass exhibits its gravitational force by \emph{curving} the geometry of the spacetime. Therefore when we think of spacetime, we shall not refer to a flat spacetime, but instead to a spacetime that can be curved, where the curvature is a measure of the intensity of the gravitational force. 

In Einstein's idea of gravity as curvature of spacetime, the gravitational force is associated with a gravitational field, which is the \emph{metric} $g_{\mu\nu}$ of the spacetime. Therefore in Einstein's idea the metric of the spacetime is a dynamical quantity, which must satisfy Einstein's equation
\begin{equation}
\notag
R_{\mu\nu} - \frac{1}{2} R g_{\mu\nu} = \frac{8\pi G}{c^4} T_{\mu\nu}  \ . 
\end{equation} 
The formulation of a theory of gravity in this way brings at least two main physical consequences: the existence of \emph{black holes} and \emph{gravitational waves}\footnote{Both of them have been recently directly confirmed with the experiments LIGO and VIRGO.}.
The concept of black holes emerges by considering a static object of mass $M$ with spherical symmetry in the vacuum. 
K. Schwarzschild solved the Einstein's equation for such configuration, and he found the celebrated Schwarzschild metric
\begin{equation}
\notag
ds^2 = - \bigg( 1 - \frac{r_*}{r} \bigg) dt^2 + \bigg( 1 - \frac{r_*}{r}\bigg)^{-1} dr^2 + r^2 d \Omega^2 \ , \qquad
r_* = \frac{2G M}{c^2} \ . 
\end{equation}
The surface $r=r_*$ defines the \emph{event horizon}, where every particle that enters in the region $r < r_*$ will never be able to reach the region $r > r_*$ again. The event horizon in this case is a 2-dimensional sphere and asymptotically the spacetime is flat. 

The Schwarzschild black hole was later generalised to the cases where the object of mass $M$ carries an electric charge $Q$ (Reissner-Nordstr\"om), or it is rotating with angular momentum $J$ (Kerr), or it is both charged and rotating (Kerr-Newman). In all cases, the event horizon is a 2-dimensional sphere and asymptotically the spacetime is flat. 
At this point, one may wonder if there exists even further types of black hole solutions than the ones just mentioned above. 
The answer is negative, because in the later years (around 1970), uniqueness theorems for black holes were discovered by Hawking et al. \cite{israel, carter, hawking, robinson1, israel2, mazur, robinson}. 
The statement is the following
\begin{center}
\emph{
``Any asymptotically flat and analytic black hole, which is solution to the 4-dimensional Einstein equations, is uniquely determined by the data $(M, Q, J)$, and its metric is the Kerr-Newman metric'' 
}
\end{center}
and it is known under the name of \emph{no-hair theorem}.

The second revolutionary discovery of the 20\textsuperscript{th} century is quantum mechanics. The triggering fact was again an inconsistency of the Maxwell equations. Consider a black body at thermal equilibrium. According to Maxwell theory, the black body emits radiations in the whole spectrum of frequencies, and the power of the radiation emitted is proportional to the square of the frequency. Therefore the total energy emitted, which is the sum of the energy in all frequencies, is infinite, in contradiction with the principle of conservation of energy. This is called the \emph{ultraviolet catastrophe}, which was solved by M. Planck in 1900 by introducing the concept that the radiation is not continuum in frequencies, but instead it comes in ``discrete packets'', called \emph{quanta}, which nowadays we call \emph{photons}. 

In 1924, L. de Broglie proposed that particles exhibit also a wave-like behaviour, which was supported by the experiment on the diffraction of electrons. This led to the concept that at quantum scales, one can describe quanta either as particles or waves (known as wave-particle duality). 
Based on this intuition, W. Heisenberg discovered the \emph{uncertainty principle}, which states that in a measuring process, position $x$ and momentum $p$ of a particle cannot be simultaneously determined with arbitrary precision. The errors $\Delta$ in the measure of $x$ and $p$ must satisfy  
\begin{equation}
\notag
\Delta x \, \Delta p \geq \frac{\hbar}{2} \ . 
\end{equation} 
From the Rydberg-Ritz combination principle of spectral lines of atoms, Heisenberg understood that at quantum level observables (e.g. position and momentum) have an interpretation as non-commutative operators. In his interpretation, position and momentum are matrices which satisfy the commutation relation
\begin{equation}
\notag
[x , p ] = i \hbar \ ,
\end{equation}
and the uncertainty principle is a consequence of this.
Criticized for being physically counterintuitive, Heisenberg theory was abandoned in favour of the wave-like description of E. Schr\"odinger. In Schr\"odinger's perspective, the particle is described by a function $\psi (t, x)$, whose square has the meaning of probability density of finding the particle at point $x$ if a position measurement is made at time $t$. The function $\psi$, called \emph{wave function}, must satisfy Schr\"odinger's equation 
\begin{equation}
\notag
i \hbar \frac{\partial \psi}{\partial t} = \bigg[ - \frac{\hbar^2}{2m} \nabla^2 + V \bigg] \psi \ .
\end{equation}
The Schr\"odinger picture of quantum mechanics was satisfactory because it can explain many physical systems with accuracy, e.g. the energy levels of the hydrogen atom. 
However the Schr\"odinger equation is not relativistic: the rest energy $m c^2$ is not included, and the equation is not Lorentz covariant. An attempt to modify the Schr\"odinger equation in order to include the rest energy is given by the Klein-Gordon and Dirac equations. 
However these two equations suffer from some problems, one of them is that the energy spectrum is unbounded from below (it extends down to $- \infty$) and therefore there exists no ground state. 
The problems of Klein-Gordon and Dirac equations are due to the fact that we are ignoring an important relativistic phenomenon, which is \emph{particle production}, where particles can be dynamically created and annihilated.
The solution comes with the introduction of \emph{Quantum Field Theory}, where the wave function of quantum mechanics is replaced by a state which can dynamically change its particle content. In this framework, there exists a state which contains no particles, the vacuum state. 
The single particle of quantum mechanics is associated with a single excitation produced above the vacuum state.  
Such excitations are produced by fields, which are objects depending on the spacetime coordinates, and satisfy a certain dynamics. 

A way to introduce the quantum behaviour, is by imposing that the fields must satisfy commutation relations, which must be compatible with the causality principle of special relativity. This is called \emph{canonical quantization}. 
A physical quantity of interest to compute is the probability that a certain initial state evolves to a certain final state, called the amplitude. In case of a scattering process, the  amplitude is proportional to the modulus squared of the cross section, which can be measured with an experiment.
An alternative way to quantise the theory is via the \emph{path integral} formulation of R. Feynman. In this formulation, the amplitude is computed by summing the weight function $e^{i S / \hbar}$ over all possible field configurations between the initial and final states. 

A typical issue which affects a quantum field theory when computing amplitudes is the presence of ultraviolet divergences. This happens for instance in quantum electrodynamics when computing the self-energy of the electron. 
A possible way to avoid this problem is by making a redefinition of the fields, so that the divergent terms disappear. This procedure is called \emph{renormalization}, and the theories for which this procedure can be applied are called \emph{renormalizable}.   
G. 't Hooft proved that non-abelian gauge theories are renormalizable. This implies that the interactions of the Standard Model (electromagnetic, weak and strong nuclear) can be consistently described within the quantum field theory formalism. 
It turns out that General Relativity is not renormalizable. 
For this reason, it can only be regarded as an effective theory of quantum gravity. String theory represents nowadays the best candidate theory of quantum gravity.

\section{String Theory}
The first concepts in string theory were introduced by G. Veneziano for the purpose of  explaining the nuclear force (e.g. the Regge behaviour). 
String theory is based on the idea that fundamental particles are not point-like, but excitations of a string. Therefore we need to determine the dynamics of a string in spacetime. 

To begin, we recall that the dynamics of a point-like particle of mass $m$ moving in Minkowski spacetime is given by\footnote{We set $\hbar = c = 1$. }
\begin{equation}
\notag
S = - m \int d s \ .
\end{equation}
Consider now a string of length $\ell_s$ moving in Minkowski spacetime. The analogous quantity to mass for the string is the tension $T$, which has dimensions of mass per unit length of the string. The dynamics is governed by the Nambu-Goto action
\begin{equation}
\notag
S = - T \int d \mathcal{A} \ ,  
\end{equation}
where $d \mathcal{A}$ is the infinitesimal area element of the string world-sheet. 

The Nambu-Goto action is inconvenient because the background spacetime must be Minkowski. An alternative action, which is classically equivalent, is the Polyakov action, which describes a two-dimensional sigma model
\begin{equation}
\notag
S = - T \int d^2 \sigma \sqrt{-\gamma}\, \nabla_a X^{\mu} \nabla_b X^{\nu} \gamma^{ab} g_{\mu\nu} \ , 
\end{equation}
where $\{ \sigma^a \}$ and $\gamma_{ab}$ are coordinates and metric on the string world-sheet respectively, and $g_{\mu\nu}$ is the spacetime metric, which cannot be generic, but specified shortly. 
Classically, the Polyakov action is invariant under a conformal rescaling of the world-sheet metric $\gamma$. 
However, after quantization, conformal symmetry might not be preserved. We require that conformal invariance is preserved at quantum level, and therefore the Polyakov action describes a 2-dimensional conformal field theory. Cancellation of conformal anomalies imposes an equation for the metric $g_{\mu\nu}$, which at leading order is Einstein's equation, together with subleading higher order terms, which are powers and derivatives of the curvature multiplied by appropriate powers of the string scale $\ell_s$. 

Associated to each 2-dimensional conformal field theory there is a number $c$, the so-called \emph{central charge}. For a free theory, $c$ is just the number of scalar fields. Then in our case, $c$ is the dimension of the spacetime. The requirement that unphysical states with negative norm disappear imposes $c = 26$. This implies that the dimension of the spacetime must be $26$. 

At this level, the string describes only bosons. Quantum mechanically, one aims to define a vacuum state and treat the Fourier modes of the string coordinates $X^{\mu}$ as ladder type operators. In this way, the excitations created above the vacuum state describe particles with integer spin. 
In doing that, one discovers that the vacuum state is a tachyon.  
The new ingredient which removes the tachyonic behaviour of the vacuum state and also allows the inclusion of fermions in the spectrum is called \emph{supersymmetry}. 

Supersymmetry relates each particle to its corresponding superpartner, which has opposite statistic. However supersymmetry is not just a symmetry between bosons and fermions.
Supersymmetry is an extension of the Poincar\'e symmetry, which is the symmetry of Minkowski spacetime. In this sense the spacetime acquires new extra fermionic directions, which are described by Gra\textbeta mann numbers\footnote{Historically, supersymmetry was introduced first in the context of the string world-sheet, and later generalised to the full spacetime.}. Mathematically, the supersymmetry algebra is a $\mathbb{Z}_2$ graded Lie algebra, where there are odd-type generators which anti-commute between themselves, the so-called supercharges
\footnote{Supersymmetry has also found some phenomenological application from the standard model perspective, such as the resolution of the hierarchy problem, the unification of the gauge coupling of the standard model interactions, and a candidate particle for dark matter.}.

String theory, after introducing supersymmetry, becomes the so-called \emph{superstring} theory, and its action looks like
\begin{equation}
\notag
S = - T \int d^2 \sigma \sqrt{-\gamma}\, \bigg( \gamma^{ab} g_{\mu\nu} \nabla_a X^{\mu} \nabla_b X^{\nu}  - i g_{\mu\nu} \overline{\psi}^{\mu} \rho^{a} \nabla_{a} \psi^{\nu} \bigg) \ , 
\end{equation}
where $\psi^{\mu}$ is a fermion, and $\rho^{a}$ are the gamma matrices in the 2-dimensional world-sheet. Quantum mechanical consistency of superstring theory now requires that the dimension of the spacetime is $10$.

The spectrum of the closed string contains a massless particle of spin two, which has precisely the right properties to be the graviton.  
Furthermore, the spectrum of the open string contains massless spin one gauge fields, like those in the standard model, with the right gauge symmetries incorporated. 
In 1974, this observation led J. Scherk and J. H. Schwarz to propose string theory as a unified theory of all forces, including gravity \cite{Scherk:1974ca}, which goes towards Einstein's dream of a unified theory of all interactions. The connection of string theory to general relativity at low energies was described by Yoneya \cite{Yoneya:1973ca, Yoneya:1974jg}. 

In the point-like description, any attempt to quantize general relativity produces ultraviolet divergences, which makes the theory non-renormalizable at least at perturbation level. The advantage of string theory is that it is ultraviolet finite at all orders in perturbation theory. The intuitive reason is that perturbatively the point-like interaction brings short-distance singularities, while this is removed in the string-like interaction, as shown in the picture below. 

\begin{figure}[H]
\begin{center}
 \includegraphics[scale=0.50]{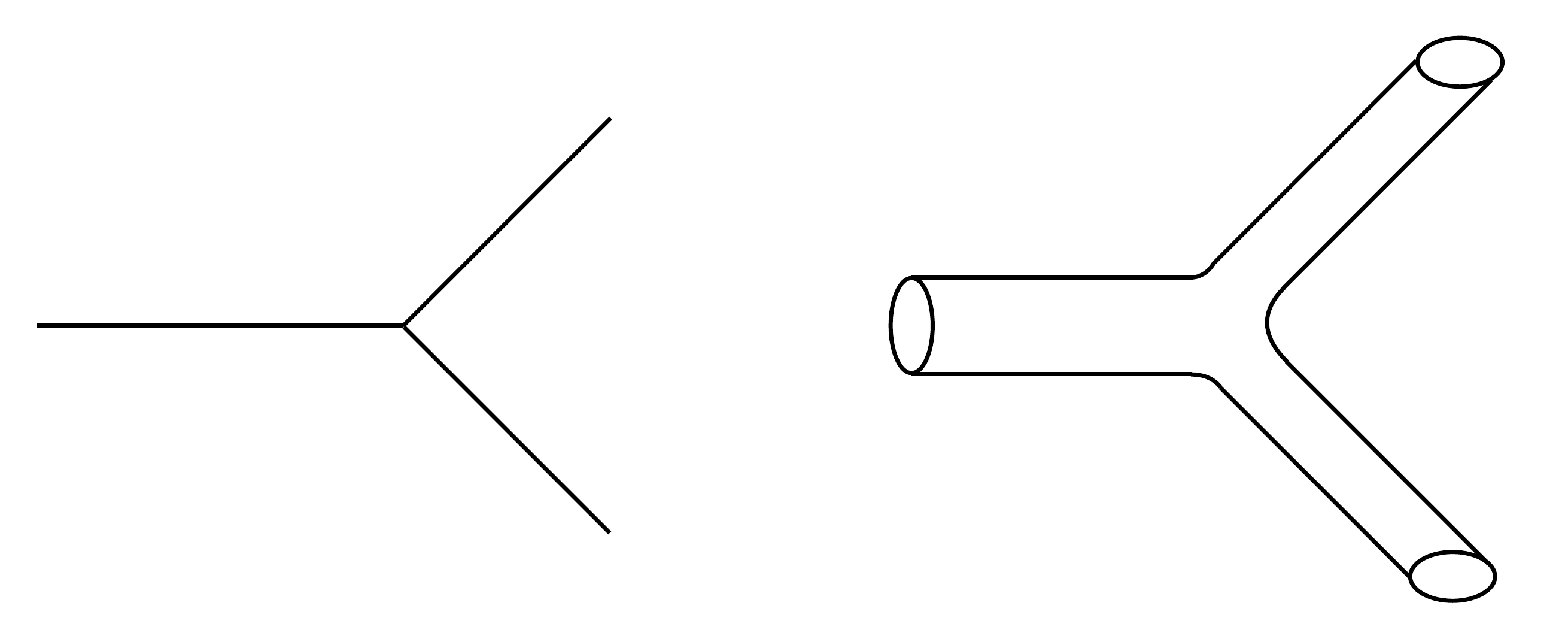}
\end{center}
\caption{Point-like vs string-like interactions.}
\end{figure}   

In a perturbative quantum field theory, scattering amplitudes are typically computed by Taylor expanding the S-matrix in the coupling constant. There is a prescription for computing the coefficients of the expansion, which consists in summing Feynman diagrams with a certain number of loops. 
The number of loops is related to the power of the coupling constant that we are considering in the expansion; though it is possible for tree level diagrams to be of higher power in the coupling constant under certain circumstances.

Computing scattering amplitudes in string theory is different. The trajectory of the string is a surface, therefore the $g$-loop, $n$-point scattering amplitude can be described in terms of \emph{Riemann surfaces} with genus $g$ and $n$ punctures. Computing scattering amplitudes in string theory consists in computing an integral over the moduli space of Riemann surfaces with $g$ and $n$ as specified.   

Though string theory is ultraviolet finite, it still suffers from the infrared divergences which affect quantum field theories. Typically in quantum field theory, infrared divergences (e.g. tadpole) appear because the vacuum state is incorrectly identified. In string theory, infrared divergences are caused by integrating over \emph{singular} Riemann surfaces. Singularities in Riemann surfaces appear when there is a degeneration, which consists in squeezing any possible handle or connection between a pair of Riemann surfaces. As explained in \cite{Sen:2015cxs}, a possible resolution of this problem consists in considering a quantum field theory which has the same amplitudes of superstring theory, the so-called \emph{superstring field theory}. Then in superstring field theory one can remove the infrared singularities in exactly the same way as in an ordinary quantum field theory.

\subsection{The first superstring revolution}
The discovery that triggered the first superstring revolution was that Type I superstring is anomaly free. This is connected to the existence of gauge, gravitational and mixed anomalies in parity-violating theories in $4 k + 2$ dimensional theories which spoils general covariance, as pointed out by L. Alvarez-Gaum\'e and E. Witten \cite{AlvarezGaume:1983ig}, based on a first result of \cite{Frampton:1983ah}. 

An \emph{anomaly} occurs when a local symmetry for the classical action is not preserved after quantization. In terms of path integral quantization, an anomalous symmetry is a symmetry of the action $S$, but not of the measure $d \mu$, such that the partition function 
\begin{equation}
\notag
\mathcal{Z} = \int d \mu \, \exp \big(i S / \hbar\big)\ , 
\end{equation}
is not invariant under the symmetry considered \cite{Fujikawa:1980eg}. In the point-like quantum field theory, a failure of gauge invariance is measured by the failure of Ward identities. 

In 4-dimensional gauge theories with parity-violating gauge couplings, an anomaly is generated by the presence of the \emph{triangle} diagram, with external  gauge bosons and a chiral fermion running in the loop. In the standard electro-weak theory such anomalies cancel out, which is important for consistency. 
The analogous question in 10-dimensions involves the \emph{hexagon} diagram, with external gauge bosons and gravitons and a chiral fermion running in the loop.

It was observed in \cite{AlvarezGaume:1983ig} that the gravitational anomaly cancel in type IIB supergravity\footnote{Type IIB superstring theory does not contain elementary Yang-Mills gauge fields, and therefore it is trivially free from gauge anomalies.}. However it appeared that type I superstring is anomalous for any choice of the gauge group. 
This problem was solved by M. Green and J. H. Schwarz \cite{Green:1984sg} in $D=10$, $\mathcal{N} =1$ supergravity coupled to a Yang-Mills gauge theory. The hexagon diagram has external bosons which are combinations of the 2-form field strength and Ricci tensor, $F^6$, $F^4 R^2$, $F^2 R^4$, $R^6$.

\begin{figure}[h]

\begin{tikzpicture}
  \begin{feynman}
    \vertex (a);
    \vertex [above right=of a] (b);
    \vertex [above right=of b] (c);
    \vertex [above =of c] (d);
         \vertex [below right=of c] (e);
    \vertex [below right=of e] (f);
 
    \diagram* {
      (a) -- [boson] (b), 
       (c) -- [boson] (d),
       (e) -- [boson] (f),
      (b) --  (c),
      (b) --  (e), 
      (c) --  (e),  
    };
  \end{feynman} 

\hspace{8cm}

 \begin{feynman}
    \vertex (a);
    \vertex [above right=of a] (b);
    \vertex [above left=of b] (c);
    \vertex [left =of c] (d);
     \vertex [above right =of c] (e);
       \vertex [above left =of e] (f);  
       \vertex [right =of e] (g);
       \vertex [above right =of g] (h);
       \vertex [below right =of g] (i);
       \vertex [right =of i] (l);
       \vertex [below left =of i] (m);
    \vertex [below right=of m] (n);
 
    \diagram* {
      (a) -- [boson] (b), 
       (c) -- [boson] (d),
       (e) -- [boson] (f),
        (g) -- [boson] (h),
         (i) -- [boson] (l),
          (m) -- [boson] (n),
      (b) --  (c),
      (c) --  (e), 
      (e) --  (g),  
         (g) --  (i),  
            (i) --  (m),
               (m) --  (b),    
    };
  \end{feynman} 

\end{tikzpicture}

\caption{The triangle and hexagon diagrams, responsible for anomalies in four and ten dimensions respectively.}

\end{figure}
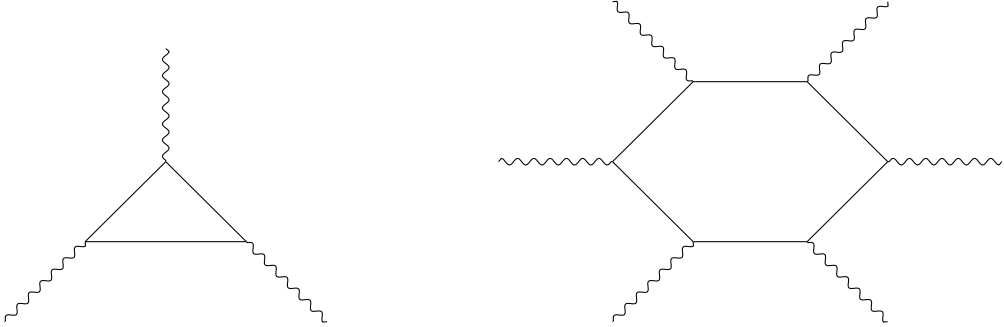

\noindent Green and Schwarz realized that the anomalies cancel if 
\begin{enumerate}
\item[(I)] one imposes that the 2-form Kalb-Ramond field $B$ transforms in a certain way under the gauge and Lorenz groups, and
\item[(II)] one couples the classical action to the counterterm
\begin{equation}
\notag
S_{GS} = \int B \wedge X_8 \ , 
\end{equation} 
where $X_8$ is an exact 8-form, which is a combination\footnote{Here the powers stand for wedge product, and not for spacetime indices contraction. The trace $\mathrm{tr} M$ stands for the trace of $M$ in the fundamental representation.} of $\mathrm{tr}F^4$, $\mathrm{tr}F^2 \mathrm{tr}R^2$, $\mathrm{tr}R^4$, and
\item[(III)] the Yang-Mills gauge group has 496 generators.
\end{enumerate} 
In particular, condition (I) implies that the field strength of $B$, in order to be gauge invariant, must be modified as
\begin{equation}
\notag
H = dB + CS (\omega) - CS (A) \ ,  
\end{equation} 
where $CS(\omega)$ and $CS(A)$ are the Lorentz and Yang-Mills Chern-Simons forms. By taking the exterior derivative, this equation implies the famous anomaly corrected Bianchi identity for $H$, 
\begin{equation}
\notag
dH = \mathrm{tr} R^2 - \mathrm{tr} F^2 \ . 
\end{equation}
The fact that the gauge group must have dimension $496$ restricts its choice down to only two possibilities: $SO(32)$ or $E_8 \times E_8$. This led Green and Schwarz to construct an anomaly-free type I superstring theory with gauge group $SO(32)$. 
Ten dimensional anomaly-free chiral string theory with gauge group $SO(16) \times SO(16)$ which is tachyon-free and without spacetime supersymmetry has also been constructed \cite{Dixon:1986iz , AlvarezGaume:1986jb}.  
Furthermore, Green and Schwarz constructed the type IIA and type IIB superstring theories, which were based on a earlier work of Gliozzi, Scherk and Olive \cite{Gliozzi:1976qd}. Type IIB is a chiral theory, which has vanishing anomaly as pointed out previously, while type IIA is a non-chiral theory, and therefore the anomalies cancel straightforwardly. However these theories were not considered from the phenomenological perspective because they do not contain Yang--Mills gauge fields, and therefore exclude the possibility of obtaining a chiral theory in four dimensions via a compactification\footnote{To obtain an effective 4-dimensional theory containing chiral fermions via a compactification of a theory in $D >4$, one needs to require that the higher dimensional theory must contain chiral fermions and elementary Yang-Mills gauge fields, \cite{Witten:1983ux}.}. Later with the discovery of D-branes, also type II theories became of phenomenological interest. 

A new type of anomaly-free string theory was constructed by Gross, Harvey, Martinec and Rohm, which is called \emph{heterotic} \cite{Gross:1984dd, Gross:1985fr, Gross:1985rr}. 
Anomaly cancellation allows two types of heterotic superstring theory, with gauge group either $SO(32)$ or $E_8 \times E_8$. 

A remarkable fact, is that all these superstring theories have no adjustable parameters. All dimensionless parameters arise either dynamically as expectation values of scalar fields, or as integers which count some quantities (e.g. topological invariants, number of branes or quantized fluxes).

The $E_8 \times E_8$ heterotic theory was particularly interesting from the phenomenological perspective. Candelas, Horowitz, Strominger and Witten found a vacuum solution where the spacetime has six compact directions, and the remaining four describe a  Minkowski spacetime, which resembles a connection with the standard model \cite{Candelas:1985en}. 
Such a vacuum solution preserves $\mathcal{N}=1$ supersymmetry in the 4-dimensional Minkowski spacetime. This condition imposes mathematical restrictions on the geometry of the 6-dimensional compact manifold, which must be Ricci flat and K\"ahler, and therefore a Calabi-Yau manifold. 
As was pointed out in \cite{Lerche:1989uy}, there are two possible choices of Calabi-Yau manifolds that give rise to the same effective $\mathcal{N} = 1$ theory in 4-dimensional flat spacetime. Such pairs of Calabi-Yau manifolds are known as mirror pairs, and this led to the discovery of \emph{mirror symmetry}. Mirror symmetry opened a new connection with mathematics after the work of Candelas, de la Ossa, Green and Parkes \cite{Candelas:1990rm}, which shows how to use mirror symmetry to solve problems in enumerative geometry\footnote{The problem solved consists in counting the number of rational curves of degree $k$ in a particular Calabi-Yau space.}.

\subsection{The second superstring revolution}
The second superstring revolution, which happened in the mid-1990s, is associated with two important discoveries: the \emph{string dualities}, and the existence of \emph{branes}.

The five types of string theories are not independent, but they are related between each other via dualities. Hull and Townsend \cite{Hull:1994ys} and Witten \cite{Witten:1995ex} pointed out that all string theories are different limits of a unique theory defined in eleven dimensions, called \emph{M-theory}, which does not contain strings. 
The string dualities consist of T-duality, S-duality, U-duality and mirror symmetry. 

\begin{figure}[h]
\begin{center}
 \includegraphics[scale=0.50]{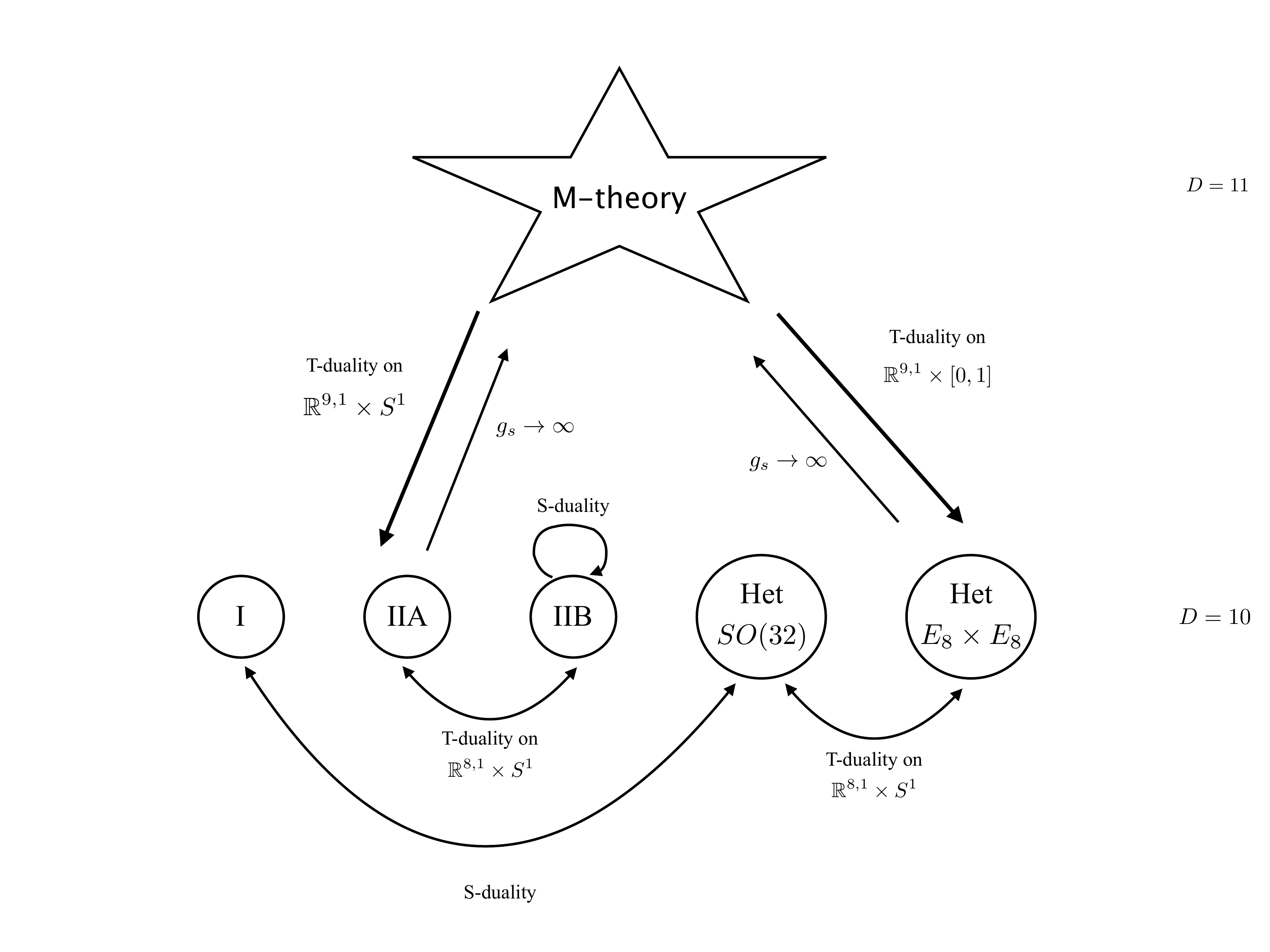}
\end{center}
\caption{Web of string dualities.}
\end{figure}   

\begin{itemize}
\item {\bf T-duality} ($R \rightarrow \ell_s^2/R$). The abelian version of T-duality consists in considering a spacetime compactified on a circle of radius $R$ and to substitute it with another circle of radius $\ell_s^2 / R$. One can show that this transformation leaves the spectrum of the bosonic string theory invariant, provided that momentum modes and winding modes are swapped\footnote{T-duality can be formulated in a path integral approach, where the $U(1)$ isometry associated to $S^1$ plays a central role. One may apply the same procedure in the case the isometry group is non-abelian. In this way one obtains the \emph{non-abelian} T-duality. In contrast to the abelian version, the non-abelian T-duality is not necessary a symmetry of string theory.}. In the case of superstring theories, such transformation swaps the type, and therefore it can be used as a duality.  For instance, type IIA theory in $\mathbb{R}^{8, 1} \times S^1$, with circle radius $R$, is equivalent to type IIB theory again in 9-dimensional Minkowski spacetime times a circle, but with radius $\ell_s^2 / R$. 

\item {\bf S-duality} ($g_s \rightarrow 1/ g_s$). By inverting the string coupling, type I superstring theory is equivalent to $SO(32)$ heterotic theory, while type IIB is equivalent to itself. This means that if we know the perturbative behaviour of these three theories, i.e. when $g_s$ is small, then we also know how they behave when $g_s \gg 1$ \cite{Schwarz:1993vs, Sen:1992ch}. The string coupling is not a free parameter, but it is given by the exponential of the vacuum expectation value of the dilaton field, i.e. $g_s = e^{\langle \Phi \rangle}$. Therefore S-duality, as well as T-duality, is a field transformation, given by $\Phi \rightarrow - \Phi$.
\end{itemize}

The U-duality was proposed by Hull and Townsend \cite{Hull:1994ys}, and it is a combination of T and S dualities in the context of toroidal compactifications. U-duality relates theories compactified on a space of large (or small) volume, to theories at large (or small) coupling.

The second important discovery of the 1990s is the fact that string theory is not just a theory of strings, but other extended objects called \emph{branes} exist. Historically the name brane is related to \emph{membrane}, which is a 2+1 dimensional surface. Branes can be of various dimensions, for instance a 0-brane is a point, a 1-brane is a string, and so on. Branes appear as non perturbative objects in string theory, and their tension is inversely related to powers of $g_s$. 

One type of common brane is called D-brane, where D stands for \emph{Dirichlet}, where the name is because open stings with Dirichlet boundary conditions can end on D-branes.  They were introduced in string theory by Polchinski \cite{Polchinski:1995mt}, where he discovered that D-branes are sources of the Ramond-Ramond $p$-forms. The idea is that the endpoints of open strings can freely move inside the D-brane, but can never leave, unless they join together to form a closed string. 
This gives rise to a new way to view extra dimensions in string theory. 
Instead of thinking of extra dimensions as tiny tubes, one can imagine that they are infinitely extended, and the universe at human scales is reproduced by a $3+1$ dimensional brane which can move inside the 10-dimensional spacetime.  Particles of the standard model are reproduced by the vibration modes of the open strings attached to the brane. 
Outside the brane, only gravity can exist since the graviton emerges from the spectrum of the closed string. 

This model, also called \emph{large extra dimensions} (LED) was proposed by Arkani-Hamed, Dimopoulos and Dvali \cite{ArkaniHamed:1998rs}, and it has a dramatic consequence.  
In the LED model the fundamental scale of energy, namely the highest energy which can be measured, is lower than the Planck scale. In particular the fundamental scale can range between few TeV to $10^{16}$ TeV. This  means that if the LED model is correct, quantum gravity and string effects can be measurable at the Large Hadron Collider of CERN\footnote{The most recent experiment at LHC started in 2015 and will stop on 3 December 2018 with a collision energy of $13$ TeV. Evidence of the model is still missing at this scale.}. 

The fact that extra dimensions in string theory can be large, and not necessarily compactified, motivated the study of black holes in higher dimensions. If the LED model is realized in nature, higher-dimensional black holes could be created at the LHC, or observed in the universe (for a review, see e.g. \cite{Kanti:2004nr}).

Furthermore, black hole uniqueness theorems do not hold in higher dimensions, and therefore one expects a richer family of black holes. Determining all possible black hole solutions in higher dimensions represents an interesting and challenging problem in general relativity.

The coexistence on the horizon of quantum effects and a strong gravitational field\footnote{This is not always the case, since large black holes can have quite small curvature on the horizon. This happens for instance to supermassive black holes placed at the centre of galaxies, where the gravitational field on the horizon is almost negligible. However in string theory we are not trying to model supermassive black holes.}, makes black holes an important theoretical laboratory where to test quantum gravity. 
Bekenstein and Hawking argued that a black hole must have an \emph{entropy}, which is proportional to the area $A$ of the event horizon, 
\begin{equation}
\notag
S_{BH} = \frac{A}{4 G} \ .
\end{equation}
According to the microscopical interpretation of entropy, this suggests that the black hole must have some microstates, whose number $d_{micro}$ is given by 
\begin{equation}
\notag
d_{micro} \sim e^{S_{BH}}  \ . 
\end{equation}
An explanation of the microscopical interpretation of the black hole entropy was first found by Strominger and Vafa \cite{Strominger:1996sh} in the context of string theory. First of all, we recall the notion that black holes in string theory can be described as a superposition of branes (we shall give some details later in section \ref{sec:BH_higher}). 
Therefore one may expect that black hole microstates are given by all possible brane configurations which macroscopically reproduce the black hole. 
In general this is difficult to check, however one may consider a black hole which saturates the BPS bound, and therefore protected by quantum corrections in $g_s$. In the weak coupling limit, one can show that the ``Schwarzchild radius'' for the system of branes becomes smaller than the string length $\ell_s$. This means that the gravitational effects can be neglected, and this allows the computation of the degeneracy of states. Supersymmetry allows this result to be extended to strong coupling, where the system of branes properly describes a black hole. 
In this way one can compute the statistical entropy, which is given as
\begin{equation}
\notag
S_{stat} (Q) = \ln d_{deg} (Q) \ , 
\end{equation}
where $Q$ is a set of charges of the BPS state, and $d_{deg} (Q)$ is the number of BPS states which carry the same set of charges $Q$. 
In the Strominger and Vafa computation, they found that 
\begin{equation}
\notag
S_{BH} (Q) =  S_{stat} (Q) \ , 
\end{equation}
which is a confirmation of the fact that 
\begin{equation}
\notag
d_{micro} = d_{deg} (Q) \ . 
\end{equation}
For a review on this topic, see e.g. \cite{Sen:2007qy, David:2002wn}.

\section{Black holes in higher dimensions}
\label{sec:BH_higher}
In higher dimensions than four, uniqueness theorems for asymptotically flat black holes lose their validity. 
The known black hole solutions in four dimensions have been generalised to higher dimensions. For instance, there exists a solution to the Einstein equations in any dimension $D$, which is a generalization of the Schwarzchild metric. This was discovered by Tangherlini \cite{Tangherlini:1963bw}, and the metric is 
\begin{equation}
\notag
ds^2 = - \bigg( 1- \frac{r_*^{D-3}}{r^{D-3}} \bigg) dt^2 + \bigg( 1- \frac{r_*^{D-3}}{r^{D-3}} \bigg)^{-1} dr^2 + r^2 d\Omega^2_{D-2} \ . 
\end{equation}
Furthermore, the Kerr metric, which describes a rotating black hole in four dimensions, can be generalised to higher dimensions. This was found by Myers and Perry \cite{Myers:1986un}. The Myers-Perry solution is specified by the mass $M$ and a set of angular momenta $J_{r}$, where $r = 1, ... , \mbox{rank}\, [SO(D-2)]$, and the horizon topology is $S^{D-2}$. In $D=5$, for solutions which have only one non-zero angular momentum $J_1 = J \neq 0$, they found the bound  $J^2 \leq 32 G M^3 / (27\pi)$, which is a generalisation of the known four dimensional Kerr bound $J \leq GM^2$. However for $D > 5$, the momentum is unbounded, and the black hole can be ultra-spinning. It was argued in \cite{Emparan:2003sy} that the Myers-Perry black hole in six or higher dimensions is unstable when the angular momentum becomes sufficiently large, which reproduces a sort of dynamical Kerr bound. The instability was a Gregory-Laflamme instability, and it suggested that there exists new stationary black holes with \emph{rippled} horizon of spherical topology.

The first evidence of the violation of the no-hair theorem is due to the discovery of the \emph{black ring} of Emparan and Reall \cite{Emparan:2001wn}. 
The black ring is an asymptotically flat black hole in a pure gravity theory in five dimensions. The topology of the black ring event horizon is $S^1 \times S^2$. The violation of the black hole uniqueness theorem occurs because for a certain range of mass and angular momentum, both Myers-Perry and Emparan-Reall solutions exist, whose horizons have different topology. This means that the asymptotic data does not uniquely determine the black hole solution.

We remark that uniqueness theorems have been proven for higher dimensional \emph{static} asymptotically flat black holes \cite{Gibbons:2002av, Gibbons:2002ju, Rogatko:2002bt}. In the static case and in any dimension, the only asymptotically flat black hole is the Schwarzschild-Tangherlini solution. The violation of the uniqueness theorem occurs when considering rotating black holes. 

A class of black holes which are particularly relevant in string theory are the \emph{supersymmetric} black holes. 
Supersymmetric black holes are black hole solutions to a certain supergravity theory which preserve some fraction of supersymmetry. This is equivalent to requiring that the solution admits at least one Killing spinor. Killing spinors are spinors which satisfy a set of conditions imposed by preservation of supersymmetry, called Killing spinor equations. 
Supersymmetric black holes must be \emph{extremal}, i.e. with vanishing surface gravity or equivalently with zero temperature.

In $\mathcal{N} =2$, and $D = 4$ supergravity, whose bosonic sector is Einstein-Maxwell theory, there are no supersymmetric and rotating  asymptotically flat single-black holes which are non-singular on the horizon, e.g. \cite{Townsend:2002yf}. This means that when one takes the supersymmetric limit of the Kerr black hole, at the same time one has to set the angular momentum to zero in order to preserve the regular horizon.

This is not true in minimal $D=5$ supergravity, where asymptotically flat supersymmetric and rotating single-black holes, with a regular horizon, exists. 
The supersymmetric version of the rotating Myers-Perry black hole in $D=5$ is parametrised by the mass $M$ and the angular momenta $J_1$ and $J_2$, where no conditions are imposed by the BPS bound. However, in order for the solution to be non-singular, a linear combination of the two angular momenta must vanish, and the solution is described by only two parameters, $M$ and $J$. This was first discovered by Breckenridge, Myers, Peet and Vafa \cite{Breckenridge:1996is}, also known as the \emph{BMPV black hole}. The event horizon of the BMPV black hole is not rotating, and the effect of the black hole rotation is to deform the horizon geometry from $S^3$ to a squashed $S^3$ \cite{Gauntlett:1998fz}. 

Based on the result of \cite{Gauntlett:2002nw}, which provides a classification of all supersymmetric bosonic solutions of minimal supergravity in five dimensions, a supersymmetric version of the black ring was constructed \cite{Elvang:2004rt}.
Asymptotically flat multi-center black holes in four dimensions which are supersymmetric and rotating have been found via a dimensional reduction of the supersymmetric black ring in a Taub-NUT background \cite{Elvang:2005sa}. 
Later it was found that even more exotic solutions can appear. It is possible indeed to superpose an arbitrary number of concentric black rings and place a spherical black hole at the center. This configuration is called \emph{black Saturn}, and it was first discovered by Gauntlett and Gutowski \cite{Gauntlett:2004wh} when some supersymmetry is preserved.

\begin{figure}[h]
\begin{center}
 \includegraphics[scale=0.50]{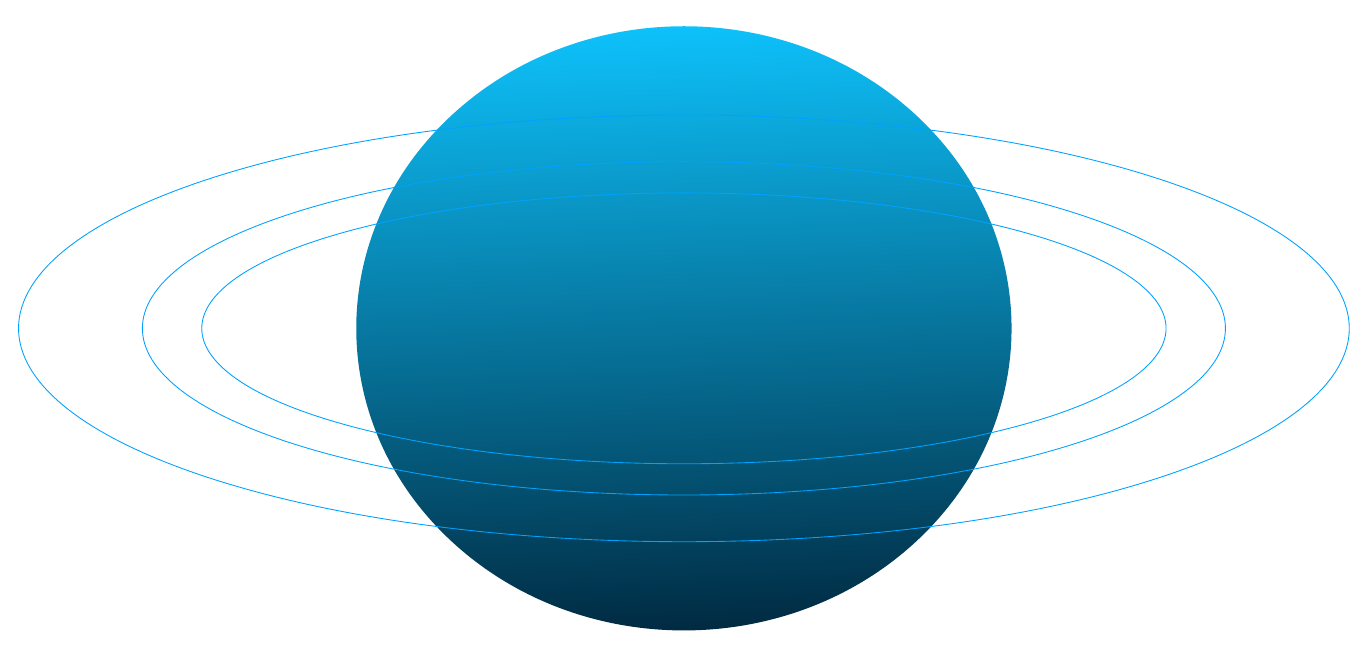}
\end{center}
\caption{Black Saturn.}
\end{figure}

Supersymmetry was instrumental in the construction of black Saturn solutions, because the conditions imposed by supersymmetry are first order equations, and therefore the multi-black hole solution is constructed as a superposition of harmonic functions. This is not the case for non-supersymmetric black holes, where the second order Einstein equations must be solved. A non-supersymmetric black Saturn solution was found by Elvang and Figueras \cite{Elvang:2007rd}, where they used the inverse scattering method.

Another example of the black hole uniqueness theorem violation is given by the construction of a 5-dimensional asymptotically flat black hole with a 2-cycle in the exterior. This solution was constructed by Kunduri and Lucietti \cite{Kunduri:2014iga}, which is characterised by a horizon with spherical topology and a ``bubble'' in the exterior. This solution represents a further violation of the uniqueness theorem because it has the same conserved charges of the BMPV black hole. 
Once again, the role of supersymmetry was instrumental to construct this solution. From the fuzzball literature one can find examples of supersymmetric smooth solitons with bubbles \cite{Bena:2005va}. Because of the linearity of the supersymmetry conditions, one can superpose a black hole solution to such geometries with bubbles, to obtain the ``bubbling'' black hole described. 

In string theory, black holes can be constructed as systems of intersecting branes. This was found for instance in \cite{Papadopoulos:1996uq}, where black hole solutions of five dimensional ungauged supergravity can be uplifted to solutions of $D=11$ supergravity which correspond to systems of intersecting branes. Black holes in $D=5$ gauged supergravity can also be uplifted to solutions in type IIB supergravity \cite{Chamblin:1999tk}. 
We also mention that when uplifted to higher dimensions, the supersymmetric black ring becomes a black supertube \cite{Elvang:2004ds}, and is reproduced in M-theory as a system of intersecting M2- and M5-branes
\begin{center}
\begin{tabular}{c@{\hspace{5mm}}ccccccccccc}
      &  0 & $x$ & $y$ & $\phi$ & $\psi$ & 5 & 6 & 7 & 8 & 9 & 10 \\
M2  & $\times$ & &  &  &  & $\times$ & $\times$ & & &  &   \\
M2 & $\times$ & &  &  &  &  &  & $\times$ & $\times$ &  &  \\
M2 & $\times$ & &  &  &  &  &  &  &  & $\times$ & $\times$  \\
M5 & $\times$ & &  &  & $\times$ &  &  & $\times$ & $\times$  & $\times$ & $\times$  \\
M5 & $\times$ & &  &  & $\times$  & $\times$ & $\times$ &  &  & $\times$ & $\times$  \\
M5 & $\times$ & &  &  & $\times$ & $\times$ & $\times$ & $\times$ & $\times$ & &   \\
\end{tabular}
\end{center}
where the supersymmetric black ring is described by the coordinates $x^0, x, y, \phi, \psi$. The M5-branes wrap a common circle described by $\psi$, and $z^i$, with $i=5, ... , 10$, are coordinates that span a 6-dimensional torus.

\section{Near-horizon geometries}

The richness of gravity in higher dimensions is responsible for the existence of several new types of black holes. Motivated by string and M theories, one is led to consider gravitational systems in ten and eleven dimensions. In such higher dimensional spacetimes, one should expect even more exotic black holes than the one described in $D=5$. This is because the number of axes around which the black hole can spin increases when the dimensions increase, and therefore one has to deal with more degrees of freedom.

The family of black holes most relevant to investigate quantum gravity properties of string theory, such as the microscopical interpretation of entropy, is represented by the \emph{extremal black holes}.
Extremal black holes, as well as any degenerate Killing horizon\footnote{A Killing horizon with vanishing surface gravity is called degenerate.}, admit a well defined notion of \emph{near-horizon limit} \cite{Reall:2002bh}, which allows the consistent decoupling of the full spacetime geometry from the horizon. In this way, one obtains the so-called notion of \emph{near-horizon geometry}. 

An interesting feature of near-horizon geometries is that they admit a scale symmetry associated with the invariance of the metric under a coordinate rescaling. This extra symmetry, which is not a symmetry of the full spacetime, makes it simpler to study analytically near-horizon geometries, and in particular to attempt a classification \cite{Kunduri:2013ana}. To study near-horizon geometries, one has in general to assume some conditions. 
There exists three approaches which are based on the following assumptions
\begin{itemize}
\item Assume isometries on the near-horizon geometry, e.g. \cite{Kunduri:2008rs},
\item Assume conditions on the stress-energy tensor $T_{\mu\nu}$ (the \emph{blackfold} approach), e.g. \cite{Emparan:2009cs}, 
\item Assume that the near-horizon geometry is \emph{supersymmetric}, e.g. \cite{hethor}.   
\end{itemize}   
For the purpose of this thesis, we shall describe the supersymmetric approach.

\subsection{Supersymmetric near-horizon geometries}

In the supersymmetric approach one assumes that on a neighbourhood of the event horizon there exists at least one Killing spinor, which is a spinor that satisfies the Killing spinor equations. Since the Killing spinor equations involve the bosonic fields of the theory, assuming supersymmetry represents a way to impose further conditions on the bosonic solution. 
For instance, near-horizon geometries in heterotic theory, without string corrections, have been classified by the number of Killing spinors \cite{hethor}. 

In many cases, supersymmetric near-horizon geometries exhibit the so-called \emph{supersymmetry enhancement}, which is here explained. If one assumes that the near-horizon geometry admits one Killing spinor, then it is possible to construct a map in terms of the bosonic fields, such that if applied to the given Killing spinor, it generates a second Killing spinor linearly independent from the first one. Therefore supersymmetry is \emph{enhanced}, because the near-horizon geometry experiences a doubling of the number of preserved supersymmetries. 

Supersymmetry enhancement implies \emph{symmetry enhancement}. The whole family of Killing spinors generated by supersymmetry enhancement can be paired to construct bilinears, which are isometries of the near-horizon geometry. Typically the Killing vectors generated in this way do not commute, but they generate an algebra which is at least $\mathfrak{sl}(2, \mathbb{R})$. The $\mathfrak{sl}(2, \mathbb{R})$ generated in this way is a \emph{dynamical} symmetry, since one has to use the equations of motion to prove it.

The supersymmetry enhancement and the $\mathfrak{sl}(2, \mathbb{R})$ isometry algebra have been demonstrated for near-horizon geometries in $D=11$ supergravity \cite{lichner11}, type IIB \cite{lichneriib}, type IIA \cite{lichneriia1}, massive type IIA \cite{lichneriia2}, uncorrected heterotic \cite{hethor}, minimal gauged $D=5$  supergravity \cite{Grover:2013ima}, $\mathcal{N} =2$, $D=4$ gauged supergravity \cite{Gutowski:2016gkg}.
This led to the formulation of the \emph{horizon conjecture}. The first part of the conjecture states a condition on the number of preserved supersymmetries. In particular, if the index of a twisted Dirac operator vanishes, the number of preserved supersymmetries is always even. To prove this part one needs to construct a generalised version of the Lichnerowicz theorem, together with an index theory argument.  
The second part of the conjecture states that every near-horizon geometry with non-trivial fluxes admit the $\mathfrak{sl}(2, \mathbb{R})$ isometry algebra, independently of the supergravity theory considered.  To prove the second part, one needs to generate new linearly independent spinors by using the map mentioned above, and to construct bilinears out of them. 

There are known examples of black holes and branes which exhibit a  supersymmetry enhancement on the horizon, for instance the Reissner-Nordstr\"om and BMPV black holes, the D3-, M2- and M5-branes, see e.g. \cite{Gibbons:1993sv}. In all these cases, the solution is half-maximally supersymmetric in the bulk, but its near-horizon geometry is maximally supersymmetric. It is interesting to investigate if this is a general behaviour or if there are other types of black holes for which there is supersymmetry enhancement away from the horizon. To answer this question one would need to extend the black near-horizon geometry into the bulk and to formulate a Lichnerowicz type theorem. So far this question is still open.

From the physical perspective, the supersymmetry enhancement has an application in investigating properties of black holes, like the entropy microstate counting e.g. \cite{Sen:2011cn}, and in AdS/CFT, where the enhancement of the isometry group to $\mathfrak{sl}(2, \mathbb{R})$ is the minimal required to assert that the dual field theory is conformal. 
From the geometric point of view, the supersymmetry enhancement implies a refinement of the $G$-structure of the horizon spatial cross section. This in turn implies further geometric conditions, which are useful for a possible classification of the near-horizon geometries. 

Finally, we mention that many supersymmetric backgrounds of supergravity theories, such as anti-de Sitter warped product solutions which are largely studied for their importance in string theory, can be described as near-horizon geometries. Therefore general features of near-horizon geometries, like the (super)symmetry enhancement and generalised Lichnerowicz theorems, are also useful to classify these type of supersymmetric backgrounds, see e.g. \cite{lichnerads1, lichnerads2, lichnerads3, lichnerads4, Gran:2017qus}.

\section{AdS/CFT duality}

The AdS/CFT duality is arguably the most striking discovery made in the context of string theory. It represents the first example of the \emph{holographic principle}, which informally states that 
\begin{center}
\emph{
``Any theory of gravity on a spacetime $\mathcal{M}$ is equivalent to a gauge theory, with no gravity, defined on $\partial \mathcal{M}$''}
\end{center}
The holographic principle compares different theories, one in $D$ dimensions and the other in $D-1$ dimensions. Therefore the gauge theory, which appears in the lower dimensional side, must develop ``enough'' degrees of freedom in order to be equivalent to the dual higher dimensional gravity theory. This is because theories in higher dimensions generically show a higher number of degrees of freedom, e.g. the one-particle state in $D$ dimensions is labelled by an additional momentum parameter in comparison to states in $D-1$ dimensions.

The solution to this issue is given by an observation of G. 't Hooft, who considered a Yang-Mills theory with gauge group $SU(N)$ in the limit where $N$ is large and the 't Hooft coupling $g_{YM}^2 N \equiv \lambda$ is fixed. This is also called the \emph{'t Hooft limit}. In this limit, 't Hooft \cite{tHooft:1973alw} showed that only planar diagrams, with quarks on the edge, dominate, and he pointed out that the partition function in this limit resembles the one of an interacting dual string. 
In the 't Hooft limit, the partition function admits a $1 / N$ perturbative expansion, where powers of $1/ N$ are related to the genus of the Feynman diagrams associated to that coefficient.  In the perturbative expansion, the 't Hooft coupling $\lambda$ also appears, whose powers enumerate the number of quantum loops. 
Schematically, the partition function is 
\begin{equation}
\notag
\mathcal{Z}_{YM} = \sum_{g=0}^{\infty} \frac{1}{N^{2g - 2}} \sum_{\ell = 0}^{\infty} c_{g, \ell} \lambda^{\ell} \ ,
\end{equation}
where $c_{g, \ell}$ are suitable coefficients evaluated at fixed genus $g$ and loop order $\ell$. The form in which the partition function is written resembles the partition function of a string theory, which is written as a perturbative expansion in $g_s$, whose powers are related to the genus of the associated string diagrams, schematically
\begin{equation}
\notag
\mathcal{Z}_{string} = \sum_{g=0}^{\infty} g_s^{2g - 2} \mathcal{F}_g \ , 
\end{equation}
for some suitable coefficients $\mathcal{F}_g $ which depend by the genus $g$. The question now is which string theory one must consider? The concrete idea was given by J. Maldacena \cite{Maldacena:1997re}, who conjectured that 
\begin{center}
\emph{
``Type IIB superstring on $AdS_5 \times S^5$ \hspace{5mm} $=$ \hspace{5mm} $\mathcal{N} = 4$, $SU(N)$ super Yang-Mills on $\partial AdS_5$''}
\end{center}
On the string side, the $AdS_5 \times S^5$ background arises as the near-horizon limit of a stack of $N$ parallel D3-branes. In the duality, the number $N$ of D3-branes corresponds to the rank $N$ of the $SU(N)$ Yang-Mills gauge group. The $\mathcal{N} = 4$ super Yang-Mills theory is a \emph{quantum conformal} theory, as its beta function vanishes exactly. 
For this reason, Maldacena's conjecture is also called the AdS/CFT duality.
This duality comes equipped with a dictionary, which translates each physical quantity of one side of the duality to quantities defined on the other side.  
A comprehensive review on the AdS/CFT correspondence is in \cite{Aharony:1999ti}.   

\begin{figure}[h]
\begin{center}
 \includegraphics[scale=0.34]{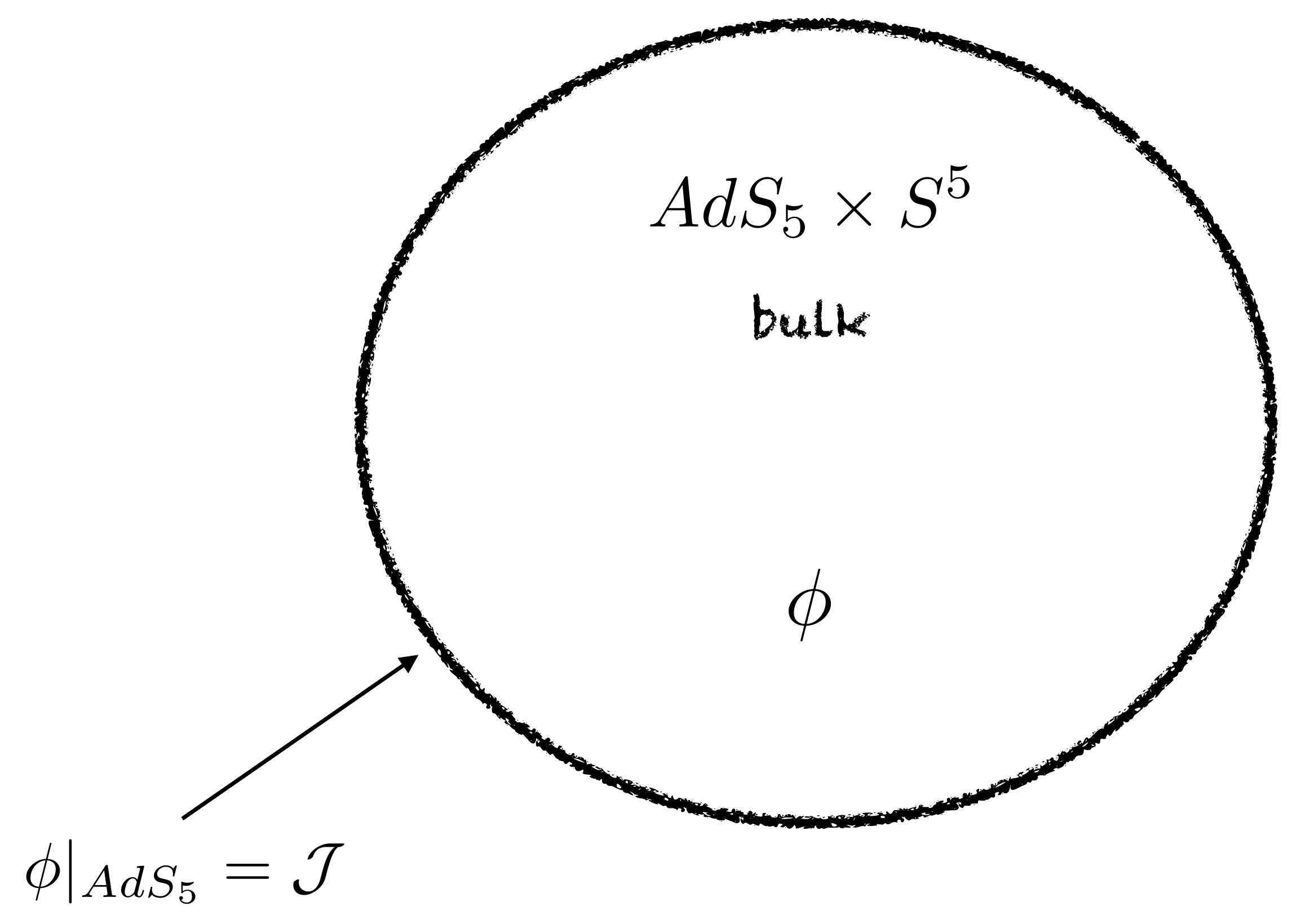}
\end{center}
\caption{Sources generation in the dual CFT.}
\end{figure}   

\noindent The fields defined on the string side in $AdS_5 \times S^5$, which we denote collectively by $\phi$, have a meaning for the dual CFT. The fields evaluated at the boundary of the spacetime, i.e. $\phi |_{AdS_5}$, are the sources $\mathcal{J}$, or equivalently conserved currents, of the dual CFT.

 In the 't Hooft limit and in the supergravity approximation where string corrections can be ignored, the equivalence between the two theories can be stated as
\begin{equation}
\notag
\mathcal{Z}_{CFT} [ \mathcal{J} ] = e^{- S_{sugra}( \phi, \mathcal{J}= \phi |_{AdS_5} ) } \ . 
\end{equation}  
The $AdS_5 \times S^5$ string sigma model contains the parameters $g_s$ and $R_{AdS}$, which is the radius of $AdS_5$. On the other side of the duality, $\mathcal{N} = 4$ super Yang-Mills theory contains the parameters $N$ and $g_{YM}$, or equivalently $\lambda$. The AdS/CFT duality imposes relations between the parameters of the two theories
\begin{equation}
\notag
\frac{4\pi \lambda}{N} = g_s \ , \qquad\qquad
\sqrt{\lambda} = \frac{R_{AdS}^2}{\ell_s^2}\ . 
\end{equation} 
The main idea of the AdS/CFT duality is that the two theories describe the \emph{same} physics, which is not obvious since a priori gravitational and gauge theories look differently. In particular, the AdS/CFT is a strong/weak duality, in the sense that highly coupled gauge fields correspond to classical weak supergravity, and weakly coupled gauge fields correspond to strings propagating in highly curved backgrounds. 
This is showed in the diagram below, where every point in the box is a physical configuration which can be either described via the ``gauge'' or the ``gravity'' axes. 
Therefore the AdS/CFT duality gives a prescription on how to compute correlators on strongly coupled gauge theories. 
\begin{figure}[H]
\begin{center}
 \includegraphics[scale=0.305]{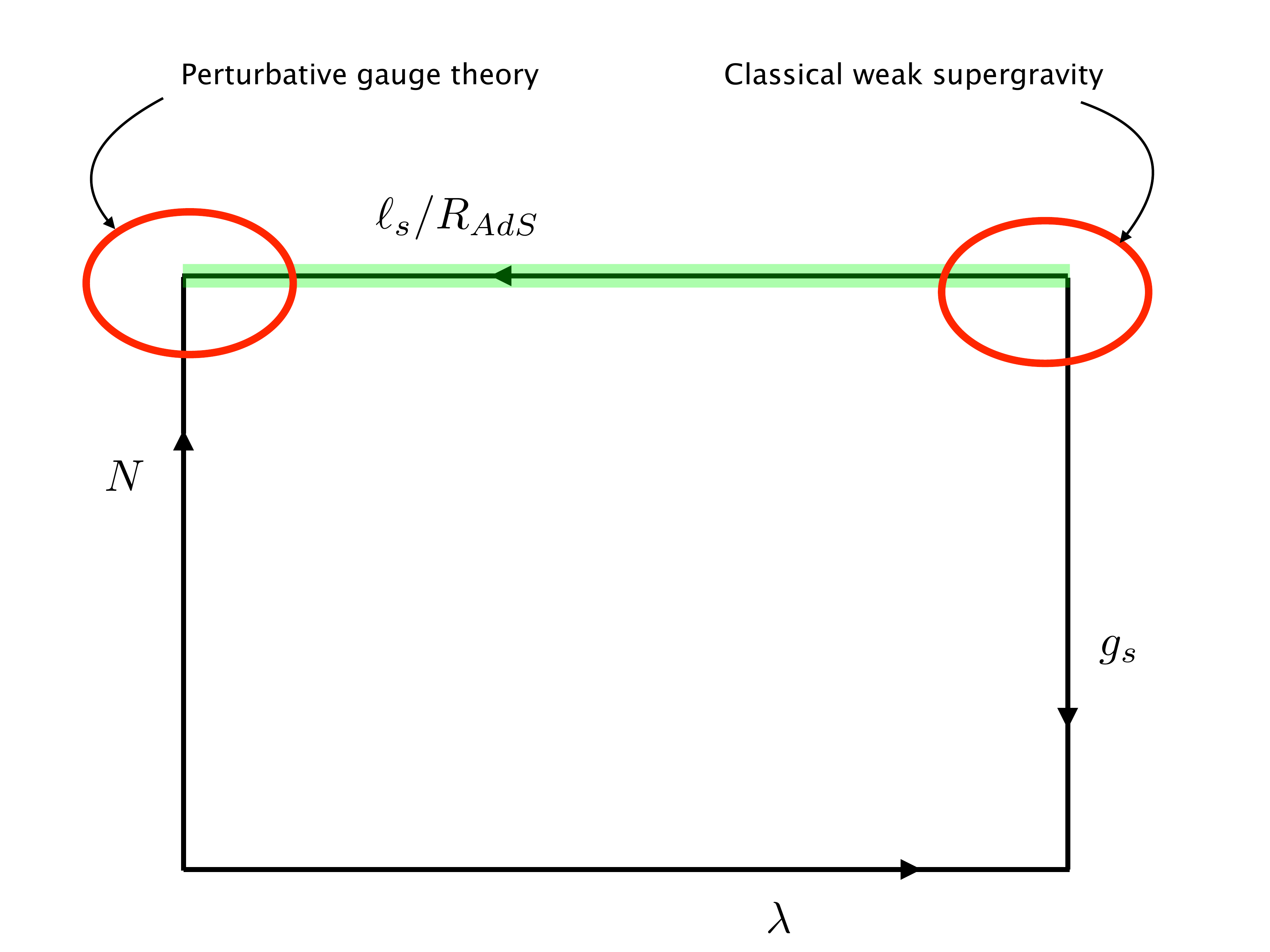}
\end{center}
\caption{The box of physics in the AdS/CFT duality.}
\label{fig:box}
\end{figure}   

\subsection{The spectrum conjecture and integrable spin chains}

A concrete way to test the AdS/CFT duality consists in computing the spectrum on both theories and check if there is any matching, for a review see \cite{rev}.  
We collectively denote string excitations in $AdS_5 \times S^5$ by $| \mathcal{O}_A \rangle$, with $A$ being a multi index, which are responsible to generate fields in the bulk. They are eigenstates of the world-sheet Hamiltonian $\mathcal{H}^{ws}$, with eigenenergies $E_A$ which depend on $\ell_s / R_{AdS}$ and $g_s$, 
\begin{equation}
\notag
\mathcal{H}^{ws} | \mathcal{O}_A \rangle = E_A   | \mathcal{O}_A \rangle \ . 
\end{equation}
On the CFT side, the analogous of the string excitations $| \mathcal{O}_A \rangle$ are composite gauge invariant operators, e.g. $\mathcal{O}_A = \mathrm{Tr} [ \phi_{\ell_1} \cdots \phi_{\ell_n} ]$, where each field $(\phi_{\ell_i})_{ab}$ is an elementary gauge field of $\mathcal{N} = 4$ super Yang-Mills theory in the adjoint representation of $SU(N)$. In a conformal field theory, the two-point function of gauge invariant operators is fully specified by the scaling dimension $\Delta$ of the operators. In the case that $\mathcal{O}_A$ is a scalar operator,
\begin{equation}
\notag
\langle \mathcal{O}_A (x) , \mathcal{O}_B (y) \rangle = \frac{\delta_{AB}}{(x - y)^{2\Delta_A}} \ , 
\end{equation}
where $\Delta_A$ depends on $\lambda$ and $1 / N$. In the AdS/CFT duality, it is conjectured that 
\begin{equation}
\notag
E_A  = \Delta_A  \ . 
\end{equation}
To check this equality exactly, one has to determine the string spectrum at all orders in the genus and $\ell_s / R_{AdS}$ and the complete dependence of the scaling dimensions on $\lambda$ and $1 / N$. This is still a difficult problem, however there is a regime where some simplification occurs. This happens when the string theory contains non-interacting strings (i.e. $g_s = 0$) and the dual gauge theory is in the planar limit. In the diagram \ref{fig:box}, this regime corresponds to the green line. 
In this regime, one has a certain control on the classical weak supergravity and the weakly coupled gauge theory. However the two theories, which are both in their perturbative regime, are not dual to each other. In particular, a classical weak supergravity is dual to a strongly coupled gauge theory and a weakly coupled gauge theory is dual to a string theory on a highly curved background. 

A solution comes from Berenstein, Maldacena and Nastase \cite{Berenstein:2002jq}, who considered a point-like configuration of the string which is spinning with large angular momentum $J$ around the equator of $S^5$. In the limit $J \rightarrow \infty$ and $J^2 / N$ is fixed, which is called \emph{BMN limit}, the background geometry detected by the fast moving point particle is the gravitational plane-wave. String fluctuations around this background can be exactly quantised in the light-cone gauge \cite{Metsaev:2001bj, Metsaev:2002re}. The eigenenergies of the string fluctuations have been computed exactly at all loops, which gives a prediction of the scaling dimensions, \begin{equation}
\notag
\Delta_n = J + 2 \sqrt{1 + \frac{\lambda}{J^2} n^2} \ ,
\end{equation}
where $\lambda / J^2$ becomes a loop counter for the effective gauge theory in the BMN limit. From the gauge theory side, the scaling dimensions in the formula above have been correctly reproduced up to the three loop order \cite{Beisert:2003tq, Beisert:2003ys, Bern:2005iz, Eden:2004ua}. 

We shall briefly review how to compute scaling dimensions in a conformal field theory. In general, the two-point function $\langle \mathcal{O}_A (x) , \mathcal{O}_B (y) \rangle$ is not proportional to $\delta_{AB}$, but it may contain off-diagonal terms. This is known as the \emph{operator mixing problem}. Therefore one has to find a suitable basis for the operators $\mathcal{O}_A$ which diagonalizes the two-point function. The superconformal group of $\mathcal{N} = 4$ super Yang-Mills theory is $PSU(2, 2 | 4)$, and one of its generators is the \emph{dilatation} generator $\mathcal{D}$. The dilatation generator acts as an operator on $\mathcal{O}_A$ and its eigenvalues are the scaling dimensions, i.e.
\begin{equation}
\notag
\mathcal{D}  \, \mathcal{O}_A = \Delta_A  \, \mathcal{O}_A \ . 
\end{equation}
Diagonalizing the dilatation operator suffices to solve the operator mixing problem. 
The dilatation operator receives quantum corrections, and perturbatively it can be expanded out as\footnote{In principle this is a double expansion, where the second perturbative parameter is $1 / N$ whose powers depend on the genus of the string diagrams. Since we are in the regime of non-interacting strings (i.e. $g_s = 0$), this second expansion do not appear.}
\begin{equation}
\notag
\mathcal{D} = \sum_{n=0}^{\infty} \lambda^{n} \mathcal{D}^{(n)} \ . 
\end{equation}
This in turn implies that the scaling dimensions also gain quantum corrections
\begin{equation}
\notag
\Delta = \Delta^{(0)} + \sum_{\ell = 1}^{\infty} \lambda^{\ell} \Delta^{(\ell)} \ , 
\end{equation}
such that 
\begin{equation}
\notag
\mathcal{D}^{(n)} \mathcal{O} = \Delta^{(n)} \mathcal{O} \ ,
\end{equation}
where $\Delta^{(0)}$ is the bare scaling dimensions, which is obtained by summing up the dimensions of the elementary fields inside the operator $\mathcal{O}$, and $\Delta^{(\ell)}$ are the anomalous scaling dimensions. 

In the field content of $\mathcal{N} = 4$ super Yang-Mills theory there are six real scalar fields $\phi_1, \cdots, \phi_6$ which transform under the $SO(6)$ R-symmetry inherited from the $S^5$ internal space of the dual string theory. These scalar fields can be combined into three complex scalar fields
\begin{equation}
\notag
Z = \frac{1}{\sqrt{2}}(\phi_1 + i \phi_2 )\ , \qquad
W =  \frac{1}{\sqrt{2}}( \phi_3 + i \phi_4 ) \ , \qquad
U = \frac{1}{\sqrt{2}}( \phi_5 + i \phi_6 ) \ . 
\end{equation}
The dual string is moving around the equator of $S^5$, specifically on the submanifold $S^3$. This means that the R-symmetry reduces from $SO(6)$ to $SU(2)$, and the corresponding closed $SU(2)$ subsector of scalar fields is $\{ Z , W \}$. Composite gauge invariant operators are built up in terms of $Z$ and $W$, e.g. 
\begin{equation}
\notag
\mathrm{Tr}[ Z  \cdots Z ] \ , \qquad\qquad
\mathrm{Tr} [ Z \cdots Z W Z \cdots Z ] \ .
\end{equation}
Minahan and Zarembo \cite{Minahan:2002ve} showed that $\mathcal{D}^{(1)}$ acts on composite gauge invariant operators, built up in terms of $Z$ and $W$, as the periodic Heisenberg XXX quantum spin chain Hamiltonian, 
\begin{equation}
\notag
\mathcal{D}^{(1)} = \frac{1}{8 \pi^2} \mathcal{H}_{\scriptsize \mbox{XXX}\normalsize} = \frac{1}{4 \pi^2} \sum_{i = 1}^L \bigg( \frac{1}{4} - \vec{\sigma}_i \cdot \vec{\sigma}_{i + 1} \bigg) \ , 
\end{equation}
where $\vec{\sigma}_i$ are the three Pauli matrices acting on the spin at the $i^{th}$ site. 
Therefore the gauge invariant operators shown above can be interpreted as excitations of a ferromagnetic $1/2$-spin chain, where $W = + 1/2$ and $Z = - 1/2$, 
\begin{eqnarray}
\notag
\mathrm{Tr}[ Z  \cdots Z ] &\equiv & | \downarrow \cdots \downarrow \rangle \ ,  \\
\notag
\mathrm{Tr} [ Z \cdots Z W Z \cdots Z ] &\equiv & | \downarrow \cdots \downarrow \uparrow \downarrow \cdots \downarrow \rangle \ , 
\end{eqnarray}
Consider a spin chain of length $L$, i.e. the total number of elementary fields $Z$ and $W$ in the trace is $L$. 
The state $\mathrm{Tr}[ Z  \cdots Z ]$ is the ground state of the ferromagnetic spin chain, while $\mathrm{Tr} [ Z \cdots Z W Z \cdots Z ] $ corresponds to the first excitation, the so-called \emph{magnon}, 
\begin{equation}
\notag
| p \rangle \equiv \frac{1}{\sqrt{L}} \sum_{n} e^{i p n}\,  \mathrm{Tr}[  Z \cdots Z W Z \cdots Z ] \ , 
\end{equation}
where $W$ is in the $n^{th}$ position. 
The magnon $|p \rangle$ is a wave propagating around the spin chain with momentum $p$. 

Magnon excitations of the spin chain are analogous to string excitations in the BMN limit. 
Since the bare scaling dimension $\Delta^{(0)}$ is the sum of the dimensions of the elementary fields, which in our case are the scalar fields $Z$ and $W$ in the spin chain, we have that $\Delta^{(0)} = L$. 
The anomalous scaling dimensions can be found diagonalizing the Heisenberg XXX spin chain Hamiltonian, which was done in 1931 by H. Bethe with a technique named after his name, called \emph{Bethe ansatz} \cite{Bethe:1931hc}. This allowed the verification of the spectrum conjecture at one-loop. The fact that the gauge theory side of the duality can be described in terms of the XXX spin chain implies that the theory is \emph{quantum integrable} at one-loop expansion. 

This argument can also be generalised to $n$-loop \cite{Beisert:2003tq, Beisert:2003ys, Bern:2005iz, Eden:2004ua}. The one-loop dilatation operator describes a spin chain with only nearest interactions. The two-loop operator acts on the closed $SU(2)$ sector introducing next-to-nearest neighbours interactions. In general, the $n$-loop operator introduces interactions at most with the $n^{th}$ nearest neighbour. Therefore the dilatation operator can be perturbatively rewritten as a Hamiltonian acting on a spin chain with the range of interaction given by the order in perturbation theory considered.
In practice, analysing properties of the full Hamiltonian is difficult, which motivates consideration of the S-matrix.
Since all these spin chains are quantum integrable, there is a strong evidence that $\mathcal{N} = 4$ super Yang-Mills theory is exactly quantum integrable in the planar limit.
However, this is a highly non-trivial result to prove.
 
This argument is also supported from the string theory side. 
Bena, Polchinski and Roiban \cite{Bena:2003wd} proved that the Green-Schwarz string classical action on $AdS_5 \times S^5$ is integrable.
This motivates to further explore the holographic principle for string backgrounds which are not $AdS_5 \times S^5$, but whose sigma model is still classically integrable.

\subsection{Black holes in AdS/CFT}

The AdS/CFT duality provides a prescription on how to compute the entropy of asymptotically anti-de Sitter (AdS) black holes: the microstates of the asymptotically AdS black hole are states of a dual CFT. 
Classical AdS black holes are dual to strongly coupled CFTs \cite{Witten:1998zw}, and if the black hole is non-extremal, the dual CFT thermalize. 
For the asymptotically $AdS_3$ Ba\~nados-Teitelboim-Zanelli (BTZ) black hole, it is possible in the dual 2-dimensional CFT to count states with the same quantum numbers of the black hole by using the Cardy formula \cite{Strominger:1997eq}. However this argument cannot applied in higher dimensions, and one has little control on the dual CFT since strongly coupled quantum field theories are poorly understood. 

A way to overcome this issue consists in considering \emph{supersymmetric} asymptotically AdS black holes. The states of the dual strongly coupled CFT are in a short representation, and it is not expected that the number of states with the same quantum numbers change when the CFT coupling varies. Therefore for this class of black holes, one can count the degeneracy of BPS states of the dual CFT at \emph{weak} coupling. 

By using localization techniques, this counting was possible for a class of supersymmetric black holes which are asymptotically $AdS_4 \times S^7$, and whose dual field theory is a topologically twisted ABJM theory \cite{Benini:2015eyy}. 
Although much is known about BPS states in $\mathcal{N} = 4$ super Yang-Mills theory, a microscopic description of the entropy for the Gutowski-Reall black hole \cite{Gutowski:2004ez, Gutowski:2004yv} and its extension \cite{Chong:2005hr, Kunduri:2006ek, Markeviciute:2018yal}, which are asymptotically $AdS_5 \times S^5$, is still an open problem. 
This is also the case for the class of black holes which are asymptotically $AdS_7 \times S^4$.

The AdS/CFT, or more generically the holographic principle, provides a potential way to define a theory of \emph{quantum gravity}. The AdS/CFT relates string theories on a certain background to some quantum mechanics or quantum field theories, which we know how to quantize. However there are still some outstanding limitations to this idea. For instance the AdS/CFT requires that the string theory is defined on backgrounds with special boundary conditions, which is not suitable to cosmological spacetimes or realistic compactifications. 

A big puzzle of modern theoretical physics, is the \emph{information loss} problem of black holes, for reviews see e.g. \cite{Polchinski:2016hrw, Harlow:2014yka}. The problem consists in the fact that if an external observer throws bits of information in a pure state inside the black hole, after the black hole evaporation, the external observer will only detect states which are mixed. This transition from pure to mixed states means that we are losing information, or equivalently, the process is not unitary.
However if we consider asymptotically AdS black holes, the AdS/CFT tells us that this process admits a description on the dual quantum field theory in absence of gravity, which we know to be unitary, and therefore there is no information loss. 
Of course this argument applies only to black holes which are asymptotically AdS, and to extend it to the most general black hole one would need the holographic principle. 

Several solutions have been proposed to the information loss paradox. 
Recently Hawking, Perry and Strominger \cite{Hawking:2016msc} have shown that four dimensional black holes do have \emph{soft hairs}, which are relevant at the quantum level. Soft hairs appear as a consequence of the invariance of the black hole spacetime at future and past null infinity under the Bondi, van der Burg, Metzner and Sachs (BMS) group, which includes in addition to the Poincar\'e transformations an infinite set of diffeomorphisms, known as supertranslations. This is further discussed in \cite{Averin:2016ybl, Carlip:2017xne}. The authors of \cite{Hawking:2016msc} argue that soft hairs store the information associated to particles crossing the event horizon, and this might lead to a solution of the information loss paradox.

\section{Plan of this thesis}

This thesis is organised as follows:

\noindent {\bf Chapter 2} is an introduction to supersymmetric near-horizon geometries, with some examples (the BMPV black hole and the supersymmetric black ring). The horizon conjecture will be stated and proved schematically in the context of type IIA supergravity.

\noindent {\bf Chapter 3} aims to extend the horizon conjecture outside the supergravity approximation. The analysis presented is in the context of heterotic supergravity with $\alpha'$ corrections and based on \cite{Fontanella:2016aok}. A consequence of our analysis is the no-go theorem \ref{th:no_AdS2}. In the case of supersymmetry enhancement,  the spacetime near-horizon geometry will be described. Finally, horizons with broken supersymmetry will be considered and the geometry described. 

\noindent {\bf Chapter 4} describes how to extend a prescribed near-horizon geometry into the bulk, in order to recover the black hole solution. This work is based on \cite{Fontanella:2016lzo} and consists in a first order expansion in the radial coordinate of the near-horizon fields. The theories considered are heterotic supergravity in absence of string corrections and $D=11$ supergravity. The main result obtained is that the space of radial deformations of the near-horizon geometry is finite dimensional. Stability issues of the extended solution will be discussed.

\noindent {\bf Chapter 5} introduces the idea of exploring the holographic principle for integrable theories. The notion of classical integrability will be introduced, with some examples regarding the KdV and Sine-Gordon equations, and supercoset sigma-models. Properties of the scattering matrix in integrable theories will be given, together with their formulation in the mathematical framework of Hopf algebras. Finally, the algebraic Bethe ansatz will be reviewed, in the context of models which admit an $\mathfrak{su}(2)$ algebra.

\noindent {\bf Chapter 6} is devoted to the study of massless modes of $AdS_2 \times S^2 \times T^6$ type IIB background and is based on \cite{Andrea2}. A set of solutions for the relativistic massless R-matrix will be determined and connected to known non-relativistic massless solutions. This will allow us to provide a further evidence of the Zamolodchikov's conjecture \ref{Zam_conj}. 
The asymptotic Bethe ansatz is formulated by using a technique based on the free-fermion condition in both relativistic and non-relativistic cases.

\noindent {\bf Chapter 7} focusses on massless modes of $AdS_3 \times S^3 \times T^4$ type IIB background and is based on \cite{Andrea}. A specific non-relativistic massless R-matrix is chosen, and its invariance under the $q$-deformed super Poincar\'e symmetry is investigated. The pseudo-invariance of the R-matrix under the boost generator leads us to rewrite the algebraic scattering problem in a geometric language, based on fibre bundles. In this new language, a candidate universal R-matrix is proposed. Finally, we show that the boost invariance of the R-matrix admits a second interpretation as an auxiliary Sch\"odinger equation. This quantum approach is supported by a coordinate Bethe ansatz, where the Hamiltonian involved in the Schr\"odinger problem is constructed.

\noindent {\bf Appendices.} This thesis ends with several appendices. In appendix A, we list our spinorial geometry convention for heterotic near-horizon geometries. 
In appendix B, we list the spin connection and Ricci tensor components of a generic horizon metric. 
In appendix C, we list relevant bosonic field equations and their decomposition in Gaussian null coordinates.  
In appendix D, we report the detailed analysis for the simplification of the heterotic reduced KSEs.  
In appendix E we show that in heterotic theory one can always construct additional spacetime parallel spinors,  without making any assumption on the spacetime geometry.  
In appendix F, we report the detailed proof of the Lichnerowicz theorem in heterotic theory including $\alpha'$ corrections. 
In appendix G, we show how to write an $AdS_{n+1}$ space as a warped fibration over $AdS_n$, and we comment its implications. 
In appendix H, we show that the integral expression of the $AdS_3$ massless R-matrix correctly reproduces the known R-matrix solution, via integrating along a straight line.  
In appendix I, we speculate about alternative interpretations of the boost pseudo-invariance of the R-matrix.

\clearpage{\pagestyle{empty}\cleardoublepage} 

\part{Black Holes in String Theory}
\chapter{\textbf{Introduction}}
\label{chap:intro_BH}

\section{Near-horizon geometries}
\label{sec:NHG}
We are interested in studying the geometry of black hole horizons. Therefore we assume that the $D$-dimensional spacetime admits a \emph{Killing horizon}. 
\begin{definition}[Killing horizon]
A null hypersurface $\mathcal{H}$ is a Killing horizon if there exists a Killing vector field $V$, defined in a neighbourhood of the spacetime which contains $\mathcal{H}$, such that $V$ is normal to $\mathcal{H}$.  
\end{definition}
A set of coordinates $(u, r, y^I)$ can be adapted in a neighbourhood of the spacetime which contains $\mathcal{H}$, the so-called \emph{Gaussian null coordinates}, where $u, r \in \mathbb{R}$ and $y^I$ are spatial coordinates with $I= 1, ... , D-2$ \cite{isen, gnull}.
In these coordinates, the Killing vector field $V$ is simply 
\begin{equation}
\label{iso_V}
V = \frac{\partial}{\partial u} \ , 
\end{equation}  
and the Killing horizon $\mathcal{H}$ is located at $r=0$. 

In a neighbourhood of the spacetime which contains $\mathcal{H}$, the metric decomposes as follows:
\begin{equation}
\label{Gauss_metric}
ds^2 = - rf du^2 + 2 du dr + 2rh_I du dy^I + \gamma_{IJ} dy^I dy^J \ , 
\end{equation}
where $f$ is a scalar, $h$ a 1-form and $\gamma$ is the metric of the \emph{spatial cross section} $\cS$, which is a co-dimension 2 manifold. 
The metric components $f, h, \gamma$ are independent of $u$, because of the Killing vector $V$, and also we assume that they are \emph{analytic} in $r$.
We also assume that the spatial cross section $\cS$ is \emph{smooth}, \emph{compact} and \emph{connected without boundary}. 

Since we assumed that $f, h, \gamma$ are analytic in $r$, we Taylor expand them at $r=0$:
\begin{eqnarray}
\notag
f(r, y) &=& \sum_{n=0}^{\infty} \frac{r^n}{n!} \partial_r^n f\big|_{r=0} \ , \\
\notag
h_I(r, y) &=& \sum_{n=0}^{\infty} \frac{r^n}{n!} \partial_r^n h_I\big|_{r=0} \ , \\
\gamma_{IJ}(r, y) &=& \sum_{n=0}^{\infty} \frac{r^n}{n!} \partial_r^n \gamma_{IJ}\big|_{r=0} \ .
\end{eqnarray}
We remark that the surface gravity\footnote{Here we use the same symbol for the Killing vector $V$ and its dual 1-form.}
\begin{equation}
i_V dV \big|_{r=0} = f(0, y) V\big|_{r=0} \ , 
\end{equation}
is proportional to $f(0, y)$. Thus if $f(0, r)$ is not zero, the black hole has temperature. 

Since we are interested in the black hole horizon, we must decouple the horizon from the rest of the spacetime. A consistent way to make it is via the \emph{near-horizon limit} \cite{Reall:2002bh}, which consists in the following substitution
\begin{equation}
\label{NH_limit}
u \rightarrow \frac{u}{\varepsilon} \ , \qquad r \rightarrow \varepsilon r \ , \qquad y^I \rightarrow y^I \ , \qquad \varepsilon \rightarrow 0 \ . 
\end{equation}  
This limit would produce a divergent term in (\ref{Gauss_metric}) arising from $-r f du^2$. However if we impose the vanishing of $f(0, y)$, which means that the black hole has no temperature, then the divergent term disappears. Therefore we shall consider only \emph{extremal} horizons. 

After taking the limit (\ref{NH_limit}), the metric (\ref{Gauss_metric}) becomes
\begin{equation}
\label{NHG}
ds^2 = - r^2 \Delta du^2 + 2 du dr + 2rh_I du dy^I + \gamma_{IJ} dy^I dy^J \ , 
\end{equation}
where now
\begin{equation}
h_I = h_I(0, y) \ , \qquad \gamma_{IJ} = \gamma_{IJ}(0, y) \ , \qquad \Delta = - \partial_r f \big|_{r=0} \ . 
\end{equation} 
The metric obtained in (\ref{NHG}) is the \emph{near-horizon geometry}, and its components $\{ \D , h_I, \gamma_{IJ}\}$ are the so-called \emph{near-horizon data}, which depend only on $y^I$. 
It turns out to be convenient to introduce the following non-coordinate basis, the so-called \emph{light-cone frame}
\begin{equation}
\label{nhbasis}
\mathbf{e}^+ = du \ , \qquad
\mathbf{e}^- = dr + rh - \frac{1}{2} r^2 \D du \ , \qquad
\mathbf{e}^i = e^i{}_J dy^J \ , 
\end{equation}
where $e^i{}_J$ is a vielbein transformation, such that the metric (\ref{NHG}) becomes
\begin{equation}
\label{NHG_frame}
ds^2= 2 \mathbf{e}^+ \mathbf{e}^- + \delta_{ij} \mathbf{e}^i \mathbf{e}^j \ . 
\end{equation}
The near-horizon geometry (\ref{NHG}), in addition to $V= \partial_u$, admits an extra symmetry associated with the scaling symmetry $u \rightarrow u/\alpha$ and $r \rightarrow \alpha r$, with $\alpha\in \mathbb{R}$. This isometry is generated by the Killing vector
\begin{equation}
\label{iso_D}
D = - r \partial_r + u\partial_u \ , 
\end{equation} 
which does not commute with $V$, i.e.
\begin{equation}
[ V, D ] = V \ . 
\end{equation}
In chapter \ref{ch:het_NHG} we aim to investigate whether a (supersymmetric) near-horizon geometry admits even more extra symmetries, while in chapter \ref{chap:bulk} we shall consider the inverse problem of extending a near-horizon geometry into the bulk.

\subsection{Examples of near-horizon geometries}
In this section we shall illustrate some concrete examples of near-horizon geometries associated with some black holes: the BMPV black hole \cite{Breckenridge:1996is, Gauntlett:1998fz} and the supersymmetric black ring \cite{Elvang:2004rt}.

\subsubsection{BMPV black hole}
The BMPV (Breckenridge, Myers, Peet and Vafa) black hole appears in the bosonic sector of $D=5$ gauged supergravity, which is Einstein-Maxwell theory with a Chern-Simons term. 
The metric is given in terms of a Gibbons-Hawking space
\begin{equation}
\label{Hawking_Gibbons}
ds^2 = - f^2 (dt + \omega )^2 + f^{-1} ds^2 (\mathbb{R}^4) \ , 
\end{equation} 
where $f$ and $\omega$ are a scalar and 1-form on $\mathbb{R}^4$ respectively, given by
\begin{equation}
f^{-1} = 1 + \frac{\mu}{\rho^2} \ , \qquad\qquad
\omega = \frac{j}{2 \rho^2} \sigma^3 \ , 
\end{equation}
where $\mu$ and $j$ are parameters proportional to the mass and the angular momentum, $\rho \in [0, + \infty [$ is the radius,  and $\sigma^3$ is defined as follows
\begin{eqnarray}
\notag
\sigma^1 &=& - \sin \psi d\theta + \cos \psi \sin \theta d\phi \ , \\ 
\notag
\sigma^2 &=& \cos \psi d\theta + \sin\psi \sin\theta d \phi \ , \\
\sigma^3 &=& d\psi + \cos\theta d \phi \ , 
\end{eqnarray}
where $\sigma_i$ are left-invariant 1-forms on $SU(2) \simeq S^3$ given in terms of the Euler angles $\theta, \phi, \psi$, where $\theta \in [0, \pi]$, $\phi \in [0, 2\pi]$, $\psi \in [0, 4\pi]$. 
In these coordinates, the metric on $\mathbb{R}^4$ becomes
\begin{equation}
ds^2(\mathbb{R}^4 ) = d\rho^2 + \frac{\rho^2}{4} \bigg( (\sigma^1 )^2 + (\sigma^2 )^2  + (\sigma^3 )^2  \bigg) \ . 
\end{equation}
The horizon is located at $\rho =0$. Then we perform the following coordinate transformation
\begin{eqnarray}
dt &=& du + \mathcal{Z}(\rho) d\rho \ , \\
d \psi &=& d \hat{\psi} + \mathcal{W}(\rho) d\rho \ , 
\end{eqnarray}
where $\mathcal{Z}$ and $\mathcal{W}$ are fixed by imposing that the terms $d\rho^2$ and $d\rho (d\hat{\psi} + \cos \theta d \phi )$ vanish. In this way, one obtains 
\begin{eqnarray}
\notag
\mathcal{Z}(\rho) &=& \frac{1}{\rho^3} \sqrt{(\rho^2 + \mu )^3 - j^2} \ , \\
\mathcal{W}(\rho) &=& \frac{2j}{\rho \sqrt{(\rho^2 + \mu )^3 - j^2}} \ . 
\end{eqnarray}
Then we set
\begin{equation}
r \equiv \rho^2 \ , 
\end{equation}
and finally we perform the transformation 
\begin{equation}
- \frac{r+\mu}{\sqrt{(r + \mu )^3 - j^2}} dr = 2 dR \ , 
\end{equation}
The metric becomes
\begin{eqnarray}
\label{BMPV_metric}
\notag
ds^2 &=& - \bigg( 1 + \frac{\mu}{R} \bigg)^{-2} du^2 
+ 2 du dR \\
\notag
&-& \frac{j R}{(R + \mu)^2}  du (d \hat{\psi} + \cos \theta d \phi )  
+ \frac{1}{4} (R + \mu) (d \theta^2 + \sin^2 \theta d\phi^2 )  \\
&+& \bigg[ \frac{1}{4} (R + \mu) - \frac{1}{4} \frac{j^2}{(R + \mu )^2} \bigg] (d \hat{\psi} + \cos \theta d \phi )^2  + .... \ ,  
\end{eqnarray}
where the ellipses are subleading terms in $R$. 

From (\ref{BMPV_metric}), we can read off the spatial cross section of the horizon 
\begin{equation}
ds^2_{\scriptsize\mbox{horizon}\normalsize} = \frac{\mu}{4} (d \theta^2 + \sin^2 \theta d\phi^2 )  
+ \bigg( \frac{\mu}{4} - \frac{j^2}{4 \mu^2} \bigg) (d \hat{\psi} + \cos \theta d \phi )^2 \ ,
\end{equation}
which describes the geometry of a squashed $S^3$. 

The metric (\ref{BMPV_metric}) in the near-horizon limit
\begin{equation}
u \rightarrow \frac{u}{\varepsilon} \ , \qquad
R \rightarrow \varepsilon R \ , \qquad
\varepsilon \rightarrow 0 \ , 
\end{equation}
becomes 
\begin{eqnarray}
\notag
ds^2 &=& - \frac{R^2}{\mu^2} du^2 + 2 du dR 
- \frac{j}{\mu^2} R du (d \hat{\psi} + \cos \theta d \phi ) \\
&-& \frac{\mu}{4} ( d\theta^2  + \sin^2 \theta d\phi^2 ) 
+ \bigg( \frac{\mu}{4} - \frac{j^2}{4 \mu^2} \bigg) 
(d \hat{\psi}^ + \cos \theta d \phi)^2 \ , 
\end{eqnarray}
which is known as the \emph{BMPV near-horizon geometry}.

\subsubsection{Supersymmetric black ring}
The theory considered is the bosonic sector of $D=5$ minimal supergravity, which is Einstein-Maxwell theory with a Chern-Simons term. The solution for the metric field is 
\begin{equation}
\label{Hawking_Gibbons}
ds^2 = - f^2 (dt + \omega )^2 + f^{-1} ds^2 (\mathbb{R}^4) \ , 
\end{equation} 
where $f$ and $\omega$ are a scalar and 1-form on $\mathbb{R}^4$ respectively. The metric on $\mathbb{R}^4$ is written as 
\begin{equation}
ds^2(\mathbb{R}^4) = \frac{R^2}{(x - y)^2} \bigg[  \frac{dy^2}{y^2 - 1} + (y^2 - 1) d\psi^2 + \frac{dx^2}{1 - x^2} + (1 - x^2) d\phi^2 \bigg] \ , 
\end{equation}
where $x \in [-1 , 1]$, $y\in ] -\infty , -1]$ and $\phi, \psi \in [0, 2\pi ]$. The scalar function $f$ is 
\begin{equation}
f^{-1} = 1 + \frac{Q - q^2}{2R^2} (x-y) - \frac{q^2}{4 R^2} (x^2 - y^2) \ , 
\end{equation}
and the non-vanishing components of the 1-form $\omega$ are
\begin{eqnarray}
\notag
\omega_{\phi} &=& - \frac{q}{8 R^2} (1 - x^2) \big[ 3 Q - q^2 (3 + x + y) \big] \ , \\
\omega_{\psi} &=& \frac{3}{2} q (1 + y) + \frac{q}{8 R^2} (1 - y^2) \big[ 3 Q - q^2 (3 + x + y) \big] \ , 
\end{eqnarray}
where $Q$ and $q$ are positive constants, which are proportional to the net charge and the local dipole charge of the ring, respectively. 

We shall introduce Gaussian null coordinates to describe the horizon, which is located at $y = -\infty$. First we define
\begin{equation}
r \equiv - \frac{R}{y} \ , 
\end{equation}
such that the horizon is located at $r = 0$. Then we perform the following coordinate transformation
\begin{eqnarray}
\notag
dt &=& du - B(r) dr \ , \\
\notag
d\phi &=& d\phi' - C(r) dr \ , \\
d\psi &=& d\psi' - C(r) dr \ , 
\end{eqnarray}
where
\begin{eqnarray}
\notag
B(r) &=& \frac{B_2}{r^2} + \frac{B_1}{r} + B_0 \ , \\
C(r) &=& \frac{C_1}{r} + C_0 \ , 
\end{eqnarray}
and the constants $\{ B_i, C_i \}$ are chosen such that the metric components are finite at $r=0$. They are given as follows
\begin{eqnarray}
\notag
B_2 &=& q^2 \frac{L}{4R} \ , \qquad\qquad
C_1 = - \frac{q}{2L} \ , \\
\notag
B_1 &=& \frac{Q + 2q^2}{4L} + L \frac{Q - q^2}{3 R^2} \ , \\
\notag
C_0 &=& - \frac{(Q - q^2)^3}{8 q^3 R L^3} \ , \\
B_0 &=& \frac{q^2 L}{8 R^3} + \frac{2 L}{3 R} - \frac{R}{2 L} + \frac{3 R^3}{2 L^3} + \frac{3(Q - q^2 )^3}{16 q^2 R L^3} \ , 
\end{eqnarray}
where 
\begin{equation}
L \equiv \sqrt{\frac{3 (Q - q^2)^2}{4 q^2} - 3R^2} \ . 
\end{equation}
In these coordinates, the metric becomes
\begin{eqnarray}
\label{black_ring_Gauss}
\notag
ds^2 &=& - \frac{16 r^4}{q^4} du^2 + \frac{2R}{L} du dr 
+ \frac{4 r^3 \sin^2 \theta}{R q} du d\phi' 
+ \frac{4R}{q} r du d\psi' 
+ \frac{3qr \sin^2 \theta}{L} dr d\phi' \\
\notag
&+& 2 \bigg[ \frac{qL}{2R} \cos \theta + \frac{3qR}{2L} 
+ \frac{(Q - q^2)(3R^2 - 2L^2)}{3qRL} \bigg] drd\psi' \\
&+& L^2 d\psi'^2 + \frac{q^2}{4} \bigg[ d\theta^2 + \sin^2 \theta (d\phi' - d\psi' )^2 \bigg] + .... \ , 
\end{eqnarray}
where we set $x = \cos \theta$, and the ellipsis are subleading terms which contains integer powers of $r$. 

The geometry of the \emph{event horizon} spatial cross section can be read off from (\ref{black_ring_Gauss}), which is 
\begin{equation}
ds^2_{\scriptsize\mbox{horizon}\normalsize} = L^2 d \psi'^2  + \frac{q^2}{4} (d\theta^2 + \sin^2\theta d\chi^2 ) \ , 
\end{equation}
where $\chi \equiv \phi' - \psi' = \phi - \psi$. This metric describes the geometry of $S^1 \times S^2$. 

Next, we rescale the radial coordinate as
\begin{equation}
r \equiv \frac{L}{R} \tilde{r} \ , 
\end{equation}
and we take the near-horizon limit
\begin{equation}
\label{nearhor_limit_black_ring}
u \rightarrow \frac{u}{\varepsilon} \ , \qquad
\tilde{r} \rightarrow \varepsilon \tilde{r} \ ,  \qquad
\varepsilon \rightarrow 0 \ , 
\end{equation}
where the remaining spatial coordinates $\theta, \psi', \phi'$ are unchanged. In this limit, the metric (\ref{black_ring_Gauss}) becomes
\begin{equation}
ds^2 = 2 du d \tilde{r} + \frac{4L}{q} \tilde{r} du d \psi' + L^2 d\psi'^2 + \frac{q^2}{4} (d\theta^2 + \sin^2 \theta d\chi^2 ) \ .
\end{equation}
which describes the geometry of $AdS_3 \times S^2$. 
We mention that this is also the near-horizon geometry of the black string \cite{Chamseddine:1999qs}. Therefore we expect that at certain point the bulk extension of the near-horizon geometry $AdS_3 \times S^2$ shows a bifurcation. 

The BMPV black hole and the supersymmetric black ring are half-maximal supersymmetric in the bulk (4 out of 8 preserved real supersymmetries), but their near-horizon geometries are maximally supersymmetric (8 out of 8 preserved real supersymmetries). This is due to the mechanism of \emph{supersymmetry enhancement} which occurs on the horizon. 
Moreover, it is known that all $D=4$ and $D=5$ supergravity black holes undergo supersymmetry enhancement in the near-horizon limit 	\cite{Kallosh:1992ta, Ferrara:1996dd}. Also, the supersymmetric asymptotically $AdS_5$ black hole found in \cite{Gutowski:2004ez} undergoes a supersymmetry enhancement from $N=2$ to $N=4$ (half-maximal supersymmetric) in the near-horizon limit.

\section{The horizon conjecture}
\label{sec:hor_conj}
We are interested in supersymmetric near-horizon geometries, i.e. which admit \emph{at least} one Killing spinor, well defined on the Killing horizon $\cH$. 
Supersymmetric near-horizon geometries which admit one Killing spinor generically experience an \emph{enhancement of supersymmetry}, which means that there exists a second Killing spinor, linearly independent from the first one. This feature has been demonstrated in a case by case basis
\cite{hethor, lichner11, lichneriib, lichneriia1, lichneriia2, Grover:2013ima, Gutowski:2016gkg}, and it implies a \emph{symmetry enhancement} of the isometry algebra to (at least) the $\mathfrak{sl}(2, \mathbb{R})$ algebra. This enhanced symmetry is dynamical since in the proof the equations of motion are required.
This leads to 
\begin{cnj}[Horizon conjecture]
\label{hor_conj}
A near-horizon geometry in supergravity theories has the following properties 
\begin{enumerate}
\item The number of preserved supersymmetries $N$ is always
\begin{equation}
N = 2 N_+ - {\rm Index}(\mathcal{D}_{\lambda})  \ , 
\end{equation}
or equivalently
\begin{equation}
N = 2 N_- + {\rm Index}(\mathcal{D}_{\lambda})  \ , 
\end{equation}
where $N_{\pm} \in \mathbb{N}$ is non-vanishing\footnote{This is the  number of positive or negative light-cone chirality spinors that one assumes for the supersymmetric near-horizon geometry to preserve initially, and we shall define it later.}, and $\mathcal{D}_{\lambda}$ is the Dirac operator twisted by the fluxes on the spatial cross section $\cS$, which depends on the supergravity theory considered. 
\item when the fluxes are not trivial and $N_{\pm} \neq 0$, the near-horizon geometry admits a $\mathfrak{sl}(2, \mathbb{R})$ isometry algebra, which is also a symmetry of the fluxes.  
\end{enumerate}
\end{cnj}
The first part of the horizon conjecture is a direct consequence, together with an index theory argument, of the fact that generalised Lichnerowicz theorems can be constructed for near-horizon geometries.   
In what follows, we shall review the key ideas necessary for the proof of the horizon conjecture in the context of type IIA supergravity \cite{lichneriia1}.  

The bosonic field content of IIA supergravity is the spacetime metric $g$, the real scalar dilaton field $\Phi$, the 2-form NS-NS gauge potential $B$, the 1- and 3-forms RR gauge potentials $A$ and $C$ respectively.
In addition, there are non-chiral fermionic fields, which are the gravitino $\psi$ and dilatino $\lambda$. 
We are interested in supersymmetric vacuum solutions, which means that $\psi = \lambda = 0$, and their supersymmetric variations\footnote{We remind that the expectation value of \emph{any} fermion in \emph{any} vacuum is zero. Here we give the reasoning. 
Consider a generic Lagrangian theory of bosons and fermions. In the quantum description, consider a generic vacuum state $|0\rangle$. If the vacuum $|0\rangle$ has fermionic number $\alpha$, then the state $\psi |0\rangle$, where $\psi$ is a fermionic field, has fermionic number $\alpha +1$. 
Since the fermionic number is conserved in any process (because the Lagrangian is a boson), then the following object
\begin{equation}
\notag
\langle 0| \psi |0\rangle = 0 \ , 
\end{equation} 
which is vanishing for the superselection rule of conservation of fermionic number. This proves the initial statement. 

At the supergravity level, all fields are classical, therefore the statement above implies that all fermionic fields are set to zero.
(We gratefully thank Alessandro Torrielli for a useful discussion over this point). 
} 
\begin{equation}
\label{KSE_intro}
\delta_{\epsilon} \psi = 0 \ , \qquad\qquad \delta_{\epsilon} \lambda = 0 \ , 
\end{equation}
admit a solution for the spinor $\epsilon$. The equations in (\ref{KSE_intro}) are the so-called \emph{Killing Spinor Equations} (KSEs), and the solution $\epsilon$ is called \emph{Killing spinor}. 

The field strengths of the gauge potentials are 
\begin{equation}
F = dA \ , \qquad H = dB \ , \qquad G = dC - H \w A \ ,  
\end{equation}
with associated Bianchi identities 
\begin{equation}
dF= 0 \ , \qquad dH = 0 \ . \qquad dG = F \w H \ . 
\end{equation}
We do not list here the equations of motion for the bosonic fields, which can be found in \cite{lichneriia1}. 

We consider a Killing horizon in the near-horizon limit, therefore the metric is given in (\ref{NHG_frame}). 
Furthermore we assume that the IIA fields are well-defined in the near-horizon limit, and the isometry $V = \partial_u$ is a symmetry of the full solution, i.e.
\begin{equation}
\cL_{\partial_u} \Phi = \cL_{\partial_u} H = \cL_{\partial_u} F = \cL_{\partial_u} G = 0 \ . 
\end{equation}
The IIA fields decompose in the near-horizon limit as
\begin{equation}
\Phi = \Phi (y) \ , 
\end{equation}
and 
\begin{equation}
H = \le^+ \w \le^- \w L + r \le^+ \w M + \tilde{H} \ , 
\end{equation}
and
\begin{equation}
F = S \le^+\w \le^-  + r \le^+ \w T + \tilde{F} \ , 
\end{equation}
and
\begin{equation}
G = \le^+ \w \le^- \w X + r \le^+ \w Y + \tilde{G} \ ,  
\end{equation}
where $S$ is a scalar function, $T, L$ are 1-forms , $\tilde{F}, M, X$ are 2-forms, $\tilde{H}, Y$ are 3-forms and $\tilde{G}$ is a 4-form, which are all defined on $\cS$ and $u, r$-independent. 
We assume that all these fields components, including the near-horizon data $\{ \D, h, \gamma \}$, are at least $\mathscr{C}^2$ differentiable. 

At this stage the $u,r$-dependence of all bosonic fields is known. 
Therefore the bosonic field equations and Bianchi identities decompose into a set of equations which only involves $\{ \D, h, \gamma, \Phi, L, M, \tilde{H}, S, T, \tilde{F}, X, Y, \tilde{G}\}$. 
We do not write here these equations, which can be found in \cite{lichneriia1}. 

The Killing spinor equations (\ref{KSE_intro}) for type IIA supergravity are:
\begin{eqnarray}
\notag
\label{KSE:gravitino}
\mathcal{D}_{\mu} \epsilon &\equiv & \nabla_{\mu} \epsilon +  \bigg( \frac{1}{8} H_{\mu\la_1\la_2} \G^{\la_1\la_2} \G_{11} + \frac{1}{16} e^{\Phi} F_{\la_1\la_2}\G^{\la_1\la_2} \G_{\mu} \G_{11} \\
&& \qquad\qquad\qquad + \frac{1}{8 \cdot 4!} e^{\Phi} G_{\la_1\la_2\la_3\la_4} \G^{\la_1\la_2\la_3\la_4} \G_{\mu} \bigg) \epsilon = 0 \ , \\
\notag
\label{KSE:algebraic}
\cA \epsilon & \equiv & \bigg( \nabla_{\mu} \Phi \G^{\mu} + \frac{1}{12} H_{\mu_1\mu_2\mu_3}\G^{\mu_1\mu_2\mu_3} G_{11} + \frac{3}{8} e^{\Phi} F_{\mu_1\mu_2} \G^{\mu_1\mu_2} \G_{11} \\
&& \qquad\qquad\qquad + \frac{1}{4 \cdot 4!} e^{\Phi} G_{\mu_1\mu_2\mu_3\mu_4\mu_5} \G^{\mu_1\mu_2\mu_3\mu_4\mu_5} \bigg) \epsilon = 0 \ , 
\end{eqnarray}
where $\epsilon$ is a Majorana, but not Weyl, spinor of $Spin(1, 9)$. 
We assume that the near-horizon geometry is supersymmetric, i.e. there exists at least one $\epsilon \neq 0$ solution of (\ref{KSE:gravitino}) and (\ref{KSE:algebraic}). 

It is a standard approach to supersymmetric solutions to assume the KSEs and to use the integrability conditions to imply the bosonic field equations and Bianchi identities. This allows one to analyse the KSEs only, which are first order PDEs instead of the bosonic field equations, which are second order PDEs. 
Our approach is in the other way around: by assuming the bosonic field equations and Bianchi identities we reduce the KSEs to a minimal set of necessary and sufficient conditions for the spinor $\epsilon$. This approach allows us to identify the minimal set of equations in order to establish supersymmetry, which will be useful later when considering the supersymmetry enhancement mechanism. 

The gravitino KSE (\ref{KSE:gravitino}) can be integrated along the $\le^+, \le^-$ directions, since after taking the near-horizon limit the $u, r$-dependence of all bosonic fields is known.
To do this, it is useful first to split the spinor $\epsilon$ into \emph{light-cone chiralities}\footnote{This notion of chirality is unrelated to the one induced by $\Gamma_{11}$. }. The projectors associated with such splitting are
\begin{equation}
P_{\pm} \equiv \frac{\mathds{1} \pm \G_{+-}}{2} \ , \qquad
P_{\pm}^2 = P_{\pm} \ , \qquad P_{\pm}\cdot  P_{\mp} = 0 \ ,
\qquad \epsilon_{\pm} \equiv P_{\pm} \epsilon\ ,
\end{equation}
where $\epsilon_+, \epsilon_-$ are the positive and negative light-cone chiralities spinors respectively.
Therefore we have that
\begin{equation}
\label{epsilon+-}
\epsilon = \epsilon_+ + \epsilon_- \ , 
\end{equation} 
where $\epsilon_{\pm}$ satisfies 
\begin{equation}
\G_{+-} \epsilon_{\pm} = \pm \epsilon_{\pm} \ , \qquad\qquad
\G_{\pm} \epsilon_{\pm} = 0 \ .
\end{equation}
Integrating first the $\mu = -$ and then the $\mu= +$ components of (\ref{KSE:gravitino}), one obtains
\begin{equation}
\label{epsilon+-_1}
\epsilon_+ = \phi_+ \ , \qquad \epsilon_- = \phi_- + r \G_- \Theta_+ \phi_+ \ , 
\end{equation}
where $\partial_r \phi_{\pm} = 0$, and
\begin{equation}
\label{epsilon+-_2}
\phi_- = \eta_- \ , \qquad \phi_+ = \eta_+ + u \G_+ \Theta_- \eta_- \ , 
\end{equation}
where $\partial_r \eta_{\pm} = \partial_u \eta_{\pm} = 0$, and
\begin{eqnarray}
\notag
\Theta_{\pm} &=& \frac{1}{4} h_i \G^i \mp \frac{1}{4} \G-{11} L_i \G^i - \frac{1}{16} e^{\Phi} \G_{11} (\pm 2S + \tilde{F}_{ij} \G^{ij} ) \\ 
&-& \frac{1}{8 \cdot 4!} e^{\Phi} ( \pm 12 X_{ij} \G^{ij} + \tilde{G}_{\ell_1\ell_2\ell_3\ell_4}\G^{\ell_1\ell_2\ell_3\ell_4}) \ .
\end{eqnarray}
The \emph{reduced spinors} $\eta_{\pm}$ arise as constants of integrations, and they depend only on $y$. The fact that the spinor (\ref{epsilon+-}) must satisfy the KSEs (\ref{KSE:gravitino}) and (\ref{KSE:algebraic}) implies that the reduced spinors $\eta_{\pm}$ must satisfy a set of equations, the so-called \emph{reduced} KSEs, which can be found in \cite{lichneriia1}. 

Assuming the bosonic field equations and the Bianchi identities, one can show that some of the reduced KSEs are implied by the others. In \cite{lichneriia1} the authors show that the minimal set of necessary and sufficient reduced KSEs is the following
\begin{equation}
\label{KSE_minimal}
\tilde{\nabla}^{(\pm)}_i \eta_{\pm} = 0 \ , \qquad\qquad
\cA^{(\pm)} \eta_{\pm} = 0 \ , 
\end{equation}
where 
\begin{equation}
\tilde{\nabla}^{(\pm)}_i = \tilde{\nabla}_i + \Psi^{(\pm)}_i \ , 
\end{equation}
with 
\begin{eqnarray}
\notag
\Psi^{(\pm)}_i &=& \bigg( \mp \frac{1}{4} h_i \mp \frac{1}{16} e^{\Phi} X_{\ell_1\ell_2} \G^{\ell_1\ell_2} \G_i + \frac{1}{8 \cdot 4!} e^{\Phi} \tilde{G}_{\ell_1\ell_2\ell_3\ell_4} \G^{\ell_1\ell_2\ell_3\ell_4} \G_i \bigg) \\
&+& \G_{11} \bigg( \mp \frac{1}{4} L_i + \frac{1}{8} \tilde{H}_{i\ell_1\ell_2} \G^{\ell_1\ell_2} \pm \frac{1}{8} e^{\Phi} S \G_i - \frac{1}{16} e^{\Phi} \tilde{F}_{\ell_1\ell_2} \G^{\ell_1\ell_2} \G_i \bigg) \ , 
\end{eqnarray}
and
\begin{eqnarray}
\notag
\cA^{(\pm)} &=& \tilde{\nabla}_i\Phi \G^i + \bigg( \mp \frac{1}{8} e^{\Phi} X_{\ell_1\ell_2} \G^{\ell_1\ell_2} + \frac{1}{4\cdot 4!} e^{\Phi} \tilde{G}_{\ell_1\ell_2\ell_3\ell_4} \G^{\ell_1\ell_2\ell_3\ell_4} \bigg) \\
&+& \G_{11} \bigg( \pm \frac{1}{2} L_i \G^i - \frac{1}{12} \tilde{H}_{\ell_1\ell_2\ell_3} \G^{\ell_1\ell_2\ell_3} \mp \frac{3}{4} e^{\Phi} S + \frac{3}{8} e^{\Phi} \tilde{F}_{ij} \G^{ij} \bigg) \ . 
\end{eqnarray}
Therefore the necessary and sufficient reduced KSEs (\ref{KSE_minimal}) turns out to be the naive restriction of the spacetime KSEs (\ref{KSE:gravitino}) and (\ref{KSE:algebraic}) to the spatial cross section $\cS$. 
The conditions for the near-horizon geometry to be supersymmetric reduces to the existence of at least one non-vanishing $\eta_+$ or $\eta_-$ which satisfies (\ref{KSE_minimal}), and the spacetime spinor $\epsilon$ is then given as a function $\epsilon = \epsilon (u, r, \eta_{\pm})$ via (\ref{epsilon+-}), (\ref{epsilon+-_1}) and (\ref{epsilon+-_2}).

\subsection{Generalised Lichnerowicz Theorems and Index Theory}
\label{sec:gen_lich_th}
In this section we shall prove the first part of the conjecture \ref{hor_conj}. First of all, we shall recall the Lichnerowicz Theorem. In what follows $\nabla$ is the Levi-Civita connection, $\G^i \nabla_i$ is the associated Dirac operator, $\langle \cdot , \cdot \rangle$ is the Dirac inner product and $R$ is the Ricci scalar. 
\begin{theorem}[Lichnerowicz Theorem]
On any compact without boundary spin manifold $\mathcal{M}$:
\begin{equation}
\label{identity_spin}
\int_{\mathcal{M}} \langle \G^i \nabla_i \epsilon, \G^j \nabla_j \epsilon \rangle = \int_{\mathcal{M}}  \langle \nabla_i \epsilon, \nabla^i \epsilon \rangle + \int_{\mathcal{M}} \frac{R}{4} \langle\epsilon , \epsilon \rangle \ , 
\end{equation}
Then:
\begin{itemize}
\item If $R>0$, there are no zero modes for the Dirac operator.
\item If $R=0$, the zero modes of the Dirac operator are in a one-to-one correspondence with spinors parallel with respect to $\nabla$.  
\end{itemize}
\end{theorem}
\begin{proof}
Consider the identity 
\begin{equation}
\label{identity2}
\int_{\mathcal{M}} \langle \G^i \nabla_i \epsilon, \G^j \nabla_j \epsilon \rangle =  \int_{\mathcal{M}} \langle \nabla_i \epsilon , \nabla^i \epsilon \rangle - \int_{\mathcal{M}} \langle \Gamma^{ij} \nabla_i \nabla_j \epsilon , \epsilon \rangle \ , 
\end{equation}
where we used the fact that $(\Gamma^i)^{\dagger} = \Gamma^i$, and $\Gamma^i \Gamma^j = \Gamma^{ij} + \delta^{ij}$. On the RHS of (\ref{identity2}), there is also a total divergence term which disappeared because $\mathcal{M}$ has no boundary. By using the following identity
\begin{equation}
[ \nabla_i , \nabla_j ] \epsilon = - \frac{1}{4} R_{ij, mn} \Gamma^{mn} \epsilon \ , 
\end{equation}
we find that 
\begin{equation}
\label{identity3}
\Gamma^{ij} \nabla_i \nabla_j \epsilon = - \frac{1}{8} R_{ij, mn} \Gamma^{ijmn} \epsilon 
- \frac{1}{2} R_{ij} \Gamma^{ij} \epsilon 
- \frac{1}{4} R \epsilon \ . 
\end{equation} 
The first two terms on the RHS of (\ref{identity3}) vanish because of the Bianchi identity and the symmetry property of the Ricci tensor. This proves (\ref{identity_spin}). 

Suppose that $R >0$. Since $\langle \nabla_i \epsilon, \nabla^i \epsilon \rangle$ is semi-positive definite, the RHS of (\ref{identity_spin}) is always strictly positive. This implies that $\langle \G^i \nabla_i \epsilon, \G^j \nabla_j \epsilon \rangle$ can never vanish, and therefore there are no zero modes for the Dirac operator $\Gamma^i \nabla_i$.

Suppose that $R = 0$ and $\nabla \epsilon = 0$. Then   (\ref{identity_spin}) implies that $\langle \G^i \nabla_i \epsilon, \G^j \nabla_j \epsilon \rangle$ must also vanish, and therefore $\Gamma^i \nabla_i \epsilon$ vanishes, i.e. $\epsilon$ is a zero mode of the Dirac operator $\Gamma^i \nabla_i$. 

Suppose that $R = 0$ and $\Gamma^i \nabla_i \epsilon = 0$. Then   (\ref{identity_spin}) implies that $\langle \nabla_i \epsilon, \nabla^i \epsilon \rangle$ vanishes, and therefore $\nabla \epsilon = 0$. This concludes the proof. 
\end{proof}

We are interested in generalising the Lichnerowicz Theorem to Killing spinors in type IIA supergravity, and identifying them with the zero modes of certain Dirac operators. These type of theorems have been proved in various theories \cite{lichner11, lichneriib, lichneriia1, lichneriia2, Grover:2013ima, Gutowski:2016gkg}. 

For this purpose, we introduce the \emph{modified connection} $\hat{\nabla}^{(\pm)}$: 
\begin{equation}
\hat{\nabla}^{(\pm)}_i \equiv \tilde{\nabla}^{(\pm)}_i + \kappa \G_i \cA^{(\pm)} \ ,  
\end{equation}
and the \emph{modified Dirac operator} $\mathscr{D}^{(\pm)}$:
\begin{equation}
\mathscr{D}^{(\pm)} \equiv \G^i \tilde{\nabla}^{(\pm)}_i - \cA^{(\pm)} \ , 
\end{equation}
where $\kappa \in \mathbb{R}$. Then in \cite{lichneriia1} the authors show
\begin{theorem}[Generalised Lichnerowicz Theorem for IIA horizons]
Killing spinors are in a one-to-one correspondence with zero modes of the modified Dirac operator, i.e. 
\begin{equation}
\tilde{\nabla}^{(\pm)} \eta_{\pm} = 0 \ , \qquad \cA^{(\pm)} \eta_{\pm} = 0\quad \Longleftrightarrow\quad \mathscr{D}^{(\pm)} \eta_{\pm} = 0 \ , 
\end{equation}
provided that $-\frac{1}{4} < \kappa < 0$. 
\end{theorem}
\begin{proof}
Suppose that $\eta_{\pm}$ is a Killing spinor, i.e. a solution of (\ref{KSE_minimal}). Then $\eta_{\pm}$ is straightforwardly also a solution of the Dirac equation $\mathscr{D}^{(\pm)} \eta_{\pm} = 0$. 

Suppose that $\mathscr{D}^{(\pm)} \eta_{\pm} = 0$, i.e. $\eta_{\pm}$ is a solution of the Dirac equation. Then one can compute the Laplacian acting on $\lVert \eta_{\pm} \rVert^2 \equiv \langle \eta_{\pm} , \eta_{\pm} \rangle$ and use the bosonic field equations and Bianchi identities to simplify it. For the case involving $\eta_+$, after using $\mathscr{D}^{(+)} \eta_{+} = 0$ one can establish the equality
\begin{equation}
\label{Lich_IIA}
\tilde{\nabla}^i \tilde{\nabla}_i \lVert \eta_+ \rVert^2 
- (2 \tilde{\nabla}^i \Phi + h^i) \tilde{\nabla}_i \lVert \eta_+ \rVert^2 = 2 \lVert \hat{\nabla}^{(+)} \eta_+ \rVert^2
+ (- 4 \kappa - 16 \kappa^2 ) \lVert \cA^{(+)} \eta_+ \rVert^2 \ , 
\end{equation}
If we impose that $\kappa$ lies in the interval $-\frac{1}{4} < \kappa < 0$, then the RHS of (\ref{Lich_IIA}) consists of a sum of two semi-positive quantities, which is never negative.  The Hopf maximum principle \cite{maxp} can be applied, since we assumed that all fields are smooth, and $\cS$ compact. This implies that $\lVert \eta_+ \rVert^2 = const.$, which in turns implies that $\eta_+$ satisfies \emph{both} gravitino and algebraic KSEs in (\ref{KSE_minimal}), i.e. $\eta_+$ is a Killing spinor. For the case involving $\eta_-$, after imposing $\mathscr{D}^{(-)} \eta_{-} = 0$ one still finds that $\eta_-$ is a Killing spinor, though not the condition $\lVert \eta_- \rVert^2 = const.$. The details can be found in \cite{lichneriia1}.
This ends the proof of the generalised Lichnerowicz theorem for IIA horizons. 
\end{proof}

\vspace{5mm}
We remark that an alternative way to prove the generalised Lichenrowicz theorem is by considering the following functional
\begin{equation}
\cI^{(\pm)} \equiv \int_{\cS} e^{c\Phi} \bigg( \langle \hat{\nabla}^{(\pm)}_i \eta_{\pm}, \hat{\nabla}^{(\pm)\, i} \eta_{\pm} \rangle 
- \langle \mathscr{D}^{(\pm)} \eta_{\pm} , \mathscr{D}^{(\pm)} \eta_{\pm} \rangle \bigg) \ , 
\end{equation}
where $c\in \mathbb{R}$. One can use the bosonic field equations and Bianchi identities to simplify $\cI^{(\pm)}$ and obtain an equality where under the restrictions on $\kappa$ made above, and the fact that $\mathscr{D}^{(\pm)} \eta_{\pm} = 0$, one implies that $\eta_{\pm}$ is also a Killing spinor.

\vspace{5mm}
Let $N$ be the total number of preserved supersymmetries. One can assume that the solution preserves a certain number of linearly independent $\eta_+$ and $\eta_-$ spinors, which we denote by $N_+$ and $N_-$ respectively. Then 
\begin{equation}
\label{N_N_+_N_-}
N = N_+ + N_- \ . 
\end{equation}
A corollary of the generalised Lichnerowicz theorem is that
\begin{equation}
N_+ = {\rm dim\, Ker}  \mathscr{D}^{(+)} \ , \qquad
N_- = {\rm dim\, Ker}  \mathscr{D}^{(-)} \ .  
\end{equation}
It turns out that there is a relation between $\mathscr{D}^{(-)}$ and $\big( \mathscr{D}^{(+)}\big)^{\dagger}$, which is 
\begin{equation}
\big( e^{2\Phi} \G_- \big) \big(\mathscr{D}^{(+)}\big)^{\dagger}  =  \mathscr{D}^{(-)} \big( e^{2\Phi} \G_- \big)
\ , 
\end{equation}
which implies that
\begin{equation}
{\rm dim\, Ker}  \mathscr{D}^{(-)}  = {\rm dim\, Ker}  \big( \mathscr{D}^{(+)} \big)^{\dagger} \ . 
\end{equation}
The index of the Dirac operator $\mathscr{D}^{(+)}$ is \cite{atiyah1}
\begin{equation}
{\rm Index} (\mathscr{D}^{(+)}) = {\rm dim\, Ker}  \mathscr{D}^{(+)} - {\rm dim\, Ker}  \big( \mathscr{D}^{(+)}\big)^{\dagger} \ , 
\end{equation}
which allows us to rewrite (\ref{N_N_+_N_-}) as 
\begin{equation}
N= 2N_+ - {\rm Index} (\mathscr{D}^{(+)}) \ , 
\end{equation}
or equivalently 
\begin{equation}
N = 2 N_- + {\rm Index} (\mathscr{D}^{(+)}) \ . 
\end{equation}
This proves the first part of the horizon conjecture \ref{hor_conj}. 

Since type IIA supergravity is non-chiral, the spin bundle associated with $\cS$ splits into $S = S_+ \oplus S_-$, where $S_+$ and $S_-$ are isomorphic $Spin (8)$ bundles associated with the Majorana, but not Weyl, ${\bf 16}$ representation.  
The Dirac operator $\mathscr{D}^{(+)}$ is a map $\mathscr{D}^{(+)} : \G (S_{+}) \rightarrow \G (S_{+})$, where $\G (S_{+})$ is a generic section of $S_{+}$. 
Moreover the Dirac operator $\mathscr{D}^{(+)}$ and the standard Dirac operator $\G^i \tilde{\nabla}_i$ have the same principal symbol. Since the index of $\G^i \tilde{\nabla}_i$ acting on Majorana, but not Weyl, spinors vanishes \cite{atiyah1}, the index of $\mathscr{D}^{(+)}$ also vanishes. 
This implies that 
\begin{equation}
\label{evenSUSY}
N = 2 N_{\pm} \ , 
\end{equation}
and therefore the number of preserved supersymmetries is always \emph{even}. Hence if we assume that the supersymmetric near-horizon geometry has only one Killing spinor, i.e. $N=1$, then this turns out to be incorrect, because (\ref{evenSUSY}) tells us that $N=2$ at least, and therefore there must exist at least a second Killing spinor. This is called \emph{supersymmetry enhancement}, and in the next section we shall show how to construct the second Killing spinor.

\subsection{The map $\eta_- \rightarrow \eta'_+$}
\label{sec:susy_enh_intro}
In \cite{lichneriia1} the authors observe that if $\eta_-$ is a Killing spinor solution to the ``$-$" set of (\ref{KSE_minimal}), then 
\begin{equation}
\eta'_+ \equiv \G_+ \Theta_- \eta_- \ , 
\end{equation}
is also a solution to the ``$+$" set of (\ref{KSE_minimal}), and hence a new Killing spinor. 

Therefore $\G_+\Theta_-$ is the map that generates the supersymmetry enhancement. This is the case provided that
the new Killing spinor $\eta'_+$ is non-vanishing. However this does not happen because in \cite{lichneriia1} the authors show that
\begin{equation}
\label{trivial_kernel}
{\rm Ker}\  \Theta_- = \{ 0 \} \ .
\end{equation}  
We remark that the new Killing spinor $\eta'_+$ is always linearly independent from the parental $\eta_-$, since it is of the opposite chirality. 
Moreover, since $\G_+\Theta_-$ is a linear map, (\ref{trivial_kernel}) states that $\G_+\Theta_-$ is injective, and therefore in the case where we have several $\eta^{(k)}_-$, the generated Killing spinors $\eta'^{(k)}_+$ are also linearly independent between each other. 
This implies that the Killing spinors in the set $\{ \eta^{(k)}_{\pm}, \eta'^{(k)}_{\mp}\}$, with $k = 1, ... , n < 16$, are all linearly independent. 

We remark that if one would be able to prove that there exists a map also in the other way around, i.e. a given $\eta_+$ Killing spinor implies the existence of a second Killing spinor $\eta'_-$, then this would be enough to establish the supersymmetry enhancement, without the need of using index theory arguments together with generalised Lichenrowicz theorems.
This happens in the case of uncorrected heterotic near-horizon geometries \cite{hethor}, but it does not in the cases of type IIA, massive IIA, IIB and $D=11$ supergravities \cite{lichneriia1, lichneriia2, lichneriib, lichner11}. 

\subsection{The $\mathfrak{sl}(2, \mathbb{R})$ symmetry}
In this section we show that the supersymmetry enhancement previously discussed implies a \emph{symmetry enhancement}. 
We assume that there exists a non-vanishing Killing spinor $\eta_-$, from which one can generate a second new Killing spinor $\eta'_+ = \G_{+}\Theta_{-} \eta_-$. 

The spacetime spinors $\epsilon_1 = \epsilon(u, r, \eta_-)$ and $\epsilon_2 = \epsilon(u, r, \eta'_+)$ are 
\begin{eqnarray}
\epsilon_1 &=& \eta_- + u \G_+ \Theta_- \eta_- + ur \G_- \Theta_+ \G_+ \Theta_- \eta_- \ , \\
\epsilon_2 &=& \eta'_+ + r \G_- \Theta_+ \eta'_+ \ . 
\end{eqnarray}
Given two generic Killing spinors $\zeta_1, \zeta_2$, one can construct the 1-form bilinear:
\begin{equation}
K(\zeta_1, \zeta_2) = \langle (\G_+ - \G_-) \zeta_1,  \G_A \zeta_1 \rangle \, \le^A \ , 
\end{equation}
where $\le^A = \{ \le^+, \le^-, \le^i\}$, such that $K$ is a a symmetry of all bosonic fields of IIA supergravity. 

In \cite{lichneriia1} the authors consider the 1-form bilinears associated with $\epsilon_1, \epsilon_2$ given by $K_1(\epsilon_1, \epsilon_2)$, $K_2(\epsilon_2, \epsilon_2)$, $K_3(\epsilon_1, \epsilon_1)$, and show that their dual vector fields are\footnote{We use the same symbol for the 1-forms and their dual vector fields.} 
\begin{eqnarray}
\label{vector_fields}
\notag
K_1 &=& - 2 u \lVert \eta_+ \rVert^2 \partial_u + 2 r \lVert \eta_+ \rVert^2 \partial_r + V^i \tilde{\partial}_i \ , \\
\notag
K_2 &=& - 2 \lVert \eta_+ \rVert^2 \partial_u \ , \\
K_3 &=& - 2 u^2 \lVert \eta_+ \rVert^2 \partial_u + (2 \lVert \eta_- \rVert^2 + 4 ru \lVert \eta_+ \rVert^2)\partial_r + 2 u V^i \tilde{\partial}_i \ , 
\end{eqnarray}
where
\begin{equation}
V_i \equiv \langle \G_+ \eta_- , \G_i \eta_+ \rangle \ .  
\end{equation}
The vector fields (\ref{vector_fields}) satisfy the following commutation relations
\begin{equation}
[K_1, K_2 ] = 2 \lVert \eta_+ \rVert^2 K_2 \ , \qquad
[K_2, K_3 ] = - 4 \lVert \eta_+ \rVert^2 K_1 \ , \qquad
[K_3, K_1 ] = 2 \lVert \eta_+ \rVert^2 K_3 \ ,  
\end{equation}
which are the commutation relations defining the $\mathfrak{sl}(2, \mathbb{R})$ algebra. This ends the proof of the second part of the horizon conjecture \ref{hor_conj}.

\begin{figure}[H]
\begin{center}
\hspace{-22 mm}
 \includegraphics[scale=0.50]{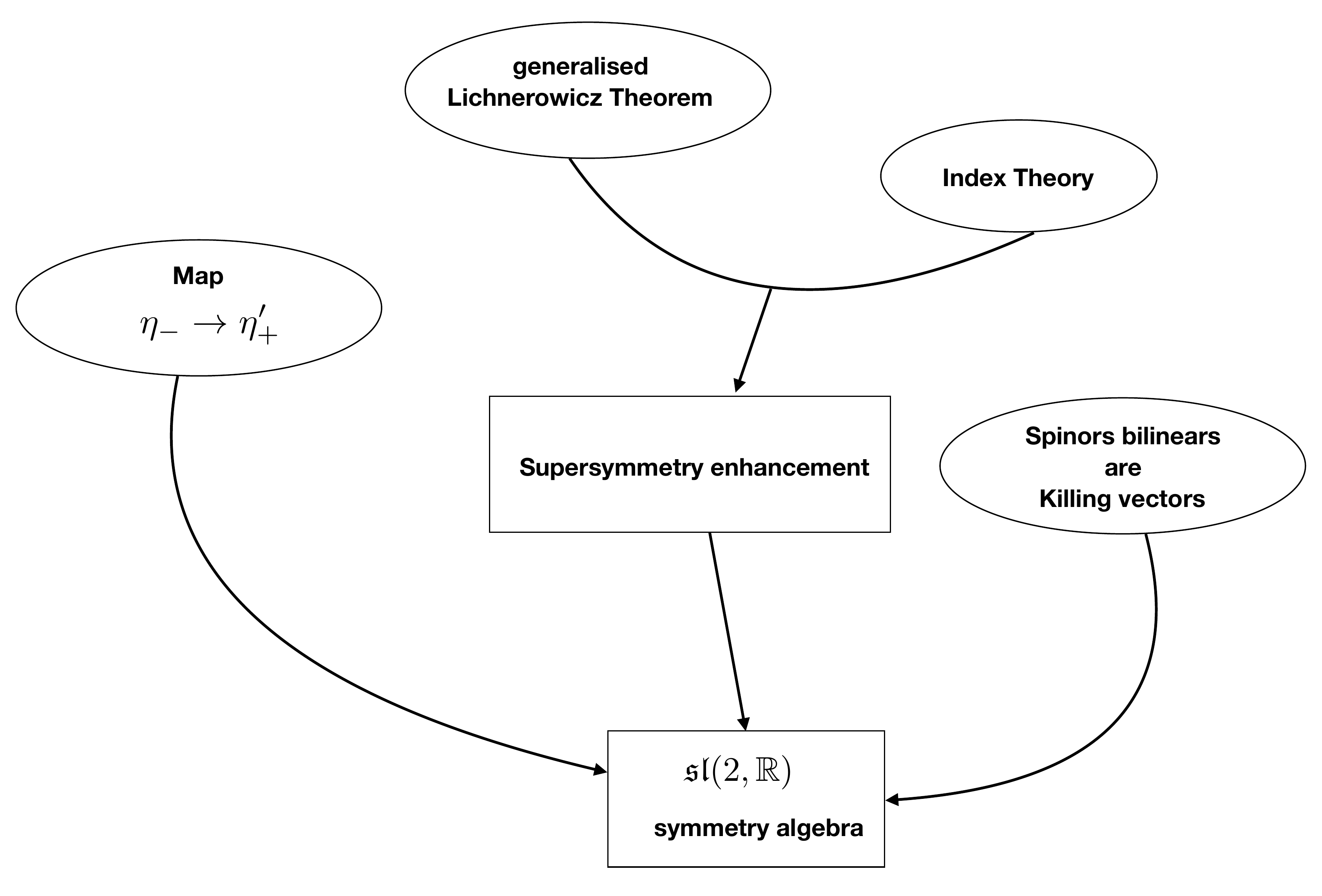}
\end{center}
\caption{The horizon conjecture.}
\end{figure}

\clearpage{\pagestyle{empty}\cleardoublepage} 

\chapter{\textbf{String corrected near-horizon geometries}}
\label{ch:het_NHG}

In this chapter we shall investigate the horizon conjecture \ref{hor_conj} beyond the supergravity approximation, by including the string corrections.
The first systematic classification of supersymmetric near-horizon geometries in a higher derivative theory \cite{Hanaki:2006pj}, was done in \cite{Gutowski:2011nk}, where the only assumption made was that the solution should preserve the minimal amount of supersymmetry. At the uncorrected level, the theory reduces to ungauged $D=5$ supergravity. In this limit, it is known that every near-horizon geometry is maximally supersymmetric with constant scalars \cite{Gutowski:2004bj}, which is consistent with the standard picture of the attractor mechanism. 
However, when higher corrections are considered, the list of near-horizon geometries fond in \cite{Gutowski:2011nk} includes the maximally supersymmetric solutions (which were classified in \cite{Castro:2008ne}), but also a class of solutions which are half-maximal supersymmetric. Although it is unclear if these solutions can be extended to a full black hole solution, this is a first evidence that string corrections can alter the mechanism of supersymmetry enhancement which generically occurs for near-horizon geometries. 
The five dimensional theory considered is not the most general, which suggests to further investigate this issue in a ten dimensional theory. Among the various possibilities, the heterotic theory is the most promising choice for at least two reasons. First, in the limit $\alpha' \rightarrow 0$, much more is known about general supersymmetric solutions and supersymmetric near-horizon geometries of heterotic supergravity, when compared to the type II theories. In particular, as a consequence of the spinoral classification techniques developed in \cite{class1, class2}, all supersymmetric near-horizon geometries have been classified \cite{hethor}. The second reason is that $\alpha'$ corrections in heterotic theory are significantly simpler compared to the other theories, and also better understood in terms of geometric quantities, e.g. the $\alpha'$ corrections to the Bianchi identity for the 3-form flux $H$ are described in terms of Pontryagin forms. 

Our conventions for the $\alpha'$ corrected heterotic theory are consistent with those of \cite{hetpap}, except with relabelling $\nabla^{(+)}$ with $\nabla^{(-)}$, and vice versa. 
 
\section{Heterotic near-horizon geometries}
The heterotic field content is the metric $g$, the real 3-form $H$, the real scalar dilaton field $\Phi$, and the non-abelian gauge potential $A$.
We assume that the bosonic fields and Killing spinors admit a perturbative Taylor series expansion in $\alpha'$, i.e. 
\begin{equation}
\xi = \xi^{[0]} + \alpha' \xi^{[1]} + \mathcal{O}(\alpha'^2) \ ,
\end{equation}
where $\xi$ is a generic field\footnote{This means that the perturbed solutions is sufficiently close to the unperturbed one, and therefore a set of Gaussian null coordinates can still be adapted on a neighbourhood of the horizon.}. We shall consider only terms which are first order in $\alpha'$. 

We assume that the spacetime contains a Killing horizon, to which we adapt Gaussian null coordinates and take the near-horizon limit, as explained in section \ref{sec:NHG}. The metric $g$ is then given in (\ref{NHG_frame}). 

We assume that all heterotic fields admit a well-defined near-horizon limit, and that $V = \partial_u$ is a symmetry of the full solution:
\begin{eqnarray}
{\cal{L}}_{\partial_u} \Phi=0 \ , \qquad {\cal{L}}_{\partial_u} H = 0 \ , \qquad {\cal{L}}_{\partial_u}A=0 \ .
\end{eqnarray}
The bosonic fields decompose in Gaussian null coordinates and in the near-horizon limit as
\begin{equation}
\Phi=\Phi(y) \ , 
\end{equation}
and
\begin{eqnarray}
\label{threef}
H = \mathbf{e}^+ \wedge \mathbf{e}^- \wedge N+r \mathbf{e}^+ \wedge Y+W \ ,
\end{eqnarray}
and \begin{eqnarray}
A= r {\cal{P}} \mathbf{e}^+ + {\cal{B}} \ ,
\end{eqnarray}
where $N$, $Y$ and $W$ are $u,r$-independent 1-, 2- and 3-forms on ${\cal{S}}$ respectively, while ${\cal{P}}$ and ${\cal{B}}$ are a $u,r$-independent $G$-valued\footnote{Where $G = SO(32)$ or $E_8 \times E_8$.} scalar and 1-form on ${\cal{S}}$ respectively.

The Bianchi identity for $H$ includes $\alpha'$ corrections coming from the anomaly cancellation mechanism, and is given in (\ref{bian}). 
The non-abelian 2-form field strength $F$ is given by
\begin{eqnarray}
F = dA + A \wedge A\ .
\end{eqnarray}
with associated Bianchi identity
\begin{equation}
dF + A \w F - F \w A = 0  \ .
\end{equation}

Since in our investigation we keep $\alpha'$ corrections, we treat the 3-form flux $H$ as a fundamental field, which must satisfy the anomaly corrected Bianchi identity (\ref{bian}). This is not the case for the non-abelian 2-form field strength $F$, whose Bianchi identity is solved in terms of a non-abelian 1-form $A$. In this case, $A$ is the fundamental field and the Bianchi identity does not imply further conditions on the $A$ components. 

For the supersymmetric solutions,  we shall assume
that there is at least one zeroth order in $\alpha'$ Killing spinor,  $\epsilon^{[0]} \neq 0$.

\section{Killing Spinor Equations}
\label{sec:KSE}
We are interested in \emph{supersymmetric} near-horizon geometries, so we assume that there exists at least one Majorana-Weyl Killing spinor $\epsilon$, which is non-vanishing at zeroth order in $\alpha'$, i.e. $\epsilon^{[0]} \neq 0$. 

We remark that the KSEs of
heterotic supergravity have been solved in \cite{class1}
and \cite{class2} for a generic class of backgrounds which admits at least one Killing vector. Therefore the solutions to the KSEs which we consider here correspond to a subclass of the solutions
in \cite{class1, class2}. However in the case of near-horizon geometries we have that the global assumptions on the spatial cross section ${\cal S}$, like compactness, allow us to derive additional conditions on the spinors and on the geometry.  So it is particularly useful to re-solve the KSEs in the way explained in section \ref{sec:hor_conj}.

We shall decompose the spinor $\epsilon$ into
positive and negative light-cone chiralities, $\epsilon=\epsilon_+ + \epsilon_-$, where
\begin{eqnarray}
\Gamma_\pm \epsilon_\pm =0, \qquad \Gamma_{+-} \epsilon_\pm
= \pm \epsilon_\pm \ .
\end{eqnarray}
and extract from the KSEs conditions on $\epsilon_{\pm}$. These conditions will be useful to carry on global analysis on $\cS$ and obtain further conditions on the bosonic fields.

\subsection{Gravitino KSE}

The \emph{gravitino} equation is the following
\begin{eqnarray}
\label{grav}
\nabla^{(-)}_{\mu}\epsilon\equiv\nabla_{\mu} \epsilon -{1 \over 8}H_{\mu \lambda_1 \lambda_2} \Gamma^{\lambda_1 \lambda_2}
\epsilon= \mathcal{O}(\alpha'^2) \ ,
\end{eqnarray}
where $\nabla^{(-)}$ is the connection with torsion defined in (\ref{connection_torsion}).
Since in the near-horizon limit the $u, r$-dependence of all bosonic fields is known, the equation (\ref{grav}) can be straightforwardly integrated along the $\le^+$ and $\le^-$ directions. 

First, by integrating the $\mu=-$ component of ({\ref{grav}}), 
we find that
\begin{eqnarray}
\label{mu-1}
\epsilon_+ = \phi_+ + \mathcal{O}(\alpha'^2) \ , \qquad
\epsilon_- = \phi_- + {1 \over 4} r (h-N)_i \Gamma_- \Gamma^i \phi_+ + \mathcal{O}(\alpha'^2)~,
\label{grav2}
\end{eqnarray}
where $\partial_r \phi_\pm=0$.
Next, by integrating the $\mu=+$ component of ({\ref{grav}}),
we find
\begin{eqnarray}
\label{mu-2}
\phi_- = \eta_- + \mathcal{O}(\alpha'^2)\  , \qquad \phi_+ = \eta_+ + {1 \over 4}u (h+N)_i
\Gamma_+ \Gamma^i \eta_- + \mathcal{O}(\alpha'^2)~,
\label{grav3}
\end{eqnarray}
where $\partial_r \eta_\pm = \partial_u \eta_\pm=0$.
In additon, the $\mu=+$ component of ({\ref{grav}}) implies a number of algebraic conditions\footnote{The $\mu = +$ component of ({\ref{grav}}) can be schematically written as 
\begin{equation}
\label{sumA}
\sum r^n \cA^+_n + \sum r^m \cA^-_m = 0 \ , 
\end{equation}
where $\cA^+_n, \cA^-_m$ are expressions on the bosonic fields and the spinors, which are of positive and negative chirality respectively. In the near-horizon limit, the coordinate $r$ is small, therefore every power of $r$ in (\ref{sumA}) must vanish separately. Furthermore, terms which are of different chirality must also vanish separately. This implies that
\begin{equation}
\cA^+_n = 0 \ , \qquad \cA^-_m = 0 \ , \qquad\forall\ n, m \ , 
\end{equation}
which in turn implies the algebraic conditions (\ref{alg1}), (\ref{alg2}), (\ref{alg3}). }:
\begin{eqnarray}
\label{alg1}
\bigg({1 \over 2} \Delta +{1 \over 8}(h^2-N^2)
-{1 \over 8}(dh+Y+h \wedge N)_{ij} \Gamma^{ij} \bigg) \phi_+= \mathcal{O}(\alpha'^2)~,
\end{eqnarray}
and
\begin{eqnarray}
\label{alg2}
\bigg(-{1 \over 2} \Delta -{1 \over 8}(h^2-N^2)
-{1 \over 8}(dh+Y+ h \wedge N)_{ij} \Gamma^{ij} \bigg) \eta_-= \mathcal{O}(\alpha'^2)~,
\end{eqnarray}
and
\begin{eqnarray}
\label{alg3}
\bigg({1 \over 4} (\Delta h_i - \partial_i \Delta)\Gamma^i
-{1 \over 32} (dh+Y)_{ij}\Gamma^{ij} (h-N)_k \Gamma^k \bigg)
\phi_+= \mathcal{O}(\alpha'^2)~.
\end{eqnarray}
We remark that ({\ref{alg1}}) and ({\ref{alg2}}) are equivalent
to\footnote{We take the inner product of (\ref{alg1}) with $\phi_+$, and we use the fact that $\langle \phi_+, \phi_+ \rangle$ is real, while $\langle \phi_+, \G_{ij} \phi_+ \rangle$ is purely imaginary, due to the fact that $\G_{ij}$ is anti-hermitian. This implies that the two terms must vanish separately. However $\langle \phi_+, \phi_+ \rangle$ never vanishes, which implies (\ref{alg4a}) and consequently (\ref{alg4b}). The same argument applies to ({\ref{alg2}}). }
\begin{eqnarray}
\label{alg4a}
{1 \over 2} \Delta +{1 \over 8}(h^2-N^2)= \mathcal{O}(\alpha'^2)~,
\end{eqnarray}
\begin{eqnarray}
\label{alg4b}
(dh+Y+ h \wedge N)_{ij} \Gamma^{ij} \phi_+= \mathcal{O}(\alpha'^2)~,
\end{eqnarray}
and
\begin{eqnarray}
\label{alg5b}
(dh+Y+ h \wedge N)_{ij} \Gamma^{ij} \eta_-= \mathcal{O}(\alpha'^2)~,
\end{eqnarray}
respectively. Furthermore, using these conditions,
({\ref{alg3}}) can also be rewritten as
\begin{eqnarray}
\label{alg6}
\bigg({1 \over 4} (\Delta h_j - \partial_j \Delta)
-{1 \over 8}(h-N)^k \big(dh+Y+2 h \wedge N)_{jk} \bigg) \Gamma^j \phi_+= \mathcal{O}(\alpha'^2)~.
\end{eqnarray}

Next, we consider the $\mu=i$ components of ({\ref{grav}}).
This implies
\begin{eqnarray}
\label{par1}
\tilde{\nabla}_i \phi_+ + \bigg({1 \over 4}(N-h)_i -{1 \over 8} W_{ijk}
\Gamma^{jk} \bigg) \phi_+= \mathcal{O}(\alpha'^2)~,
\end{eqnarray}
and
\begin{eqnarray}
\label{par2}
\tilde{\nabla}_i \eta_- + \bigg({1 \over 4}(h-N)_i -{1 \over 8} W_{ijk}
\Gamma^{jk} \bigg) \eta_-= \mathcal{O}(\alpha'^2)~,
\end{eqnarray}
together with the algebraic condition
\begin{eqnarray}
\label{alg7}
\bigg(\tilde{\nabla}_i (h-N)_j + {1 \over 2}(h_i N_j - h_j N_i)
-{1 \over 2}(h_i h_j -N_i N_j)
\nonumber \\
-(dh-Y)_{ij} -{1 \over 2} W_{ijk}(h-N)^k \bigg)
\Gamma^j \phi_+= \mathcal{O}(\alpha'^2)~.
\end{eqnarray}
These conditions exhaust the content of ({\ref{grav}}).

\subsection{Dilatino and Gaugino KSEs}
We consider the \emph{dilatino} KSE:
\begin{eqnarray}
\label{akse1}
\bigg(\Gamma^{\mu} \nabla_{\mu} \Phi -{1 \over 12}H_{\lambda_1 \lambda_2 \lambda_3}
\Gamma^{\lambda_1 \lambda_2 \lambda_3} \bigg) \epsilon = \mathcal{O}(\alpha'^2)~.
\end{eqnarray}
On making use of the conditions coming from the analysis of the gravitino equation, it is straightforward
to show that the dilatino KSE is equivalent to the following three conditions
\begin{eqnarray}
\label{aksecon1}
\bigg(\Gamma^i \tilde{\nabla}_i \Phi +{1 \over 2} N_i \Gamma^i -{1 \over 12} W_{ijk} \Gamma^{ijk} \bigg) \phi_+= \mathcal{O}(\alpha'^2)~,
\end{eqnarray}
and
\begin{eqnarray}
\label{aksecon2}
\bigg(\Gamma^i \tilde{\nabla}_i \Phi -{1 \over 2} N_i \Gamma^i -{1 \over 12} W_{ijk} \Gamma^{ijk} \bigg) \eta_-= \mathcal{O}(\alpha'^2)~,
\end{eqnarray}
and
\begin{eqnarray}
\label{aksecon2b}
\bigg( \big(\Gamma^i \tilde{\nabla}_i \Phi -{1 \over 2} N_i \Gamma^i -{1 \over 12} W_{ijk}
\Gamma^{ijk} \big) (h-N)_\ell \Gamma^\ell + Y_{ij} \Gamma^{ij} \bigg)
\phi_+= \mathcal{O}(\alpha'^2)  \ .
\end{eqnarray}

Then we consider the \emph{gaugino} KSE:
\begin{eqnarray}
\label{akse2}
F_{\mu\nu} \Gamma^{\mu\nu} \epsilon = \mathcal{O}(\alpha')~.
\end{eqnarray}
This implies the following conditions
\begin{eqnarray}
\label{akseconaux1}
\bigg(2 {\cal{P}} + {\tilde{F}}_{ij} \Gamma^{ij} \bigg) \phi_+= \mathcal{O}(\alpha')~,
\end{eqnarray}
and
\begin{eqnarray}
\label{akseconaux2}
\bigg(-2 {\cal{P}}+{\tilde{F}}_{ij} \Gamma^{ij} \bigg) \eta_- = \mathcal{O}(\alpha')~,
\end{eqnarray}
and
\begin{eqnarray}
\label{akseconaux2b}
\notag
\bigg( {1 \over 4}\big(-2 {\cal{P}} &+& {\tilde{F}}_{ij} \Gamma^{ij}\big)
(h-N)_\ell \Gamma^\ell \\
&+&2\big(h {\cal{P}}+ {\cal{P}} {\cal{B}}
- {\cal{B}} {\cal{P}}-d {\cal{P}}\big)_i  \Gamma^i \bigg) \phi_+ = \mathcal{O}(\alpha')~,
\end{eqnarray}
where we have introduced 
\begin{eqnarray}
{\tilde{F}}=d {\cal{B}} + {\cal{B}} \wedge {\cal{B}} \ .
\end{eqnarray}
The conditions ({\ref{akseconaux1}}) and ({\ref{akseconaux2}}) imply that
\begin{eqnarray}
{\cal{P}}= \mathcal{O}(\alpha')~,
\end{eqnarray}
and so $F={\tilde{F}} + \mathcal{O}(\alpha')$. Therefore ({\ref{akse2}}) is equivalent
to
\begin{eqnarray}
\label{aksecon3}
{\tilde{F}}_{ij} \Gamma^{ij} \phi_+= \mathcal{O}(\alpha')~,
\end{eqnarray}
and
\begin{eqnarray}
\label{aksecon4}
{\tilde{F}}_{ij} \Gamma^{ij} \eta_-=\mathcal{O}(\alpha')~,
\end{eqnarray}
and
\begin{eqnarray}
\label{aksecon4b}
{\tilde{F}}_{ij} \Gamma^{ij} (h-N)_\ell \Gamma^\ell \phi_+ = \mathcal{O}(\alpha')~.
\end{eqnarray}
This exhaust the content of the dilatino and gaugino KSEs. 

We have shown that the heterotic KSEs for the spacetime spinor $\epsilon$ decompose into a set of differential and algebraic conditions for the near-horizon spinors $\eta_{\pm}$, which are sections of the positive and negative chirality spin bundle associated with $\cS$. These conditions are the so-called \emph{reduced} KSEs, which can be reduced to a minimal set of necessary and sufficient conditions for the spinors $\eta_{\pm}$.

\section{Necessary and sufficient reduced KSEs}
\label{sec:necess_suff_KSEs}

The reduced KSEs can be simplified.
The details of this simplification are delegated to appendices \ref{apx:simp_KSEs_1}, \ref{apx:simp_KSEs_2} and \ref{ind}. 
The analysis consists in considering the two cases for which either $\phi_+^{[0]} \equiv 0$ or $\phi_+^{[0]} \not \equiv 0$, and the general idea is to compute the Laplacian acting on the norm of the spinors and apply the Hopf maximum principle to extract conditions on the bosonic near-horizon fields. In appendix \ref{ind} only local analysis is needed, therefore the bosonic field equations together with the Bianchi identities are sufficient.

It turns out that the independent reduced KSEs are:
\begin{eqnarray}
\label{gravsimp}
\tilde{\nabla}^{(-)}\eta_\pm &\equiv& \tilde{\nabla}_i \eta_\pm - {1 \over 8} W_{ijk} \Gamma^{jk} \eta_\pm = \mathcal{O}(\alpha'^2) \ , \\
\label{algsimpmax}
\cA \eta_{\pm} &\equiv& \bigg(\Gamma^i \tilde{\nabla}_i \Phi \pm {1 \over 2} h_i \Gamma^i -{1 \over 12} W_{ijk} \Gamma^{ijk} \bigg) \eta_\pm = \mathcal{O}(\alpha'^2) \ .
\end{eqnarray}
This result holds irrespectively on whether $\phi_+^{[0]} \equiv 0$ or $\phi_+^{[0]} \not= 0$. 
This simplification turns out to be useful when we will consider the supersymmetry enhancement. To establish supersymmetry on the horizon, one needs only to check that there exists either $\eta_+$ or $\eta_-$ that satisfies the conditions (\ref{gravsimp}) and (\ref{algsimpmax}), instead of the full set of reduced KSEs.

Furthermore, we obtained the following conditions on the bosonic fields
\begin{equation}
\label{Delta_zero}
\D = \mathcal{O}(\alpha'^2) \ , \qquad
N = h + \mathcal{O}(\alpha'^2) \ , \qquad
Y = dh + \mathcal{O}(\alpha'^2) \ , 
\end{equation}
which implies that the 3-form flux can be written as
\begin{equation}
H = d (\mathbf{e}^- \wedge \mathbf{e}^+) +W + \mathcal{O}(\alpha'^2) \ .
\end{equation}

\section{Supersymmetry enhancement at zeroth order in $\alpha'$: A review}
\label{sec:review}
In this section we shall review the supersymmetry enhancement mechanism which hold for the uncorrected heterotic supergravity \cite{hethor}. In this section only, we shall neglect $\alpha'$ corrections and set $\alpha'=0$.  

We calculate the Laplacian acting on $h^2$. To avoid the trivial case when $h^2 = 0$, in which it can be shown by considering the dilaton field equation that $\Phi = const.$, $H = 0$ and $\cS$ is Ricci flat, we assume that $h \neq 0$. We obtain
\begin{equation}
\label{lap_h_zero}
\tilde{\nabla}^i \tilde{\nabla}_i h^2 + (h-2 d \Phi)^j \tilde{\nabla}_j h^2
= 2 \tilde{\nabla}^{(i} h^{j)} \tilde{\nabla}_{(i} h_{j)}
+{1 \over 2}(dh - i_h W)_{ij} (dh-i_h W)^{ij} \ , 
\end{equation}
In computing this expression, we made use of the Einstein equation ({\ref{einsp}})
together with the gauge field equations ({\ref{geq1a}}) and
({\ref{geq1b}}), neglecting the $\alpha'$ corrections. Since all fields are smooth, $\cS$ is compact and the RHS of (\ref{lap_h_zero}) is semi-positive definite, one can apply the Hopf maximum principle to (\ref{lap_h_zero}) and imply the following conditions:
\begin{equation}
\label{Hopf_h^2}
h^2 = \mbox{const} \ , \qquad
\tilde{\nabla}_{(i} h_{j)} = 0 \ , \qquad
dh - i_h W = 0 \ . 
\end{equation}
which in turn imply the following condition
\begin{equation}
\tilde{\nabla}^{(-)}_ih_j\equiv \tilde{\nabla}_i h_j - {1 \over 2} W_{ijk} h^k = 0 \ ,  
\end{equation}
i.e. $h$ is parallel with respect to the connection with torsion $\tilde{\nabla}^{(-)}$. Moreover the conditions (\ref{Hopf_h^2}) imply 
\begin{eqnarray}
\label{cond_1}
i_h dh &=& 0 \ ,  \\
\label{cond_2}
\cL_h W &=& 0  \ , \\
\label{cond_3}
\cL_h \Phi &=& 0 \ ,
\end{eqnarray}
where (\ref{cond_1}) is a consequence of the third condition in (\ref{Hopf_h^2}), (\ref{cond_2}) is implied by the third condition in (\ref{Hopf_h^2}) and the fact that $dW = 0$, (\ref{cond_3}) is implied by the second equation in (\ref{Hopf_h^2}) and the $+-$ component of the gauge field equation (\ref{gaugeH}). 

The necessary and sufficient reduced KSEs are:
\begin{eqnarray}
\label{gravsimp_zero}
\tilde{\nabla}^{(-)}\eta_\pm&=& \tilde{\nabla}_i \eta_\pm - {1 \over 8} W_{ijk} \Gamma^{jk} \eta_\pm = 0 \\
\label{algsimpmax_zero}
\cA \eta_{\pm} &=& \bigg(\Gamma^i \tilde{\nabla}_i \Phi \pm {1 \over 2} h_i \Gamma^i -{1 \over 12} W_{ijk} \Gamma^{ijk} \bigg) \eta_\pm =  0  \ .
\end{eqnarray}
Assume that there exists one spinor $\eta_+ \neq 0$ which satisfies (\ref{gravsimp_zero}) and (\ref{algsimpmax_zero}) with the upper sign. Then one can show by using the conditions (\ref{Hopf_h^2}) - (\ref{cond_3}) that the following spinor 
\begin{equation}
\eta'_- \equiv \G_- \slashed h \eta_+ \ , 
\end{equation}
also satisfies (\ref{gravsimp_zero}) and (\ref{algsimpmax_zero}) with the lower sign. 

It is true also the other way around, i.e. suppose that there exists $\eta_- \neq 0$ which satisfies (\ref{gravsimp_zero}) and (\ref{algsimpmax_zero}) with the lower sign, then 
\begin{equation}
\eta'_+ \equiv \G_+ \slashed h \eta_- \ , 
\end{equation}
also satisfies (\ref{gravsimp_zero}) and (\ref{algsimpmax_zero}) with the upper sign. 

This means that if we assume the existence of one Killing spinor, then we are able to construct a second Killing spinor, which is linearly independent because of opposite chirality. This is the case provided that $\slashed h \eta_{\pm} \neq 0$, which is true since we assumed $h \neq 0$. 
This is how the supersymmetry enhancement takes place in the uncorrected heterotic supergravity. In the next section we shall reintroduce again the $\alpha'$ corrections.

\section{Global Properties}

\subsection{Maximum principle on  $h^2$} \label{hsq}
\label{sec: max h^2}
In this section we shall investigate if the argument made in section \ref{sec:review} to imply the supersymmetry enhancement at zeroth order also applies to the corrected case. 
To avoid the trivial case when $h^2=\mathcal{O}(\alpha'^2)$, we take $h^{[0]} \neq 0$.

We calculate the Laplacian of $h^2$ to find
that
\begin{eqnarray}
\label{lap1}
&&\tilde{\nabla}^i \tilde{\nabla}_i h^2 + (h-2 d \Phi)^j \tilde{\nabla}_j h^2
= 2 \tilde{\nabla}^{(i} h^{j)} \tilde{\nabla}_{(i} h_{j)}
+{1 \over 2}(dh - i_h W)_{ij} (dh-i_h W)^{ij}
\nonumber \\
&&~~-{\alpha' \over 4} h^i h^j
\bigg(-2 dh_{i \ell}
dh_j{}^\ell + \tilde{R}^{(+)}{}_{i \ell_1 \ell_2 \ell_3}
\tilde{R}^{(+)}{}_j{}^{\ell_1 \ell_2 \ell_3}
- {\tilde{F}}_{i\ell}{}^{ab} {\tilde{F}}_j{}^\ell{}_{ab} \bigg) + \mathcal{O}(\alpha'^2) \ , 
\end{eqnarray}
where in computing this expression, we made use of the Einstein equation ({\ref{einsp}})
together with the gauge field equations ({\ref{geq1a}}) and
({\ref{geq1b}}).
The $\alpha'$ terms
in ({\ref{lap1}}) originate from the $\alpha'$ terms in
$2 h^i h^j {\tilde{R}}_{ij}$ of the Einstein equation.

To begin, we consider ({\ref{lap1}}) to zeroth order in $\alpha'$.
We then re-obtain the conditions found in \ref{sec:review} via a maximum principle argument, i.e.
\begin{eqnarray}
\label{firstiso}
h^2= {\rm const} + \mathcal{O}(\alpha') \ , \qquad
\tilde{\nabla}_{(i} h_{j)}=  \mathcal{O}(\alpha') \ ,\qquad
dh - i_h W = \mathcal{O}(\alpha') \ . 
\end{eqnarray}
and
\begin{equation}
\label{hpar_first}
\tilde{\nabla}^{(-)}_ih_j\equiv \tilde{\nabla}_i h_j - {1 \over 2} W_{ijk} h^k =  \mathcal{O}(\alpha')   \ ,  
\end{equation}
and 
\begin{eqnarray}
\label{firstordercond}
i_h dh= \mathcal{O}(\alpha') \ , \qquad
{\cal{L}}_h \Phi = \mathcal{O}(\alpha') \ , \qquad {\cal{L}}_h W = \mathcal{O}(\alpha') \ .
\end{eqnarray}

Consider now (\ref{lap1}) at first order in $\alpha'$, which by using the conditions (\ref{firstiso}) and (\ref{firstordercond}) simplifies to 
\begin{eqnarray}
\tilde{\nabla}^i \tilde{\nabla}_i h^2 + (h-2 d \Phi)^j \tilde{\nabla}_j h^2 = -{\alpha' \over 4} h^i h^j
\bigg( \tilde{R}^{(+)}{}_{i \ell_1 \ell_2 \ell_3}
\tilde{R}^{(+)}{}_j{}^{\ell_1 \ell_2 \ell_3}
- {\tilde{F}}_{i\ell}{}^{ab} {\tilde{F}}_j{}^\ell{}_{ab} \bigg) + \mathcal{O}(\alpha'^2) \ , 
\end{eqnarray}
In what follows, we shall prove that the $i_h \tilde{R}^{(+)}$ term on the RHS disappears.
As we showed in section \ref{sec:review}, if $\eta_+$ satisfies
({\ref{gravsimp}}) and (\ref{algsimpmax}), then $\eta'_- \equiv \Gamma_- h_i \Gamma^i \eta_+$ also satisfies
({\ref{gravsimp}}) and (\ref{algsimpmax}) to zeroth order in $\alpha'$. The integrability conditions of the gravitino equation for $\eta_+$ and $\eta'_-$ imply that\footnote{One can show that if $\hat{\nabla}$ is a generic connection, with connection symbols $\hat{\Gamma}$, then according to our convention in appendix \ref{apx:A} we have
\begin{equation}
\notag
[\hat{\nabla}_i, \hat{\nabla}_j ] \eta_+ = - \frac{1}{4} R(\hat{\Gamma})_{ij, mn} \Gamma^{mn} \eta_+ - T_{ij}{}^k \hat{\nabla}_k \eta_+ \ , 
\end{equation}
where $T$ and $R(\hat{\Gamma})$ are the torsion and the curvature of $\hat{\nabla}$ respectively. In our case $\hat{\nabla} = \tilde{\nabla}^{(-)}$, and since $\eta_+$ satisfies the gravitino KSE the torsion term disappears. }
\begin{eqnarray}
\tilde{R}^{(-)}_{ijmn} h^m \Gamma^n \eta_+ = \mathcal{O}(\alpha')~,
\end{eqnarray}
and by using (\ref{curvcross}), this in turn implies\footnote{This is a consequence of spinorial geometry. }
\begin{eqnarray}
\tilde{R}^{(+)}{}_{mnij} h^m = \mathcal{O}(\alpha') \ .
\end{eqnarray}

On substituting these conditions back into ({\ref{lap1}}) one finds that
the remaining content of ({\ref{lap1}}), after rewriting the Laplacian term conveniently, is
\begin{eqnarray}
\label{lap2}
\tilde{\nabla}^i \bigg( e^{-2 \Phi} \tilde{\nabla}_i h^2 \bigg) + e^{-2 \Phi} h^j \tilde{\nabla}_j h^2
= {\alpha' \over 2} e^{-2 \Phi} h^i h^j {\tilde{F}}_{i\ell}{}^{ab} {\tilde{F}}_j{}^\ell{}_{ab}
+ \mathcal{O}(\alpha'^2)\ .
\end{eqnarray}
On integrating both sides of ({\ref{lap2}}) over the zeroth order
horizon section, and by using the gauge field equation (\ref{geq1a}), one finds that
\begin{eqnarray}
i_h {\tilde{F}} = \mathcal{O}(\alpha') \ ,
\end{eqnarray}
which in turn implies
\begin{eqnarray}
h^2 = {\rm const} + \mathcal{O}(\alpha'^2) \ .
\end{eqnarray}
Therefore we do find the condition $h^2 =$ const $+\, \mathcal{O}(\alpha'^2)$, where the $\mathcal{O}(\alpha'^2)$ terms do not necessary need to be constant, however we do \emph{not} get 
\begin{equation}
\label{cond_import}
\tilde{\nabla}_{(i} h_{j)} = \mathcal{O}(\alpha'^2) \ , \qquad\qquad
dh - i_h W = \mathcal{O}(\alpha'^2) \ . 
\end{equation}
We recall that (\ref{cond_import}) are crucial properties of the $\slashed h$ map in order to generate the supersymmetry enhancement at right order in $\alpha'$, and therefore at this stage we can only establish the enhancement of supersymmetry at zeroth order in $\alpha'$.

\subsection{Lichnerowicz Type Theorem}
In section \ref{sec: max h^2} we showed that including $\alpha'$ corrections in heterotic supergravity spoils the $\slashed h$ map that generates the doubling of number of preserved supersymmetries. 
In this section we shall investigate if one can construct a generalised Lichenerowicz theorem including $\alpha'$ corrections. As explained in the introduction \ref{sec:hor_conj}, this would be useful because together with an index theory argument it would imply that the number of preserved supersymmetries is even, and therefore show the supersymmetry enhancement.

We introduce the \emph{modified connection with torsion} ${\nabla}^{(\kappa)}$
\begin{equation}
{\nabla}^{(\kappa)}_{i} \equiv \tilde{\nabla}^{(-)}_{i} + \kappa \, \Gamma_{i} \mathcal{A}  \ , 
\end{equation}
and the \emph{modified Dirac operator} $\mathcal{D}$
\begin{equation}
\mathcal{D}  \equiv \Gamma^{i} \tilde{\nabla}^{(-)}_{i} + q \, \mathcal{A} \ ,
\end{equation}
where $\kappa, q \in \mathbb{R}$, and $\tilde{\nabla}^{(-)}$ and $\cA$ are defined by the independent reduced KSEs (\ref{gravsimp}) and (\ref{algsimpmax}) respectively. 
Then we have the following
\begin{theorem} If $\eta_{\pm}$ is a Killing spinor up to second order corrections in $\alpha'$, then $\eta_{\pm}$ is also a solution to the Dirac equation $\mathcal{D} \eta_{\pm} = \mathcal{O}(\alpha'^2)$. 

Conversely, if $\eta_{\pm}$ is a solution to the Dirac equation $\mathcal{D} \eta_{\pm} = \mathcal{O}(\alpha'^2)$, then $\eta_{\pm}$ is a Killing spinor at zeroth order in $\alpha'$ only, i.e.
\begin{equation}
\label{killsp1}
\tilde{\nabla}^{(-)}_i \eta_{\pm} = \mathcal{O}(\alpha') \ , \qquad \qquad \mathcal{A} \eta_\pm = \mathcal{O}(\alpha') \ .
\end{equation}
and also satisfies 
\begin{eqnarray}
\label{killsp2}
{{dh}}_{ij} \Gamma^{ij} \eta_\pm = \mathcal{O}(\alpha') \ , \qquad \qquad {{\tilde{F}}}^{ab}_{ij}\Gamma^{ij} \eta_\pm
= \mathcal{O}(\alpha') \ .
\end{eqnarray}
\end{theorem}
\begin{proof}
The first part of the theorem is straightforward, since the fact that $\eta_{\pm}$ satisfies $\tilde{\nabla}^{(-)}_i \eta_{\pm} = \mathcal{O}(\alpha'^2)$ and $\mathcal{A} \eta_\pm = \mathcal{O}(\alpha'^2)$ implies that $\mathcal{D} \eta_{\pm} = \mathcal{O}(\alpha'^2)$. 
For the second part of the proof we consider the following functional:
\begin{eqnarray}
\label{I functional}
\mathcal{I} \equiv  \int_{\mathcal{S}} e^{c\Phi} \bigg( \langle {\nabla}^{(\kappa)}_{i} \eta_{\pm} , {\nabla}^{(\kappa)i} \eta_{\pm} \rangle
-  \langle\mathcal{D} \eta_{\pm} , \mathcal{D} \eta_{\pm} \rangle \bigg) \ ,
\end{eqnarray}
where $c \in \mathbb{R}$, and we assume all the  field equations. After some algebra, which is described in appendix \ref{calcId}, we find
\begin{align}
\label{final_I}
\notag
\mathcal{I}  = &\left(8\kappa^2 - \frac{1}{6} \kappa \right) \int_{\mathcal{S}} e^{-2 \Phi} \parallel \mathcal{A}\,  \eta_{\pm} \parallel^2
+ \int_{\mathcal{S}} e^{-2\Phi} \langle \eta_{\pm}, \Psi \mathcal{D} \eta_{\pm} \rangle \\
&- \frac{\alpha'}{64} \int_{\mathcal{S}} e^{-2\Phi} \left( 2 \parallel \slashed{dh}\, \eta_{\pm} \parallel^2 + \parallel \slashed{\tilde{F}} \eta_{\pm} \parallel^2 - \langle \tilde{R}^{(+)}{}_{\ell_1\ell_2,\, ij}\Gamma^{\ell_1\ell_2}\eta_{\pm}, \tilde{R}^{(+)}{}^{ ij}_{\ell_3\ell_4,}\Gamma^{\ell_3\ell_4}\eta_{\pm}\rangle \right) + \mathcal{O}(\alpha'^2)\ ,
\end{align}
which is true if and only if $q= \frac{1}{12} + \mathcal{O}(\alpha'^2)$ and $c = -2 +\mathcal{O}(\alpha'^2)$, and the  $\Psi$ is defined as follows
\begin{eqnarray}
\Psi \equiv 2\left(\kappa - \frac{1}{12}\right) \mathcal{A}^{\dagger} -2 \Gamma^{i}\tilde{\nabla}_{i} \Phi - \frac{1}{6} \Gamma^{\ell_1\ell_2\ell_3}W_{\ell_1\ell_2\ell_3} + \mathcal{O}(\alpha'^2) \ .
\end{eqnarray}
The values of $q$ and $c$ are fixed by requiring that certain terms in the
functional ({\ref{I functional}}), which cannot be rewritten in terms of
the Dirac operator ${\cal{D}}$, or ${\cal{A}}^\dagger {\cal{A}}$, and which have no fixed sign, should vanish.
Consider the zeroth order of (\ref{final_I}). If $0 < \kappa < \frac{1}{48}$, then the RHS is a sum of two semi-positive quantities, which implies that
\begin{eqnarray}
\label{Dirac->Killing}
\label{gravitino + alg}
\mathcal{D} \eta_{\pm} = \mathcal{O}(\alpha'^2)  \quad \Longrightarrow  ({\ref{killsp1}})
\end{eqnarray}
and establishes the first part of the theorem. Next
the integrability condition of $\tilde{\nabla}^{(-)}\eta_{\pm} = \mathcal{O}(\alpha')$ is
\begin{eqnarray}
\tilde{R}^{(-)}_{mn, \ell_1\ell_2}\Gamma^{\ell_1\ell_2}\eta_{\pm} = \mathcal{O}(\alpha') \ ,
\end{eqnarray}
which in turn implies that
\begin{eqnarray}
\tilde{R}^{(+)}{}_{\ell_1\ell_2, mn}\Gamma^{\ell_1\ell_2} \eta_{\pm} = \mathcal{O}(\alpha') \ .
\end{eqnarray}
Hence we shall neglect the term in (\ref{final_I}) which is quadratic in
$\tilde{R}^{(+)}$, as this term is ${\cal{O}}(\alpha'^3)$.
Then, by using (\ref{Dirac->Killing}), the part of (\ref{final_I})
which is first order in $\alpha'$ further implies ({\ref{killsp2}}).
This completes the proof.
\end{proof}

Therefore we proved that generalised Lichenrowicz theorems are spoiled by the presence of string corrections, because $\mathcal{D} \eta_{\pm} = \mathcal{O}(\alpha'^2)$ does not imply the KSEs up to ${\cal{O}}(\alpha'^2)$ corrections.

\section{A sufficient condition for supersymmetry enhancement}
In the previous section we showed that standard theorems involving global properties of the horizon do not hold when string corrections are considered. 

In this section we shall show that one can establish supersymmetry enhancement, including $\alpha'$ corrections, provided that there exists at least one non-vanishing at zeroth order negative light-cone chirality spinor, i.e. $\eta_-^{[0]} \neq 0$.

To prove this, it suffices to demonstrate that $h$ leaves all fields invariant and that it is covariantly constant with respect
to the connection with torsion $\tilde{\nabla}^{(-)}$ on ${\cal S}$.  Indeed, first note that if there exists at least one $\eta_-^{[0]} \neq 0$, then one can consider ({\ref{udepa}}), which implies that
\begin{eqnarray}
\label{niceh}
\tilde{\nabla}^{(-)}_ih_j\equiv \tilde{\nabla}_i h_j - {1 \over 2} W_{ijk} h^k= \mathcal{O}(\alpha'^2)~.
\end{eqnarray}
In particular, to both zeroth and first order in $\alpha'$,
$h$ defines an isometry on ${\cal{S}}$, with the following properties implied by (\ref{niceh}):
\begin{equation}
h^2= {\rm const} + \mathcal{O}(\alpha'^2) \ , \qquad
\tilde{\nabla}_{(i} h_{j)}=  \mathcal{O}(\alpha'^2) \ ,\qquad
dh - i_h W = \mathcal{O}(\alpha'^2) \ . 
\end{equation} 
Then the gauge equation ({\ref{geq1a}})
implies
\begin{eqnarray}
\label{phlie}
{\cal{L}}_h \Phi = \mathcal{O}(\alpha'^2)~.
\end{eqnarray}
Also, the $u$-dependent part of ({\ref{auxalg1c}}) implies
\begin{eqnarray}
\label{extraalg3}
(i_h {\tilde{F}})_i \Gamma^i \eta_-= \mathcal{O}(\alpha')~,
\end{eqnarray}
which implies that $i_h {\tilde{F}}= \mathcal{O}(\alpha')$. So
in the gauge for which $i_h {\cal{B}}=0$, one has
\begin{eqnarray}
{\cal{L}}_h {\tilde{F}} = \mathcal{O}(\alpha') \ .
\end{eqnarray}
Next we consider ${\cal{L}}_h W$, where
\begin{eqnarray}
\label{lie3}
{\cal{L}}_h W = -{\alpha' \over 2}   \bigg( {\rm tr}\big( (i_h R^{(+)}) \wedge R^{(+)}\big) \bigg)+\mathcal{O}(\alpha'^2)~,
\end{eqnarray}
because $dh=i_h W+\mathcal{O}(\alpha'^2)$. To evaluate this expression, note first that because of (\ref{niceh}) the spinors $\eta_-$ and $\slashed h \eta_-$ satisfy 
\begin{eqnarray}
\tilde{\nabla}^{(-)}_i \eta_-=\mathcal{O}(\alpha'^2), \qquad \tilde{\nabla}^{(-)}_i(h_\ell \Gamma^\ell \eta_-)=\mathcal{O}(\alpha'^2)
\end{eqnarray}
for which the associated integrability conditions are
\begin{eqnarray}
\tilde{R}^{(-)}_{ijpq} \Gamma^{pq} \eta_-=\mathcal{O}(\alpha'^2), \qquad
\tilde{R}^{(-)}_{ijpq} \Gamma^{pq} (h_\ell \Gamma^\ell \eta_-)=\mathcal{O}(\alpha'^2)
\end{eqnarray}
from which we obtain the condition
\begin{eqnarray}
h^\ell \tilde{R}^{(-)}_{ij\ell q} =\mathcal{O}(\alpha'^2)~,
\end{eqnarray}
and hence, as a consequence of ({\ref{curvcross}}),
\begin{eqnarray}
h^\ell \tilde{R}^{(+)}{}_{\ell qij} =\mathcal{O}(\alpha')~.
\end{eqnarray}
Moreover,
\begin{eqnarray}
h^\ell  R^{(+)}{}_{\ell q+-} = h^i (dh)_{i q} =\mathcal{O}(\alpha'^2)~.
\end{eqnarray}
It follows that the contribution of $i_h R^{(+)}$ to the RHS of ({\ref{lie3}}) is of at least $\mathcal{O}(\alpha')$, and hence
\begin{eqnarray}
{\cal{L}}_h W=\mathcal{O}(\alpha'^2) \ .
\end{eqnarray}
So, we have shown that to both zero and first order in $\alpha'$,
the Lie derivative of the metric on ${\cal{S}}$, as well as $h, \Phi$ and $W$ with respect to $h$
vanishes, and the Lie derivative of ${\tilde{F}}$ with respect to $h$ vanishes to zeroth order
in $\alpha'$.

Supersymmetry is therefore enhanced, because  if $\eta_+$ satisfies ({\ref{gravsimp}})
and ({\ref{algsimpmax}}), then so does $\eta_-' = \Gamma_- h_i \Gamma^i \eta_+$.  Conversely, if $\eta_-$ satisfies
({\ref{gravsimp}})
and ({\ref{algsimpmax}}), then so does $\eta_+'= \Gamma_+ h_i \Gamma^i \eta_-$.
This establishes  a 1-1 correspondence between
spinors $\eta_+$ and $\eta_-$ satisfying ({\ref{gravsimp}})
and ({\ref{algsimpmax}}), so the number of supersymmetries preserved is always even. Thus we have found a sufficient condition which guarantees the supersymmetry enhancement, namely the existence of at least one $\eta_-^{[0]} \neq 0$.

\section{A No-Go Theorem}
In this section we show an implication from the facts that $\D = \mathcal{O}(\alpha'^2)$, as explained in (\ref{Delta_zero}), and that $h$ is an isometry at zeroth and first order in $\alpha'$, provided that there exists at least one $\eta_-^{[0]} \neq 0$. 
\begin{theorem}
\label{th:no_AdS2}
There are no $AdS_2\times_w \cS $ backgrounds in uncorrected heterotic supergravity, for which all fields are smooth and the internal space $\cS$ is smooth, compact and without boundary. The result extends up to second order corrections in $\alpha'$, provided that there exists at least one $\eta_-^{[0]} \neq 0$.
\end{theorem}
\begin{proof}
Consider the case where there exists at least one $\eta_-^{[0]} \neq 0$. The uncorrected case does not need such requirement, and it follows immediately by mathematically setting $\alpha' = 0$ in the following formul\ae.
The condition (\ref{Delta_zero}) implies that we can set $\D = \mathcal{O}(\alpha'^2)$ in the near-horizon metric (\ref{NHG}). 
As a consequence of (\ref{niceh}), the vector field $h$ is an isometry. 
We introduce local coordinates such that $h$ can be written as
\begin{equation}
\label{h_psi}
h = \frac{\partial}{\partial \psi} \ , 
\end{equation}
where $\psi$ is a local spatial coordinate. The metric (\ref{NHG}) becomes
\begin{equation}
\label{NH_metric_h_iso}
ds^2 = 2 du dr + 2 r h du + \frac{1}{k^2} h \otimes h + ds^2_{(7)} + \mathcal{O}(\alpha'^2) \ ,  
\end{equation}
where we set $h^2 = k^2 + \mathcal{O}(\alpha'^2)$, with $k$ a constant, and the 1-form dual to (\ref{h_psi}), which we indicate with the same symbol $h$, is given by
\begin{equation}
\label{h_1form}
h = k^2 ( d \psi + \alpha ) \ , 
\end{equation}
where $\alpha$ is a 1-form only locally defined, while $d\alpha$ is a 2-form globally defined. 
The metric (\ref{NH_metric_h_iso}) can be written as 
\begin{equation}
ds^2 = 2 du dr + 2 k^2 r du (d\psi + \alpha) + k^2 (d\psi + \alpha )^2 + ds^2_{(7)} \ . 
\end{equation}
This metric locally describes an $AdS_3$ fibration over a 7-dimensional manifold $\mathcal{B}^7$, and therefore there are no $AdS_2\times_w \cS$ backgrounds.  
\end{proof}
We shall remark that the existence of near-horizon geometries which are locally $AdS_3 \times \mathcal{B}^7$ does not contradict theorem \ref{th:no_AdS2}. 
This is because if one attempts to write 
\begin{equation}
AdS_3 = AdS_2 \times_w Y \ , 
\end{equation}
where $Y$ is a 1-dimensional manifold, then one violates one or more conditions on $\cS$, which we assumed from the beginning. In particular one finds the following possibilities for $Y$:
\begin{itemize}
\item $Y = \mathbb{R}$, which violates \emph{compactness},
\item $Y = [0, 1]$, which violates the \emph{no boundary} condition, 
\item $Y = S^1$ with discontinuous warp factor, which violates \emph{smoothness}.  
\end{itemize}
This result can be found in detail in appendix \ref{apx:AdS_n}. Of course this is not the case in type II supergravities, where $AdS_2$ backgrounds, with global conditions on $\cS$ as assumed, exist.

\section{Description of the spacetime geometry}
\label{sec:geometry}

In the uncorrected case \cite{hethor}, horizons with non-trivial fluxes preserve an even number of supersymmetries. Furthermore horizons with more than 8 supersymmetries are trivial, i.e. $h$ vanishes. Therefore the non-trivial heterotic horizons preserve 2, 4, 6 and 8 supersymmetries.

Up to ${\cal O}(\alpha')$, the investigation of the horizon geometry is identical to \cite{hethor} for heterotic horizons
with $H$ closed. Here we shall describe the geometry of the horizons that admit a $\eta_-$ Killing spinor up to ${\cal O}(\alpha'^2)$. We have seen that for such horizons $h$ is parallel with respect to the connection with torsion up to ${\cal O}(\alpha'^2)$. Because of this, the geometry of such horizons is very similar to that of horizons with closed 3-form flux. The only difference between the geometries of the two cases are solely located in the modified Bianchi identity for the 3-form flux.

In what follows we shall describe the geometry of horizons preserving 2 and 4 supersymmetries, including $\alpha'$ corrections, for which there exists at least one Killing spinor $\eta^{[0]}_- \neq 0$. For the cases which preserve 6 and 8 supersymmetries, see \cite{hethor, Fontanella:2016aok}. 


\subsection{Horizons with $G_2$ structure}

These horizons admit two supersymmetries up to ${\cal O}(\alpha'^2)$ corrections, which are 
\begin{equation}
\eta_+^1 = 1 + e_{1234} \ , \qquad\qquad
\eta_-^2 = \G_- \G^i h_i (1 + e_{1234} )  \  , 
\end{equation}
where $h$ satisfies (\ref{niceh}). The associated spacetime Killing spinors are
\begin{equation}
\label{G_2spinors}
\epsilon^1 = 1 + e_{1234} \ ,  \qquad\qquad
\epsilon^2 = - k^2 u (1 + e_{1234}) + \G_- \G^i h_i (1 + e_{1234}) \ . 
\end{equation}
The isotropy group of both Killing spinors is $G_2 \subset Spin(9, 1)$. Since both $\epsilon^1$ and $\epsilon^2$ satisfy the gravitino KSE, the holonomy of $\nabla^{(-)}$ is also a subgroup of $G_2$. 

One can construct 1-form (mixed) bilinears in terms of the spacetime spinors (\ref{G_2spinors}), which are as follows
\begin{eqnarray}
\label{g2vbi}
\lambda^- &=& \mathbf{e}^-~,~~~
\lambda^+ = \mathbf{e}^+ - {1 \over 2} k^2 u^2 \mathbf{e}^- -u h~,~~~
\lambda^1 = k^{-1} \big(h+ k^2 u \mathbf{e}^-\big)~,
\end{eqnarray}
where $k^2 \equiv h^2$ is constant up to ${\cal O}(\alpha'^2)$. 
One can check that the vector fields associated to $\lambda^-, \lambda^+, \lambda^1$ satisfy the $\mathfrak{sl}(2,\mathbb{R})$ algebra. 

The spacetime metric can be written as
\begin{eqnarray}
ds^2=\eta_{ab} \lambda^a \lambda^b+d\tilde s_{(7)}^2+{\cal O}(\alpha'^2) \ ,
\end{eqnarray}
where the non-vanishing components of $\eta$ are $\eta_{+-} = 2$ and $\eta_{11} = 1$. 
The spacetime can be locally described as a principle bundle with a $SL(2, \mathbb{R})$ fibre over a 7-dimensional base space manifold $B^7$. We indicate the metric and 3-form flux data on $B^7$ by $d\tilde s_{(7)}^2$ and $\tilde H_{(7)}$ respectively. The connection $\tilde{\nabla}^{(-)}_{(7)}$ with torsion $\tilde H_{(7)}$ has holonomy contained in $G_2$.
The spacetime 3-form flux can be written as
\begin{eqnarray}
H=CS(\lambda)+\tilde H_{(7)}+{\cal O}(\alpha'^2) \ ,
\end{eqnarray}
where the 3-form $CS(\lambda)$ is the Chern-Simons form of the principal bundle connection $\lambda^a$,
\begin{equation}
CS(\lambda) = \frac{1}{3} \eta_{ab} \lambda^a \wedge d \lambda^b + \frac{2}{3} \eta_{ab} \lambda^a \wedge \mathcal{F}^b \ , 
\end{equation}
and $\mathcal{F}$ is the curvature of $\lambda$, given by 
\begin{equation}
\mathcal{F}^a = d\lambda^a - \frac{1}{2} ( \lambda \wedge \lambda )^a \ . 
\end{equation}
A direct computation shows that 
\begin{equation}
\label{CS_exp}
CS(\lambda)= du\wedge dr\wedge h+r du\wedge dh+k^{-2} h\wedge dh \ .
\end{equation}
Moreover, the 3-form flux on $B^7$ can be described in terms of the $G_2$ structure as 
\begin{eqnarray}
\tilde H_{(7)}=k  \varphi+ e^{2\Phi} \star_7d\big( e^{-2\Phi} \varphi\big)+\mathcal{O}(\alpha'^2) \ .
\end{eqnarray}
where $\varphi$ is the $G_2$ fundamental 3-form, which descends from the fundamental $Spin(7)$ 4-form $\phi$ associated with the existence of one Killing spinor $\eta_+$, via 
\begin{equation}
\varphi=\frac{1}{k} i_h\phi+\mathcal{O}(\alpha'^2)\ .
\end{equation}
The algebraic KSE implies that
\begin{equation}
\theta_{\varphi} = 2 d \Phi \ , \qquad\qquad
\partial_{a} \Phi = \mathcal{O}(\alpha'^2) \ , 
\end{equation}
where $\theta_{\varphi}$ is the Lee form of $\varphi$, defined for a generic $p$-form $\xi$ as $\theta_{\xi} = - \frac{1}{6}\star (\star d \xi \w \xi)$. 
Therefore the dilaton $\Phi$ depends only on the coordinates of $B^7$.
The system of equations for the geometric data of $G_2$ holonomy heterotic near-horizon geometries is
\begin{eqnarray}
\label{g2cons}
&&d[e^{-2\Phi}\star_7\varphi]={\cal O}(\alpha'^2)\ , \\
\label{g2cons1}
&&k^{-2}\,dh\wedge dh+ d\tilde H_{(7)}=-{\alpha'\over4} \bigg(-2 dh\wedge dh+ \mathrm {tr}( R^{(+)}_{(8)}\wedge R^{(+)}_{(8)}- F\wedge F)\bigg)+{\cal O}(\alpha'^2) \ , \qquad\\
\label{g2cons2}
&&(dh)_{ij}\varphi^{ij}{}_k = {\cal O}(\alpha'^2) \ , \qquad\qquad\qquad
\tilde{F}_{ij}\varphi^{ij}{}_k = {\cal O}(\alpha') \ .
\end{eqnarray}
The condition (\ref{g2cons}) is the $G_2$ holonomy condition, i.e. the condition required for $B^7$ to admit a $G_2$ structure compatible with a connection with torsion. The condition (\ref{g2cons1})
is the anomalous Bianchi identity of the 3-form field strength written in terms of $B^7$ data, where the curvature $R^{(+)}_{(8)}$ is associated with the near-horizon section ${\cal S}$ with metric and 3-form flux
\begin{eqnarray}
d\tilde s_{(8)}^2= k^{-2} h\otimes h+d\tilde s_{(7)}^2+\mathcal{O}(\alpha'^2)~,~~~\tilde H_{(8)}= k^{-2}  h\wedge dh+\tilde H_{(7)}+\mathcal{O}(\alpha'^2)~.
\end{eqnarray}
Finally, the two equations in (\ref{g2cons2}), which descend from (\ref{auxalg2}) and (\ref{auxalg2c}), 
imply that both $dh$ and the gauge connection are $\mathfrak{g}_2$ instantons on $B^7$.

\subsection{Horizons with $SU(3)$ structure}

These horizons preserve 4 supersymmetries up to ${\cal O}(\alpha'^2)$ corrections, which are 
\begin{eqnarray}
\label{spinors_su(3)}
\notag
&&\eta_+^1 = 1 + e_{1234} \ , \qquad\qquad
\eta_-^2 = \G_- \G^i h_i (1 + e_{1234} )  \  , \\
&&\eta_+^3 = i (1 - e_{1234} )  \ , \qquad\quad
\eta_-^4 = i \G_- \G^i h_i (1 - e_{1234} ) \ , 
\end{eqnarray}
where again $h$ satisfies (\ref{niceh}). The associated spacetime spinors are 
\begin{eqnarray}
\label{SU3_spinors}
\notag
&&\epsilon^1 = 1 + e_{1234} \ ,  \qquad\qquad
\epsilon^2 = - k^2 u (1 + e_{1234}) + \G_- \G^i h_i (1 + e_{1234}) \ , \\
&& \epsilon^3 = i (1 - e_{1234} ) \ , \qquad\quad
\epsilon^4 = - i k^2 u (1 - e_{1234} ) + i \G_- \G^i h_i (1 - e_{1234}) \ . 
\end{eqnarray}
The isotropy group of all these spinors is $SU(3) \subset G_2$. Since the spinors $\epsilon^1, ... , \epsilon^4$ satisfy the gravitino KSE, the holonomy of $\nabla^{(-)}$ is a subgroup of $SU(3)$. 

One can construct 1-form spinor bilinears in terms of (\ref{SU3_spinors}). The linearly independent ones are again $\lambda^+, \lambda^-, \lambda^1$ as in (\ref{g2vbi}), and an additional $\lambda^6$, 
\begin{eqnarray}
\label{u1_gen}
\lambda^6=k^{-1} \ell \ , \qquad\qquad
\ell_i = h_j I^j{}_i \ , 
\end{eqnarray}
where $h^2=k^2$ is constant up to ${\cal O}(\alpha'^2)$, and $I$ is the complex structure $I^2 = - \delta$, given as a mixed bilinear of Killing spinors (\ref{spinors_su(3)}), which in holomorphic coordinates becomes $I_{\bar{\alpha}\beta} =  i \delta_{\bar{\alpha}\beta}$. 
The vector field associated with $\lambda^6$ generates a $\mathfrak{u}(1)$ Lie algebra, and commutes with the remaining vector fields associated with $\lambda^+, \lambda^-, \lambda^1$, which generates the $\mathfrak{sl}(2, \mathbb{R})$ algebra. 

The spacetime metric can be written as
\begin{eqnarray}
ds^2 =\eta_{ab} \lambda^a \lambda^b+ d\tilde s^2_{(6)}+{\cal O}(\alpha'^2) \ , 
\end{eqnarray}
where $\lambda^a$, $a=+,-,1,6$ are given in (\ref{g2vbi}) and (\ref{u1_gen}), and the non-vanishing components of $\eta$ are $\eta_{+-}=2$, and $\eta_{11} = \eta_{66} = 1$. 
Locally the spacetime is a principal bundle with a $SL(2, \mathbb{R})\times U(1)$ fibre over a K\"ahler with torsion manifold $B^6$
with Hermitian form $\omega_{(6)}$ \cite{hethor}.
By using the algebraic KSE, one can show that 
\begin{equation}
N ( I ) = 0 \ , 
\end{equation}
where $N$ is the Nijenhuis tensor of $I$. The fact that $I$ is a bilinear in terms of Killing spinors implies that $I$ is an isometry of $B^6$. 
The Hermitian form $\omega_{(8)}$ generated as mixed bilinear decomposes in terms of $\omega_{(6)}$ as
\begin{equation}
\omega_{(8)} = k^{-2} h \w \ell + \omega_{(6)} \ .
\end{equation}
The condition $\tilde{\nabla}^{(-)}_{(8)} \omega_{(8)} = {\cal O}(\alpha'^2)$ reduces in terms of data of $B^6$ as 
\begin{equation}
\tilde{\nabla}^{(-)}_{(6)} \omega_{(6)} = {\cal O}(\alpha'^2) \ . 
\end{equation}
The spacetime torsion can be written as
\begin{eqnarray}
H=CS(\lambda)+\tilde H_{(6)}+{\cal O}(\alpha'^2)~,
\end{eqnarray}
where the Chern-Simons 3-form is given as in (\ref{CS_exp}), and the 3-form flux on $B^6$ is 
\begin{eqnarray}
\tilde H_{(6)}=-i_I d\omega+\mathcal{O}(\alpha'^2) =e^{2 \Phi} \star_6 d [e^{-2\Phi} \omega_{(6)}]+\mathcal{O}(\alpha'^2) \ .
\end{eqnarray}
In addition, the algebraic KSE implies that the K\"ahler with torsion manifold $B^6$ is conformally balanced, i.e.
\begin{eqnarray}
\theta_{\omega_{(6)}}=2d\Phi+\mathcal{O}(\alpha'^2)~,
\end{eqnarray}
where $\theta_{\omega_{(6)}}$ is the Lee form of $\omega_{(6)}$, together with the condition
\begin{equation}
\partial_{a} \Phi = \mathcal{O}(\alpha'^2) \ , 
\end{equation}
i.e. the dilaton $\Phi$ depends only on the coordinates of $B^6$. 
The system of equations for the geometric data of $SU(3)$ holonomy heterotic near-horizon geometries are 
\begin{eqnarray}
\label{SU3_cond_2}
\notag
&&k^{-2} dh\wedge dh+k^{-2} d\ell\wedge d\ell+ d\Big(e^{2 \Phi}\star_6 d [e^{-2\Phi} \omega]\Big)=\\
&&\qquad\qquad-{\alpha'\over4} \bigg(-2 dh\wedge dh+ \mathrm {tr}( R^{(+)}_{(8)}\wedge R^{(+)}_{(8)}- F\wedge F)\bigg)+{\cal O}(\alpha'^2) \ , \\
\label{SU3_cond_3}
&&dh^{2,0}=d\ell^{2,0}=\mathcal{O}(\alpha'^2)~,~~~dh_{ij} \omega_{(6)}^{ij}=\mathcal{O}(\alpha'^2)~,~~~d\ell_{ij} \omega_{(6)}^{ij}=-2 k^2+\mathcal{O}(\alpha'^2) \ , \qquad \\
\label{SU3_cond_4}
&&F^{2,0}=\mathcal{O}(\alpha')~,~~~F_{ij} \omega_{(6)}^{ij}=\mathcal{O}(\alpha')~. \\
\label{SU3_cond_1}
&&\tilde{R}^{(-)}_{(6)}{}_{ij} \omega_{(6)}^{ij}=-2 k^2 d\ell+\mathcal{O}(\alpha'^2) \ , 
\end{eqnarray}
The condition (\ref{SU3_cond_2}) is the anomalous Bianchi identity, where $R^{(+)}_{(8)}$ is the curvature of the connection with torsion on ${\cal S}$. The metric and torsion on $\cS$ are given by
\begin{eqnarray}
d\tilde s^2&=&k^{-2} (h\otimes h+\ell\otimes\ell)+d\tilde s_{(6)}^2+\mathcal{O}(\alpha'^2)~,
\nonumber \\
\tilde H&=& k^{-2}  (h\wedge dh+\ell\wedge d\ell)+\tilde H_{(6)}+\mathcal{O}(\alpha'^2)~.
\end{eqnarray}
Note that $\nabla^{(-)}_{(8)}$ has holonomy contained in $SU(3)$ and so $R^{(+)}_{(8)}$ is a well defined form on $B^6$.
The conditions (\ref{SU3_cond_3}) and (\ref{SU3_cond_4}), which are implied by (\ref{auxalg2}) and (\ref{auxalg2c}), states that both $h$ and the gauge connection are $\mathfrak{su}(3)$ instantons on $B^6$, while $\ell$ is a $\mathfrak{u}(3)$ instanton on $B^6$. 
Finally, the condition (\ref{SU3_cond_1}) ensures that the $U(3)$ structure on $B^6$ lifts to a $SU(3)$ structure on the spacetime, or equivalently, on the spatial horizon section ${\cal S}$.

\section{Nearly supersymmetric horizons with $G_2$ structure}

In this section, we shall consider solutions for which the
supersymmetry is explicitly partially broken, in the sense that such near-horizon geometries admit spinors which satisfy the
gravitino KSE ({\ref{gravsimp}}) but do not satisfy the algebraic KSE ({\ref{algsimpmax}}). 
Our assumptions are the following:
\begin{enumerate}
\item[\emph{(i)}]  there exists exactly \emph{one} $\eta_+$, with $\eta_+^{[0]} \neq 0$, solution to the gravitino KSE
\begin{eqnarray}
\label{covcon1}
\tilde{\nabla}^{(-)} \eta_+ = \mathcal{O}(\alpha'^2) \ ,
\end{eqnarray}
\item[\emph{(ii)}] the spinor $\eta_+$ does not satisfy the algebraic KSE, i.e.
\begin{equation}
\big({\cal{A}} \eta_+\big)^{[0]} \neq 0 \ . 
\end{equation}
\item[\emph{(iii)}] The fields $\D$ and $H$ satisfy
\begin{eqnarray}
\Delta = \mathcal{O}(\alpha'^2), \qquad\quad 
H = d (\mathbf{e}^- \wedge \mathbf{e}^+) +W + \mathcal{O}(\alpha'^2) \ .
\label{nearh}
\end{eqnarray}
\end{enumerate}
We recall that the condition \emph{(i)} does not exclude the possibility to have a second spinor which is $\mathcal{O}(\alpha')$ and satisfies (\ref{covcon1}). 
The condition \emph{(iii)} was previously obtained via the supersymmetry analysis; here we shall assume it.
In particular, all of the conditions obtained from
the global analysis of the Laplacian of $h^2$ in section \ref{sec: max h^2}
remain true. 

Let us define the spinor $\tau_+$
\begin{equation}
\label{tau_+}
\tau_+ \equiv \mathcal{A} \eta_+  \ , 
\end{equation}
which is not vanishing at zeroth order in $\alpha'$. 
Then we have 
\begin{theorem}
\label{th:extra_parallel}
Suppose there exists one $\eta_+$ which satisfies (i) and (ii). Then $\tau_+$ defined in (\ref{tau_+}) is $\tilde{\nabla}^{(-)}$-parallel, i.e. 
\begin{equation}
\tilde{\nabla}^{(-)} \tau_+ = \mathcal{O}(\alpha'^2) \ . 
\end{equation}
\end{theorem}
\begin{proof}
First, we note the following useful identity 
\begin{eqnarray}
\tilde{\nabla}_i W_{\ell_1 \ell_2 \ell_3} \Gamma^{\ell_1 \ell_2 \ell_3} \eta_+ &=&
\tilde{\nabla}_i (\mathcal{A} \eta_+) -{1 \over 8} W_{i \ell_1 \ell_2} \Gamma^{\ell_1 \ell_2}
(\mathcal{A} \eta_+)
\nonumber \\
&+&3 W_{\ell_1 \ell_2 q} W_{i \ell_3}{}^q \Gamma^{\ell_1 \ell_2 \ell_3} \eta_+
-\big(6 \tilde{\nabla}^m \Phi+3 h^m\big) W_{mi\ell} \Gamma^\ell \eta_+
\nonumber \\
&+&\big(12 \Gamma^\ell \tilde{\nabla}_i \tilde{\nabla}_\ell \Phi +6 \tilde{\nabla}_i h_\ell \Gamma^\ell\big) \eta_+ \ .
\end{eqnarray}
The integrability conditions of ({\ref{covcon1}}) imply that
\begin{eqnarray}
\label{ksenil1}
{1 \over 6} \bigg(\tilde{\nabla}_i (\mathcal{A} \eta_+) -{1 \over 8} W_{i \ell_1 \ell_2}
\Gamma^{\ell_1 \ell_2} (\mathcal{A} \eta_+) \bigg)
-{\alpha' \over 8} ({\tilde{F}}_{i \ell})_{ab} \Gamma^\ell ({\tilde{F}}_{q_1 q_2})^{ab}
\Gamma^{q_1 q_2} \eta_+
\nonumber \\
-{\alpha' \over 16}  dh_{i \ell} \Gamma^\ell dh_{q_1 q_2} \Gamma^{q_1 q_2} \eta_+ = \mathcal{O}(\alpha'^2)~,
\end{eqnarray}
and hence
\begin{eqnarray}
{1 \over 6} \langle \eta_+, \Gamma^i \tilde{\nabla}_i(\mathcal{A} \eta_+) -{1 \over 8} W_{\ell_1 \ell_2 \ell_3}
\Gamma^{\ell_1 \ell_2 \ell_3} (\mathcal{A} \eta_+) \rangle
+{\alpha' \over 8} \langle ({\tilde{F}}_{\ell_1 \ell_2})_{ab} \Gamma^{\ell_1 \ell_2} \eta_+
, ({\tilde{F}}_{q_1 q_2})^{ab} \Gamma^{q_1 q_2} \eta_+ \rangle
\nonumber \\
+{\alpha' \over 16} \langle dh_{\ell_1 \ell_2} \Gamma^{\ell_1 \ell_2} \phi_+,
dh_{q_1 q_2} \Gamma^{q_1 q_2} \eta_+ \rangle = \mathcal{O}(\alpha'^2) \ .
\nonumber \\
\end{eqnarray}
Integrating this expression over ${\cal{S}}$ yields the conditions
\begin{eqnarray}
\label{F_dh_cond}
{\tilde{F}}_{ij} \Gamma^{ij} \eta_+ = \mathcal{O}(\alpha'), \qquad
dh_{ij} \Gamma^{ij} \eta_+ = \mathcal{O}(\alpha')~,
\end{eqnarray}
and substituting these conditions back into ({\ref{ksenil1}})  implies that
\begin{eqnarray}
\tilde{\nabla}_i (\mathcal{A} \eta_+) -{1 \over 8} W_{i \ell_1 \ell_2} \Gamma^{\ell_1 \ell_2} (\mathcal{A} \eta_+)=\mathcal{O}(\alpha'^2) \ .
\label{naeta}
\end{eqnarray}
\end{proof}
The theorem \ref{th:extra_parallel} establishes that if there exists one spinor parallel with respect to $\tilde{\nabla}^{(-)}$, which do not satisfy the algebraic KSE, then one can constract a second spinor which is also $\tilde{\nabla}^{(-)}$-parallel. This important feature will allow us to describe the geometry of $\cS$. 
We remark that the presence of an additional $\tilde{\nabla}^{(-)}$-parallel spinor is not in contradiction with the assumption \emph{(i)} because $\tau_+$ is an odd spinor, while $\eta_+$ is even\footnote{We recall that the spinor $\G_{\ell_1 ... \ell_n} \eta_+$
is \emph{even} (resp. \emph{odd}) if $n$ is even (resp. odd).}.

The existence of the spinors $\eta_+$ and $\tau_+$ implies that at zeroth order in $\alpha'$ the spatial cross section ${\cal{S}}^{[0]}$ admits a $G_2$ structure. Furthermore, as a corollary of the theorem \ref{th:extra_parallel}, the $G_2$ fundamental 3-form is $\tilde{\nabla}^{(-)}$-parallel, and therefore the connection $\tilde{\nabla}^{(-)}$ of ${\cal{S}}^{[0]}$ has $G_2$ holonomy. 

Next, we show that the existence of extra $\tilde{\nabla}^{(-)}$-parallel spinors implies the existence of extra isometries of the spatial cross section.  
We define the following vector field
\begin{eqnarray}
\label{vecV}
V_i \equiv \langle \eta_+, \Gamma_i \tau_+ \rangle \ , 
\end{eqnarray}
where $\tau_+$ is defined in (\ref{tau_+}). Then we have
\begin{theorem}
\label{th:V_symmetry}
$V$ is a symmetry of all bosonic fields on $\cS$ up to $\mathcal{O}(\alpha'^2)$ terms.  
\end{theorem}

\begin{proof}
As $\tau_+^{[0]} \neq 0$, this  implies that $V^{[0]} \neq 0$.
In addition, as $\eta_+$ and $\tau_+$ satisfy
\begin{eqnarray}
\label{eta_tau}
\tilde{\nabla}^{(-)} \eta_+ = \mathcal{O}(\alpha'^2)~, \qquad \tilde{\nabla}^{(-)} \tau_+
=\mathcal{O}(\alpha'^2)~,
\end{eqnarray}
it follows that
\begin{eqnarray}
\label{parallelV}
\tilde{\nabla}^{(-)} V = \mathcal{O}(\alpha'^2)~,
\end{eqnarray}
so that $V^2=const. + \mathcal{O}(\alpha'^2)$, and $V$ is an isometry of
${\cal{S}}$ to both zero and first order in $\alpha'$.

Next, we consider the relationship of $V$ to $h$. In particular, the
spinors $h_i \Gamma^i {\cal{A}} \eta_+$ and $V_i \Gamma^i {\cal{A}} \eta_+$
are both parallel with respect to $\tilde{\nabla}^{(-)}$ at zeroth order in $\alpha'$. As we have assumed that ({\ref{covcon1}}) admits only one
solution, there must be a nonzero constant $c$ such that
\begin{eqnarray}
V = ch+ \mathcal{O}(\alpha')~.
\end{eqnarray}

Next we consider $\cL_V W$, and we have
\begin{eqnarray}
\mathcal{L}_V W = i_V dW + \mathcal{O}(\alpha'^2)~,
\end{eqnarray}
because $dV=i_V W + \mathcal{O}(\alpha'^2)$ from (\ref{parallelV}). Also, as $V=ch+\mathcal{O}(\alpha')$ it follows that
\begin{eqnarray}
\mathcal{L}_V W = c i_h dW + \mathcal{O}(\alpha'^2)~.
\end{eqnarray}
As a consequence of $\tilde{\nabla}^{(-)}_i h_j =\mathcal{O}(\alpha')$, one has that $i_h dh=\mathcal{O}(\alpha')$,
and from the global analysis of the Laplacian of $h^2$, we find
$i_h {\tilde{F}}=\mathcal{O}(\alpha')$ as well as  $\tilde{R}^{(+)}{}_{mnij} h^m = \mathcal{O}(\alpha')$.
These conditions imply that
\begin{eqnarray}
\mathcal{L}_V W = \mathcal{O}(\alpha'^2)~,
\end{eqnarray}
and so $W$ is invariant.

Next we consider $\mathcal{L}_V \Phi$. As $V = c h + \mathcal{O}(\alpha')$ it follows that
\begin{eqnarray}
\mathcal{L}_V dh = c \mathcal{L}_h dh + \mathcal{O}(\alpha') = \mathcal{O}(\alpha')~.
\end{eqnarray}
Also we have
\begin{eqnarray}
\mathcal{L}_V {\tilde{R}}_{ij,pq} = \mathcal{O}(\alpha'^2)~,
\end{eqnarray}
and
\begin{eqnarray}
\big(\mathcal{L}_V {\tilde{F}}\big)_{ij}{}^a{}_b \tilde{F}^{ijb}{}_a
= \mathcal{O}(\alpha')~,
\end{eqnarray}
which follows from
\begin{eqnarray}
\mathcal{L}_V {\tilde{F}} = c [{\tilde{F}}, i_h {\cal{B}}]+ \mathcal{O}(\alpha')~.
\end{eqnarray}
Hence we have
\begin{eqnarray}
\mathcal{L}_V \bigg( \alpha' \big(-2dh_{ij} dh^{ij} + \tilde{R}^{(+)}{}_{ij,pq}
\tilde{R}^{(+)}{}^{ij,pq} - ({\tilde{F}}_{ij})^{ab} ({\tilde{F}}^{ij})_{ab} \big) \bigg) = \mathcal{O}(\alpha'^2)~.
\end{eqnarray}
So, on taking the Lie derivative of the trace of ({\ref{einsp}}) with respect to $V$ we find
\begin{eqnarray}
\label{L_trEins}
\mathcal{L}_V \bigg( \tilde{\nabla}^i h_i +2 \tilde{\nabla}_i \tilde{\nabla}^i \Phi \bigg) =\mathcal{O}(\alpha'^2)~,
\end{eqnarray}
and hence, as a consequence of the field equation ({\ref{geq1a}}), we find
\begin{eqnarray}
\label{liex1}
\mathcal{L}_V \bigg( h^i \tilde{\nabla}_i \Phi + \tilde{\nabla}^i \tilde{\nabla}_i \Phi \bigg) = \mathcal{O}(\alpha'^2)~.
\end{eqnarray}
Also, on taking the Lie derivative of the dilaton field equation
({\ref{deqsimp1}}), we get
\begin{eqnarray}
\label{liex2}
\mathcal{L}_V \bigg(-h^i \tilde{\nabla}_i \Phi -2 \tilde{\nabla}_i \Phi \tilde{\nabla}^i \Phi + \tilde{\nabla}^i \tilde{\nabla}_i \Phi \bigg) = \mathcal{O}(\alpha'^2) \ .
\end{eqnarray}
On taking the sum of ({\ref{liex1}}) and ({\ref{liex2}}), we find
\begin{eqnarray}
\mathcal{L}_V \bigg( \tilde{\nabla}^i \tilde{\nabla}_i \Phi - \tilde{\nabla}^i \Phi \tilde{\nabla}_i \Phi \bigg) = \mathcal{O}(\alpha'^2) \ ,
\end{eqnarray}
and hence if $f= \mathcal{L}_V \Phi$ we have
\begin{eqnarray}
\label{laplx3}
\tilde{\nabla}_i \tilde{\nabla}^i f -2 \tilde{\nabla}^i \Phi \tilde{\nabla}_i f = \mathcal{O}(\alpha'^2)~.
\end{eqnarray}
We know $\mathcal{L}_h \Phi = \mathcal{O}(\alpha')$ as a consequence of the analysis
of the Laplacian of $h^2$, so $f=\alpha' f^{[1]}+ \mathcal{O}(\alpha'^2)$.
Then, on integrating, ({\ref{laplx3}}) implies that
\begin{eqnarray}
\int_{{\cal{S}}^{[0]}} e^{-2 \Phi^{[0]}} \tilde{\nabla}_i f^{[1]} \tilde{\nabla}^i f^{[1]} = 0~,
\end{eqnarray}
so $f^{[1]}=\beta$ for constant $\beta$, and so
\begin{eqnarray}
\mathcal{L}_V \Phi = \beta \alpha' + \mathcal{O}(\alpha'^2)~.
\end{eqnarray}
As we require that $\Phi$ must attain a global maximum on ${\cal{S}}$,
at this point $\mathcal{L}_V \Phi=0$ to all orders in $\alpha'$, for any $V$.
This fixes $\beta=0$, so
\begin{eqnarray}
\label{L_Phi}
\mathcal{L}_V \Phi = \mathcal{O}(\alpha'^2)~,
\end{eqnarray}
which proves the invariance of $\Phi$.

Next, we consider $\mathcal{L}_V h$. On taking the Lie derivative of the field  equation of the 2-form
gauge potential ({\ref{geq1c}}) we find
\begin{eqnarray}
\label{Lie_gauge}
d (\mathcal{L}_V h)_{ij} - (\mathcal{L}_V h)^k W_{ijk} = \mathcal{O}(\alpha'^2)~,
\end{eqnarray}
and on taking the Lie derivative of the Einstein equation
({\ref{einsp}}) we get
\begin{eqnarray}
\label{Lie_einst}
\tilde{\nabla}_{(i} (\mathcal{L}_V h)_{j)} = \mathcal{O}(\alpha'^2)~,
\end{eqnarray}
where we have used
\begin{eqnarray}
\mathcal{L}_h \bigg( {\tilde{F}}_{i\ell}{}^{ab} {\tilde{F}}_j{}^\ell{}_{ab} \bigg) = \mathcal{O}(\alpha')~.
\end{eqnarray}
It follows that
\begin{eqnarray}
\tilde{\nabla}^{(-)}_i (\mathcal{L}_V h)_j = \mathcal{O}(\alpha'^2)~.
\end{eqnarray}
As $V=ch+ \mathcal{O}(\alpha')$, it is convenient to write
\begin{eqnarray}
\mathcal{L}_V h = \alpha' \Lambda+\mathcal{O}(\alpha'^2)~,
\end{eqnarray}
where
\begin{eqnarray}
\tilde{\nabla}^{(-)} \Lambda = \mathcal{O}(\alpha')~.
\end{eqnarray}
As $\Lambda_j \Gamma^j {\cal{A}} \eta_+$ and $h_j \Gamma^j {\cal{A}} \eta_+$
are both parallel with respect to $\tilde{\nabla}^{(-)}$ at zeroth order in $\alpha'$, and since we assumed that there is only one spinor parallel with respect to $\tilde{\nabla}^{(-)}$ at zeroth order in $\alpha'$, we must have
\begin{eqnarray}
\Lambda = b h + \mathcal{O}(\alpha')~,
\end{eqnarray}
for constant $b$. It is also useful to compute
\begin{eqnarray}
\label{hL_h}
h^i (\mathcal{L}_V h_i) = h^i \bigg( V^j \tilde{\nabla}_j h_i + h_j \tilde{\nabla}_i V^j \bigg)
= {1 \over 2} \mathcal{L}_V h^2 + h^i h^j \tilde{\nabla}_i V_j = \mathcal{O}(\alpha'^2)~,
\end{eqnarray}
which follows because $h^2 = \mathrm{const} + \mathcal{O}(\alpha'^2)$, and $\tilde{\nabla}^{(-)} V = \mathcal{O}(\alpha'^2)$.
This implies  that $b=0$, and hence
\begin{eqnarray}
\mathcal{L}_V h = \mathcal{O}(\alpha'^2)~.
\end{eqnarray}
So $V$ is a symmetry of the full solution to both zeroth and first order in
$\alpha'$.
\end{proof}

\subsection{Description of the geometry of $\cS$}
In this section we use the results of the theorems \ref{th:extra_parallel} and \ref{th:V_symmetry} to describe the geometry of $\cS$.  
The existence of one spinor $\eta_+$ which is parallel with respect to $\tilde{\nabla}^{(-)}$ implies that the structure group $Spin(8)$ of $\cS$ reduces to $Spin(7)$, i.e. the isotropy group of $\eta_+$, and the fundamental self-dual 4-form $\phi$ of $Spin(7)$ in $\cS$ satisfies 
\begin{equation}
\tilde{\nabla}^{(-)} \phi = \mathcal{O}(\alpha'^2) \ . 
\end{equation} 
The torsion $W$ is uniquely determined in terms of the metric on $\cS$ and the self-dual 4-form $\phi$, without any additional condition on the $Spin(7)$ structure on $\cS$ \cite{ivanovspin7}. 

The theorem \ref{th:extra_parallel} states that there exists an additional $\tilde{\nabla}^{(-)}$-parallel spinor, i.e. $\tau_+$, and together with $\eta_+$, it generates the vector field $V$, which is an isometry, symmetry of all bosonic fields on $\cS$, and $\tilde{\nabla}^{(-)}$-parallel. This implies that the $Spin(7)$  structure group of $\cS$ reduces to $G_2$, where the fundamental 3-form is given by 
\begin{equation}
\label{varphi}
\varphi = \frac{1}{\ell} i_V \phi \ , \qquad
\tilde{\nabla}^{(-)}  \varphi = \mathcal{O}(\alpha'^2) \ , 
\end{equation}
where we set $V^2 = \ell^2 + \mathcal{O}(\alpha'^2)$, with $\ell$ constant. This also implies that we can decompose the metric and the 3-form $W$ as follows
\begin{eqnarray}
d\tilde s^2={1\over\ell^2} V\otimes V+ds^2_{(7)}+\mathcal{O}(\alpha'^2)\ , \qquad 
W=\ell^{-2}V\wedge dV+W_{(7)}+\mathcal{O}(\alpha'^2) \ ,
\end{eqnarray}
where $ds^2_{(7)}$ is (locally) the metric on $M^7 \subset \cS$, which is the space of orbits of $V$ in $\cS$, and it is orthogonal to $V$. The 3-form $W_{(7)}$ is the torsion 3-form flux projected on $M^7$, which is orthogonal to $V$, i.e. $i_V W_{(7)} = \mathcal{O}(\alpha'^2)$. This follows from the fact that $dV = i_V W + \mathcal{O}(\alpha'^2)$. 

The question now is whether $M^7$ inherits the $G_2$ structure inside $\cS$. 
In what follows we shall prove that this is the case. To prove this we need to show that $\cL_V \varphi = \mathcal{O}(\alpha'^2)$. 
First we notice that (\ref{varphi}) implies $i_V \varphi = \mathcal{O}(\alpha'^2)$. Then we consider $dV$, which decomposes under the splitting induced from the $G_2$ 3-form $\varphi$ as 
\begin{equation}
dV = dV^{{\bf 7}} + dV^{{\bf 14}} + \mathcal{O}(\alpha'^2) \ , 
\end{equation}
where we used the fact that $i_V dV=\mathcal{O}(\alpha'^2)$. 
Then we use (\ref{bianx}) together with
$\tilde{\nabla}^{(-)} \varphi=\tilde{\nabla}^{(-)} V=\mathcal{O}(\alpha'^2)$ and $i_V dW={\cal O}(\alpha'^2)$ to show that
\begin{eqnarray}
\tilde{\nabla}^{(-)} dV^{\bf 7}=\mathcal{O}(\alpha'^2)~.
\end{eqnarray}
As $dV^{\bf 7}$ is a vector\footnote{$dV^{\bf 7}$ is a 2-form dual to a vector via the contraction with the 3-form $\varphi$. } in ${\cal S}$ orthogonal to $V$, if it does not vanish, it will generate an additional $\tilde{\nabla}^{(-)}$-parallel spinor on ${\cal S}$ of the same chirality as $\eta_+$.  As we have restricted
the number of such spinors to one, we have to set 
\begin{equation}
dV^{\bf 7}=\mathcal{O}(\alpha'^2) \ . 
\end{equation}
Consider now a generic $k$-form $\zeta$. One can show that if $U$ is a generic vector field such that
\begin{equation}
\tilde{\nabla}^{(-)} U = \mathcal{O}(\alpha'^2)	 \ , 
\end{equation} 
then the statement
\begin{equation}
\cL_U \zeta = \mathcal{O}(\alpha'^2) \ , 
\end{equation}
is equivalent to
\begin{equation}
i_U W_{[\ell_1}{}^q \zeta_{|q| \ell_2 ... \ell_k]} = \mathcal{O}(\alpha'^2) \ , 
\end{equation}
which means that $\zeta$ is invariant under the infinitesimal rotation generated by $i_U W$.

In our case $U=V$ and $\zeta = \varphi$. 
Since $i_V W=dV+\mathcal{O}(\alpha'^2)$, and $dV$ takes values in $\mathfrak{g}_2$, by definition of $\varphi$ we conclude that
\begin{eqnarray}
{\mathcal{L}}_V \varphi=\mathcal{O}(\alpha'^2) \ .
\end{eqnarray}
Therefore $M^7$ admits a $G_2$ structure compatible with connection with skew-symmetric torsion given by the data $(ds^2_{(7)}, W_{(7)})$.  In such a
case $W_{(7)}$ can be determined uniquely in terms of $\varphi$ and $ds^2_{(7)}$, provided
a particular geometric constraint is satisfied \cite{ivanovg2}.

This concludes the description of the geometry of $\cS$ of nearly supersymmetric horizons in the case where there exists one $\tilde{\nabla}^{(-)}$-parallel spinor $\eta_+$. We do not report here the description of the cases where one assumes the existence of more than one $\tilde{\nabla}^{(-)}$-parallel spinors. These cases can be found in \cite{Fontanella:2016aok} and they follow a similar type of analysis presented here for the $G_2$ structure case.

\chapter{\textbf{Bulk extension of a near-horizon geometry}}
\label{chap:bulk}

\section{Radial deformation of a near-horizon geometry}
We are interested in the inverse problem of determining all (extremal) black holes associated with a prescribed near-horizon geometry. The metric of a near-horizon geometry is fully specified via the data $\{ \D(y) , h(y), \gamma(y) \}$, which do only depend on the internal coordinates on $\cS$. In order to extend the horizon into the bulk, one has to recover the $r$-dependence of the near-horizon data which was suppressed by taking the near-horizon limit. 
Therefore our problem can be formulated as follows: given the near-horizon data $\{ \D(y) , h(y), \gamma(y) \}$, can we extrapolate the data $\{ \D(r, y) , h(r, y), \gamma(r, y) \}$? 
Our approach to this problem consists in Taylor expanding in $r$ the horizon data, and to show that the first order deformations of the horizon fields must satisfy ellliptic PDEs. We shall also couple matter fields to the metric field, by considering radial deformations of near-horizon geometries in heterotic supergravity and $D=11$ supergravity. 

Furthermore, we have also considered radial deformations of a near-horizon geometry in Einstein-Maxwell-Dilaton theory in any dimension, including topological terms in $D=4, 5$. The analysis and the final result obtained is very similar to the heterotic and $D=11$ supergravity cases, and the details can be found in \cite{Fontanella:2016lzo}. 
The aim of this chapter is to prove the following
\begin{theorem}
\label{th:finiteness_moduli}
The moduli space of radial deformations of a given near-horizon geometry in heterotic and $D=11$ supergravities is finite dimensional. 
\end{theorem}

\subsection{Metric moduli and gauge freedom}
 
In this section we shall describe the metric moduli, which are common to both heterotic and $D=11$ supergravities. 
Consider a near-horizon geometry with metric given in (\ref{NHG}). We assume that such near-horizon geometry belongs to a black hole, whose event horizon $\cH$ is a Killing horizon, and therefore on a neighbourhood of $\cH$ one can adapt Gaussian null coordinates, where the metric is given in (\ref{Gauss_metric}). 
To determine the components of (\ref{Gauss_metric}), we Taylor expand the horizon data in $r$, 
\begin{eqnarray}
\Delta &=& {\buildrel \circ \over \Delta}(y)+ r \delta \Delta (y) + {{\cal{O}}(r^2)},
\nonumber \\
h &=& {\buildrel \circ \over h} (y) + r \delta h (y) + {{\cal{O}}(r^2)},
\nonumber \\
\gamma &=& {\buildrel \circ \over \gamma}(y)+ r \delta \gamma(y) + {{\cal{O}}(r^2)}
\end{eqnarray}
where $\{ {\buildrel \circ \over \Delta}, {\buildrel \circ \over h}, {\buildrel \circ \over \gamma}\}$ are the zeroth order terms, which are determined by the data of the given near-horizon geometry. The first order radial deformations of the horizon data, which we call \emph{metric moduli}, are given by $\{\delta \Delta, \delta h, \delta \gamma\}$. 
In our analysis, we shall neglect higher order terms, i.e. terms which are not linear in the radial deformation $\delta$, e.g. $\d \Delta \d h$.  
We shall assume that the near-horizon spatial cross section ${\buildrel \circ \over{{\cal{S}}}}$, equipped with metric ${{\buildrel \circ \over \gamma}}$, is compact and without boundary.

\subsubsection{Gauge fixing}
The near-horizon metric (\ref{NHG}) admits two obvious isometries, generated by the vector fields $V$ in (\ref{iso_V}) and $D$ in (\ref{iso_D}).

The horizon metric (\ref{Gauss_metric}) truncated at first order in $r$ admits a 1-parameter family of isometries, generated by the vector fields \cite{Li:2015wsa}
\begin{eqnarray}
K_f = {1 \over 2} f \bigg(dr+r {\buildrel \circ \over h}-{1 \over 2} r^2 {\buildrel \circ \over \Delta} du \bigg)
-{1 \over 4} r^2 \bigg({\buildrel \circ \over \Delta} f + {\cal{L}}_{{\buildrel \circ \over h}} f \bigg) du -{1 \over 2} r df
\end{eqnarray}
where $f$ is an arbitrary smooth function on ${\cal{S}}$, such that $\cL_{K_f} ({\buildrel \circ \over g} + \delta g) = 0$.
Notice that when $f = const$ or $f=u \cdot const$, the vector field $K_f$ is proportional to $V$ and $D$ respectively. 
We remark that the existence of a 1-parameter family of isometries is an artefact of the first order expansion. This is because Gaussian null coordinates are unique, up to fixing coordinates on $\cS$. However this artificial redundancy is expected to disappear once the contributions of all orders in $r$ are summed up in the metric.

Under the diffeomorphisms generated by $K_f$, the horizon data transform as 
\begin{eqnarray}
\label{gtrans}
\delta {{\gamma}}_{ij} &\rightarrow& \delta \gamma_{ij} + {\buildrel \circ \over \nabla}_i {\buildrel \circ \over \nabla}_j f
- {\buildrel \circ \over h}_{(i} {\buildrel \circ \over \nabla}_{j)} f
\nonumber \\
\delta {{h}}_i &\rightarrow& \delta h_i +{1 \over 2} {\buildrel \circ \over \Delta} {{\buildrel \circ \over \nabla}}_i f
-{1 \over 4} ({{\buildrel \circ \over \nabla}}_i {\buildrel \circ \over h}_j) {{\buildrel \circ \over \nabla}}^j f
-{1 \over 4} {\buildrel \circ \over h}_i {\buildrel \circ \over h}_j {{\buildrel \circ \over \nabla}}^j f
+{1 \over 2} ({{\buildrel \circ \over \nabla}}_j {\buildrel \circ \over h}_i) {{\buildrel \circ \over \nabla}}^j f
+{1 \over 4} {\buildrel \circ \over h}_j {{\buildrel \circ \over \nabla}}_i {{\buildrel \circ \over \nabla}}^j f
\nonumber \\
\delta {{\Delta}} &\rightarrow& \delta \Delta +{1 \over 2} {{\buildrel \circ \over \nabla}}^i f
\bigg({{\buildrel \circ \over \nabla}}_i {\buildrel \circ \over \Delta} - {\buildrel \circ \over h}_i {{\buildrel \circ \over \Delta}} \bigg) \ ,
\end{eqnarray}
where indices $i,j,\dots$ are with respect to the orthonormal basis ${\bf{e}}^i\big|_{r=0}$
on ${\buildrel \circ \over{{\cal{S}}}}$, and ${\buildrel \circ \over \nabla}$ denotes the Levi-Civita connection on ${\buildrel \circ \over{{\cal{S}}}}$.

As we shall see later, in order to construct an elliptic PDE for $\d \gamma$ we need to fix the trace modulus $\delta \gamma_k{}^k$. We shall show that this can be achieved via a gauge fixing. 
To see this note that under the transformation ({\ref{gtrans}})
\begin{eqnarray}
\label{trtrans}
\delta \gamma_k{}^k \rightarrow \delta \gamma_k{}^k + {\cal{D}} f
\end{eqnarray}
where ${\cal{D}}$, and its adjoint ${\cal{D}}^\dagger$, are given by
\begin{eqnarray}
{\cal{D}} \equiv {\buildrel \circ \over \nabla}^2 - {{\buildrel \circ \over h}}^i {\buildrel \circ \over \nabla}_i \ , \qquad {\cal{D}}^\dagger = {\buildrel \circ \over \nabla}^2 + {{\buildrel \circ \over h}}^i {\buildrel \circ \over \nabla}_i + {\buildrel \circ \over \nabla}^i {{\buildrel \circ \over h}}_i \ .
\end{eqnarray}
We decompose $\delta \gamma_k{}^k$ as a sum of two terms, $\phi \in {\rm Im} {\cal{D}}$, and $\phi^\perp \in \big({\rm Im} {\cal{D}} \big)^\perp$ as
\begin{eqnarray}
\delta \gamma_k{}^k = \phi + \phi^\perp \ .
\end{eqnarray}
By definition, $\phi$ can be written in terms of an arbitrary smooth function $\tau$, i.e.
\begin{eqnarray}
\phi = {\cal{D}} (\tau) \ ,
\end{eqnarray}
and since $\phi$ is orthogonal to $\phi^{\perp}$, we have
\begin{equation}
0 = (\phi^{\perp}, \phi) = ( {\cal{D}}^\dagger  \phi^{\perp}, \tau ) \ .
\end{equation}
As $\tau$ is arbitrary (because $f$ is arbitrary), this implies 
\begin{equation}
{\cal{D}}^\dagger \phi^\perp =0 \ . 
\end{equation}
Now we fix the gauge: on setting $f=-\tau$ in ({\ref{trtrans}}), we have
\begin{eqnarray}
\delta \gamma_k{}^k = \phi^\perp
\end{eqnarray}
and hence
\begin{eqnarray}
\label{trelliptic}
\bigg({\buildrel \circ \over \nabla}^2 + {{\buildrel \circ \over h}}^i {\buildrel \circ \over \nabla}_i + {\buildrel \circ \over \nabla}^i {{\buildrel \circ \over h}}_i \bigg) \delta \gamma_k{}^k=0 \ .
\end{eqnarray}
This is an elliptic PDE which depends on the near-horizon data only.  This equation can be solved for $\d \gamma_k{}^k$, and admits a finite number of solutions. This fixes the trace modulus. This condition is independent of the matter content of the
theory which we couple to gravity, and is common to heterotic and $D=11$ supergravity theories. 
We shall make use of this result in the analysis of the
metric moduli in the following sections, in particular the linearized Einstein equations include
a Hessian term in $\delta \gamma_k{}^k$. Without the gauge fixing condition, this term
would destroy the ellipticity of the associated equation. However, as the trace modulus
is fixed by ({\ref{trelliptic}}), the linearized Einstein equation acting on the traceless part of $\delta \gamma$ will be elliptic.

\section{Heterotic Supergravity}
In this section we consider the bulk extension of a given near-horizon geometry in heterotic supergravity without $\alpha'$ corrections. The dynamic of the non abelian 2-form $F$ decouples from the rest of the heterotic fields, and therefore without loss of generality we shall set $F=0$. 

We assume that the Killing vector $V = \partial_u$ is a symmetry of the full solution,
\begin{equation}
\cL_{\partial_u} \Phi = \cL_{\partial_u} H = 0 \ . 
\end{equation}
In Gaussian null coordinates the heterotic fields decompose as 
\begin{equation}
\Phi = \Phi ( r, y ) \ , 
\end{equation}
and 
\begin{equation}
H = du \w dr \w N + r du \w Y + dr \w Z + W \ ,
\end{equation}
where $N, Y$ and $W$ are $u$-independent 1- 2- 3-forms on $\cS$ respectively, which survive if one would take the near-horizon limit. Here there is a novel term: $Z$ is a $u$-independent 2-form on $\cS$ which disappears if one would take the near-horizon limit. We assume that all fields are analytic in $r$. 

The Bianchi identity $dH = 0$ decomposes as 
\be
\label{dH=0bulk}
\tilde{d} Y = 0 \ , \qquad \tilde{d} W = 0 \ , \qquad  \tilde{d} N - Y - r \dot{Y} = 0 \ , \qquad \tilde{d} Z - \dot{W} = 0 \ ,  
\ee
where we denote by $\tilde{d}$ the exterior derivative restricted to
hypersurfaces $r=const$, and by $\dot{\xi}$ the Lie derivative of $\xi$ along the vector field ${\partial \over \partial r}$, i.e. 
\begin{eqnarray}
\dot{\xi} \equiv {{\cal L}}_{{\partial \over \partial r}} \xi  \ .
\end{eqnarray}
We Taylor expand the heterotic fields $\Phi, N, Y, W$ as follows
\be
\notag
\Phi &=& \hPhi + r \d \Phi + \rsq  \ , \\
\notag
N &=& \hN + r \d N + \rsq \ , \\
\notag
Y &=& \hY + r \d Y + \rsq \ , \\
W &=& \hW + r \d W + \rsq \ .
\ee
The zeroth order terms $\{ \hPhi, \hN, \hY, \hW \}$ are fixed by the given heterotic near-horizon geometry.
The $Z$ field is peculiar because it enters in $H$ as $dr \w Z$, which scales linearly with $r$, therefore it must be expanded as 
\be
\label{Z_exp}
Z = \d Z + \mathcal{O} (r) \ .
\ee
This is consistent the $\varepsilon$-expansion instead of the $r$-expansion, which requires that under the one parameter family of diffeomorphism
\be
\label{epsilon_exp}
(u, r, y^I ) \longrightarrow (\varepsilon^{-1} u, \varepsilon r , y^I ) \ , \qquad\qquad \varepsilon \in \mathbb{R}_{>0} \ , 
\ee
all moduli scales linearly in $\varepsilon$. 
Moreover this is also consistent with the $--$ component of the Einstein equation, which implies the expansion (\ref{Z_exp}). 
This can be interpreted as a consequence of the fact that the $Z$ term disappears in the near-horizon limit, and therefore ${\buildrel \circ \over Z} = 0$.
The flux moduli are $\d N, \d Y, \d Z, \d W$ and the dilaton modulus is $\d \Phi$.

The moduli of a given heterotic near-horizon geometry are $\{ \d \Delta, \d h, \d \gamma, \d N, \d Y, \d Z, \d W , \d \Phi \}$. 
In what follows we shall show that not all moduli are independent, but some of them can be fixed in terms of the others by using the linearised Bianchi identities and field equations, which are listed in appendix \ref{appx:unc_het}. 
  
The Bianchi identity (\ref{dH=0bulk}) implies that
\begin{equation}
\label{BianchiZW}
\d Y = \frac{1}{2}\tilde{d} \d N   \ ,  \qquad\qquad
\d W  = \tilde{d} \d Z   \ ,
\end{equation}
which we use to fix $\d Y$ and $\d W$ in terms of $\d N$ and $\d Z$. 

The $-i$ component of the gauge field equation for $H$ given in (\ref{gaugeH}) allows us to fix $\d N$ as 
\be
\label{Nmodulus_fix}
\d N_i = \hhn^j \d Z_{ji} +  \d \Phi \hN_i + \d \gamma_{ik} \hN^k - \frac{1}{2} \d \gamma_k{}^k \hN_i + \hh^j \d Z_{ij} + \d Z_{ij} \hhn^j \hPhi  \ .
\ee
Moreover, considering the $-i$ and $+-$ components of the Einstein equations (\ref{-i ein}) and (\ref{+- ein}), we fix $\d h$ and $\d \D$ respectively as follows
\be
\label{hmodulus_fix}
\notag
\d h_i &=& - \frac{1}{2} \hn^j \d \gamma_{ji} + \frac{1}{2} \hhn_i \d \gamma_k{}^k - \frac{1}{4} \hh_i \d \gamma_k{}^k + \frac{1}{2} \d \gamma_{ij} \hh^j \\
&+& \frac{1}{2} e^{-\hPhi} \hN^j \d Z_{ij} + \frac{1}{4} e^{-\hPhi} \d Z_{jk} \hW_i{}^{jk} + \frac{1}{2} \d \Phi \hhn_i \hPhi\ , 
\ee
and 
\be
\label{Dmodulus_fix}
\notag
\d \D &=&  - \frac{1}{6} \d \gamma_{ij} \hhn^i \hh^j + \frac{1}{3} \hhn^i \d h_i - \frac{1}{6} \hh^j\hhn^i \d \gamma_{ij}  + \frac{1}{12} \hh_i \hhn^i \d \gamma_k{}^k \\
\notag
&+& \frac{1}{3} \d \gamma_{ij} \hh^i \hh^j -  \hh^i \d h_i - \frac{1}{6} \hD \d \gamma_k{}^k - \frac{1}{12} \d \gamma_k{}^k \hh_i \hh^i + \frac{1}{4} \d N_i \hN^i e^{-\hPhi}  \\
\notag
&-& \frac{1}{8} \d \gamma_{ij} \hN^i \hN^j e^{-\hPhi} - \frac{1}{8} \d \Phi e^{-\hPhi} ( \hN_i \hN^i - \frac{1}{18} \hW_{ijk} \hW^{ijk} ) + \frac{1}{72} \d W_{ijk} \hW^{ijk} \e\\
\notag
&+& \frac{1}{24} e^{-\hPhi} (\hh \w \hN )^{ij} \d Z_{ij} - \frac{1}{24} e^{-\hPhi} \hY^{ij} \d Z_{ij} - \frac{1}{48} \d \gamma_{ij} \e \hW^i{}_{\ell_1\ell_2} \hW^{j\ell_1\ell_2} \\
&-& \frac{1}{72} \e (\hh \w \d Z )_{ijk} \hW^{ijk} \ . 
\ee
This allows us to reduce the full set of moduli down to $\{ \d Z, \d \gamma, \d \Phi \}$, which we treat as independent.  

In what follows we shall show that the independent moduli $\d Z, \d \gamma$ and $\d \Phi$ must satisfy elliptic PDEs. 
We start with the $\d Z$ moduli.
By taking the divergence of the Bianchi identity (\ref{BianchiZW}), we obtain
\be
\label{ellipticPDE_Z}
\hhn^{\ell} \hhn_{\ell} \d Z_{ij} + 2 \hhn_{[i} \hhn^{\ell} \d Z_{j] \ell} - 2 {\bcc R}_{\ell [i} \d Z^{\ell}{}_{j]} - \hhn^{\ell} \d W_{\ell ij} = 0 \ , 
\ee
where $\buildrel \circ \over{R}$ denotes the Ricci tensor of ${\buildrel \circ \over{{\cal{S}}}}$.  
This is not exactly an elliptic PDE for $\d Z$, except for the Hessian term $\hhn_{[i} \hhn^{\ell} \d Z_{j] \ell}$. However by using the $ij$ component of the gauge field equation (\ref{ijgauge}), we can rewrite $\hhn^{\ell} \d W_{\ell ij}$ as
\be
\hhn^{\ell} \d W_{\ell ij} = 2 \hhn_{[i} \hhn^{\ell} \d Z_{j] \ell}  + \mathcal{F}_{ij} \ , 
\ee
where 
\begin{eqnarray}
\notag
\mathcal{F}_{ij} &\equiv & - 2 \hhn_{[i} ( \hN_{j]}\d \Phi  ) - 2 \hhn_{[i} ( \d \gamma_{j] k} \hN^k) + \hhn_{[i} (\hN_{j]} \d \gamma_k{}^k ) \\
\notag
&-& 2 \hhn_{[i} ( \d Z_{j]k} \hh^k ) - 2 \hhn_{[i} ( \d Z_{j]k} \hhn^k \hPhi ) + 2(\d h\w \hN)_{ij} + 2(\hh \w \d N )_{ij}  \\
\notag
&+& \d h_k \hW^k{}_{ij} + \hh^k \d W_{kij} - \d \gamma_{k\ell} \hh^k \hW^{\ell}{}_{ij} -  \d \gamma_{k\ell} \hhn^k \hPhi \hW^{\ell}{}_{ij} + \hW_{kij}  \hhn^k \d \Phi \\
\notag
&+&  \d W_{kij} \hhn^k \hPhi + \d \gamma_{k\ell} \hhn^k \hW^{\ell}{}_{ij} + \hW^{\ell}{}_{ij} \hhn^k \d \gamma_{k \ell} - \frac{1}{2} \hW_{\ell ij} \hhn^{\ell} \d \gamma_k{}^k \\
\notag
&+& \hhn_k \d \gamma_{\ell [i} \hW_{j]}{}^{k\ell} + \hW^{k\ell}{}_{[j} \hhn_{i]} \d \gamma_{k\ell} - \hhn_{\ell} \d \gamma_{k [i} \hW_{j]}{}^{k \ell} +  \d \Phi \hY_{ij} -  \d \Phi ( \hh\w \hN)_{ij}  \\
\notag
&-& 2 \hD \d Z_{ij} + 2 \d \gamma_{k[i} ( \hY - \hh \w \hN )_{j]}{}^k  - \frac{1}{2} \d \gamma_k{}^k ( \hY - \hh\w\hN)_{ij} - 2 \hh^k (\hh\w \d Z)_{ijk} \\
\notag
&+& \d h_k \hW^k{}_{ij} -  \hhn^k \hPhi ( \hh \w \d Z)_{kij} -  \d \Phi \hh^k \hW_{kij} + \hhn^k ( \hh\w \d Z)_{kij} - \hh^k \d W_{kij}  \\
&-&  \d \gamma_{\ell k} \hh^k \hW^{\ell}{}_{ij} + \frac{1}{2} \d \gamma_k{}^k \hh^{\ell} \hW_{\ell ij} - 2 \d \gamma^{\ell}{}_{[i} \hW_{j]k\ell} \hh^k \ . 
\end{eqnarray}
Then we substitute this expression into ({\ref{ellipticPDE_Z}}); the
$\hhn_{[i} \hhn^{\ell} \d Z_{j] \ell}$ terms cancel out. Furthermore, the terms linear in
$\delta W$, $\d N$, and $\delta h$ are rewritten using 
the Bianchi identity ({\ref{BianchiZW}}) and
({\ref{Nmodulus_fix}}), ({\ref{hmodulus_fix}}), producing
terms linear in $\delta Z, \delta \gamma, \delta \Phi, {\buildrel \circ \over \nabla} \delta Z, {\buildrel \circ \over \nabla} \delta \gamma, {\buildrel \circ \over \nabla} \delta \Phi$. The resulting expression produces
an elliptic PDE for $\delta Z$, with principal symbol generated by ${\buildrel \circ \over \nabla}^2$.

Next we consider the $\d \Phi$ modulus. The linearised dilaton field equation (\ref{lin_dilaton}) provides
\be
\label{elliptic_dilaton}
\notag
&&\hhn^2 \d \Phi - \d \gamma_{ij} \hhn^i \hhn^j \hPhi - \hhn^i \d \gamma_{ij} \hhn^j \hPhi - \frac{1}{2} \hhn^i \d \gamma_k{}^k \hhn_i \hPhi  + \d \gamma_{ij} \hh^i \hhn^j \hPhi \\
\notag
&&- \d h_i \hhn^i \hPhi - \hh_i \hhn^i \d \Phi + \frac{1}{2} \d \Phi \e \hN_i \hN^i - \e \d N_i \hN^i + \frac{1}{2} \e \d \gamma_{ij} \hN^i \hN^j  \\
\notag
&&- \frac{1}{12} \d \Phi \e \hW_{ijk} \hW^{ijk} + \frac{1}{6} \e \d W_{ijk} \hW^{ijk} - \frac{1}{4} \e \d \gamma_{ij} \hW^i{}_{\ell_1\ell_2} \hW^{j\ell_1\ell_2} \\
\notag
&&+ 2 \hD \d \Phi - 2 \hh^i \hhn_i \d \Phi - \d \Phi \hhn_i \hh^i + 2 \hh_i \hh^i \d \Phi + \d \gamma_{ij} \hh^i \hhn^j \hPhi - \d h_i \hhn^i \hPhi \\
\notag
&&- \frac{1}{2} \d \gamma_k{}^k \hh_i \hhn^i \hPhi + \frac{1}{2} \e \hY^{ij} \d Z_{ij} - \frac{1}{2} \e (\hh\w \hN)^{ij} \d Z_{ij} \\
&&- \frac{1}{6} \e \hW^{ijk} (\hh\w\d Z)_{ijk} = 0  \ . 
\ee
After rewriting the terms linear in $\delta W$, $\d N$, and $\delta h$ by using ({\ref{BianchiZW}}),
({\ref{Nmodulus_fix}}) and ({\ref{hmodulus_fix}}), which produces terms linear in $\delta Z, \delta \gamma, \delta \Phi, {\buildrel \circ \over \nabla} \delta Z, {\buildrel \circ \over \nabla} \delta \gamma, {\buildrel \circ \over \nabla} \delta \Phi$, we obtain an elliptic PDE for $\delta \Phi$, with principal symbol generated by ${\buildrel \circ \over \nabla}^2$.

Finally, we consider the metric moduli $\d \gamma$. The linearised $ij$ component of the Einstein equation produces the following equation
\be
\label{ij_Einst_moduli}
\hhn^2 \d \gamma_{ij}  - \hhn_i \hhn_j \d \gamma_k{}^k + (\hhn_j \hhn_k- \hhn_k \hhn_j ) \d \gamma^k{}_i + ( \hhn_i \hhn_k - \hhn_k \hhn_i ) \d \gamma^k{}_j = \mathcal{H}_{ij} \ , 
\ee
where 
\be
\notag
\mathcal{H}_{ij} &\equiv &  - 2 \hh^k \hhn_{(i} \d \gamma_{j) k} + \hh_{(i} \hhn_{j)} \d \gamma_k{}^k + 2 \d \gamma_{k (i} \hhn_{j)} \hh^k - 2 \d Z_{k(i} \hhn_{j)} ( \e \hN^k)\\
\notag 
&+& 2 \e \hN^k \hhn_{(i} \d Z_{j) k} + \hhn_{(i} ( \e \hW_{j)}{}^{k\ell} ) \d Z_{k\ell} + \e \hW^{k\ell}{}_{(i} \hhn_{j)} \d Z_{k\ell}  \\
\notag
&+& 2 \hhn_{(i} \d \Phi \hhn_{j)} \hPhi + 2 \d \Phi \hhn_i \hhn_j \hPhi - 8 \d h_{(i} \hh_{j)} + 2 ( - \hD + \frac{1}{2} \hhn_k \hh^k - \hh_k \hh^k ) \d \gamma_{ij} \\
\notag
&-& 2 \d \gamma_{k (i} \hhn^k \hh_{j)} + 3 \hh^k \hhn_k \d \gamma_{ij} - 2 \hh_{(i} \hhn^k \d \gamma_{j)k} + 4 \hh^k \hh_{(i} \d \gamma_{j) k} - \hh_i \hh_j \d \gamma_k{}^k  \\
\notag
&-& \d \Phi \e \hN_i \hN_j + 2 \e \d N_{(i} \hN_{j)} + \frac{1}{2} \d \Phi \e \hW_{i\ell_1\ell_2} \hW_j{}^{\ell_1\ell_2} - \e \d W_{\ell_1\ell_2(i} \hW_{j)}{}^{\ell_1\ell_2}  \\
\notag
&+& \e \d \gamma_{\ell_1\ell_2} \hW_{ik}{}^{\ell_1} \hW_j{}^{k\ell_2} + \frac{1}{4} \d \Phi \e \h\gamma_{ij} \hN_k \hN^k - \frac{1}{4} \e \d \gamma_{ij} \hN_k \hN^k - \frac{1}{2} \e \h\gamma_{ij} \d N_k \hN^k  \\
\notag
&+& \frac{1}{4} \e \h\gamma_{ij} \d \gamma_{k\ell} \hN^k \hN^{\ell} - \frac{1}{24} \d \Phi \e \h\gamma_{ij} \hW_{\ell_1\ell_2\ell_3} \hW^{\ell_1\ell_2\ell_3} + \frac{1}{24} \e \d \gamma_{ij} \hW_{\ell_1\ell_2\ell_3} \hW^{\ell_1\ell_2\ell_3} \\
\notag
&+& \frac{1}{12} \e \h\gamma_{ij} \d W_{\ell_1\ell_2\ell_3} \hW^{\ell_1\ell_2\ell_3} - \frac{1}{8} \e \h\gamma_{ij} \d \gamma_{mn} \hW^m{}_{\ell_1\ell_2} \hW^{n \ell_1\ell_2}  - 2 \hhn_{(i} \d \Phi \hhn_{j)} \hPhi \\
\notag
&+& 2 \e \hY^k{}_{(i} \d Z_{j) k} - 2 \e (\hh\w \hN )^k{}_{(i} \d Z_{j)k} + \e \hW^{\ell_1\ell_2}{}_{(i} (\hh\w \d Z)_{j)\ell_1\ell_2}  \\
\notag
&+& \frac{1}{4} \h\gamma_{ij} \hY^{\ell_1\ell_2} \d Z_{\ell_1\ell_2} - \frac{1}{4} \h\gamma_{ij} ( \hh\w \hN)^{\ell_1\ell_2} \d Z_{\ell_1\ell_2} - \frac{1}{12} \h\gamma_{ij} ( \hh\w \d Z)_{\ell_1\ell_2\ell_3} \hW^{\ell_1\ell_2\ell_3}  \\
&+& 2 \d \Phi \hh_{(i} \hhn_{j)} \hPhi \ . 
\ee
which is a linear expression in $\delta Z, \delta \Phi, \delta \gamma, {\buildrel \circ \over \nabla} \delta Z, {\buildrel \circ \over \nabla} \delta \Phi, {\buildrel \circ \over \nabla} \delta \gamma$. Furthermore, in ({\ref{ij_Einst_moduli}}), terms of the form $( {\buildrel \circ \over \nabla}_{\ell} {\buildrel \circ \over \nabla}_j - {\buildrel \circ \over \nabla}_j {\buildrel \circ \over \nabla}_{\ell} ) \delta g^{\ell}{}_i$ can be rewritten as terms linear in $\delta \gamma$ and the Riemann tensor ${\buildrel \circ \over R}$, hence can be incorporated into the algebraic term on the RHS. The trace term $\delta \gamma_k{}^k$
is fixed by the elliptic condition ({\ref{trelliptic}}), so ({\ref{ij_Einst_moduli}})
is an elliptic set of PDEs for the traceless part of $\delta \gamma$, with principal symbol generated by ${\buildrel \circ \over \nabla}^2$.

Taken together the conditions ({\ref{ellipticPDE_Z}}), ({\ref{elliptic_dilaton}}), ({\ref{ij_Einst_moduli}})
and ({\ref{trelliptic}}) constitute elliptic PDEs on the moduli $\delta Z$, $\delta \Phi$, $\delta \gamma$. The remaining moduli $\{\delta \Delta, \delta h, \delta N, \delta Y, \delta W \}$ are fixed in terms of $\{\delta Z, \delta \Phi, \delta \gamma \}$
by ({\ref{BianchiZW}}), ({\ref{Nmodulus_fix}}), ({\ref{hmodulus_fix}}) and ({\ref{Dmodulus_fix}}). From standard Fredholm theory, it follows that the moduli space is finite dimensional.

\section{$D=11$ Supergravity}
\label{sec:mod_space_11D}
In this section we shall investigate the bulk extension of a given near-horizon geometry in $D=11$ supergravity, and the procedure that we will follow is similar to the heterotic case. 
The bosonic field content of $D=11$ supergravity is the $D=11$ metric $g$, and a 4-form $F$, $F=dC$. 
The field equations and their decomposition in Gaussian null coordinates is given in appendix \ref{appx:D=11}. Since our analysis is purely bosonic, the coefficient $q$ of the Chern-Simons topological term is kept arbitrary. 

We shall assume that the Killing vector ${\partial \over \partial u}$ is a symmetry of the full solution, i.e.
\begin{eqnarray}
{{\cal L}}_{{\partial \over \partial u}} F = 0 \ ,
\end{eqnarray}
We decompose $F$ in Gaussian null co-ordinates as
\begin{eqnarray}
\label{4form}
F = du \wedge dr \wedge \Psi +r du \wedge W + dr \wedge Z + X
\end{eqnarray}
where $\Psi$ is a $u$-independent 2-form, $W, Z$ are $u$-independent 3-forms, and $X$ is a $u$-independent 4-form on ${{\cal S}}$, which are all assumed to be analytic in $r$.

The Bianchi identity $dF=0$ decomposes as
\begin{eqnarray}
{\tilde{d}} \Psi -W -r \dot{W}=0, \qquad {\tilde{d}} W =0, \qquad {\tilde{d}} Z - {\dot{X}}=0, \qquad {\tilde{d}} X=0 \ .
\end{eqnarray}
We Taylor expand in $r$ the $F$ components as
\begin{eqnarray}
\Psi &=& {\buildrel \circ \over{{\Psi}}} + r \delta \Psi + {\cal{O}}(r^2)
\nonumber \\
W &=&  {\buildrel \circ \over{{W}}} + r \delta W + {\cal{O}}(r^2)
\nonumber \\
X &=& {\buildrel \circ \over{{X}}}+ r \delta X + {\cal{O}}(r^2) \ .
\end{eqnarray}
As well as in the heterotic case, the $Z$ term appears in $F$ as $dr \wedge Z$, which scales linearly with $r$. Therefore we shall expand $Z$ as
\begin{eqnarray}
Z= \delta Z + {\cal{O}}(r) \ , 
\end{eqnarray}
which is consistent with both the $\varepsilon$-expansion explained in (\ref{epsilon_exp}), and with the $--$ component of the Einstein equation. 

The moduli of $D=11$ supergravity are therefore
$\{ \d \Delta, \d h, \d \gamma, \delta \Psi, \delta W, \delta X, \delta Z \}$, and we shall show that not all of them are independent by using the linearised Bianchi identities and field equations, which are listed in \ref{appx:D=11}. 

The Bianchi identity provides the following conditions:
\begin{eqnarray}
\label{Bianchi2}
\delta W   = \frac{1}{2}\tilde{d} \delta \Psi   \ , \qquad\qquad
\delta X = \tilde{d}\delta Z  \ ,
\end{eqnarray}
which we use to fix $\delta W$ and $\delta X$ in terms of $\delta \Psi$ and $\delta Z$. 
By using the $-k_1k_2$ component of the gauge field equations (\ref{g4eq}), we further fix $\delta \Psi$ as
\begin{eqnarray}
\nonumber
\label{psifix2}
\delta \Psi_{k_1k_2} &=& {\buildrel \circ \over \nabla}^{\ell} \delta Z_{\ell k_1k_2} - {\buildrel \circ \over h}^{\ell} Z_{\ell k_1k_2} + \delta \gamma_{k_1}{}^{\ell} {\buildrel \circ \over \Psi}_{\ell k_2} - \delta \gamma_{k_2}{}^{\ell} {\buildrel \circ \over \Psi}_{\ell k_1} \\
&-& \frac{1}{2} \delta \gamma_{\ell}{}^{\ell} \Psi_{k_1k_2} - \frac{q}{72} \epsilon_{k_1k_2}{}^{\ell_1\ell_2\ell_3\ell_4\ell_5\ell_6\ell_7} \delta Z_{\ell_1\ell_2\ell_3} {\buildrel \circ \over X}_{\ell_4\ell_5\ell_6\ell_7} \ . 
\end{eqnarray}
Also, by using the $-i$ and $+-$ components of the Einstein equations (\ref{-iein}) and (\ref{+-ein}), we fix $\delta h$ and $\delta \Delta$ respectively as follows
\begin{eqnarray}
\label{hfix2}
\nonumber
\delta h_i &=& \frac{1}{2} {\buildrel \circ \over \nabla}_i \delta \gamma_k{}^k - \frac{1}{2} {\buildrel \circ \over \nabla}^j \delta \gamma_{ji} + \frac{1}{2} \delta \gamma_{ij} {\buildrel \circ \over h}^j - \frac{1}{4} {\buildrel \circ \over h}_i \delta \gamma_k{}^k \\ 
&+& \frac{1}{4} {\buildrel \circ \over \Psi}_{\ell_1\ell_2}\delta Z_i{}^{\ell_1\ell_2} + \frac{1}{12} \delta Z_{\ell_1\ell_2\ell_3} {\buildrel \circ \over X}_i{}^{\ell_1\ell_2\ell_3} \ , 
\end{eqnarray}
and
\begin{eqnarray}
\label{deltfix2}
\nonumber
\delta \Delta &=& \frac{1}{3} {\buildrel \circ \over \nabla}_i \delta h^i + \frac{1}{12} {\buildrel \circ \over h}^i {\buildrel \circ \over \nabla}_i \delta \gamma_k{}^k - {\buildrel \circ \over h}^i \delta h_i - \frac{1}{6} {\buildrel \circ \over \Delta} \delta \gamma_k{}^k - \frac{1}{12} {\buildrel \circ \over h}_i {\buildrel \circ \over h}^i \delta \gamma_k{}^k  + \frac{1}{3} \delta \gamma_{ij} {\buildrel \circ \over h}^i{\buildrel \circ \over h}^j \\
\nonumber
&-& \frac{1}{6} \delta \gamma_{ij} {\buildrel \circ \over \nabla}^i {\buildrel \circ \over h}^j - \frac{1}{6} {\buildrel \circ \over h}^j {\buildrel \circ \over \nabla}^i \delta \gamma_{ij} + \frac{1}{9} \delta \Psi_{\ell_1\ell_2} {\buildrel \circ \over \Psi}^{\ell_1\ell_2}  - \frac{1}{9}{\buildrel \circ \over \Psi}^{\ell_1}{}_m {\buildrel \circ \over \Psi}^{\ell_2 m} \delta \gamma_{\ell_1\ell_2} \\
\nonumber
&-& \frac{1}{108} ({\buildrel \circ \over W} - {\buildrel \circ \over h}\wedge {\buildrel \circ \over \Psi} )_{\ell_1\ell_2\ell_3} \delta Z^{\ell_1\ell_2\ell_3} + \frac{1}{216} \delta X_{\ell_1\ell_2\ell_3\ell_4} {\buildrel \circ \over X}^{\ell_1\ell_2\ell_3\ell_4} \\
&-& \frac{1}{108}{\buildrel \circ \over X}^m{}_{\ell_1\ell_2\ell_3} {\buildrel \circ \over X}^{n\ell_1\ell_2\ell_3} \delta \gamma_{mn} - \frac{1}{216} ({\buildrel \circ \over h}\wedge \delta Z)_{\ell_1\ell_2\ell_3\ell_4} {\buildrel \circ \over X}^{\ell_1\ell_2\ell_3\ell_4} \ . 
\end{eqnarray}
We remark that the expressions for $\delta \Psi$, $\delta X$ and $\delta h$ are linear in
$\delta Z, \delta \gamma, {\buildrel \circ \over \nabla} \delta Z, {\buildrel \circ \over \nabla} \delta \gamma$, whereas the expressions for
$\delta W$ and $\delta \Delta$ involve some second order derivatives acting
on $\delta \gamma$, $\delta Z$. The remaining (unfixed) moduli are $\delta \gamma$ and $\delta Z$.

We shall now show that the independent moduli $\d Z$ and $\d \gamma$ must satisfy elliptic PDEs. 
We consider first the $\delta Z$ moduli. By taking the divergence of the Bianchi identity (\ref{Bianchi2}), we obtain 
\begin{eqnarray}
\label{divBianchi2}
{\buildrel \circ \over \nabla}^i {\buildrel \circ \over \nabla}_i \delta Z_{k_1k_2k_3} - 3{\buildrel \circ \over \nabla}_{[k_1} {\buildrel \circ \over \nabla}^{\ell} \delta Z_{ k_2 k_3] \ell} - 3{\buildrel \circ \over R}_{\ell [k_1} \delta Z^{\ell}{}_{k_2k_3]} - {\buildrel \circ \over \nabla}^{\ell} \delta X_{\ell k_1k_2k_3} = 0 \ ,
\end{eqnarray}
and by using the $k_1 k_2 k_3$ component of the gauge field equation (\ref{g4eq4}), we express the divergence of $\delta X$ as follows
\begin{eqnarray}
{\buildrel \circ \over \nabla}^{\ell} \delta X_{\ell k_1k_2k_3} = - 3 {\buildrel \circ \over \nabla}_{[k_1} {\buildrel \circ \over \nabla}^{\ell} \delta Z_{k_2k_3] \ell} + \mathcal{B}_{ij}
\end{eqnarray}
where
\begin{eqnarray}
\notag
\mathcal{B}_{ij} &\equiv &3 {\buildrel \circ \over \nabla}_{[k_1} ( {\buildrel \circ \over h}^{\ell} \delta Z_{k_2k_3] \ell}) + 6 {\buildrel \circ \over \nabla}_{[k_1} ( \delta \gamma_{k_2}{}^{\ell} {\buildrel \circ \over{\Psi}}_{k_3] \ell})   \\
\nonumber
&+& \frac{3}{2} {\buildrel \circ \over \nabla}_{[k_1} ( {\buildrel \circ \over{\Psi}}_{k_2k_3]} \delta \gamma_{\ell}{}^{\ell} ) + \frac{q}{72} \epsilon_{[k_1 k_2}{}^{\ell_1\ell_2\ell_3\ell_4\ell_5\ell_6\ell_7} {\buildrel \circ \over \nabla}_{k_3]} ( \delta Z_{\ell_1\ell_2\ell_3} {\buildrel \circ \over X}_{\ell_4\ell_5\ell_6\ell_7} ) \\
\nonumber
&+& \frac{3}{2} ({\buildrel \circ \over \nabla}_m \delta \gamma^{\ell}{}_{[k_1} + {\buildrel \circ \over \nabla}_{[k_1} \delta \gamma_{|m|}{}^{\ell} - {\buildrel \circ \over \nabla}^{\ell}\delta \gamma_{m [k_1} ) {\buildrel \circ \over X}^m{}_{k_2k_3]\ell} + \delta \gamma_{ij} {\buildrel \circ \over \nabla}^i {\buildrel \circ \over X}^j{}_{k_1k_2k_3}  \\
\nonumber
&+&   {\buildrel \circ \over \nabla}^m \delta \gamma_{m\ell} {\buildrel \circ \over X}^{\ell}{}_{k_1k_2k_3} - \frac{1}{2} {\buildrel \circ \over \nabla}^{\ell} \delta \gamma_m{}^m {\buildrel \circ \over X}_{\ell k_1k_2k_3} + {\buildrel \circ \over \nabla}^{\ell}( {\buildrel \circ \over h} \wedge \delta Z )_{\ell k_1k_2k_3}\\
\nonumber
&-& 2 {\buildrel \circ \over \Delta} \delta Z_{k_1k_2k_3} -\delta \gamma_{ij} {\buildrel \circ \over h}^i {\buildrel \circ \over X}^j{}_{k_1k_2k_3} + 2 \delta h^{\ell} {\buildrel \circ \over X}_{\ell k_1k_2k_3} + {\buildrel \circ \over h}^{\ell} \delta X_{\ell k_1k_2k_3}  \\
\nonumber
&+& 2 (\delta h\wedge {\buildrel \circ \over \Psi} )_{k_1k_2k_3} + 2 ({\buildrel \circ \over h} \wedge \delta \Psi )_{k_1k_2k_3} + 3 \delta \gamma^{\ell}{}_{[k_1} ( {\buildrel \circ \over W} - {\buildrel \circ \over h} \wedge {\buildrel \circ \over \Psi} )_{k_2k_3]\ell} \\
\nonumber
&+& {\buildrel \circ \over h}^{\ell} \delta X_{\ell k_1k_2k_3} - 2{\buildrel \circ \over h}^{\ell} ( {\buildrel \circ \over h}\wedge\delta Z)_{\ell k_1k_2k_3} - \frac{1}{2} \delta \gamma_{\ell}{}^{\ell} ( {\buildrel \circ \over W} - {\buildrel \circ \over h}\wedge {\buildrel \circ \over \Psi} )_{k_1k_2k_3} \\
\nonumber
&+& \frac{1}{2} ( -2{\buildrel \circ \over h}^{\ell} \delta \gamma_{\ell}{}^q + {\buildrel \circ \over h}^q\delta \gamma_{\ell}{}^{\ell} ) {\buildrel \circ \over X}_{q k_1k_2k_3} + 3 {\buildrel \circ \over h}^q \delta \gamma_{[k_1}{}^{\ell} {\buildrel \circ \over X}_{k_2k_3]\ell q} \\
\nonumber
&-& \frac{q}{24} \epsilon_{k_1k_2k_3}{}^{\ell_1\ell_2\ell_3\ell_4\ell_5\ell_6} ( \delta \Psi_{\ell_1\ell_2} {\buildrel \circ \over X}_{\ell_3\ell_4\ell_5\ell_6} + {\buildrel \circ \over \Psi}_{\ell_1\ell_2} \delta X_{\ell_3\ell_4\ell_5\ell_6} 
\\ \nonumber 
&-& {\buildrel \circ \over \Psi}_{\ell_1\ell_2} ( {\buildrel \circ \over h}\wedge \delta Z )_{\ell_3\ell_4\ell_5\ell_6} ) 
+ \frac{q}{12} \delta \gamma_{\ell_1m} \epsilon_{k_1k_2k_3}{}^{\ell_1\ell_2\ell_3\ell_4\ell_5\ell_6} {\buildrel \circ \over \Psi}^m{}_{\ell_2} {\buildrel \circ \over X}_{\ell_3\ell_4\ell_5\ell_6}  
\\ \nonumber 
&+& \frac{q}{6} \delta \gamma_{\ell_3m} \epsilon_{k_1k_2k_3}{}^{\ell_1\ell_2\ell_3\ell_4\ell_5\ell_6} {\buildrel \circ \over \Psi}_{\ell_1\ell_2} {\buildrel \circ \over X}^m{}_{\ell_4\ell_5\ell_6} \\
&+& \frac{q}{18} \epsilon_{k_1k_2k_3}{}^{\ell_1\ell_2\ell_3\ell_4\ell_5\ell_6} ( {\buildrel \circ \over W} - {\buildrel \circ \over h}\wedge {\buildrel \circ \over \Psi} )_{\ell_1\ell_2\ell_3} \delta Z_{\ell_4\ell_5\ell_6} \ .
\end{eqnarray}
On substituting this expression back into ({\ref{divBianchi2}}), 
the ${\buildrel \circ \over \nabla}_{[k_1} {\buildrel \circ \over \nabla}^{\ell} \delta Z_{k_2k_3] \ell} $ terms cancel out. Furthermore, there
are a number of terms which are linear in $\delta X$, $\delta h$, $\delta \Psi$
which are eliminated on using the Bianchi identity ({\ref{Bianchi2}}), together with
({\ref{hfix2}}) and ({\ref{psifix2}}), producing terms which are linear in $\delta Z, \delta \gamma, {\buildrel \circ \over \nabla} \delta Z, {\buildrel \circ \over \nabla} \delta \gamma$. The resulting PDEs are an elliptic system for 
$\delta Z$, with principal symbol generated by ${\buildrel \circ \over \nabla}^2$.

Next, we consider the metric moduli $\delta \gamma$. The linearized $ij$ component of the Einstein equations is
\begin{eqnarray}
\label{11dlin_ij}
{\buildrel \circ \over \nabla}^2 \delta \gamma_{ij} - {\buildrel \circ \over \nabla}_i {\buildrel \circ \over \nabla}_j \delta \gamma_k{}^k - ( {\buildrel \circ \over \nabla}_{\ell} {\buildrel \circ \over \nabla}_j - {\buildrel \circ \over \nabla}_j {\buildrel \circ \over \nabla}_{\ell} ) \delta \gamma^{\ell}{}_i - ( {\buildrel \circ \over \nabla}_{\ell} {\buildrel \circ \over \nabla}_i - {\buildrel \circ \over \nabla}_i {\buildrel \circ \over \nabla}_{\ell} ) \delta \gamma^{\ell}{}_j = \mathcal{C}_{ij} \ , 
\nonumber \\
\end{eqnarray}
where
\begin{eqnarray}
\nonumber
\mathcal{C}_{ij} &\equiv & {\buildrel \circ \over \nabla}_{(i} (\delta Z_{j)}{}^{\ell_1\ell_2} {\buildrel \circ \over \Psi}_{\ell_1\ell_2}) + \frac{1}{3} {\buildrel \circ \over \nabla}_{(i} ( {\buildrel \circ \over X}_{j)}{}^{\ell_1\ell_2\ell_3} \delta Z_{\ell_1\ell_2\ell_3}) \\
\nonumber
&-& 8 {\buildrel \circ \over h}_{(i} \delta h_{j)} + 2 ( - {\buildrel \circ \over \Delta} + \frac{1}{2} {\buildrel \circ \over \nabla}_k {\buildrel \circ \over h}^k - {\buildrel \circ \over h}_k {\buildrel \circ \over h}^k ) \delta \gamma_{ij}  \\
\nonumber
&+& 2 \delta \gamma_{k(i} {\buildrel \circ \over \nabla}_{j)} {\buildrel \circ \over h}^k + {\buildrel \circ \over h}_{(i} {\buildrel \circ \over \nabla}_{j)} \delta \gamma_k{}^k   - 2 \delta \gamma_{k(i} {\buildrel \circ \over \nabla}^k {\buildrel \circ \over h}_{j)} - 2 {\buildrel \circ \over h}^k {\buildrel \circ \over \nabla}_{(i} \delta \gamma_{j)k}\\
\nonumber
&+& 3 {\buildrel \circ \over h}^k {\buildrel \circ \over \nabla}_k \delta \gamma_{ij} - 2 {\buildrel \circ \over h}_{(i} {\buildrel \circ \over \nabla}^k \delta \gamma_{j)k} + 4 {\buildrel \circ \over h}_k {\buildrel \circ \over h}_{(i} \delta \gamma_{j)}{}^k - \delta \gamma_k{}^k {\buildrel \circ \over h}_i {\buildrel \circ \over h}_j - 2 \delta \Psi_{\ell (i} {\buildrel \circ \over \Psi}_{j)}{}^{\ell} \\
\nonumber
&-& {\buildrel \circ \over \Psi}_i{}^m {\buildrel \circ \over \Psi}_j{}^n \delta \gamma_{mn} + ({\buildrel \circ \over h}\wedge {\buildrel \circ \over \Psi} - {\buildrel \circ \over W})_{\ell_1\ell_2 (i} \delta Z_{j)}{}^{\ell_1\ell_2} + \frac{1}{3} \delta X_{\ell_1\ell_2\ell_3 (i} {\buildrel \circ \over X}_{j)}{}^{\ell_1\ell_2\ell_3}  \\
\nonumber
&+& \frac{1}{2} {\buildrel \circ \over X}_{i\ell_1\ell_2}{}^m {\buildrel \circ \over X}_j{}^{\ell_1\ell_2 n} \delta \gamma_{mn}  - \frac{1}{3} ({\buildrel \circ \over h}\wedge \delta Z)_{\ell_1\ell_2\ell_3 (i} {\buildrel \circ \over X}_{j)}{}^{\ell_1\ell_2\ell_3} - \frac{1}{3} {\buildrel \circ \over \gamma}_{ij} \delta \Psi_{\ell_1\ell_2} {\buildrel \circ \over \Psi}^{\ell_1\ell_2} 
\\ \nonumber
&+&  {1 \over 3} {\buildrel \circ \over \gamma}_{ij} {\buildrel \circ \over \Psi}_m{}^\ell {\buildrel \circ \over \Psi}^{mq} \delta \gamma_{\ell q}
+{1 \over 72} \delta \gamma_{ij} {\buildrel \circ \over X}_{\ell_1 \ell_2 \ell_3 \ell_4} {\buildrel \circ \over X}^{\ell_1 \ell_2 \ell_3 \ell_4}
+ \frac{1}{36} {\buildrel \circ \over \gamma}_{ij} \delta X_{\ell_1\ell_2\ell_3\ell_4} {\buildrel \circ \over X}^{\ell_1\ell_2\ell_3\ell_4}
\\ \nonumber
 &-& \frac{1}{18} {\buildrel \circ \over \gamma}_{ij} {\buildrel \circ \over X}^m{}_{\ell_1\ell_2\ell_3} {\buildrel \circ \over X}^{n \ell_1\ell_2\ell_3} \delta \gamma_{mn} 
+ \frac{1}{9} {\buildrel \circ \over \gamma}_{ij} ({\buildrel \circ \over W} - {\buildrel \circ \over h}\wedge {\buildrel \circ \over \Psi})_{\ell_1\ell_2\ell_3} \delta  Z^{\ell_1\ell_2\ell_3}
\\
 &-& \frac{1}{36} {\buildrel \circ \over \gamma}_{ij} ({\buildrel \circ \over h} \wedge \delta Z)_{\ell_1\ell_2\ell_3\ell_4} {\buildrel \circ \over X}^{\ell_1\ell_2\ell_3\ell_4} -  {1 \over 6} \delta \gamma_{ij} {\buildrel \circ \over \Psi}_{mn} {\buildrel \circ \over \Psi}^{mn} \ , 
\end{eqnarray}
which is linear in $\delta Z, \delta \gamma, {\buildrel \circ \over \nabla} \delta Z, {\buildrel \circ \over \nabla} \delta \gamma$. 
Furthermore, in ({\ref{11dlin_ij}}), terms of the form $( {\buildrel \circ \over \nabla}_{\ell} {\buildrel \circ \over \nabla}_j - {\buildrel \circ \over \nabla}_j {\buildrel \circ \over \nabla}_{\ell} ) \delta \gamma^{\ell}{}_i$ can be rewritten as terms linear in $\delta \gamma$ and the Riemann tensor ${\buildrel \circ \over R}$, hence can be incorporated into the algebraic term on the RHS. The trace term $\delta \gamma_k{}^k$
is again fixed by the elliptic condition ({\ref{trelliptic}}), so ({\ref{11dlin_ij}})
is an elliptic set of PDEs for the traceless part of $\delta \gamma$, with principal symbol generated by ${\buildrel \circ \over \nabla}^2$.

Taken together, the conditions ({\ref{divBianchi2}}) and ({\ref{11dlin_ij}}) and ({\ref{trelliptic}}) constitute elliptic PDEs on the moduli $\{\delta Z, \delta \gamma \}$,
with the remaining moduli $\{\delta W, \delta \Psi, \delta X, \delta h, \delta \Delta \}$
fixed in terms of $\{\delta Z,  \delta \gamma \}$
by ({\ref{Bianchi2}}), ({\ref{psifix2}}), ({\ref{hfix2}}) and ({\ref{deltfix2}}). The moduli space is therefore finite dimensional.

\section{Stability conditions for the extended solution}
A priori, there is no guarantee that a given near-horizon geometry belongs to a genuine black hole. For example, there are known near-horizon geometries with toroidal topology. However a generalization to higher dimensions of the Hawking's horizon topology theorem \cite{Hawking:1971vc}, found by Galloway and Schoen \cite{Galloway:2005mf}, shows that for asymptotically flat solutions the spatial topology of the horizon must be such that it admits a metric with positive Ricci scalar (or equivalently, positive Yamabe invariant). However general criteria which allows to systematically exclude near-horizon geometries, like necessary and sufficient conditions for a horizon to belong to a black hole, are still missing.  

There are various necessary conditions that one can choose to impose on a near-horizon geometry in order to belong to a black hole, for a review see \cite{Booth:2005qc}.  
One of these is that the spatial cross section $\cS$ be a \emph{marginally trapped surface}. Intuitively, a marginally trapped surface is a surface such that every geodesic which intersects it cannot escape, but evolves inside the surface. A marginally trapped surface describes well the idea of a black hole.
Requiring that $\cS$ be a marginally trapped surface translates into geometric conditions for the metric deformations $\delta \gamma_{ij}$, and they have been discussed in \cite{Li:2015wsa}. 

\begin{figure}[h]
\begin{center}
 \includegraphics[scale=0.50]{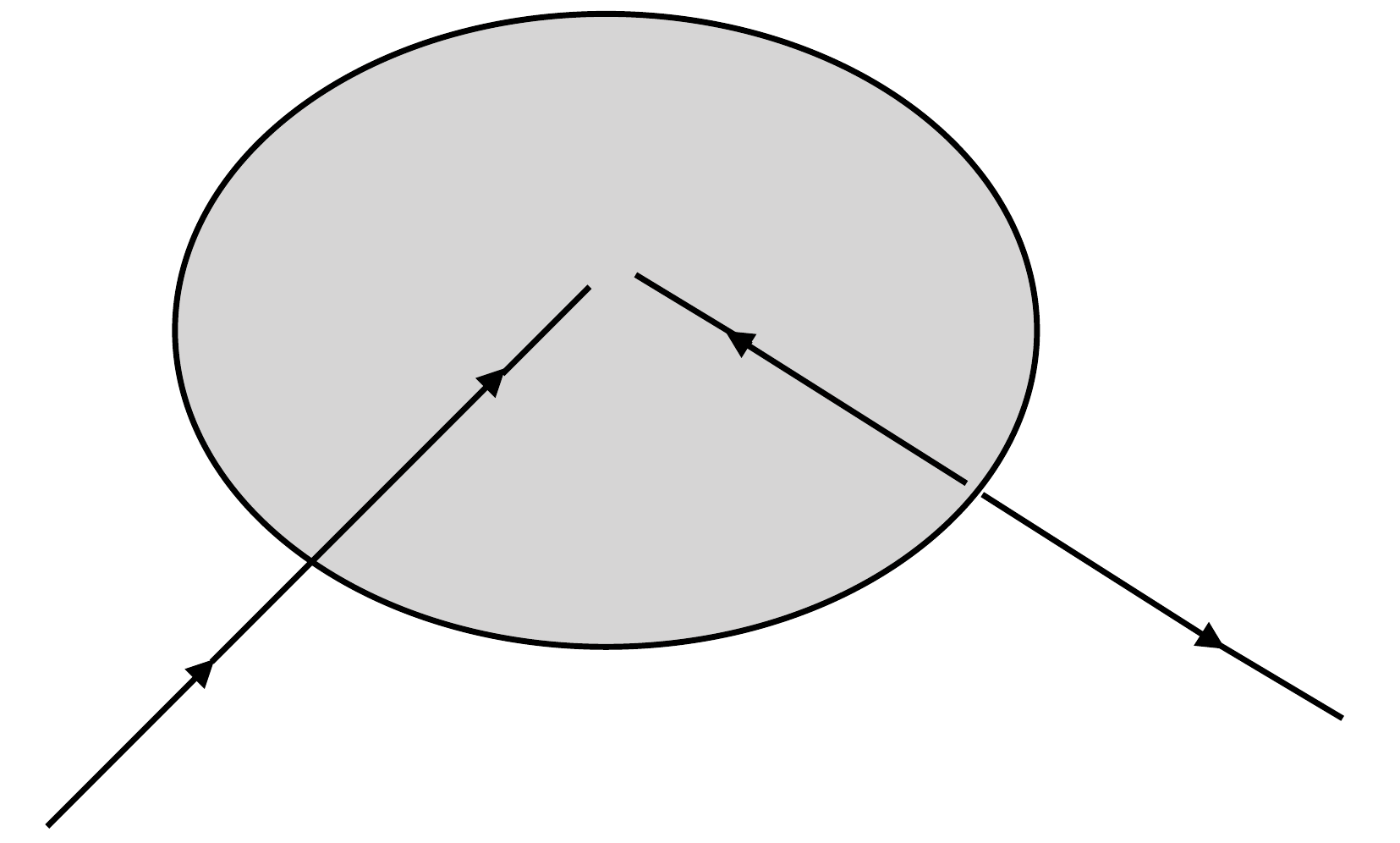}
\end{center}
\caption{The grey region is an example of marginally trapped surface.}
\end{figure}

Moreover, in \cite{Li:2015wsa} it has been considered radial deformations of the extreme Kerr near-horizon geometry, such that $\cS$ is a marginally trapped surface, and explicit solutions for the PDE which involves the metric modulus have been found. In order to render the problem tractable, the authors \emph{assumed} that the isometries of the extreme Kerr near-horizon geometry are \emph{also} isometries of the deformed solution. 
They found that the solution for the metric modulus is unique, and reproduces the first order data of the extreme Kerr black hole solution. 
Although this sounds like a uniqueness theorem, one has to remember that the logic here is different from the idea behind the black hole uniqueness theorem of Hawking et al.. In the latter case, one has a control on the asymptotic data of the black hole, such as the geometry of the spacetime, mass, charge and angular momenta of the black hole. But in the case considered here, the idea is to begin with a horizon and try to get away from it, and for this reason the asymptotic data is unknown. 
Nevertheless, we comment that it would be interesting to determine which extra conditions the moduli must satisfy, instead of assuming that the isometries of the near-horizon solution are inherited by the deformed one.

One may notice that if we choose a near-horizon geometry which belongs to a multi-black hole, then the first order deformations might not carry the information about the various black hole centres. From this perspective, a multi-black hole would be indistinguishable from a single-black hole\footnote{This was questioned by G. W. Gibbons in a seminar which I gave at DAMTP, Cambridge.}. However we remark that in higher dimensions than four, multi-black holes do not have a smooth horizon. For instance, in five dimensions the horizon metric components contain fractional powers of $r$, e.g. $r^{5/2}$, which renders the metric coefficients $\mathscr{C}^2$, but not smooth \cite{Candlish:2007fh}. Since our approach contemplates only horizons which are analytic in $r$, multi-black hole horizons are excluded, and this issue does not appear.

\part{Integrable structures of String Theory}
\chapter{{\bf Introduction}}
\label{ch:Intro_Integra}

\section{Gauge/Gravity duality in integrable backgrounds}
The first formulation of the gauge/gravity duality appears in the context of $AdS_5 \times S^5$ background in Type IIB Superstring \cite{Aharony:1999ti}. 
The duality was first tested for BPS operators, which are protected from quantum corrections \cite{DHoker:2002nbb}, however in the later years, the duality was also tested for non-BPS operators. The key ingredient of the success is the fact that the string on $AdS_5 \times S^5$ background defines a classically integrable sigma-model \cite{Minahan:2002ve, Bena:2003wd}. 

We are interested in understanding how general is the gauge/gravity duality outside the context of $AdS_5 \times S^5$. However at the same time we also desire to retain a certain control in such exploration. Therefore we are interested in studying the duality on backgrounds which still give rise to  \emph{classically integrable} string sigma-model.

In this context, the scattering matrix must satisfy a set of fundamental equations which completely determine its entries, and integrability imposes one of them, the so-called Yang-Baxter equation. The set of fundamental equations is explained in detail in section \ref{Smatrix}. 

Once the explicit solution for the scattering matrix is obtained, the next step consists in formulating a \emph{Bethe ansatz}. There exists a consolidated technique which uses the scattering matrix to find the eigenvalues and eigenstates of the Hamiltonian governing the dynamics, and therefore it solves the spectrum. 
This technique relies on the concept of the monodromy matrix associated with the scattering matrix, and on the construction of a pseudo-vacuum state, which allows to construct eigenstates by acting on it via ladder-type operators. A brief overview of this technique is given in section \ref{Bethe}. 

The program which we shall describe is schematically the following:
\begin{figure}
\centering
\begin{tikzpicture}[scale=0.85]
\node (A) at (0,0)  {Supercoset background};
\node (M) at (0, -1) {\small Section \ref{supercoset}\normalsize};
\node (B) at (6,0)  {Classically integrable};
\node (N) at (6, -1) {\small Section \ref{class_integ}\normalsize};
\node (C) at (11,0) {S-matrix};
\node (L) at (11, -1)  {\small Section \ref{Smatrix}\normalsize};
\node (D) at (15, 0) {Bethe ansatz};
\node (I) at (15, -1)  {\small Section \ref{Bethe}\normalsize};
\draw[->] (A) -- (B);
\draw[->] (B) -- (C);
\draw[->] (C) -- (D);
\end{tikzpicture}
\caption{Gauge/Gravity duality outside $AdS_5 /CFT_4$.}
\end{figure}
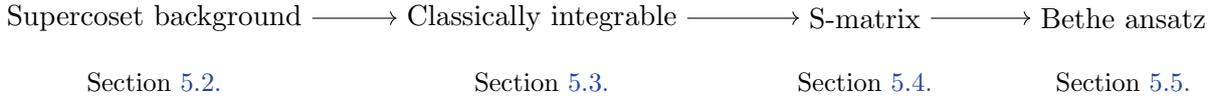

\section{Supercoset spaces}
\label{supercoset}
The sigma-model associated with the $AdS_5 \times S^5$ background, including also the fermionic degrees of freedom, can be described as a sigma-model associated with the following supercoset:
\begin{equation}
\label{AdS5supercoset}
AdS_5 \times S^5 = \frac{SO(4,2) \times SO(6)}{SO(4,1) \times SO(5)} \subset \frac{PSU(2,2 | 4)}{SO(4,1) \times SO(5)} \ .
\end{equation}
The embedding of $AdS_5 \times S^5$ into the supercoset (\ref{AdS5supercoset}) is such that the bosonic and fermionic degrees of freedom are balanced, according to preservation of supersymmetry. 

Motivated by exploring the gauge/gravity duality in other backgrounds different from $AdS_5 \times S^5$, it would be convenient to have a classification of all supercoset spaces from where one can potentially choose a background. 
In mathematics, there exists a classification of simple Lie algebras and symmetric spaces, which was done by W. Killing and E. Cartan. 
Similarly, a classification of simple Lie superalgebras has been done by V. G. Kac \cite{Kac:1977em}, and later V. V. Serganova also classified symmetric superspaces \cite{Serganova:1983vp}. 
However, not all symmetric superspaces from Serganova's classification provide a (classically) integrable sigma-model. The key ingredient which guarantees integrability of the sigma-model is that the supercoset admits a $\mathbb{Z}_4$ outer automorphism; in such a case a classically integrable action can be constructed (the reason is summarized in section \ref{class_integ}).

Supercosets which admit a ${\mathbb{Z}}_4$ symmetry 
compatible with the commutation relations are called \emph{semi-symmetric superspaces}, and have been classified \cite{sym}. 
It is important to note that there are additional conditions that need to be required in order for a semi-symmetric superspace to define a consistent \emph{string background}. The two extra conditions to impose are \cite{sym}: 
\begin{enumerate}
\item the beta function vanishes, and
\item the total central charge is zero, 
\end{enumerate}
or equivalently, the Killing form is identically zero. The full classification of integrable string backgrounds is still not available, although some progress has been made, see e.g. \cite{sym, sym2, sym3, sym4}. 
Some integrable string backgrounds of recent interest are listed in table \ref{tab:backgrounds}. 

\begin{table}[H]
\begin{center}
\begin{tabular}{c@{\hskip 1cm}c@{\hskip 1cm}c}
\hline
{\it Integrable string $\sigma$-model} & {\it Supercoset rep.} & {\it Dual CFT} \\
\hline
& &  \\
$AdS_5 \times S^5$ in type IIB& $\frac{PSU(2,2 | 4)}{SO(4,1) \times SO(5)}$ & $\mathcal{N}=4$ super Yang-Mills  \\
 & &  \\
$AdS_4 \times \mathbb{C}P^3$ in type IIA \cite{Arutyunov:2008if, Stefanski:2008ik} & $\frac{OSp(6 | 4)}{SO(3, 1) \times U(3)}$ & ABJM \cite{Aharony:2008ug}  \\
& &  \\
$AdS_3 \times S^3 \times S^3 \times S^1$ in type IIB & $\frac{D(2,1 ; \alpha) \times D(2,1 ; \alpha)}{SO(1,2)\times SO(3) \times SO(3)}$ & 
conjectured CFT$_2$ \cite{Gukov:2004ym, Tong:2014yna, Eberhardt:2017pty} \\
& &  \\
$AdS_3 \times S^3 \times T^4$ in type IIB & 
$\frac{PSU(1,1|2) \times PSU(1,1|2)}{SO(1,2) \times SO(3)}$ & 
$Sym^N (T^4)$ \cite{Berkovits:1999im , Sax:2014mea} \\
& &  \\
$AdS_2 \times S^2 \times T^6$ in type IIB & $\frac{PSU(1,1|2)}{U(1) \times U(1)}$ & unknown chiral CFT$_2$  \\
 & &  or Quantum Mechanics 
\end{tabular}
\end{center}
\caption{Some integrable string backgrounds and their dual CFTs.}
\label{tab:backgrounds}
\end{table}

\section{Classical integrability}
\label{class_integ}
In this section we shall review the concept of classical integrability in the context of a $1+1$ dimensional field theory, i.e. in one spatial dimension $x$ and time $t$. Lecture notes on this topic can be found in e.g. \cite{Torrielli:2016ufi, Arutyunov:lecture}. 

The standard notion of integrability is based on the Liouville theorem, which relies on the existence of conserved quantities in involution.
This works well for systems with a finite number of degrees of freedom, however it becomes inadequate when the number of degrees of freedom is infinite, which is the case for field theories.  
The more appropriate way to express the concept of integrability in field theories is based on the notion of \emph{Lax pair}
\footnote{A Lax pair definition of integrability also exists in the context of systems with finite number of degrees of freedom. In this case, definition (\ref{Lax}) becomes 
\begin{equation}
\label{Lax_finite}
\frac{dU}{dt} = [V, U] \ . 
\end{equation} 
A system of $n$ degrees of freedom is integrable \`a la Liouville if it admits $n$ independent integrals of motion in involution, from which one can locally construct action-angle variables $(I_j , \theta_j)$, with $j = 1, ... , n$, such that the equations of motion become
\begin{equation}
\label{Liouville}
\frac{dI_j}{dt} = 0 \ , \qquad\qquad
\frac{d\theta_j}{dt} = \frac{\partial H}{\partial I_j} \ , 
\end{equation}
where $H$ is the Hamiltonian. In a Liouville-integrable system, one can canonically construct a Lax pair in terms of the action-angle variables, see e.g. \cite{Babelon}.}.

Suppose that the 2-dimensional field theory dynamic is governed by a Lagrangian, whose associated Euler-Lagrange equations are known.  
Then we have:

\begin{definition}[{\bf Classical Integrability}]
A 2-dimensional field theory is called \emph{classically integrable} if the Euler-Lagrange equations can be written as
\begin{equation}
\label{Lax}
\frac{\partial U}{\partial t} - \frac{\partial V}{\partial x} + [ U, V ] = 0  \ ,
\end{equation} 
where $U, V$ are $n\times n$ matrices, which can depend on the so-called spectral parameter $\lambda$. 
\end{definition}
The two matrices $\{ U, V \}$ form the so-called \emph{Lax pair}, and they are responsible for the integrable structure as follows. 

Let us introduce the so-called \emph{monodromy matrix} $M$:
\begin{equation}
M \equiv \mathscr{P} \exp \bigg( \int_{x_A}^{x_B} U(x, t; \lambda) dx \bigg) \ , 
\end{equation} 
where $\mathscr{P}\exp$ is the path-ordered exponential, $x_A, x_B$ are some spatial points of the real line, such that $x_A < x_B$. 
One can show the following property:
\begin{equation}
\partial_t M = V(x_B, t; \lambda) M - M V(x_A, t ; \lambda) \ . 
\end{equation}
We assume that the spatial domain is compact and ranges in the interval $[0, 2\pi]$ with periodic boundary conditions on the fields. Then by taking $x_A = 0, x_B= 2\pi$ we have
\begin{equation}
\partial_t M = [V(0, t; \lambda) , M] \ . 
\end{equation}
This implies that the \emph{transfer matrix} $T$, which is the trace of the monodromy matrix, i.e.
\begin{equation}
T  \equiv {\rm tr} M \ ,
\end{equation}
is conserved for all values of $\lambda$. This allows us to construct infinitely many conserved quantities simply by Taylor expanding in $\lambda$ the transfer matrix\footnote{There is also an alternative method to construct infinitely many charges from the monodromy matrix. One can show that
\begin{equation}
H^{(n)} = {\rm tr} M^n \ , \qquad \forall n \in \mathbb{N} \ , 
\end{equation}  
are conserved quantities, which are also called \emph{conserved hamiltonians}. It turns out that every element in the set $\{ Q_n \}$ can be written as a linear combination of elements in the set $\{ H^{(n)} \}$, and therefore the two definitions are equivalent.
We should also remark that for a generic theory the conserved quantities $\{ Q_n \}$, or equivalently $\{ H^{(n)} \}$, are not necessary linearly independent. }
, i.e.
\begin{equation}
T(\lambda) = \sum \lambda^n Q_n \ , \qquad\qquad
\partial_t Q_n = 0 \ ,
\end{equation}
where we assumed analyticity of $T$ in $\lambda=0$. 

The Euler-Lagrange equations for an integrable 2-dimensional field theory, written in the language of Lax pair, admit at least two mathematical interpretations. 
\begin{enumerate}
\item \emph{Consistency of an auxiliary system of PDEs}: Consider the following linear system of PDEs of first order:
\begin{eqnarray}
\label{aux_PDEs}
\notag
\frac{\partial \Psi}{\partial x} &=& U(x, t; \lambda) \Psi \ , \\
\frac{\partial \Psi}{\partial t} &=& V(x, t; \lambda) \Psi \ , 
\end{eqnarray}  
where $\Psi$ is a rank $n$ vector. 

A consistency condition in order for (\ref{aux_PDEs}) to admit a well-defined solution is obtained by differentiating the first equation with respect to $t$ and the second with respect to $x$, and by imposing that the second order derivatives of $\Psi$ must coincide. This implies (\ref{Lax}). 
\item \emph{Flat Lax connection}: One can associate to the Lax pair the following 1-form:
\begin{equation}
\mathscr{L} = U dx + V dt \ ,
\end{equation}
which is called \emph{Lax connection}. Then the zero curvature condition for $\mathscr{L}$, which is
\begin{equation}
\label{zerocurvature}
d \mathscr{L} - \mathscr{L} \wedge \mathscr{L} = 0 \ ,
\end{equation} 
implies the Euler-Lagrange equations (\ref{Lax}). 
\end{enumerate}

\subsection{Example of classically integrable systems}
A general method to finding a solution of (\ref{zerocurvature}) in terms of $U$ and $V$ is given by the \emph{dressing method}, which consists in solving a Riemann-Hilbert problem. It is not our purpose to discuss here this method, but instead we report the Lax pairs of some known integrable models: the KdV and Sine-Gordon models. 

\begin{itemize}
\item \emph{KdV (Korteweg-de-Vries) equation}

In $1+1$ dimensions, with coordinates $(t, x)$, the KdV equation for the field $u(t, x)$ is
\begin{equation}
u_t + 6 u u_x + u_{xxx} = 0 \ , 
\end{equation}
where $u_t \equiv \partial u / \partial t$, and the same for $u_x$.  
The Lax pairs is given by 
\begin{equation}
U = \begin{pmatrix}
0 & 1 \\
\lambda + u & 0 
\end{pmatrix} \ , \qquad\qquad
V = \begin{pmatrix}
u_x & 4\lambda - 2u \\
4\lambda^2 + 2 \lambda u + u_{xx} - 2 u^2 & - u_x \\
\end{pmatrix} \ . 
\end{equation}
and the curvature of the associated Lax connection is 
\begin{equation}
\partial_t U - \partial_x V + [U, V] = \begin{pmatrix}
0 & 0 \\
u_t + 6 u u_x - u_{xxx} & 0 
\end{pmatrix} \ . 
\end{equation}
Thus the zero curvature condition for the Lax connection is equivalent to the KdV equation. 

\item \emph{Sine-Gordon equation}

Consider the Sine-Gordon equation for the field $\phi(t, x)$ in 2-dimensional Minkowski space
\begin{equation}
\phi_{tt} - \phi_{xx} + \frac{8m^2}{\beta} \sin (2 \beta \phi) = 0 \ , 
\end{equation}
where $m$ and $\beta$ are constant parameters, and the field $\phi$ is periodic, i.e. $\phi(t, x) \sim \phi (t, x) + 2\pi / \beta$. 
The Lax pair is 
\begin{eqnarray}
\notag
U &=& i \begin{pmatrix}
\frac{\beta}{2} \phi_t & m \lambda e^{i\beta \phi} - \frac{m}{\lambda} e^{-i\beta \phi} \\
m\lambda e^{-i \beta\phi} - \frac{m}{\lambda} e^{i\beta\phi} &
- \frac{\beta}{2} \phi_t 
\end{pmatrix} \ , \\
V &=& i \begin{pmatrix}
\frac{\beta}{2} \phi_x & - m\lambda e^{i\beta\phi} - \frac{m}{\lambda} e^{-i \beta \phi} \\
- m\lambda e^{-i \beta \phi} - \frac{m}{\lambda} e^{i \beta \phi} & - \frac{\beta}{2} \phi_x 
\end{pmatrix} \ ,
\end{eqnarray}
which satisfies the following condition
\begin{eqnarray}
\notag
\partial_t U - \partial_x V + [U, V]  &=& i \begin{pmatrix}
\frac{\beta}{2} \phi_{tt} - \frac{\beta}{2} \phi_{xx} + 4m^2 \sin (2\beta \phi) & 0 \\
0 & - \frac{\beta}{2} \phi_{tt} + \frac{\beta}{2} \phi_{xx} - 4 m^2 \sin (2 \beta\phi) 
\end{pmatrix} \ . 
\end{eqnarray}
Therefore the Lax connection is flat if and only if the Sine-Gordon equation holds. 
\end{itemize}

\subsection{Classically integrable supercoset sigma-models}
In this section we shall sketch the proof of the following:
\begin{thm}
A supercoset sigma-model which admits 
a ${\mathbb{Z}}_4$ outer automorphism admits a classically integrable action.
\end{thm}

To construct the action of the  $G/H_0$ sigma-model we shall consider the generic field $g(x)$, which takes values in $G$, modulo the identification 
\begin{equation}
g(x) \sim g(x) h(x) \ , 
\end{equation}
where $h(x)$ is an element of $H_0$, and $x$ are the world-sheet coordinates. We introduce the left-invariant current $J$, 
\begin{equation}
J_{\mu} = g^{-1} \partial_{\mu} g \ , 
\end{equation}
which is the Maurer-Cartan 1-form, and takes values in $\mathfrak{g}$, i.e. the Lie  algebra of $G$. 
The action of $G$ on the left is a global symmetry, i.e. $g(x) h(x) \rightarrow g' g(x) h(x)$ is a symmetry. 

The action for a supercoset sigma model is of Metsaev-Tseytlin type \cite{Metsaev:1998it}, see e.g. \cite{sym, Candu:2013cga}, and includes the kinetic and the topological Wess-Zumino terms \footnote{In some cases, which depends by the geometry of the supercoset, it is possible to introduce terms which include the $B$ field ($\theta$-terms) or the field strength of $B$ (WZW terms). For instance, if $H_0$ contains a $U(1)$ factor, then a $\theta$-term is allowed.}.  
Schematically:
\begin{equation}
\label{action_coset}
S \propto \int d^2x \, \textrm{Str} \bigg( J \wedge \star J + J \wedge J \bigg)  \ . 
\end{equation}
The $\mathbb{Z}_4$ outer automorphism:
\begin{equation}
\Omega : \mathfrak{g} \rightarrow \mathfrak{g} \ , \qquad \Omega^4 = {\rm id}  
\end{equation} 
induces a decomposition of $\mathfrak{g}$ as follows
\begin{equation}
\label{Z4decomp}
\mathfrak{g} = \mathfrak{h}_0 \oplus \mathfrak{h}_1 \oplus \mathfrak{h}_2 \oplus \mathfrak{h}_3 \ , 
\end{equation}
where $\mathfrak{h}_0$ is the Lie algebra associated with $H_0$, the bosonic subalgebra of $\mathfrak{g}$ is $\mathfrak{h}_0 \oplus \mathfrak{h}_2$, and the fermionic is $\mathfrak{h}_1 \oplus \mathfrak{h}_3$. The subspaces $\mathfrak{h}_n$ are characterised under the $\mathbb{Z}_4$ automorphism by the charges
\begin{equation}
\Omega ( \mathfrak{h}_n ) = i^n \mathfrak{h}_n \ , 
\end{equation} 
and the graded commutation relations between elements of the subspaces are
\begin{equation}
[ \mathfrak{h}_n , \mathfrak{h}_m \} \subset \mathfrak{h}_{(n+m) {\rm mod} 4} \ . 
\end{equation}
The $\mathbb{Z}_4$ decomposition (\ref{Z4decomp}) also induces a decomposition of the Maurer-Cartan 1-form $J$ as follows
\begin{equation}
J = J^{(0)} + J^{(1)} + J^{(2)} + J^{(3)} \ ,  
\end{equation}
where $J^{(i)} \in \mathfrak{h}_i$. The action (\ref{action_coset}) becomes
\begin{equation}
\label{action_coset_Z4}
S \propto \int d^2x \, \textrm{Str} \bigg( J^{(2)} \wedge \star J^{(2)} + J^{(1)} \wedge J^{(3)} \bigg) \ . 
\end{equation}
By following a procedure analogous to the case of $AdS_5 \times S^5$ in \cite{Bena:2003wd}, one can show that the equations of motion descending from varying the action (\ref{action_coset_Z4}) can be written in terms of a Lax pair. The Lax connection is of the type
\begin{equation}
\mathscr{L} = \alpha_0 J^{(0)} + \alpha_1 J^{(1)} + \alpha_2 J^{(2)} + \alpha_3 J^{(3)} \ ,  
\end{equation}
where the coefficients $\{ \alpha_i \}$ are fixed\footnote{One can show that the curvature of $\mathscr{L}$ is proportional to a sum of terms of the type $c_{ij} J^{(i)} \wedge J^{(j)}$, where $\{ c_{ij} \}$ depends on $\{ \alpha_i \}$. By imposing $c_{ij} = 0$ independently for all $(i, j)$, one obtains a system of equations, typically with large degree of redundancy, which fixes $\{ \alpha_i \}$ (usually) in terms of a 1-parameter family.} by imposing the zero curvature condition (\ref{zerocurvature}).

\section{Scattering matrix}
\label{Smatrix}
In this section we shall explain how the S-matrix can be exactly determined in a $1+1$ dimensional \emph{integrable} theory, which for our purpose is an integrable string supercoset sigma-model.   

A subgroup of the isometries of the supercoset must be symmetries of the S-matrix, which (partially) constrain its entries. 
Integrability implies that every $n$-body scattering can be factorized into a series of 2-body scatterings. This is the physical meaning of the so-called \emph{Yang-Baxter equation} (YBE), which together with the background isometries completely determines the entries of the S-matrix, up to an overall scalar factor, the so-called \emph{dressing factor}. 
\emph{Crossing symmetry}, which is a consequence of the fact that anti-particles can be viewed as particles travelling back in time, provides the last important equation which fixes the dressing factor. 

In what follows, we shall first introduce the concept of single particle state, and how the symmetry algebra acts on it.
Then we will illustrate the aforementioned equations which every S-matrix of an integrable theory must satisfy. Finally we will mention the proper mathematical framework were these equations formalise, which is in the context of the \emph{Hopf algebras}.

\subsection{Single particle representation}
\label{sec: single particle}
The particles interested in the scattering are world-sheet excitations of the string propagating in a given coset background. 
In $AdS_5 \times S^5$, there exists a duality between world-sheet string excitations and excitations of the spin-chain associated with the dual field theory, the so-called \emph{magnons}, see e.g. \cite{rev}. 

In order to define a magnon, one must introduce a vacuum state, from which all spin-chain excitations can be generated by acting with ladder-type operators. 
In terms of world-sheet excitations, defining a vacuum state implies that the full background isometry group is broken. The remaining preserved symmetry subgroup is the little group associated with the vacuum state. 
 
We shall consider the supercoset $G/H$, with isometry supergroup $G$. Let $K$, with $K \subset G$, be the little supergroup which leaves the vacuum state invariant, and let us denote by $\mathfrak{K}$ the associated Lie superalgebra.

The superalgebra $\mathfrak{K}$ decomposes in terms of the $\mathbb{Z}_2$ grading as
\begin{equation}
\mathfrak{K} = \mathfrak{K}_0 \oplus \mathfrak{K}_1 \ , 
\end{equation}
where $\mathfrak{K}_0$ and $\mathfrak{K}_1$ are the bosonic and fermionic subalgebras respectively, such that
\begin{equation}
\mathfrak{K}_m \, \mathfrak{K}_n \subset \mathfrak{K}_{m+n \, | \,  \text{mod}\ 2}  \ , \qquad\qquad  m, n = 0, 1  \ . 
\end{equation}
We denote by $\{ \mathfrak{a}^{(0)}_i \}$ and $\{ \mathfrak{a}^{(1)}_{\alpha} \}$, where $i, \alpha \in \mathbb{N}$ , the set of generators of $\mathfrak{K}_0$ and $\mathfrak{K}_1$ respectively, which satisfy
\begin{equation}
[ \mathfrak{a}^{(0)}_i , \mathfrak{a}^{(0)}_j ] \subset \mathfrak{K}_0 \ , \qquad 
[ \mathfrak{a}^{(0)}_i , \mathfrak{a}^{(1)}_{\alpha} ] \subset \mathfrak{K}_1 \ , \qquad
\{ \mathfrak{a}^{(1)}_{\alpha} , \mathfrak{a}^{(1)}_{\beta} \} \subset \mathfrak{K}_0 \ ,
\end{equation}
where $[ \cdot\, , \cdot ]$ (resp. $\{ \cdot\, , \cdot \}$) is the standard (anti)-commutator. 

As we shall see in chapters \ref{Chapter: AdS2} and \ref{chapter: AdS3}, one can represent the generators $\{ \mathfrak{a}^{(0)}_i , \mathfrak{a}^{(1)}_{\alpha} \}$ as operators acting on states  which are bosons or fermions. When $\mathfrak{a}^{(0)}_i$ acts on a bosonic (fermionic) state produces another bosonic (fermionic) state, while when $\mathfrak{a}^{(1)}_{\alpha}$ acts on a  bosonic (fermionic) state produces a fermionic (bosonic) state.

\subsection{Matrix representation}
\label{sec:matrix_rep}
The type of superalgebras defined in terms of matrices are the most relevant for the physics in $AdS_2$ and $AdS_3$ string sigma models.
In what follows, we shall consider the action of the graded linear maps on the complex superspace $\mathbb{C}^{n|m}$. 

\begin{itemize}
\item $\mathfrak{gl}(2|2)$

The elements of this group act linearly on $\mathbb{C}^{2|2}$, and can be represented as
\begin{equation}
\label{gl(2|2)_matrix}
M = \begin{pmatrix}
B & F \\
\hat{F} & \hat{B} 
\end{pmatrix} \ , 
\end{equation}
where $B, \hat{B} \in \mathfrak{gl}(2)$ are $0$-graded  (bosonic), while $F, \hat{F} \in \mathfrak{gl}(2)$ are $1$-graded  (fermionic). 

\item $\mathfrak{sl}(2|2)$

The elements of this group are the elements of $\mathfrak{gl}(2|2)$ given in (\ref{gl(2|2)_matrix}) with the further constraint
\begin{equation}
\label{supertrace_zero}
\textrm{Str} M \equiv \textrm{tr} B - \textrm{tr} \hat{B} = 0 \ , 
\end{equation}
where $\rm{Str}$ denotes the supertrace. 

\item $\mathfrak{u}(1,1|2)$

The elements of this group are elements (\ref{gl(2|2)_matrix}) of $\mathfrak{gl}(2|2)$ which satisfies the condition 
\begin{equation}
M^{\dagger} = - \eta^{-1} M \eta \ , \qquad\qquad
\eta = \eta^{-1} \equiv \textrm{diag}(1, -1, 1, 1) \ ,
\end{equation}
which admits eight $0$-graded and eight $1$-graded solutions. 
Among the $0$-graded solutions, one finds the following two central elements 
\begin{equation}
\textrm{Id} = \textrm{diag} (1, 1, 1, 1) \ , \qquad\qquad
\tilde{\textrm{Id}} = \textrm{diag} (1, 1, -1, -1) \ . 
\end{equation}

\item $\mathfrak{su}(1,1|2)$

The elements of this group are elements of $\mathfrak{u}(1,1|2)$ with the additional condition (\ref{supertrace_zero}) on the supertrace. Such condition does not admit $\tilde{\textrm{Id}}$ as a solution, but it admits $\textrm{Id}$. 

The superalgebra $\mathfrak{psu}(1,1|2)$, whose center is empty, is given by $\mathfrak{su}(1,1|2) \scriptsize \setminus \normalsize \{ \textrm{Id} \}$. 

\item $\mathfrak{su}(1|1)$

The elements of this group are linear maps in $\mathbb{C}^{1|1}$ which can be represented as
\begin{equation}
\label{su(1|1)_elements}
M = \begin{pmatrix}
b & f \\
\hat{f} & \hat{b} 
\end{pmatrix} \ , 
\end{equation}
where $b, \hat{b} \in \mathbb{C}$ are $0$-graded, and $f, \hat{f} \in \mathbb{C}$ are $1$-graded. Furthermore (\ref{su(1|1)_elements}) must satisfy
\begin{equation}
M^{\dagger} = - M \ , \qquad\qquad
\textrm{Str}M = b - \hat{b} = 0 \ . 
\end{equation}
The only central element of  $\mathfrak{su}(1|1)$ is the identity matrix, which when an appropriate quotient is taken, produces the superalgebra $\mathfrak{psu}(1|1)$.
\end{itemize}

\subsection{Fundamental equations of the integrable scattering}
\label{sec:fundamental_eq_R} 
We are interested in scatterings between string modes in a $1+1$ integrable field theory.  A generic interacting integrable theory in 2-dimensions enjoys peculiar features, which heavily constrain the scattering matrix:
\begin{enumerate}
\item The number of particles is conserved in any scattering process, i.e. 
\begin{equation}
n \longrightarrow n \ , 
\end{equation}
is the only type of allowed scattering. 
\item Every single momentum is conserved, although permutations of the momenta are allowed:
\begin{equation}
| p_1, p_2 , \cdots , p_n \rangle  \longrightarrow |p_{\pi(1)} , p_{\pi(2)} , \cdots , p_{\pi(n)} \rangle \ , \qquad \pi : \{ 1, \cdots , n \} \rightarrow \{ 1, \cdots , n \} \ . 
\end{equation}
\item Every $n$-body scattering is reducible in terms of 2-body scatterings. 
\end{enumerate}
Therefore a generic scattering in an integrable $1+1$ theory reduces to study only 2-body scatterings $2 \rightarrow 2$, which must behave like classical hard balls collisions. Quantum charges (e.g. flavours) however can change in the process. 

The two incident particles involved in the scattering are described by asymptotic states $|u\rangle$ and $|v \rangle$, where we assume that asymptotically there are no interactions. The states then evolve in time, interact between each other and again some final asymptotic states are produced. The scattering matrix is the object which tells us the linear relation between the initial and final states. Since fermionic particles are involved in the scattering, it turns out to be more convenient to talk about the R-matrix, instead of the S-matrix. 

To define the R-matrix, first let us introduce the \emph{permutation operator} $\Pi$, defined on a two-particle state $|u \rangle \otimes | v \rangle$ as follows
\begin{equation}
\label{permutation_operator}
\Pi |u \rangle \otimes | v \rangle = (-1)^{|u| |v|} |v \rangle \otimes | u \rangle \ , 
\end{equation}
where $|u|$ is the \emph{parity} of $u$, defined as
\begin{equation}
|u| = \begin{cases} 0, & \mbox{if } u = \phi \ \mbox{(boson)} \\ 1, & \mbox{if } u = \psi \ \mbox{(fermion)} \end{cases}
\end{equation}
One can also define $\Pi$ on a tensor product of superalgebra generators:
\begin{equation}
\Pi ( \mathfrak{a} \otimes \mathfrak{b} ) = (-1 )^{|\mathfrak{a}| |\mathfrak{b}|}  \mathfrak{b}\otimes \mathfrak{a} \ , 
\end{equation}
where $|\mathfrak{a}|$ is the parity of $\mathfrak{a} \in \mathfrak{K}$, defined as
\begin{equation}
|\mathfrak{a}| = \begin{cases} 0, & \mbox{if } \mathfrak{a} \in \mathfrak{K}_0  \\ 1, & \mbox{if } \mathfrak{a} \in \mathfrak{K}_1  \end{cases}
\end{equation}

The relation between the R- and S-matrix on two-particle states is the following:
\begin{equation}
\label{S_and_R}
S |u \rangle \otimes | v \rangle = \Pi \circ R \circ \Pi\,  |u \rangle \otimes | v \rangle \ , 
\end{equation}
which means that first one has to permute the initial states, then act with the R-matrix, and finally permute the final states. 

The R-matrix must satisfy a set of fundamental equations. Integrability implies that the R-matrix must satisfy the \emph{Yang-Baxter equation} (YBE), which is defined as follows. 
Consider the initial asymptotic state composed of three particles, which we denote by $|u_1 \rangle \otimes |u_2 \rangle \otimes |u_3 \rangle$, and where $p_1, p_2, p_3$ are the associated momenta. Then the YBE is\footnote{In what follows, any fermionic sign is implicit inside the definition of the formul\ae. } 
\begin{equation}
\label{YBE}
R_{12} (p_1, p_2) R_{13}(p_1, p_3) R_{23}(p_2, p_3)  = R_{23}(p_2, p_3) R_{13}(p_1, p_3) R_{12}(p_1, p_2) \ ,
\end{equation}
where both sides of (\ref{YBE}) implicitly act on $|u_1 \rangle \otimes |u_2 \rangle \otimes |u_3 \rangle$. The YBE tells us that the 3-body scattering factorizes into 2-body scatterings, and the order of the collisions is irrelevant\footnote{Recall that the space-time contains only one spatial dimension, therefore the particle 2 is always the first one involved in the collision.}. 
The action of $R_{12} (p_1, p_2)$ on the triple tensor product $|u_1 \rangle \otimes |u_2 \rangle \otimes |u_3 \rangle$ is defined as
\begin{equation}
R_{12} (p_1, p_2)  |u_1 \rangle \otimes |u_2 \rangle \otimes |u_3 \rangle \equiv ( R  \otimes \mathds{1} ) |u_1 \rangle \otimes |u_2 \rangle \otimes |u_3 \rangle \ , 
\end{equation} 
and in general $R_{ij}(p_i, p_j)$ only acts on the states $|u_i\rangle$ and $|u_j\rangle$, leaving $|u_k\rangle$, $k \neq i, j$ invariant.  
In terms of particle world-lines, the YBE is pictorially represented as follows
\begin{figure}[H]
\begin{center}
 \includegraphics[scale=0.3]{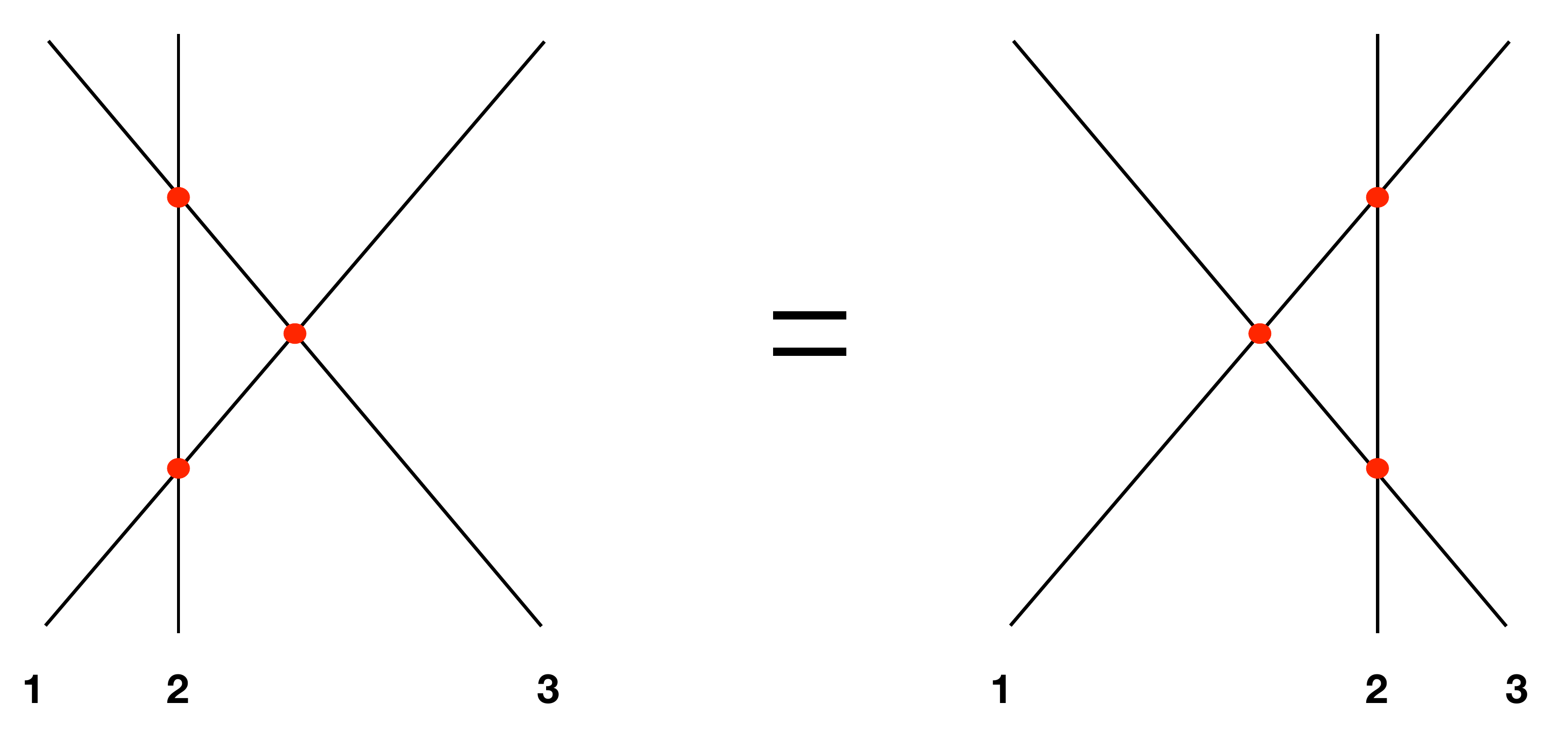}
\end{center}
\caption{Yang-Baxter equation.}
\end{figure}
Another fundamental equation which the R-matrix must satisfy is the so-called \emph{braiding unitarity}, which is different from physical unitarity, as explained in \cite{Hoare:2013hhh}, but it finds its motivation on quantum groups. On a two-particle state $|u_1 \rangle \otimes |u_2 \rangle$, braiding unitarity is
\begin{equation}
\label{braid_unitarity}
(\Pi \circ R) R = \mathds{1} \otimes \mathds{1} \ ,
\end{equation} 
or equivalently
\begin{equation}
R_{21}(p_2, p_1) R_{12}(p_1, p_2) = \mathds{1}  \ .
\end{equation}
where one defines
\begin{equation}
R_{12} (p_1, p_2) \equiv R |u_1\rangle \otimes |u_2 \rangle \ , \qquad\qquad R_{21} (p_2, p_1) \equiv \Pi \circ R |u_1\rangle \otimes |u_2 \rangle \ ,
\end{equation}
and the momenta in $R_{21}$ have been exchanged by hand. 
Braiding unitarity, in terms of particle world-lines, becomes 
\begin{figure}[H]
\begin{center}
 \includegraphics[scale=0.3]{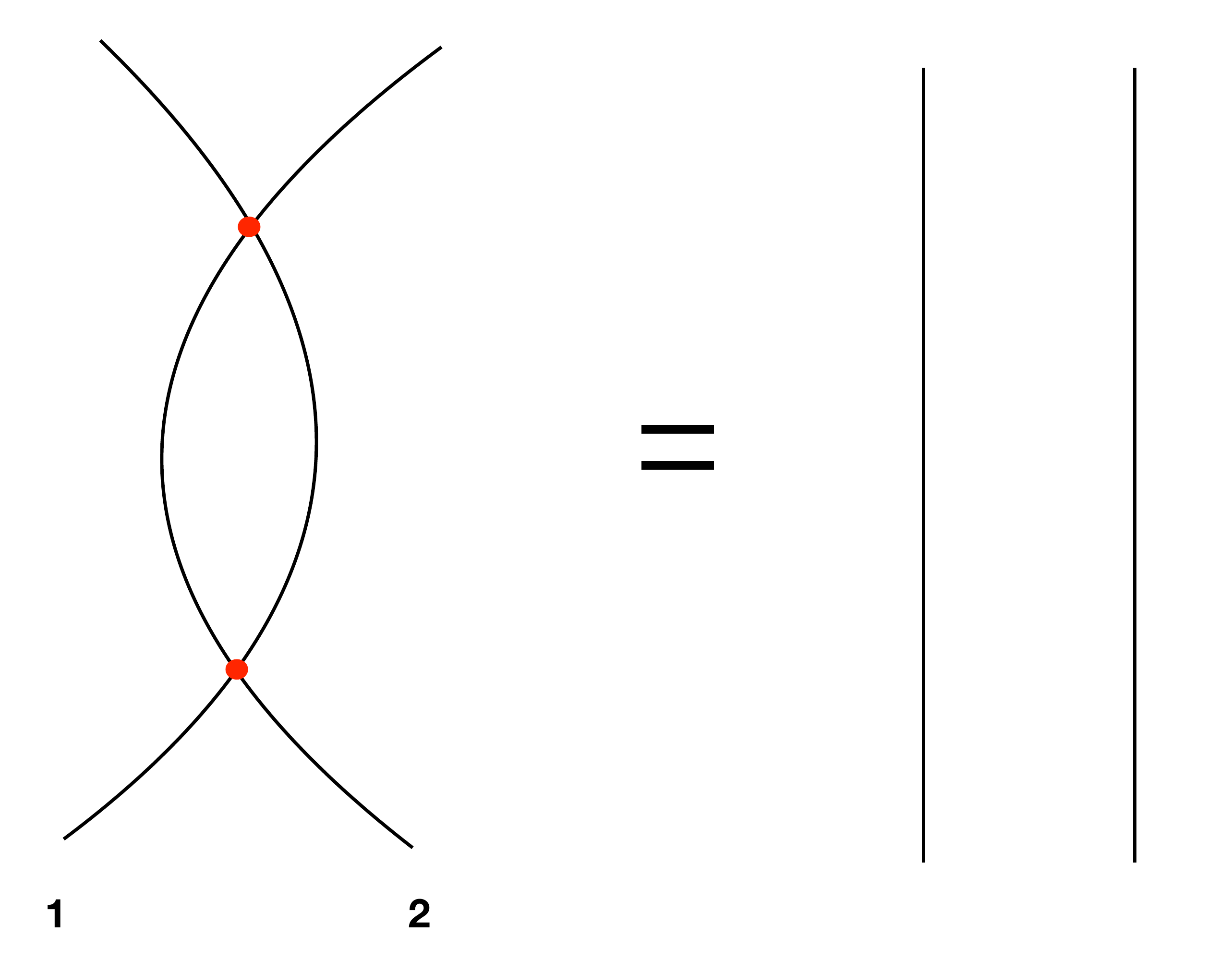}
\end{center}
\caption{Braiding unitarity equation.}
\end{figure}
\noindent which means that if we collide two particles twice between themselves is equivalent, in terms of final asymptotic states, to have no interaction.

From the physical observation that anti-particles can be interpreted as particles travelling back in time, the R-matrix must satisfy the \emph{crossing symmetry} equation, which is schematically the following 
\begin{equation}
\label{crossing}
R_{12} (p_1, p_2)  R_{\bar{1}2} (- p_1, p_2) = \mathds{1} \otimes \mathds{1} \ ,
\end{equation}
where $\bar{1}$ is the anti-particle of $1$. A more rigorous version of the crossing symmetry equation will be presented in section \ref{sec: Hopf}. 
Crossing can be represented in terms of particle world-lines as follows
\begin{figure}[H]
\begin{center}
 \includegraphics[scale=0.3]{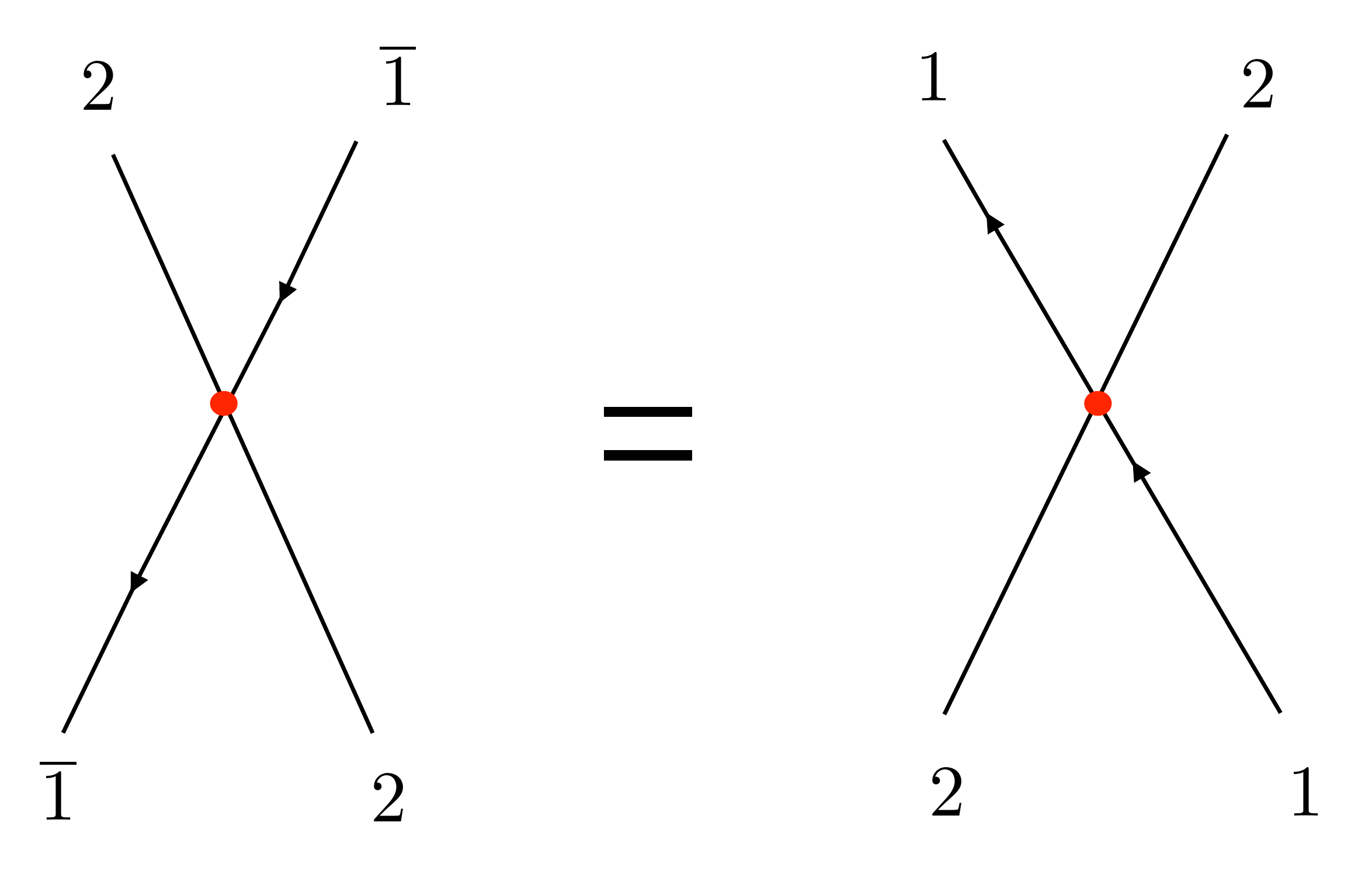}
\end{center}
\caption{Crossing equation.}
\end{figure}

Finally, the isometries of the background must be a symmetries also of the R-matrix. In particular, the residual superalgebra $\mathfrak{K}$ introduced in section \ref{sec: single particle} must be a symmetry of the R-matrix. To implement this concept, one has to define the action of $\mathfrak{K}$ on a two-particle state $|u\rangle \otimes |v\rangle$, and the formal way to make it is in the context of Hopf algebras, as we shall see in section \ref{sec: Hopf}. 
We denote by $\Delta (\mathfrak{a})$ the action of a generic generator $\mathfrak{a} \in \mathfrak{K}$ on $|u\rangle \otimes |v\rangle$, for which there are several possible definitions. 
However, a legitimate definition of $\Delta$ must preserve the Lie superalgebra graded commutation relations, i.e. if the superalgebra generators satisfy
\begin{equation}
\label{graded_algebra_rel}
[ \mathfrak{a}^{(0)}_i,  \mathfrak{a}^{(0)}_j ] = f_{ij}{}^k \mathfrak{a}^{(0)}_k \ , \qquad
[\mathfrak{a}^{(0)}_i, \mathfrak{a}^{(1)}_{\alpha} ] = f_{i\alpha}{}^{\beta} \mathfrak{a}^{(1)}_{\beta} \ , \qquad
\{ \mathfrak{a}^{(1)}_{\alpha}, \mathfrak{a}^{(1)}_{\beta} \} = f_{\alpha\beta}{}^{i} \mathfrak{a}^{(0)}_{i} \ , 
\end{equation}
where $f_{MN}{}^{P}$, $M = i, \alpha$, are the structure constants, then 
\begin{eqnarray}
\label{Delta}
\notag
&&[ \Delta(\mathfrak{a}^{(0)}_i ),  \Delta(\mathfrak{a}^{(0)}_j )] = f_{ij}{}^k \Delta (\mathfrak{a}^{(0)}_k) \ , \qquad
[\Delta( \mathfrak{a}^{(0)}_i ), \Delta (\mathfrak{a}^{(1)}_{\alpha}) ] = f_{i\alpha}{}^{\beta} \Delta(\mathfrak{a}^{(1)}_{\beta} )\ , \qquad \\
&& \qquad\qquad\qquad\qquad\quad \{\Delta(\mathfrak{a}^{(1)}_{\alpha}), \Delta(\mathfrak{a}^{(1)}_{\beta}) \} = f_{\alpha\beta}{}^{i} \Delta(\mathfrak{a}^{(0)}_{i}) \ , 
\end{eqnarray}
Let $\mathfrak{a}$ be a generic graded generator of $\mathfrak{K}$. 
The most simple definition of $\Delta$ is based on the Leibniz rule, which is
\begin{equation}
\label{trivial_coprod}
\Delta(\mathfrak{a}) = \mathfrak{a} \otimes \mathds{1} + \mathds{1} \otimes \mathfrak{a} \ , 
\end{equation}
where $\mathfrak{a}$ acts on the first (resp. second) space in the single particle representation, and leaves the second (resp. first) space invariant. In the context of $AdS$ scatterings, this coproduct appears when the particles are in the relativistic limit. 

Another possible definition, which introduces non-localities, consists in acting simultaneously on both states. For example
\begin{equation}
\Delta (\mathfrak{a}) = \mathfrak{a} \otimes\big( f(p) \mathds{1}\big) + \big( g(p) \mathds{1} \big) \otimes \mathfrak{a} \ , 
\end{equation}
where $f, g$ are functions of the momentum, which are constrained by requiring (\ref{Delta}). This happens for non-relativistic scatterings, where typically $f$ and $g$ depends exponentially on the momentum. 

We finally mention one more possible definition of $\Delta$, which is of the following type
\begin{equation}
\Delta (\mathfrak{a}) = \mathfrak{a} \otimes\big( f(p) \mathds{1}\big) + \big( g(p) \mathds{1} \big) \otimes \mathfrak{a} + \sum \big( h_1(p)\mathfrak{b} \big) \otimes \big( h_2 (p) \mathfrak{c} \big) \ , 
\end{equation}
where $\mathfrak{b}, \mathfrak{c} \in \mathfrak{K}$, and $h_1, h_2$ functions of the momentum. This definition appears when we  will consider the boost generator in chapter \ref{chapter: AdS3}. 

The invariance of the R-matrix under the residual symmetry algebra $\mathfrak{K}$ becomes
\begin{equation}
\label{algebra_R}
\Delta^{op}(\mathfrak{a}) R = R \Delta (\mathfrak{a}) \ , \qquad\qquad \forall \, \mathfrak{a} \in \mathfrak{K} \ ,  
\end{equation}
where
\begin{equation}
\Delta^{op} \equiv \Pi \circ \Delta \ . 
\end{equation}
In the context of Hopf algebras, $\Delta$ is called \emph{coproduct}, and $\Delta^{op}$ is the \emph{opposite} coproduct.

\subsection{Dressing factor and crossing symmetry}
\label{sec: dressing}
The crossing symmetry equation (\ref{crossing}) is crucial to completely solve the R-matrix. In most of the cases where the representation of the symmetry superalgebra $\mathfrak{K}$ is irreducible\footnote{This happens in e.g. $AdS_3$ sigma model, but not in $AdS_2$, where one has to use also the Yang-Baxter equation to fix the R-matrix entries.}, the symmetry invariance condition for $R$ (\ref{algebra_R}) suffices to determine the structure of the R-matrix entries, up to an overall factor. Such factor is a complex function depending on the particle momenta, which is called \emph{dressing factor} $\Phi$.

Suppose that the structure of the R-matrix is known up to the dressing factor. Then one can decompose 
\begin{equation}
\label{R_decomp}
R_{12} (p_1, p_2 ) = \Phi (p_1, p_2) \hat{R}_{12}(p_1, p_2 ) \ , 
\end{equation}
where $\hat{R}$ is known. Schematically, the crossing equation (\ref{crossing}) for (\ref{R_decomp}) becomes 
\begin{equation}
\label{cross_dress}
\Phi(p_1, p_2) \hat{R}_{12}(p_1, p_2)  \Phi (-p_1, p_2) \hat{R}_{\bar{1}2}(-p_1, p_2) = \mathds{1} \otimes \mathds{1} \ .
\end{equation}
Then one can compute the product between the two $\hat{R}$ matrices, which must be proportional to the identity, 
\begin{equation}
\hat{R}_{12}(p_1, p_2)\hat{R}_{\bar{1}2}(-p_1, p_2) = f \, \mathds{1} \otimes \mathds{1} \ , 
\end{equation}
where $f$ is a function of the particle momenta. Then (\ref{cross_dress}) implies that 
\begin{equation}
\Phi (p_1, p_2) \Phi(-p_1, p_2) = \frac{1}{f} \ , 
\end{equation}
which turns out to be a Riemann-Hilbert problem for the dressing factor.

\subsection{Fundamental equations revisited: Hopf algebras}
\label{sec: Hopf}
In this section we discuss some formal aspects of the scattering process. In particular we shall see that the appropriate mathematical language to describe the R-matrix, and the associated scattering equations, is given by the Hopf algebras. 
The concept of multiplying together generators of the Lie superalgebra $\mathfrak{K}$, which is relevant when considering the R-matrix symmetries, is encoded in the Hopf algebra formalism.

First, we shall define the \emph{universal enveloping algebra} of $\mathfrak{K}$, denoted by $\mathcal{U}[\mathfrak{K}]$. Let $\mathfrak{K}$ be a Lie superalgebra, with basis of generators $\{ \mathfrak{a}^{(0)}_i , \mathfrak{a}^{(1)}_{\alpha} \}$ which satisfy the graded commutation relations (\ref{graded_algebra_rel}). 

Then $\mathcal{U}[\mathfrak{K}]$ is constructed by considering the graded elements $x^{(0)}_i , x^{(1)}_{\alpha}$ together with the identity element $\mathds{1}$, and by building words out of them with arbitrary length, modulo the equivalence relation $\sim\,$: 
\begin{equation}
\label{equiv_univ}
\sim\ : \quad [ x^{(0)}_i , x^{(0)}_j ] = f_{ij}{}^k x^{(0)}_k \ , \qquad
[ x^{(0)}_i , x^{(1)}_{\alpha} ] = f_{i \alpha}{}^{\beta} x^{(1)}_{\beta} \ , \qquad
\{ x^{(1)}_{\alpha}, x^{(1)}_{\beta} \} = f_{\alpha\beta}{}^i x^{(0)} \ . 
\end{equation}
Therefore 
\begin{equation}
\mathcal{U}[\mathfrak{K}] \equiv \{\mathds{1}, \ \, x^{(0)}_i, \ \, x^{(1)}_{\alpha} , \ \,  x^{(1)}_1x^{(0)}_1,\ \,  x^{(0)}_2x^{(1)}_2x^{(1)}_4x^{(0)}_3 , .... \} \bigg/ \sim \ .  
\end{equation}
We remark that the equivalence relation (\ref{equiv_univ}) is the \emph{only} relation which one imposes. 
This means that $\mathcal{U}[\mathfrak{K}]$ does not inherit any additional extra relation which the generators of $\mathfrak{K}$ might satisfy when written in the matrix representation, as shown in the following example.

\emph{Example:}
Consider the simple bosonic case $\mathfrak{K} = \mathfrak{sl}(2, \mathbb{R})$. Let $X, Y, H$ be the generators, which satisfy the relations
\begin{equation}
[H, X] = 2 X \ , \qquad
[H, Y] = - 2 Y \ , \qquad
[X, Y] = H \ . 
\end{equation}
Then $X, Y, H$ can be represented in terms of matrices in the fundamental representation as:
\begin{equation}
X = \begin{pmatrix}
0 & 1 \\
0 & 0 
\end{pmatrix} \ , 
\qquad
Y = \begin{pmatrix}
0 & 0 \\
1 & 0 
\end{pmatrix} \ , \qquad
H = \begin{pmatrix}
1 & 0 \\
0 & -1 
\end{pmatrix} \ . 
\end{equation}
The universal enveloping algebra $\mathcal{U}[\mathfrak{sl}(2, \mathbb{R}) ]$ is defined as
\begin{equation}
\mathcal{U}[\mathfrak{sl}(2, \mathbb{R})] = \{ \mathds{1}, x, y, h , xyx, hyx, h^2 y^2 x^3 , .... \} \bigg/ \sim \ , 
\end{equation}
where
\begin{equation}
\sim \, : \qquad
hx - xh = 2 x \ , \qquad
hy - yh = - 2 y \ , \qquad
xy - yx = h \ . 
\end{equation}
From the fundamental matrix representation of $X$, we observe that $X^2 = 0$. However, we cannot infer this condition on the element $x \in\mathcal{U}[\mathfrak{sl}(2, \mathbb{R})]$ in the abstract algebra. 

Formally, the universal enveloping algebra admits the following properties
\begin{itemize}
\item a \emph{neutral element} $\mathds{1}$;
\item a \emph{multiplication map} 
\begin{equation}
\notag
\mu : \mathcal{U}[\mathfrak{K}] \otimes \mathcal{U}[\mathfrak{K}] \rightarrow \mathcal{U}[\mathfrak{K}] \ , 
\end{equation} 
which allows us to take words of elements of $\mathcal{U}[\mathfrak{K}]$ and generate again an element of $\mathcal{U}[\mathfrak{K}]$; 
\item a \emph{unit map}
\begin{equation}
\notag
\eta : \mathbb{C} \rightarrow \mathcal{U}[\mathfrak{K}] \ , 
\end{equation}
\end{itemize}
together with the additional maps: 
\begin{itemize}
\item a \emph{co-unit map}
\begin{equation}
\notag
\varepsilon : \mathcal{U}[\mathfrak{K}] \rightarrow \mathbb{C} \ , 
\end{equation} 
\item a \emph{co-product}
\begin{equation}
\notag
\Delta : \mathcal{U}[\mathfrak{K}] \rightarrow \mathcal{U}[\mathfrak{K}] \otimes \mathcal{U}[\mathfrak{K}] \ , 
\end{equation}
compatible with the graded commutation relations of $\mathfrak{K}$.  
\end{itemize}
As an axiom, the co-product $\Delta$ must satisfy the \emph{co-associativity} property, which is 
\begin{equation}
\big( \Delta \otimes \mathds{1}  -  \mathds{1} \otimes \Delta \big) \mathfrak{a} = 0  \ , \qquad
\forall \, \mathfrak{a} \in \mathcal{U}[\mathfrak{K}] \ . 
\end{equation}
which means that the following diagram
\begin{equation}
\begin{tikzcd}
\mathcal{U}[\mathfrak{K}] \arrow{r}{\Delta} \arrow{d}{\Delta} & \mathcal{U}[\mathfrak{K}] \otimes \mathcal{U}[\mathfrak{K}] \arrow{d}{\mathds{1} \otimes \Delta} \\
\mathcal{U}[\mathfrak{K}] \otimes \mathcal{U}[\mathfrak{K}] \arrow{r}{\Delta \otimes \mathds{1}}& \mathcal{U}[\mathfrak{K}]\otimes \mathcal{U}[\mathfrak{K}] \otimes \mathcal{U}[\mathfrak{K}]
\end{tikzcd} 
\end{equation}
commutes. 
We need just an additional map to turn a universal enveloping algebra into a Hopf algebra, which is given as follows 
\begin{definition}[Hopf algebra]
A Hopf algebra $H$ is a universal enveloping algebra with the additional property that there exists an antipode map
\begin{equation}
\notag
\Sigma : H \rightarrow H \ , 
\end{equation}
which is compatible with the graded commutation relations of $\mathfrak{K}$, and with the maps of the universal enveloping algebra as follows
\begin{equation}
\mu \circ (\Sigma \otimes \mathds{1} )\circ \Delta = \eta \circ\varepsilon \ . 
\end{equation}
\end{definition}
Suppose to introduce a matrix representation for elements $\mathfrak{a} \in H$, denoted by $\pi(\mathfrak{a})$. 
Then the antipode map allows us to define the so-called \emph{antiparticle} representation $\tilde{\pi}(\mathfrak{a})$, as follows
\begin{equation}
\label{antiparticle_rep}
\Sigma ( \pi(\mathfrak{a})) = C^{-1}  \tilde{\pi}^{str}(\mathfrak{a}) C \ , 
\end{equation}
where $C$ is the charge-conjugation matrix, and $\pi^{str}$ is the supertranspose of $\pi$. If we restrict $\Delta$ to be (\ref{trivial_coprod}) and $\mathfrak{a}$ to be just a generator of the Lie superalgebra $\mathfrak{K}$, then\footnote{The antipode, if exists, is uniquely determined from the coproduct.}
\begin{equation}
\Sigma (\mathfrak{a}) = - \mathfrak{a} \ , \qquad\qquad
\mathfrak{a} \in \mathfrak{K} \ ,
\end{equation} 
and equation (\ref{antiparticle_rep}) becomes
\begin{equation}
- \pi(\mathfrak{a}) = C^{-1}  \tilde{\pi}^{str}(\mathfrak{a}) C \ , \qquad\qquad
\mathfrak{a} \in \mathfrak{K} \ .
\end{equation}
The language of Hopf algebras turns out to be particularly suitable to describe integrable scatterings, where the algebra invariance equation (\ref{algebra_R}), braiding unitarity (\ref{braid_unitarity}), crossing symmetry (\ref{crossing}) and the Yang-Baxter equation (\ref{YBE}) formalise. 

We first notice that one can compose the co-product map $\Delta$ with the permutation operator $\Pi$, and generate a new legitimate co-product map, $\Delta^{op} \equiv \Pi \circ \Delta$. In general $\Delta$ and $\Delta^{op}$ are different (e.g. quantum groups), however there exists also the following case:
\begin{definition}[Cocommutative Hopf algebra]
A cocommutative Hopf algebra is a Hopf algebra such that
\begin{equation}
\notag
\Delta = \Delta^{op} \ . 
\end{equation}
\end{definition}
Consider a Hopf algebra which is \emph{not} cocommutative, i.e. $\Delta \neq \Delta^{op}$. Then one may wonder whether the two tensor product representations $\Delta$ and $\Delta^{op}$ are \emph{equivalent}. This leads us to the following 
\begin{definition}[Quasi-cocommutative]
A quasi-cocommutative Hopf algebra is a Hopf algebra for which there exists $R \in H \otimes H$ invertible
such that 
\begin{equation}
\label{quasi_cocom}
\Delta^{op} (\mathfrak{a}) R = R \Delta (\mathfrak{a} ) \ , \qquad\qquad
\forall \ \mathfrak{a} \in H \ , 
\end{equation}
\end{definition}
The invertible element $R$, which guarantees the equivalence between $\Delta$ and $\Delta^{op}$, is called \emph{R-matrix}, and in the context of scattering theory, it is related to the physical S-matrix via (\ref{S_and_R}). 

Inside the class of quasi-cocommutative Hopf algebras, one can identify a subclass of Hopf algebras for which the R-matrix satisfy additional properties, the so-called \emph{bootstrap} conditions. 
\begin{definition}[Quasi-triangular Hopf algebra]
A quasi-triangular Hopf algebra is a quasi-cocommutative Hopf algebra for which the R-matrix satisfy the following conditions \begin{equation}
\label{boostrap}
(\mathds{1} \otimes \Delta ) R = R_{13} R_{23} \ , \qquad
(\Delta \otimes \mathds{1} ) R = R_{13} R_{12} \ , 
\end{equation}
where $R_{ij} \equiv \phi_{ij}(R)$, and $\phi_{ij} : H \otimes H \rightarrow H \otimes H \otimes H$ is an algebra morphism given by 
\begin{equation}
\phi_{12}( a\otimes b) = a \otimes b \otimes \mathds{1} \ , \qquad
\phi_{13} (a \otimes b) = a \otimes \mathds{1} \otimes b \ , \qquad
\phi_{23} (a \otimes b) = \mathds{1} \otimes a \otimes b \ . 
\end{equation}
Conditions (\ref{boostrap}) are the bootstrap equations. 

\end{definition}
 
One can show that the boostrap equations imply the following conditions \cite{Kassel}
\begin{equation}
\label{YBE_free}
(R \otimes \mathds{1} ) ( \mathds{1} \otimes R ) (R \otimes \mathds{1} ) = ( \mathds{1} \otimes R ) (R \otimes \mathds{1}) ( \mathds{1} \otimes R ) \ , 
\end{equation}
and
\begin{equation}
\label{Cross_free}
(\Sigma\otimes \mathds{1}) R = R^{-1}\ , \qquad\qquad
(\mathds{1}\otimes \Sigma^{-1})R =  R^{-1} \ . 
\end{equation}
Equation (\ref{YBE_free}) is the \emph{Yang-Baxter equation}, while equation (\ref{Cross_free}) is the \emph{crossing equation}, which can be written formally in terms of a matrix representation as
\begin{equation}
\bigg[(C^{-1} \otimes \mathds{1} ) (\tilde{\pi}^{str}_1 \otimes \pi_2) R^{str_1}(-p_1, p_2) (C \otimes \mathds{1} ) \bigg] \bigg[(\pi_1 \otimes \pi_2)R(p_1, p_2) \bigg] = \mathds{1} \otimes \mathds{1} \ ,  
\end{equation}
where $(\pi_1 \otimes \pi_2)R $ denotes $R$ in the (tensor product) matrix representation. 
Furthermore, the definition of quasi-cocommutative Hopf algebra (\ref{quasi_cocom}) is nothing else then the \emph{algebra invariance} condition for the R-matrix and the fact that $R$ is invertible is implied by \emph{braiding unitarity}.
Therefore this shows that the mathematical framework of quasi-triangular Hopf algebra includes all the equations which describes the symmetries associated with integrable scatterings. 

An important remark is the following. The language of Hopf algebras is \emph{not} just a fancy mathematical way to rewrite the integrable scattering problem. In the context of Hopf algebras there exists the notion of \emph{universal R-matrix}, which is an abstract solution to the quasi-cocommutativity condition, and in particular it is representation-free. Of course if a particular representation is chosen the universal R-matrix reduces to the R-matrix in that representation. 
Physically, it is important to study the properties of the universal R-matrix because it allows to understand the structure of the hidden symmetry algebra of the integrable system, known as \emph{Yangian symmetry} \cite{rev}, which is relevant for the AdS/CFT spin-chain. 
However we shall not further discuss this topic here.

\section{Bethe ansatz}
\label{Bethe}
Given a generic theory, in generic dimensions, it is an interesting question to ask which is the spectrum of the theory. It is also well known that finding the complete set of eigenstates and eigenenergies for the Hamiltonian is in general a difficult problem. In quantum mechanics, for instance, this is possible for the harmonic oscillator and the hydrogen atom, which are regarded as toy models.

Classically integrable theories, however, are particularly good candidates. If one can identify the S-matrix for such theories, then there exists a well established technique to formulate the so-called \emph{Bethe ansatz}, which allows at the end to solve the spectral problem. If an integrable theory admits a Bethe ansatz formulation, then the theory is also \emph{quantum integrable}.  

Historically, Hans Bethe discovered a procedure to find the complete set of eigenstates and eigenenergies of the XXX spin chain. It was found later that this procedure, named Bethe ansatz, can also be applied to other classically integrable systems. 

In the context of scatterings in 1+1 dimensional integrable theories, one consider initial asymptotic states which will produce, after the scattering process, some final asymptotic states. The Bethe ansatz for this problem is called \emph{Asymptotic Bethe Ansatz}, and it tells us information about the continuum spectrum.  

On the other hand, one can consider the scattering problem inside a box of finite size, instead of infinite spatial directions.
In $1+1$ dimensions, this requires to consider a spacetime which topologically looks like $\mathbb{R} \times [0, 1]$. After a double Wick rotation, A. B. Zamolodchikov shows that the roles of the (infinitely extended) time and the compact space exchange between each other. This is equivalent to study scatterings in an infinitely extended space, with the new ingredient of turning on a  \emph{temperature}, since the time direction is now compact, \cite{Zamolodchikov:1989cf, Zamolodchikov:1991vh}. Dealing with a non-zero temperature implies that one has to consider systems in thermal equilibrium. The Bethe ansatz in this case is called \emph{Thermodynamic Bethe Ansatz} (TBA), and it would tell us about the spectrum of the finite-volume problem by using the data associated with the infinite-volume scattering \cite{rev} (chapter III.6). To formulate the thermodynamic Bethe ansatz, one has to first find the asymptotic Bethe ansatz, and then to take the large number of particles limit, the so-called thermodynamic limit.
An interesting feature of the thermodynamic Bethe ansatz is that it provides information about the central charge of the associated CFT at the fixed points in the RG flow\footnote{In the case of massless scatterings in AdS/CFT, it has still to be understood if the central charge obtained by TBA is associated to the world-sheet CFT at fixed points.}.

In this section we shall review the standard technique to formulate the asymptotic Bethe ansatz (see e.g. \cite{Faddeev:1996iy, Levkovich-Maslyuk:2016kfv}). Our description applies to models which are of the XXX spin chain type.

Consider a set of $n$ particles on a circle with periodic boundary conditions, where the first and $n$-th particles are identified. Suppose to take the first particle all around the circle and finally back to the original position. Consider first the case where there are no interactions between the particles. 
Then the wave function associated with the first particle will pick up a phase factor of $e^{ip_1 L}$. 
However we imposed periodicity conditions, and therefore
\begin{equation}
e^{i p_1 L} = 1 \ , 
\end{equation} 
which implies 
\begin{equation}
\label{quant}
p_1 = \frac{2\pi}{L} k \ , \qquad\qquad k \in \mathbb{Z} \ .
\end{equation}
Hence the spectrum of the first particle is determined, and in particular it is discrete. 

Consider now the case of interacting particles, such that the scattering factorises, and make again the first particle moving around the circle. The wave function will pick up again a phase $e^{ip_1 L}$ for free propagation, plus a contribution from the interactions with the other particles, $S(p_1, p_2) S(p_1, p_3) \cdots S(p_1, p_n)$. By periodicity, we have that
\begin{equation}
\label{periodicity_int}
e^{i p_1 L}S_{12}(p_1, p_2) S_{13}(p_1, p_3) \cdots S_{1n}(p_1, p_n) = \mathds{1} \otimes \mathds{1}  \ , 
\end{equation}
and in principle (\ref{periodicity_int}) fixes the spectrum of the first particle. 
Moreover this equation can be generalised to each particle in $\{ 1, \cdots , n\}$, since the above argument still holds. 
The generalisation of (\ref{periodicity_int}) is the following
\begin{equation}
\label{periodicity_gen}
e^{i p_k L}S_{k, k+1} S_{k, k+2} \cdots S_{k, n} S_{k, 1} \cdots  S_{k, k-1}  = \mathds{1} \otimes \mathds{1} \ , \qquad\quad \forall \, k = 1, ... , n \ . 
\end{equation}  
 
To solve (\ref{periodicity_gen}), one introduces an auxiliary particle, denoted by $0$ with momentum $\lambda$, living on an auxiliary Hilbert space $V_0$. Virtually one takes the particles $0$ all around the circle until reaching the original position. In this way the auxiliary particle interacts with the particles in $\{ 1, \cdots , n\}$.  
One introduces the \emph{monodromy matrix} $\mathcal{T}_0$: 
\begin{equation}
\mathcal{T}_0 (\lambda) \equiv S_{01} (\lambda, p_1)\cdot S_{02}(\lambda, p_2) \cdot ... \cdot S_{0n}(\lambda, p_n)  \ ,  
\end{equation}
where $\cdot$ is the multiplication in the auxiliary space. One introduces the \emph{transfer matrix} $\mathcal{T}$ as the trace of the monodromy matrix in $V_0$, 
\begin{equation}
\mathcal{T} (\lambda) \equiv {\rm tr}_0 \mathcal{T}_0(\lambda) \ .
\end{equation}
$S, \mathcal{T}_0$ and $\mathcal{T}$ are linear maps between tensor products of Hilbert spaces: the S-matrix $S_{ab} \in End(V_a \otimes V_b)$, the monodromy matrix $\mathcal{T}_0 \in End( V_0 \otimes V_1 \otimes \cdots \otimes V_n )$ and the transfer matrix $T \in End ( V_1 \otimes \cdots \otimes V_n )$.
The monodromy matrix can be written as a $2^{n+1} \times 2^{n+1}$ matrix, however to emphasize the action on the auxiliary space, one can write $\mathcal{T}_0$ as a $2\times 2$ block matrix acting on $V_0$, 
\begin{equation}
\mathcal{T}_0 (\lambda) = \begin{pmatrix}
A(\lambda) & B(\lambda) \\
C(\lambda) & D(\lambda) 
\end{pmatrix} \ , 
\end{equation} 
where $A, B, C, D \in End( V_1 \otimes \cdots \otimes V_n )$ are operators acting on the so-called \emph{quantum} space. 
The transfer matrix then becomes
\begin{equation}
\mathcal{T} (\lambda) = A(\lambda) + D(\lambda) \ . 
\end{equation}
The Yang-Baxter equation for the R-matrix implies the so-called RTT relations, which we write here in terms of the S-matrix:
\begin{equation}
\label{RTT}
\mathcal{T}_{0}(\lambda) \mathcal{T}_{0'}(\lambda') S_{00'}(\lambda - \lambda') = S_{00'} (\lambda - \lambda') \mathcal{T}_{0'}(\lambda') \mathcal{T}_{0}(\lambda) \ ,   
\end{equation}
where $0$ and $0'$ are two distinct auxiliary particles. 
The RTT relations can be interpreted as follows: scattering $0$ against $0'$ and then taking them all around the circle is equivalent to first taking $0$ and $0'$ all around the circle and then scattering $0$ and $0'$ against each other. 

Furthermore, the RTT relations (\ref{RTT}) imply that 
\begin{equation}
\label{T_commutes}
[ \mathcal{T} (\lambda), \mathcal{T} (\lambda') ] = 0  \ , 
\end{equation}
which allows us to construct conserved quantities 
\begin{equation}
\mathcal{Q}_n = \frac{d^n}{d \lambda^n} \log \mathcal{T}(\lambda) \bigg|_{\lambda = 0} \ , 
\end{equation}
where $\mathcal{Q}_1$ is proportional to the Hamiltonian. 

Equation (\ref{periodicity_gen}) implies the following equation for the transfer matrix
\begin{equation}
\label{T_spec}
\mathcal{T} ( \lambda ) |\Psi \rangle = e^{- i \lambda L} |\Psi \rangle \ , 
\end{equation} 
If one can find the eigenvalues for the transfer matrix
\begin{equation}
\label{eigenvalue}
\mathcal{T} |\Psi \rangle = (A + D) |\Psi \rangle =\Lambda |\Psi \rangle \ , 
\end{equation}
where $|\Psi \rangle$ is a state for the $n$ particles in the circle, then the spectrum is fixed by
\begin{equation}
\label{T_spectrum_eig}
\lambda = \frac{i}{L} \log \Lambda + \frac{2\pi}{L}k \ , \qquad \qquad k\in \mathbb{Z} \ . 
\end{equation}

We construct a reference state $|0 \rangle$, the so-called \emph{pseudo-vacuum}\footnote{This is called pseudo-vacuum because it has not necessarily to coincide with the vacuum state.}, such that it satisfies the following properties
\begin{equation}
\label{pseudovacuum_cond}
A(\lambda) |0 \rangle = a(\lambda) |0 \rangle \ , \qquad
D(\lambda) |0 \rangle = d(\lambda) |0 \rangle \ , \qquad
C(\lambda) |0 \rangle = 0 \ , 
\end{equation}
and such that 
\begin{equation}
\label{excitation}
|p_1 \rangle = B(p_1) |0 \rangle 
\end{equation}
is the first magnon excitation of the associated spin-chain, which is a propagating wave.
This describes the \emph{highest weight module} of the algebra $\mathfrak{K}$, where $|0 \rangle $ is the highest weight state and $B(\lambda)$ is a ladder-type operator. 

We remark that the construction given above in (\ref{pseudovacuum_cond}) and (\ref{excitation}) strongly depends on the entries of the S-matrix considered, and for integrable theories in general this procedure works. 
One can aim to construct eigenstates solutions to (\ref{eigenvalue}) by acting several times on the pseudo-vacuum with $B(\lambda)$:
\begin{equation}
\label{eigen_Psi}
|\Psi \rangle = \prod_{j= 1}^m B(p_j) |0 \rangle \ , \qquad\qquad m \in \mathbb{N} \ . 
\end{equation}
Imposing that (\ref{eigen_Psi}) is an eigenstate for (\ref{eigenvalue}) will lead to a set of algebraic conditions on $p_1, ... , p_n$. 
It is useful to consider the following generalised commutation relations, which follows from the RTT relations (\ref{RTT}): 
\begin{eqnarray}
\label{BB_com}
B(\lambda) B(\lambda') &=& B(\lambda') B(\lambda)  \ , \\
\label{AB_comm}
A(\lambda) B(\lambda') &=& f(\lambda, \lambda') B(\lambda') A(\lambda) + \tilde{f}(\lambda, \lambda') B(\lambda) A(\lambda') \ , \\
\label{DB_comm}
D(\lambda) B(\lambda') &=& g(\lambda, \lambda') B(\lambda') D(\lambda) + \tilde{g}(\lambda, \lambda') B(\lambda) D(\lambda') \ , 
\end{eqnarray}
where $f, \tilde{f}, g, \tilde{g}$ are generic functions of $\lambda, \lambda'$. We remark that (\ref{BB_com}), (\ref{AB_comm}), (\ref{DB_comm}) are sensitive to the S-matrix entries, but for theories of XXX spin chain type in general they capture a standard behaviour. 
Physically, (\ref{BB_com}) says that the order in which the excitations are created does not matter. 

To prove that (\ref{eigen_Psi}) is a eigenstate of the transfer matrix, we have to commute $A(\lambda)$ and $D(\lambda)$ through all the $B$'s by using the commutation relations above. Let us consider $A(\lambda)$ first. One obtains
\begin{eqnarray}
\label{ABBBBB_comm}
\notag
A(\lambda) B(p_1) ... B(p_m) |0\rangle &=& \prod_{k=1}^m f(\lambda, p_k) a(\lambda) B (p_1) ... B(p_m) |0\rangle  \\
&+& \sum_{k=1}^m M_k(\lambda | p_1, ... , p_m) B(p_1) ... \hat{B}(p_k) ... B(p_m) B(\lambda) |0\rangle \ , 
\end{eqnarray}
where $\hat{B}(p_k)$ indicates that the operator $B$ in position $k$ is missing. 
The first term on the RHS of (\ref{ABBBBB_comm}) is obtained from using the first term on the RHS of (\ref{AB_comm}), and it is in a desirable form since we reproduce the state $|\Psi \rangle$. This is a \emph{wanted term}. The second term on the RHS of (\ref{ABBBBB_comm}) does not reproduce the state $|\Psi \rangle$, and the coefficients $M_k$ can be quite involved. This is an \emph{unwanted term}. However the first coefficient $M_1$ can be easily computed. One has to commute $A(\lambda)$ with $B(p_1)$ by using the second term on the RHS of (\ref{AB_comm}), and then commute $A(p_1)$ with the remaining $B$'s by using only the first term on the RHS of (\ref{AB_comm}). One obtains 
\begin{equation}
\label{M_1}
M_1 = \tilde{f}(\lambda, p_1) \prod_{\ell = 2}^m f(p_1, p_\ell) a(p_1) \ . 
\end{equation}  
We remark that all the $B$'s commute, therefore all $M_k$ can be obtained from (\ref{M_1}) by replacing $p_1$ with $p_k$. Therefore we have
\begin{equation}
M_k = \tilde{f}(\lambda, p_k) \prod_{\ell \neq k}^m f(p_k, p_\ell) a(p_k) \ . 
\end{equation} 
The case in which $D(\lambda)$ pass thorough all $B$'s is conceptually the same, and it gives us
\begin{eqnarray}
D(\lambda) B(p_1) ... B(p_m) |0\rangle &=& \prod_{k=1}^m g(\lambda, p_k) d(\lambda)B (p_1) ... B(p_m) |0\rangle  \\
&+& \sum_{k=1}^m N_k(\lambda| p_1, ... , p_m) B(p_1) ... \hat{B}(p_k) ... B(p_m) B(\lambda) |0\rangle \ , 
\end{eqnarray}
where 
\begin{equation}
N_k = \tilde{g}(\lambda, p_k) \prod_{\ell \neq k}^m g(p_k, p_\ell) d(p_k) \ . 
\end{equation} 
The unwanted terms cancel against each other, provided that one imposes the condition $M_k = - N_k$, for all values of $k$, i.e.
\begin{equation}
\tilde{f}(\lambda, p_k) \prod_{\ell \neq k}^m f(p_k, p_\ell) a(p_k) = 
- \tilde{g}(\lambda, p_k) \prod_{\ell \neq k}^m g(p_k, p_\ell) d(p_k) \ , \qquad\forall \, k = 1, ... , m\ .
\end{equation} 
For models of XXX spin chain type, one finds that 
\begin{equation}
\label{f=-g}
\tilde{f}(\lambda, p_k) = - \tilde{g}(\lambda, p_k) \ , 
\end{equation}
which implies
\begin{equation}
\label{Bethe_eqn}
\frac{a(p_k)}{d(p_k) } = \prod_{\ell \neq k}^m \frac{g(p_k, p_\ell)}{f(p_k, p_\ell)}\ , \qquad\forall \, k = 1, ... , m\ .
\end{equation}
These are the so-called \emph{Bethe equations}, and represent a generalised version of the momentum quantisation condition for the free particle (\ref{quant}). Condition (\ref{f=-g}) also guarantees that the set of physical momenta do not depend on the momentum $\lambda$ of the auxiliary particle. 

At this stage the unwanted terms have been eliminated, and the eigenvalues problem (\ref{T_spec}), (\ref{eigenvalue}), (\ref{T_spectrum_eig}), also called \emph{momentum-carrying Bethe equation}, is solved with eigenvalues  
\begin{equation}
\Lambda = a(\lambda) \prod_{k=1}^m f(\lambda, p_k)  + d(\lambda) \prod_{k=1}^m g(\lambda, p_k) \ , 
\end{equation}
provided that the momenta $\{ p_1, ... , p_n\}$ satisfy the Bethe equations (\ref{Bethe_eqn}). 

We comment that not every integrable system admits a pseudo-vacuum state. This happens for instance for the XYZ spin chain \cite{Baxter:1972hz2} or the eight-vertex model \cite{Baxter:1972hz}, and it is a common feature of $\mathcal{N}=1$ supersymmetric integrable models \cite{Schoutens, MC}. The string sigma model on $AdS_2 \times S^2 \times T^6$ also suffers from this problem, and in chapter \ref{Chapter: AdS2} we shall present an alternative way to formulate the asymptotic Bethe ansatz, which do not rely on a pseudo-vacuum state, but instead on an algebraic relation which the R-matrix entries must satisfy, called free-fermion condition.

\clearpage{\pagestyle{empty}\cleardoublepage} 
  
\chapter{\textbf{Spectrum of $AdS_2$ massless modes}}
\label{Chapter: AdS2}

\section{The background algebra}
\label{sec:back_AdS2}
In this chapter we shall consider the spectral problem of massless string excitations in the $AdS_2 \times S^2 \times T^6$ type IIB string sigma model. In this background there are 8 preserved real supercharges, and the bosonic degrees of freedom are described by the coset space
\begin{equation}
\frac{SO(1,2)\times SO(3) \times U(1)^6}{SO(1,1) \times SO(2)} \ .
\end{equation}
This background appears as the near-horizon limit of the D3-D3-D3-D3 brane system, where the metric is
\begin{equation}
ds^2 =  ds^2 (AdS_2) +  ds^2 (S^2) + ds^2 (T^6) \ ,
\end{equation}
and the self-dual RR 5-form flux is
\begin{equation}
F^{(5)} = -\frac{e^{-\phi}}{R}\bigg( \mathrm{vol} (AdS_2) \wedge \mathfrak{Re} (\Omega_3) + \star\,  \mathrm{vol} (AdS_2) \wedge \mathfrak{Re} (\Omega_3) \bigg) \ , 
\end{equation}
where $\phi$ is the constant dilaton, $R$ is the (common) radius of $AdS_2$ and $S^2$, and $\Omega_3 = dz^1 \wedge dz^2 \wedge dz^3$ is a holomorphic 3-form on $T^6$.
To include the fermionic degrees of freedom, one has to embed the coset above into the supercoset space
\begin{equation}
\frac{PSU(1,1 | 2)}{U(1)^2} \ . 
\end{equation}
Therefore the isometry superalgebra is $\mathfrak{psu}(1,1|2)$. The fact that we have to introduce a vacuum state in order to define particle excitations will break the full isometry superalgebra down to the vacuum little group, which is $[\mathfrak{psu}(1|1)]^2$. We shall consider just one copy of the isotropy algebra, namely $\mathfrak{psu}(1|1)$, since one can obtain the representation on $[\mathfrak{psu}(1|1)]^2$ by tensor product of two representations of $\mathfrak{psu}(1|1)$. 

In order to describe physical particles which have energy, mass and momentum, we have to consider the centrally-extended $\mathfrak{psu}(1|1)$ algebra, which we shall simply denote\footnote{This is a slight abuse of notation, since in mathematics the group $\mathfrak{su}(1|1)$ is defined with the bosonic generator $\mathcal{H}$ only, and with $\mathcal{P} =  \mathcal{K} = 0$.} by $\mathfrak{su}(1|1)$. The non-vanishing algebra relations are the following
\begin{eqnarray}
\{ \mathcal{Q}, \mathcal{Q} \} = 2\mathcal{P}, \qquad 
\{ \mathcal{S}, \mathcal{S} \} = 2 \mathcal{K}, \qquad
\{ \mathcal{Q}, \mathcal{S} \} = 2 \mathcal{H},
\end{eqnarray}
where $\mathcal{P}, \mathcal{K}$ and $\mathcal{H}$ are central bosonic generators, $\mathcal{Q}$ and $\mathcal{S}$ are fermionic generators. 

We represent the $\mathfrak{su}(1|1)$ generators as $2\times 2$ matrices acting on a pair of boson-fermion $( |\phi \rangle , | \psi \rangle )^T$ as
\begin{equation}
\label{Q_S}
\mathcal{Q} = \begin{pmatrix}
0 & b \\
a & 0 
\end{pmatrix} \ , \qquad
\mathcal{S} = \begin{pmatrix}
0 & d \\
c & 0 
\end{pmatrix} \ , 
\end{equation}
and 
\begin{equation}
\label{H_P_K}
\mathcal{H} = H \begin{pmatrix}
1 & 0 \\
0 & 1 
\end{pmatrix} \ , \qquad
\mathcal{P} = P \begin{pmatrix}
1 & 0 \\
0 & 1 
\end{pmatrix} \ , \qquad
\mathcal{K} = K \begin{pmatrix}
1 & 0 \\
0 & 1 
\end{pmatrix} \ 
\end{equation}
where $a, b, c, d, H, P, K \in \mathbb{C}$ are the \emph{representation parameters}. As we shall see later, the only \emph{independent} representation parameters are $a, b, c, d$. The fact that $\mathcal{H}, 	\mathcal{P}, \mathcal{K}$ must be proportional to the identity is because the only central element of $\mathfrak{su}(1|1)$ is the identity, as described in section \ref{sec:matrix_rep}.

\section{Representations of $\mathfrak{su}(1|1)$}
\label{sec:II}
In this section we shall explicitly show the representation parameters for the $\mathfrak{su}(1|1)$ algebra in the three cases where the particle is \emph{massive}, \emph{massless} and \emph{relativistic massless}. 

\subsubsection{Massive representation}
In the boson-fermion representation (\ref{Q_S}) and (\ref{H_P_K}), we have that
\begin{align}
\label{par1_massive}
a =& \frac{\alpha \,e^{\frac{i p}{4}-\frac{i\pi}4}}{\sqrt2} \sqrt{\mathcal{E}+ m} \ ,  &\nonumber
b =& \frac{\alpha^{-1} e^{-\frac{i p}{4}+\frac{i\pi}4}}{\sqrt2} \frac{ h(1-e^{i p})}{\sqrt{\mathcal{E}+ m}} \ , \\
c =\, & \frac{\alpha\, e^{\frac{i p}{4}-\frac{i\pi}4}}{\sqrt2} \frac{ h(1-e^{-i p})}{\sqrt{\mathcal{E}+ m}} \ , &
d =\, & \frac{\alpha^{-1}e^{-\frac{i p}{4}+\frac{i\pi}4}}{\sqrt 2} \sqrt{\mathcal{E}+ m} \ ,
\end{align}
and
\begin{eqnarray}
\label{par2_massive}
H = \frac{\mathcal{E}}{2} \ , \qquad P = \frac{h}{2} (1 - e^{i  p} ) \ , \qquad K = \frac{h}{2} (1 - e^{-i  p}) \ , 
\end{eqnarray}
where $\mathcal{E}, p$ and $m$ are the energy, spatial momentum and mass respectively, $ h$ is the coupling constant and $\alpha$ is an undetermined phase, which we are free to fix later. 
The closure of the algebra implies the \emph{dispersion relation}\footnote{It is not possible to find a quadratic Casimir for the algebra $\mathfrak{su}(1|1)$ because the Killing form is vanishing. However if one introduces the secret symmetry $\mathcal{B}$, which couples non-trivially with the generator $\mathcal{H}$, then there exists a quadratic Casimir $\mathfrak{C}_2$. In this case, the condition $\mathfrak{C}_2 = m^2$ gives the dispersion relation (\ref{hence}). The same situation applies also to $\mathfrak{su}(2|2)$. }
\begin{eqnarray}
\label{hence}
\mathcal{E}^2 = m^2 + 4 h^2 \sin^2 \frac{p}{2} \ . 
\end{eqnarray}
The $AdS_2$ massive representation is a {\it long} representation, where no bound is imposed in (\ref{hence}). In particular, the ``mass" parameter $m$ is completely unconstrained. 
We attribute to $m$ the meaning of mass just by analogy with the higher-dimensional $AdS$ cases, which is not proved formally yet.

\subsubsection{Massless representation}
The (non-relativistic) massless representation of $\mathfrak{su}(1|1)$ is obtained by performing the limit
\begin{equation}
\label{massless_limit}
m \rightarrow 0 \ , \qquad\qquad h \quad \mbox{finite} \ , 
\end{equation}
of the massive representation. The representation parameters become 
\begin{eqnarray}
\label{par1_massless}
&& a =\alpha \,e^{\frac{ip}{4}-\frac{i\pi}4} \sqrt{h \sin (p / 2) },  \qquad\qquad b = \pm \frac{1}{\alpha} e^{\frac{i p}{4} -\frac{i\pi}4} \sqrt{ h \sin ( p / 2)} \ , \nonumber \\
&& c = \pm \alpha\, e^{-\frac{i p}{4}+\frac{i\pi}4} \sqrt{ h \sin ( p / 2) }, \qquad\ \  d = \frac{1}{\alpha}e^{-\frac{i p}{4}+\frac{i\pi}4} \sqrt{ h \sin ( p / 2)} \ , 
\end{eqnarray}
and
\begin{eqnarray}
\label{par2_massless}
H = \frac{\mathcal{E}}{2} \ , \qquad P = \frac{h}{2} (1 - e^{i  p} ) \ , \qquad K = \frac{h}{2} (1 - e^{-i  p}) \ , 
\end{eqnarray}
where the upper (lower) sign is for right (left) movers\footnote{For the left movers case, one also needs to account for a global factor of $\sqrt{-1} = i$ according to our choice of branch, which will matter in the mixed right-left and left-right coproducts}, and $\mathfrak{Re} p \in [0,\pi]$ for right movers, $\mathfrak{Re}  p \in [-\pi,0]$ for left movers\footnote{There is a second possible choice of left and right movers momenta, via the shift $p \rightarrow p + \pi$. In this choice, $\mathfrak{Re} p \in (0,\pi)$ for right movers and  $\mathfrak{Re} p \in (\pi, 2 \pi)$ for left movers. }

The massless dispersion relation is obtained by setting $m = 0$ in (\ref{hence}),
\begin{eqnarray}
\mathcal{E} = 2  h \, \Big\lvert\sin \frac{ p}{2}\Big\rvert ,
\end{eqnarray} 
this is a shortening condition \cite{Hoare:2014kma}, which is expected to be protected by supersymmetry from quantum corrections in the complete theory \cite{Sorokin:2011rr,Cagnazzo:2011at, Cagnazzo:2011at2}, though this has yet to be fully proven.

\begin{figure}[h]
\begin{center}
 \includegraphics[scale=0.52]{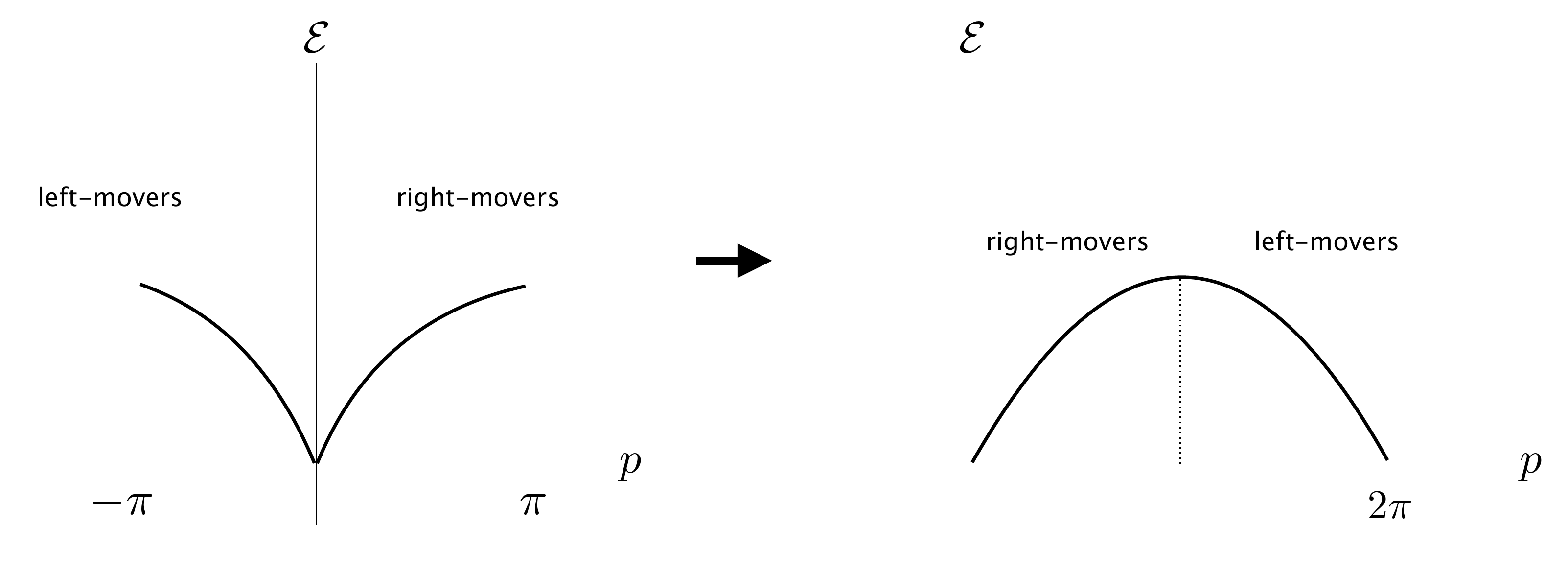}
\end{center}
\caption{Possible choices of right and left movers momenta.}
\end{figure}

\subsubsection{Relativistic massless representation}
The relativistic massless representation is obtained by taking the relativistic limit of the massless representation. The relativistic limit is the following  
\begin{eqnarray}
\label{rel_lim}
p \rightarrow \tau q, \qquad\qquad  h \rightarrow \frac{c}{\tau}, \qquad\qquad \tau\rightarrow 0^+. 
\end{eqnarray}
where $c$ is the speed of light and 
\begin{equation}
q \equiv \pm e^{\theta} \ , 
\end{equation}
where $\theta \in \mathbb{C}$ is the rapidity, and the upper (lower) sign stands for right (left) movers.
Then the representation parameters become
\begin{eqnarray}
\label{par1massless}
\nonumber
&& a = \alpha \,e^{-i\frac{\pi}{4}}\sqrt{\frac{c \, e^{\theta}}{2}}, 
\qquad b = \pm \alpha^{-1}\,e^{-i\frac{\pi}{4}}\sqrt{\frac{c \, e^{\theta}}{2}}, \\
&& c = \pm \alpha \,e^{i\frac{\pi}{4}}\sqrt{\frac{c \, e^{\theta}}{2}}, 
\qquad d = \alpha^{-1} \,e^{i\frac{\pi}{4}}\sqrt{\frac{c \, e^{\theta}}{2}},   
\end{eqnarray}
and 
\begin{eqnarray}
\label{par2_massless_rel}
C = \frac{c \, e^{\theta}}{2}, \quad\qquad P = -i\frac{c \, e^{\theta}}{2}, \quad\qquad K = i\frac{c \, e^{\theta}}{2}, 
\end{eqnarray}
where the upper (lower) sign is for right (left) movers. The relativistic massless dispersion relation is
\begin{eqnarray}
\mathcal{E} = c \, |e^{\theta}|, 
\end{eqnarray}
Relativistic invariance will guarantee that the R-matrix only depends on the difference of the rapidities of the two scattering particles.

\section{Relativistic massless R-matrix}
\label{sec:III}
In this section we consider the problem of finding the R-matrix in the relativistic massless regime. We show that the R-matrix is not unique, and we find a list of possible solutions. Of course, it is unknown if this list exhausts all possible R-matrix solutions. In \cite{MC} the authors studied the most generic R-matrix for $\mathcal{N} = 1$ supersymmetric theories, which is relevant for us. However they have imposed unitarity and crossing symmetry, which we do not do it here. The algebra invariance together with the Yang-Baxter equation will be sufficient for us to completely fix the R-matrix entries, up to the dressing factor. For this reason, the solutions obtained here are more general then the one found in \cite{MC}. Unitarity and crossing symmetry are checked only at posteriori, and we shall see that not all solutions satisfy both conditions. We will give a physical interpretation of this fact. 

The coproduct action of the $\mathfrak{su}(1|1)$ superalgebra in the relativistic massless representation is 
\begin{eqnarray}
\label{rel_coproduc}
&&\Delta(\mathcal{Q}) = \mathcal{Q} \otimes \mathds{1} + \mathds{1} \otimes \mathcal{Q} \ , \qquad 
\Delta (\mathcal{S} ) = \mathcal{S} \otimes \mathds{1} + \mathds{1} \otimes \mathcal{S} \ , 
\end{eqnarray}
and
\begin{eqnarray}
\label{rel_coproduc2}
\Delta( \mathcal{H}) = \mathcal{H} \otimes \mathds{1} + \mathds{1} \otimes \mathcal{H}\ , 
\qquad 
\Delta (\mathcal{P}) = \mathcal{P} \otimes \mathds{1} + \mathds{1} \otimes \mathcal{P} \ , 
\qquad 
\Delta ( \mathcal{K} ) = \mathcal{K} \otimes \mathds{1} + \mathds{1} \otimes \mathcal{K} \ . 
\end{eqnarray}
We notice that the non-trivial braiding factors typical of the non-relativistic representation \cite{Hoare:2014kma} tend to $\mathds{1}$ in the relativistic limit $\tau \rightarrow 0$. 

The generic R-matrix solution must be of the following form
\begin{equation}
\label{R_generic}
R = \Phi \begin{pmatrix}
1 & 0 & 0 & \omega_1 \\
0 & \omega_2 & \omega_3 & 0 \\
0 & \omega_4 & \omega_5 & 0 \\
\omega_6 & 0 & 0 & \omega_7 
\end{pmatrix} \ , 
\end{equation}
where $\Phi, \omega_i \in \mathbb{C}$, and the various zeroes are because of conservation of the total fermionic number (e.g. $\phi \phi \rightarrow \phi \psi$ is not allowed). 
We impose that (\ref{R_generic}) is invariant under the $\mathfrak{su}(1|1)$ superalgebra action in the relativistic massless representation, i.e.
\begin{eqnarray}
\label{com_massles_rel}
\Delta ( \mathfrak{a} ) R = R \Delta (\mathfrak{a}) \ , \qquad\qquad
\forall \ \mathfrak{a} \in \mathfrak{su}(1|1) \ ,  
\end{eqnarray}
where we used the fact that in the relativistic massless representation $\Delta = \Delta^{op}$. Furthermore, we impose that (\ref{R_generic}) satisfies the Yang-Baxter equation (\ref{YBE}). 
The algebra invariance condition (\ref{com_massles_rel}) together with the Yang-Baxter equation suffices to determine completely (\ref{R_generic}), up to the dressing factor $\Phi$.

Below we list the various R-matrix solutions found, where we set $\theta \equiv \theta_1 - \theta_2$.
In what follows we shall distinguish two types of scatterings: left-right and right-right (or equivalently right-left and left-left). 
A posteriori, we check whether the R-matrix solutions found satisfy or not crossing symmetry and braiding unitarity.
As we will see, the left-right R-matrix describes the zeroth order data of a physical scattering, while some of the collinear right-right (left-left) R-matrices are non-perturbative objects, which do not describe any physical scattering process.  

\vspace{2mm}
\underline{Mixed scatterings}
\begin{itemize}
\item {\bf Solution 1}:  for $\alpha$ arbitrary,
\begin{eqnarray}
\mbox{diag}(1, 1, 1, 1) \ .
\end{eqnarray}
This solution trivially satisfies crossing symmetry and braiding-unitarity, and signals a decoupling of left and right movers. 
\end{itemize}

\underline{Collinear scatterings}
\begin{itemize}
\item {\bf Solution 2 (``Fendley ${\rm p}=\frac{1}{2}$")}: set $\alpha^2 = 1$ (right-right), $\alpha^2 = -1$ (left-left),
\begin{eqnarray}
\label{Fendley p=1/2}
\begin{pmatrix}
1 & 0 & 0 & -\frac{\sinh \frac{\theta}{2}}{\cosh \theta}\\
0 & -\tanh\theta &\frac{\cosh\frac{\theta}{2}}{\cosh\theta}  & 0 \\
0 &\frac{\cosh\frac{\theta}{2}}{\cosh\theta}  & \tanh\theta & 0 \\
-\frac{\sinh \frac{\theta}{2}}{\cosh \theta} & 0 & 0 & - 1
\end{pmatrix}.
\end{eqnarray}
This solution satisfies crossing symmetry and braiding-unitarity. 

This is one of the solutions found in \cite{Fendley:1990cy}, and the ``${\rm p} = \frac{1}{2}$" parameter in the name reflects the notations of that paper - ${\rm p}$ being no momentum at all in this case. It was there obtained for massive particles, however as discussed in \cite{Borsato:2016xns}, massless relativistic R-matrices for collinear particles formally coincide with the massive one. 

\item {\bf Solution 3}: for $\alpha$ arbitrary,
\begin{eqnarray}
\begin{pmatrix}
\label{left-right}
1 & 0 & 0 & \mp\alpha^{-2} e^{-\frac{\theta}{2}}\\
0 & -1 & e^{-\frac{\theta}{2}} & 0 \\
0 & e^{-\frac{\theta}{2}} & 1 & 0 \\
\mp\alpha^2 e^{-\frac{\theta}{2}} & 0 & 0 & -1 
\end{pmatrix},
\end{eqnarray}
where the upper (lower) sign is for right-right (left-left). 

This solution satisfies {\it cross-unitarity} (\ref{cross_unitarity}), a combined relation between unitarity and crossing which we shall specify in section \ref{sec:IV}, but it does \emph{not} satisfy separately crossing and braiding-unitarity. 
\item {\bf Solution 4 (``Fendley ${\rm p}= - \frac{3}{2}$")}: set $\alpha^2 = 1$ (right-right), $\alpha^2 = -1$ (left-left),
\begin{eqnarray}
\label{Fendley p=-3/2}
\begin{pmatrix}
1 & 0 & 0 &  \frac{\sinh \frac{3}{2}\theta}{\cosh \theta}\\
0 &\tanh\theta  &\frac{\cosh\frac{3}{2}\theta}{\cosh\theta}  & 0 \\
0 & \frac{\cosh\frac{3}{2}\theta}{\cosh\theta}& - \tanh\theta   & 0 \\
\frac{\sinh \frac{3}{2}\theta}{\cosh \theta} & 0 & 0 & - 1
\end{pmatrix}.
\end{eqnarray}
This solution satisfies crossing symmetry and braiding-unitarity, and it was first found in \cite{Fendley:1990cy}.
\item {\bf Solution 5}: for arbitrary values of $\alpha$,
\begin{eqnarray}
\begin{pmatrix}
\label{right-left}
1 & 0 & 0 & \pm \alpha^{-2} e^{\frac{\theta}{2}}\\
0 & 1 & e^{\frac{\theta}{2}} & 0 \\
0 & e^{\frac{\theta}{2}} & -1 & 0 \\
\pm \alpha^2 e^{\frac{\theta}{2}} & 0 & 0 & -1 
\end{pmatrix},
\end{eqnarray}
where the upper (lower) sign is for right-right (left-left). 

This solution satisfies cross-unitarity, but does \emph{not} satisfy separately crossing and braiding-unitarity. 
\end{itemize}

Solutions 3 and 5 are reproduced as asymptotic limits of Solutions 2 and 4 as follows

\begin{table}[H]
\begin{center}
\begin{tabular}{c@{\hskip 1cm}c@{\hskip 1cm}c}
\hline
{\bf Solution} & {\bf Limit of $\theta$} & {\bf Limit solution} \\
\hline
Solution 2  & $\theta \rightarrow + \infty$ & Solution 3 (right-right) $\alpha^2 =1$\\
Solution 4 & $\theta \rightarrow - \infty$ & Solution 3 (left-left) $\alpha^2 =-1$\\
Solution 2  & $\theta \rightarrow - \infty$ & Solution 5 (right-right) $\alpha^2 =1$\\
Solution 4  & $\theta \rightarrow + \infty$ & Solution 5 (left-left) $\alpha^2 =-1$\\
\end{tabular}
\end{center}
\caption{Asymptotic limits of Solutions 2 and 4.}
\end{table}

\subsection{Connection to the non-relativistic massless scattering}
\label{connections}
In this section we show that some of the R-matrix solutions presented in section \ref{sec:III} can be reproduced by taking the relativistic limit of the (non-relativistic) massless R-matrix found in \cite{Hoare:2014kma}. In particular, Solutions 1, 3 and 5 can be obtained in this way. 

The coproduct action of the $\mathfrak{su}(1|1)$ superalgebra in the massless representation is 
\begin{eqnarray}
\label{rel_coproduc1}
&&\Delta(\mathcal{Q}) = \mathcal{Q} \otimes \mathds{1} + e^{i \frac{{ p}}{2}}\mathds{1} \otimes \mathcal{Q} \ , \qquad \Delta (\mathcal{S} ) = \mathcal{S} \otimes \mathds{1} + e^{-i \frac{{ p}}{2}}\mathds{1} \otimes \mathcal{S} \ .  
\end{eqnarray}
and 
\begin{equation}
\label{rel_coproduc3}
\Delta(\mathcal{H}) = \mathcal{H} \otimes \mathds{1} + \mathds{1}\otimes \mathcal{H} \ , \quad 
\Delta (\mathcal{P} ) = \mathcal{P} \otimes \mathds{1} + e^{i p}\mathds{1} \otimes \mathcal{P} \ , \qquad
\Delta (\mathcal{K} ) = \mathcal{K} \otimes \mathds{1} + e^{-i p}\mathds{1} \otimes \mathcal{K} \ . 
\end{equation}
Following the notation of \cite{Hoare:2014kma}, we introduce the massless Zhukovsky variables $(x_1^{\pm}, x_2^{\pm})$
\begin{eqnarray}
x_i^+ = \pm e^{i \frac{{ p}_i}{2}}\ , \qquad 
x_i^+ = \frac{1}{x_i^-}, \qquad i=1,2.
\end{eqnarray}
where the upper sign stands for right movers $\mathfrak{Re} { p}_i \in (0, \pi)$, the lower sign for left movers  $\mathfrak{Re}{  p}_i \in (-\pi,0)$.

In \cite{Hoare:2014kma}, the authors introduce the following function $f$ 
\begin{eqnarray}
f = \frac{\sqrt{\frac{x_1^+}{x_1^-}} \Big(x_1^- - \frac{1}{x_1^+}\Big) - \sqrt{\frac{x_2^+}{x_2^-}} \Big(x_2^- - \frac{1}{x_2^+}\Big)}{1 - \frac{1}{x_1^+ x_1^- x_2^+ x_2^-}}, 
\end{eqnarray}
which is responsible for various signs entering in the R-matrix entries  in the massless limit. 
In the limit where both particles are massless, the function $f$ takes the indeterminate form $\frac{0}{0}$, and in particular the limit is not well-defined. 
A possible way to take the massless limit on $f$ consists in performing the massless limit on just one of the two particles, assuming that the other one is still massive, and finally performing the massless limit also on the other particle. The order in which one performs the two limits matters \emph{only} for scattering of type right-right and left-left, while for scattering of type right-left and left-right such ambiguity does not appear (i.e. the limit is unique). In particular we have
\begin{equation}
f \rightarrow \begin{cases}
+1 & \mbox{right-left}  \\
-1 & \mbox{left-right} 
\end{cases}\ , \qquad\qquad
f \rightarrow \begin{cases}
\pm 1 & \mbox{right-right}  \\
\pm 1 & \mbox{left-left} 
\end{cases}\ .  
\end{equation}
In the right-right and left-left cases, we will show that the two values of the limit of $f$ is physically connected to the two possible choices of Fendley's relativistic R-matrix (Solutions 2 and 4). For the mixed cases, the ambiguity is absent, which guarantees that the BMN limit reproduces the trivial scattering matrix. We have verified that this pattern precisely matches the table given in section 5.2 of \cite{Hoare:2014kma}.

In what follows we take the relativistic limit of the (non-relativistic) massless R-matrix found in \cite{Hoare:2014kma} and make a connection with the solutions found in \ref{sec:III}.  

\vspace{5mm}
\noindent \underline{Right-left / left-right}

\noindent In the mixed case, the non-relativistic massless R-matrix is
\begin{eqnarray}
\label{Rnonrel}
R= \begin{pmatrix}
1 & 0 & 0 &  \pm \frac{1}{\alpha^2} \kappa({ p}_1,{ p}_2)\\
0 & \pm \delta(x_1^+)  & \tilde{\kappa}({ p}_1,{ p}_2) & 0 \\
0 & \tilde{\kappa}({ p}_1,{ p}_2) & \mp \delta(x_2^{+}) & 0 \\
\pm \alpha^2\kappa({ p}_1,{ p}_2) & 0 & 0 & - \delta(x_1^+) \delta(x_2^+)
\end{pmatrix},
\end{eqnarray}
where
\begin{eqnarray}
\delta(x_i^+) = \begin{cases} +1, & \mbox{if } x_i^+\mbox{ is right-mover} \\ -1, & \mbox{if } x_i^+\mbox{ is left-mover} \end{cases},
\end{eqnarray}
and
\begin{eqnarray}
\notag
\kappa(p_1,p_2) &\equiv & -i \sqrt[4]{\frac{x_1^{+ 2}}{x_2^{+2}}} \frac{x_2^+ \sqrt{i(x_1^- - x_1^+)} \sqrt{i(x_2^- - x_2^+)}}{1- x_1^+ x_2^+ \pm (x_1^+ - x_2^+)}\ ,  \\
 \tilde{\kappa}(p_1,p_2) &\equiv & 	\delta(x_1^+) \delta(x_2^+) \kappa(p_1, p_2) \ .
\end{eqnarray}
The upper (lower) sign in (\ref{Rnonrel}) corresponds to the right-left (left-right) case. 

In the \emph{right-left} case ($f \rightarrow 1$), the relativistic limit of (\ref{Rnonrel}) gives
\begin{eqnarray}
\label{RLrel}
\mbox{diag}(1, 1, 1, 1) \ ,
\end{eqnarray}
which is the first-order term appearing in perturbation theory \cite{amsw, amsw2}. This is in agreement with the Zamolodchikov's conjecture (discussed in section \ref{sec:Zamolod}). 

In the \emph{left-right} case ($f \rightarrow - 1$), the relativistic limit of (\ref{Rnonrel}) gives again the identity matrix (\ref{RLrel}). This is again consistent with the Zamolodchikov's conjecture. 

\vspace{5mm}
\noindent \underline{Right-right}

\noindent The non-relativistic \emph{right-right} R-matrix is
\begin{eqnarray}
\label{Rnonrel1}
R= \begin{pmatrix}
1 & 0 & 0 &  \pm \frac{1}{\alpha^2} \Big[\frac{\tan \frac{p_1}{4}}{\tan \frac{p_2}{4}}\Big]^{\pm \frac{1}{2}}\\
0 & \pm 1  & \Big[\frac{\tan \frac{p_1}{4}}{\tan \frac{p_2}{4}}\Big]^{\pm \frac{1}{2}} & 0 \\
0 & \Big[\frac{\tan \frac{p_1}{4}}{\tan \frac{p_2}{4}}\Big]^{\pm \frac{1}{2}} & \mp 1 & 0 \\
\pm \alpha^2\Big[\frac{\tan \frac{p_1}{4}}{\tan \frac{p_2}{4}}\Big]^{\pm \frac{1}{2}} & 0 & 0 & -1
\end{pmatrix}.
\end{eqnarray}
In the right-right case we have that $f$ can take both values $f \rightarrow \pm 1$, which is a consequence of the ambiguity in the order of taking the massless limits, as mentioned above. 
The upper (lower) sign in (\ref{Rnonrel1}) corresponds to $f \to +1$ ($f \to -1$). 

In the case $f \to +1$, the relativistic limit of (\ref{Rnonrel1}) gives
\begin{eqnarray}
\label{rightright5}
\begin{pmatrix}
1 & 0 & 0 & \frac{1}{\alpha^2} e^{\frac{\theta}{2}}\\
0 & 1 & e^{\frac{\theta}{2}} & 0 \\
0 & e^{\frac{\theta}{2}} & -1 & 0 \\
\alpha^2 e^{\frac{\theta}{2}} & 0 & 0 & -1 
\end{pmatrix}, \qquad f \rightarrow + 1, 
\end{eqnarray}
which reproduces the Solution 5 right-right. 

In the case $f \to -1$, the relativistic limit of (\ref{Rnonrel1}) is
\begin{eqnarray}
\begin{pmatrix}
1 & 0 & 0 & - \frac{1}{\alpha^2} e^{-\frac{\theta}{2}}\\
0 & -1 & e^{-\frac{\theta}{2}} & 0 \\
0 & e^{-\frac{\theta}{2}} & 1 & 0 \\
-\alpha^2 e^{-\frac{\theta}{2}} & 0 & 0 & -1 
\end{pmatrix}, \quad f \rightarrow - 1, 
\end{eqnarray}
which reproduces the Solution 3 right-right.

\noindent \underline{Left-left}

\noindent The non-relativistic \emph{left-left} R-matrix is
\begin{eqnarray}
\label{Rnonrel2}
R= \begin{pmatrix}
1 & 0 & 0 &  \pm \frac{1}{\alpha^2} \Big[\frac{\tan \frac{p_1}{4}}{\tan \frac{p_2}{4}}\Big]^{\mp \frac{1}{2}}\\
0 & \mp 1  & \Big[\frac{\tan \frac{p_1}{4}}{\tan \frac{p_2}{4}}\Big]^{\mp \frac{1}{2}} & 0 \\
0 & \Big[\frac{\tan \frac{p_1}{4}}{\tan \frac{p_2}{4}}\Big]^{\mp \frac{1}{2}} & \pm 1 & 0 \\
\pm \alpha^2\Big[\frac{\tan \frac{p_1}{4}}{\tan \frac{p_2}{4}}\Big]^{\mp \frac{1}{2}} & 0 & 0 & -1
\end{pmatrix}.
\end{eqnarray}
The upper (lower) sign in (\ref{Rnonrel2}) corresponds to $f \to +1$ ($f \to -1$).

In the case $f \to +1$, the relativistic limit of (\ref{Rnonrel2}) is
\begin{eqnarray}
\label{leftleft3}
\begin{pmatrix}
1 & 0 & 0 &  \frac{1}{\alpha^2}e^{-\frac{\theta}{2}}\\
0 & -1 & e^{-\frac{\theta}{2}} & 0 \\
0 & e^{-\frac{\theta}{2}} & 1 & 0 \\
 \alpha^2 e^{-\frac{\theta}{2}} & 0 & 0 & -1 
\end{pmatrix}, \qquad f \rightarrow + 1, 
\end{eqnarray}
which reproduces the Solution 3 left-left. 

In the case $f \to -1$, the relativistic limit of (\ref{Rnonrel2}) gives
\begin{eqnarray}
\begin{pmatrix}
1 & 0 & 0 & -\frac{1}{\alpha^2} e^{\frac{\theta}{2}}\\
0 & 1 & e^{\frac{\theta}{2}} & 0 \\
0 & e^{\frac{\theta}{2}} &-1 & 0 \\
-\alpha^2 e^{\frac{\theta}{2}} & 0 & 0 & -1 
\end{pmatrix}, \qquad f \rightarrow - 1, 
\end{eqnarray}
which reproduces the Solution 5 left-left.

We remark that the non-triviality of the massless collinear R-matrices in the BMN limit is consistent with the Zamolodchikov's conjecture \ref{Zam_conj}, where $R$ is non-perturbative. 
The following diagram shows the big picture about the origin of Solutions 1, 2, 3, 4, 5, and how they are related between each other. 

\begin{figure}[H]
\begin{center}
 \includegraphics[scale=0.48]{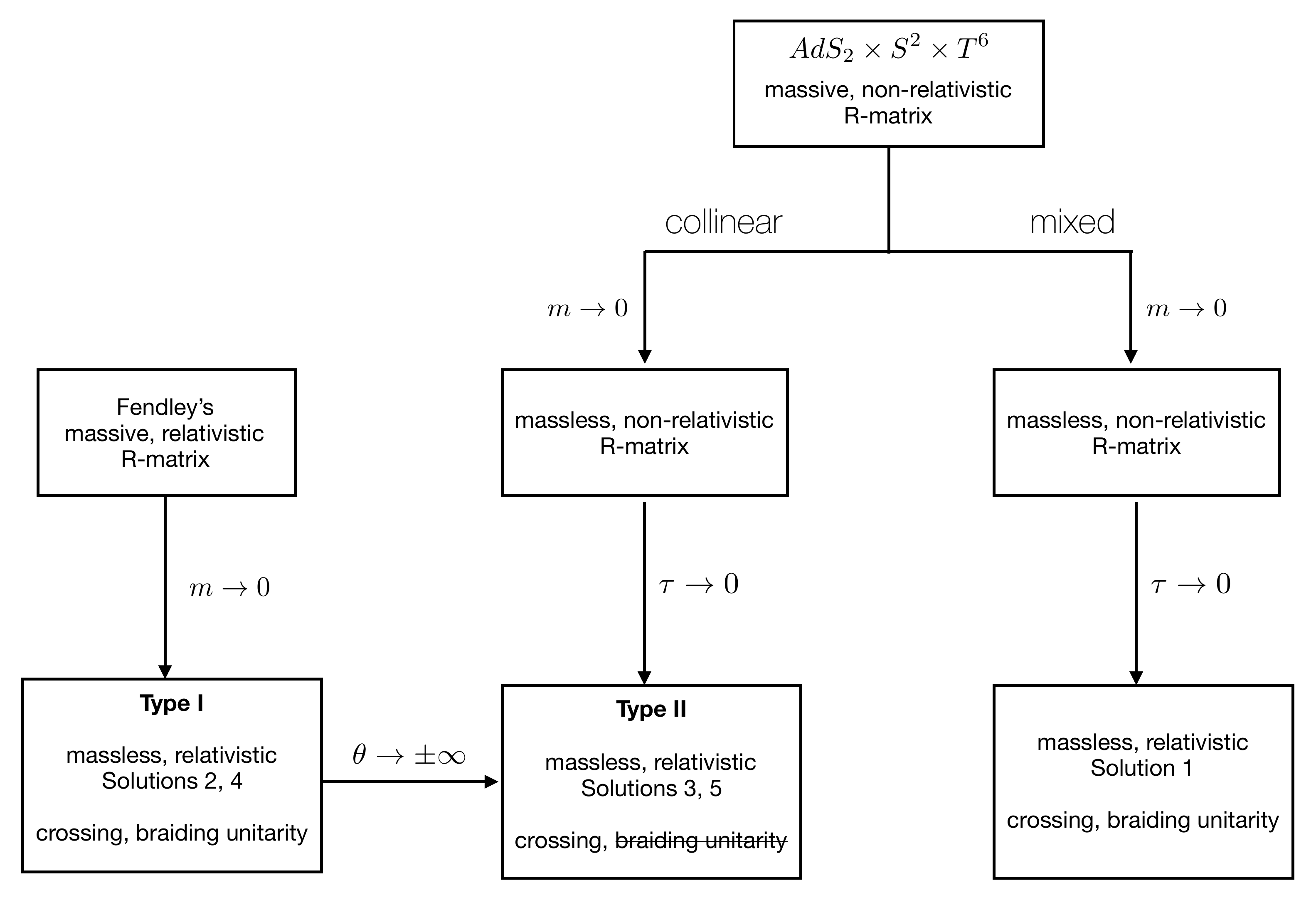}
\end{center}
\caption{Origin of Solutions 1, 2, 3, 4, 5. }
\end{figure}
In the diagram above, we denote by {\it type I} the relativistic massless R-matrices which satisfy crossing and braiding unitarity separately, while {\it type II} the relativistic massless R-matrices which satisfy cross-unitarity, but do \emph{not} satisfy braiding unitarity.

\noindent{\bf Comment} 

\noindent There are two solutions which are invariant under the relativistic massless coproduct action (\ref{rel_coproduc}) and (\ref{rel_coproduc2}), which satisfy the Yang-Baxter equation, but which do not follow by any limit of the massless non-relativistic R-matrix in \cite{Hoare:2014kma}. These solutions are: 
\begin{itemize}
\item {\bf Solution 6}: for $\alpha$ arbitrary,
\begin{eqnarray}
\begin{pmatrix}
1 & 0 & 0 & 0\\
0 & \frac{-1 + e^{\theta}}{-1 + e^{\theta} + e^{\kappa +\theta }} & \frac{e^{\kappa + \frac{\theta}{2}}}{-1 +e^{\theta} + e^{\kappa + \theta}} & 0 \\
0 & \frac{e^{\kappa + \frac{\theta}{2}}}{-1 + e^{\theta}+ e^{\kappa + \theta}} & 1 - \frac{e^{\kappa}}{-1 + e^{\theta} + e^{\kappa + \theta }} & 0 \\
0 & 0 & 0 & \frac{-1 - e^{\kappa} + e^{\theta}}{-1 + e^{\theta} + e^{\kappa + \theta}}
\end{pmatrix}.
\end{eqnarray}
where $\kappa$ is an arbitrary constant. This solution satisfies braiding-unitarity, but does \emph{not} satisfy crossing symmetry. 
\item {\bf Solution 7}: for $\alpha^2 = 1$,
\begin{eqnarray}
\begin{pmatrix}
1+  \frac{2i\sin \beta \pi  }{\sinh \theta} & 0 & 0 &  \frac{i\sin \beta \pi }{\cosh \frac{\theta}{2}}\\
0 & 1  & \frac{i \sin \beta \pi  }{\sinh \frac{\theta}{2}}  & 0 \\
0 & \frac{i \sin \beta \pi}{\sinh \frac{\theta}{2}}  & 1   & 0 \\
\frac{i \sin \beta \pi}{\cosh\frac{\theta}{2} } & 0 & 0 & 1 - \frac{2i \sin \beta \pi}{\sinh \theta}
\end{pmatrix}.
\end{eqnarray}
where $\beta$ is an arbitrary constant. This solution satisfies crossing symmetry and braiding-unitarity. This R-matrix appears as a solution for the supersymmetric Sine-Gordon model \cite{SW,Ahn:1993qa,Ahn:1990uq,HRS} (with $\beta$ related to the coupling). We remark that the Pohlmeyer reduction of the $AdS_2$ superstring is the ${\cal{N}}=2$ supersymmetric Sine-Gordon theory \cite{pr, pr2}, whose R-matrix is built from those of ${\cal{N}}=1$ supersymmetric Sine-Gordon\footnote{We thank Ben Hoare for discussions about this point.}. 
\end{itemize}

\subsection{Zamolodchikov's conjecture}
\label{sec:Zamolod}
In this section we state the Zamolodchikov's conjecture regarding the R-matrix behaviour in the massless limit \cite{Zamol2, Zamol22,Borsato:2016xns}. We remind that when we deal with massive particles, right-left scatterings can be turned into right-right scatterings, and vice versa, by using a boost transformation.
This is not possible in the case of massless particles, where such swapping does not occur because the particles are travelling at speed of light. 

\begin{cnj}[Zamolodchikov]
\label{Zam_conj}
Let $R$ be the R-matrix describing the scattering of massive particles. Consider the same system in the massless limit. Then 
\begin{itemize}
\item \underline{Mixed case}: $R$ still describes the physical scattering between particles, and it admits a perturbative series expansion. 
$R$ satisfies crossing and braiding unitarity separately. 

\item \underline{Collinear case}: $R$ does not describe the physical scattering, and it is non-perturbative.
$R$ satisfies cross-unitarity, but does not necessary satisfy crossing and braiding unitarity separately.

\end{itemize}
\end{cnj}

Here we do not prove the Zamolodchikov's conjecture, but we provide an example where one breaks braiding unitarity.

Let $R^{(II)}, R^{(III)}, R^{(V)}$ be the R-matrices of Solutions 2, 3 and 5 respectively. Then, for an appropriate value of $\alpha^2$, we know that asymptotically 
\begin{equation}
\lim_{\theta \rightarrow + \infty} R^{(II)}  = R^{(III)} \ , \qquad
\lim_{\theta \rightarrow - \infty} R^{(II)}  = R^{(V)} \ .
\end{equation}
Since $R^{(II)}$ is a solution of type I, it satisfies braiding-unitarity i.e.
\begin{eqnarray}
\label{buI}
R^{(II)}_{12} (\theta) R^{(II)}_{21} (-\theta) = \mathds{1}.
\end{eqnarray}
Now we take the limit $\theta \rightarrow + \infty$ in (\ref{buI}). In this limit, we notice that the first factor $R^{(II)}_{12} (\theta)$ reproduces $R^{(III)}$, however the second factor $ R^{(II)}_{21} (-\theta)$ must be expanded around $- \infty$, and therefore it reproduces $R^{(V)}$. Asymptotically, we obtain 
\begin{equation}
R^{(III)}_{12}(\theta) R^{(V)}_{21} (-\theta) = \mathds{1} \ , 
\end{equation}
which is \emph{not} braiding unitarity for either solutions $R^{(III)}$ or $R^{(V)}$.

The fact that type I solutions satisfy crossing and braiding unitarity separately is not in contradiction with the Zamolodchikov's conjecture, although they are relativistic massless R-matrices which recover the collinear solutions at large rapidities.  
The reason is the following. To obtain the type I solutions one first has to take the relativistic limit and second the massless limit.  
However the entries of massive relativistic R-matrices are functionally identical to the entries of the massless relativistic ones, because the relativistic limit has suppressed any mass dependence (see e.g. Fendley's solutions \cite{Fendley:1990cy}). For this reason type I solutions behave like massive solutions, which do not suffer form the issues contemplated in the Zamolodchikov's conjecture. 
This is not the case for type II solutions, which are derived by taking the massless limit first, and the relativistic limit subsequently. This is an example where \emph{physical limits do not commute}.

\subsection{Crossing and Dressing Factors}
\label{sec:IV}
In this section, we show the dressing factors for Solutions 2 and 3. In what follows, we set $\alpha^2 = 1$ and focus on right-right scattering. In the relativistic limit, crossing symmetry is implemented in the following fashion. Define the supertranspose of a matrix $M$ as
\begin{equation}
M^{str}_{ij} = (-)^{ij+i} \, M_{ji} \ ,
\end{equation}
and the charge conjugation matrix as 
\begin{equation}
C = \mbox{diag}(i,1) \ ,
\end{equation}
such that\footnote{In this light notation, $\mathcal{Q}_{q}$ stands for the matrix representation of the generator $\mathcal{Q}$, whose entries depend by $q$.}
\begin{eqnarray}
-\mathcal{Q}_{q} = C^{-1} \mathcal{Q}_{-q}^{str} \, C, \qquad -\mathcal{S}_{q} = C^{-1} \mathcal{S}_{-q}^{str} \, C \ ,\nonumber \\
\end{eqnarray}
where the crossing map is given by
\begin{equation}
q \to - q, \qquad \theta \to i \pi + \theta \ .
\end{equation}
Then unitarity and crossing reads 
\begin{equation}
\label{rel_unit}
R_{12} (\theta ) R_{21} (- \theta ) = \mathds{1} \otimes \mathds{1} \ , 
\end{equation}
and
\begin{equation}
\label{rel_cross}
R_{12} ( i \pi - \theta ) = \big[C^{-1}\otimes \mathds{1}\big] R_{21}^{str_1} (\theta ) \big[C \otimes \mathds{1}\big]  \ .
\end{equation}
A mixed \emph{cross-unitarity} relation can be written combining (\ref{rel_unit}) and (\ref{rel_cross})
\begin{equation}
\label{cross_unitarity}
R_{12} (\theta ) \big[C^{-1}\otimes \mathds{1}\big] R_{12}^{str_1} ( i\pi + \theta) \big[C \otimes \mathds{1}\big] = \mathds{1} \otimes \mathds{1}  \ . 
\end{equation}
The cross-unitarity condition for Solution 2 reads
\begin{equation}
\label{cror}
R (\theta) \, \big[C^{-1}\otimes \mathds{1}\big] \, R^{str_1}(i \pi + \theta) \big[C \otimes \mathds{1}\big] = \bigg(1+\frac{\sinh^2 \frac{\theta}{2}}{\cosh^2 \theta}\Bigg) \mathds{1} \otimes \mathds{1} \ .
\end{equation}
From here, we deduce the cross-unitarity equation for the dressing factor $\Phi$:
\begin{equation}
\label{cror}
\Phi (\theta) \Phi(\theta + i \pi) = \bigg(1+\frac{\sinh^2 \frac{\theta}{2}}{\cosh^2 \theta}\Bigg) ^{-1} \ .
\end{equation}
The dressing factor given in \cite{Fendley:1990cy} is\footnote{In the original paper \cite{Fendley:1990cy} it is missing the overall factor of $4$. }
\begin{eqnarray}
\label{Fe}
\Phi(\theta) &=& 4\bigg[\frac{1}{2} - \frac{\theta}{\pi i }\bigg]^2 \, \prod_{j=1}^\infty \frac{\Big(j-\frac{1}{2}\Big) \, \prod_{k=1}^{3} \Big(3j + \frac{1}{2} - k\Big)}{\Big(2j - \frac{1}{2}\Big)^2 \Big(2j + \frac{1}{2}\Big)^2 \Big(4 j^2 - \big[\frac{1}{2} - \frac{\theta}{\pi i }\big]^2\Big)^2}\nonumber\\
&&\times \frac{\Gamma\Big(3j-\frac{5}{2}+\frac{3}{2} \frac{\theta}{\pi i}\Big)\Gamma\Big(3j-1-\frac{3}{2} \frac{\theta}{\pi i}\Big)}{\Gamma\Big(3j-1+\frac{3}{2} \frac{\theta}{\pi i}\Big)\Gamma\Big(3j+\frac{1}{2}-\frac{3}{2} \frac{\theta}{\pi i}\Big)} \frac{\Gamma\Big(j-\frac{1}{2}+\frac{\theta}{2\pi i}\Big)\Gamma\Big(j-\frac{\theta}{2\pi i}\Big)}{\Gamma\Big(j+\frac{1}{2}-\frac{\theta}{2\pi i}\Big)\Gamma\Big(j+\frac{\theta}{2\pi i}\Big)} \ .
\end{eqnarray}
We have verified that (\ref{Fe}) solves equation (\ref{cror}). It also has the right analiticity structure, meaning no poles in the physical strip $\Im \theta \in (0,\pi)$ (as massless particles cannot form bound states).  

The minimal solution to the cross-unitarity equation satisfied by Solution 3 can also be constructed by factorising Zamolodchikov's formula for the Sine-Gordon dressing factor \cite{Zamolodchikov:1977nu}. It can be directly verified, by using properties of the Gamma function and its product-representation, that 
\begin{eqnarray}
\Omega(\theta) = \frac{e^{\frac{\gamma}{2}- \frac{\pi i }{8}+\frac{\theta}{4}}}{\sqrt{2 \pi}} \prod_{j=1}^\infty e^{-\frac{1}{2 j}} \, j \, \frac{\Gamma\Big(j-\frac{1}{2}+\frac{\theta}{2\pi i}\Big)\Gamma\Big(j-\frac{\theta}{2\pi i}\Big)}{\Gamma\Big(j+\frac{1}{2}-\frac{\theta}{2\pi i}\Big)\Gamma\Big(j+\frac{\theta}{2\pi i}\Big)} \ , 
\end{eqnarray}
satisfies
\begin{equation}
\Omega (\theta) \Omega(\theta + i \pi) = \frac{e^{\frac{\theta}{2}}}{2 \cosh \frac{\theta}{2}} \ ,
\end{equation}
where $\gamma$ is Euler's constant. The factor $\Omega(\theta)$ has no poles in the physical strip. Hence, no CDDs are necessary, and we therefore expect it to be associated with the limit of the massless $AdS_2$ dressing factor in the corresponding relativistic limit, shadowing an analogous phenomenon occurring in the $AdS_3$ case \cite{Bogdan}.

\subsection{Differential equations}
\label{sec:V}
In this section we show that the R-matrices Solutions 2, 3, 4, 5 given in (\ref{Fendley p=1/2}), (\ref{left-right}), (\ref{Fendley p=-3/2}),  (\ref{right-left}) satisfy the following type of partial differential equations 
\begin{eqnarray}
\label{PDE1}
\left[\frac{\partial}{\partial \theta} + \Gamma^{(i)}_{\theta}\right] R^{(i)} = 0 \ ,
\end{eqnarray}
where $R^{(i)}$ stands for the R-matrix Solution $i$, and $\Gamma^{(i)}$ is an algebraic term.
In what follows we set $\alpha^2=1$ and we consider the right-right scattering only. 

By computing $\partial R^{(i)}/\partial \theta$ we find that the algebraic terms $\Gamma^{(i)}$ are 
\begin{eqnarray}
\Gamma^{(3 / 5)}_{\theta} = \pm \frac{1}{2(1 + e^{\pm \theta})} \mathds{1} \otimes \mathds{1} - \frac{1}{4 \cosh(\frac{\theta}{2})} (\mathsf{E}_{12} + \mathsf{E}_{21}) \otimes (\mathsf{E}_{12} + \mathsf{E}_{21})\ ,
\end{eqnarray}
(with $3 / 5$ associated with the upper/ lower sign), and
\begin{eqnarray}
\notag
\Gamma^{(2)}_{\theta} &=& \frac{\tanh\frac{\theta}{2}(2 - \cosh\theta)}{2\cosh\theta(1 - 2 \cosh\theta)} \mathds{1}\otimes \mathds{1} +\frac{2 - \cosh\theta}{2\cosh\frac{3\theta}{2}} (\mathsf{E}_{12} \otimes \mathsf{E}_{12} + \mathsf{E}_{21}\otimes\mathsf{E}_{21} )\\ 
&+&\frac{\cosh\frac{\theta}{2}}{1-2\cosh\theta} (\mathsf{E}_{12}\otimes\mathsf{E}_{21} + \mathsf{E}_{21}\otimes\mathsf{E}_{12} )\ , 
\end{eqnarray}
and 
\begin{eqnarray}
\notag
\Gamma^{(4)}_{\theta} &=& \bigg( \frac{\sinh\theta - 2 \sinh 2\theta}{1-2\cosh\theta + 2 \cosh 2\theta} - \frac{1}{2} \tanh\frac{\theta}{2} + \tanh\theta \bigg) ( \mathsf{E}_{11}  \otimes\mathsf{E}_{11} + \mathsf{E}_{22} \otimes \mathsf{E}_{22}) \\
\notag
&&- \frac{4\sinh\theta + \sinh 3\theta + 2\tanh\theta}{2(\cosh 2\theta + \cosh 3\theta)} (\mathsf{E}_{11} \otimes \mathsf{E}_{22} + \mathsf{E}_{22} \otimes \mathsf{E}_{11} ) \\
\notag
&& -  \frac{\cosh\frac{\theta}{2}(-3 + 2\cosh\theta)}{1- 2\cosh\theta + 2\cosh 2\theta} ( \mathsf{E}_{12} \otimes \mathsf{E}_{21} + \mathsf{E}_{21} \otimes \mathsf{E}_{12} ) \\
&&- \frac{3 - \cosh\theta + \cosh 2\theta}{2\cosh \frac{5\theta}{2}} (\mathsf{E}_{12}\otimes\mathsf{E}_{12} + \mathsf{E}_{21}\otimes \mathsf{E}_{21} ) \ ,
\end{eqnarray}
where 
\begin{equation}
\label{E_ij_basis}
\mathsf{E}_{11} \equiv \begin{pmatrix}1&0\\0&0\end{pmatrix}\ ,
\qquad 
\mathsf{E}_{22} \equiv \begin{pmatrix}0&0\\0&1\end{pmatrix} \ , \qquad
\mathsf{E}_{12} \equiv \begin{pmatrix}0&1\\0&0\end{pmatrix}\ ,
\qquad
\mathsf{E}_{21} \equiv \begin{pmatrix}0&0\\1&0\end{pmatrix}\ .
\end{equation}
The fact that all $\Gamma^{(i)}$ are not trivial suggests that there might be a possible geometric interpretation in terms of a connection on a fibre bundle, in analogy to the work presented in Chapter \ref{chapter: AdS3}. 
The (would-be) connections $\Gamma^{(i)}$ are meromorphic with poles in the complex $\theta$-plane. In the spirit of \cite{Joakim,Andrea}, we conjecture that the above represents the relativistic limit of a ``non-relativistic" fibre bundle, with a 2D torus as a base space - which is then decompactified and complexified to $\mathbb{C}^2 \ni (p_1,p_2)$ because of the analytic continuation of the dressing factor - and a $\mathfrak{su}(1 | 1)$ fibre. We conjecture that the relativistic limit might be responsible for the shrinking of the base space $T^2$ to $S^1$, decompactified and complexified to $\mathbb{C} \ni \theta$.

We have checked that the connections for Solution 3 / 5 respectively coincide with those for Solution 2 / 4 in the asymptotic large-$\theta$ regime, as expected from the discussion in section \ref{sec:II}. 

\section{Bethe ansatz}
\label{sec:VI}
In this section we study the Bethe-ansatz for the massless sector of $AdS_2$ superstrings. The technique which we adapt here do not rely on a definition of a pseudo-vacuum state.   
We approach the problem in both the relativistic and non-relativistic regimes. In the relativistic case, we study the Bethe ansatz for both Solution 2 (Fendley's) and Solution 3, which is obtained expanding at large $\theta$ the Solution 2. 
Finally, we show that the same technique can be applied to the non-relativistic R-matrix, and we show that we match the auxiliary Bethe equations conjectured in \cite{Sorokin:2011rr}. 
The technique itself relies on an algebraic condition for the S-matrix entries, the so-called \emph{free-fermion} condition, and it was introduced in \cite{Felderhof, Felderhof2, Felderhof3} and subsequently applied in \cite{MC,Ahn:1993qa}.

\subsection{Free-fermion condition and basis-change}
\label{sec:VIA}
To formulate the Bethe ansatz, it turns out to be convenient to switch from the R-matrix language to the S-matrix, and after that, to proceed by {\it ignoring any further fermionic sign}\footnote{The S-matrix solutions still satisfy the Yang-Baxter equations, and the spectrum is not altered by this switch.}, as done in \cite{MC,Ahn:1993qa}. 
We also suppresses the dressing factor, which can be easily reinstated at the end of the procedure.

We begin with writing the S-matrix as  
\begin{eqnarray}
S = \begin{pmatrix}A&B\\C&D\end{pmatrix}\ ,
\end{eqnarray}
where the $2 \times 2$ block S-matrix acts on the {\it auxiliary} space, while the operators $A$, $B$, $C$ and $D$ act on the {\it quantum} space, and they are 
\begin{eqnarray}
&&A = a_+ \, \mathsf{E}_{11} \, + \, b_+ \, \mathsf{E}_{22} \ , \qquad B = d_+ \, \mathsf{E}_{12} \, + \, c_- \, \mathsf{E}_{21}\ , \nonumber \\
&&C = c_+ \, \mathsf{E}_{12} \, + \, d_- \, \mathsf{E}_{21}\ , \qquad D = b_- \, \mathsf{E}_{11} \, + \, a_- \, \mathsf{E}_{22}\ ,
\end{eqnarray}
where $\mathsf{E}_{ij}$ is defined in (\ref{E_ij_basis}).
For Solution 2, we have that
\begin{eqnarray}
&& a_+ = a_- = 1 \ , \qquad\qquad\  b_- = - b_+ = \tanh \theta\ , \nonumber\\
&& c_+ = c_- = \frac{\cosh \frac{\theta}{2}}{\cosh \theta}\ , \qquad d_+ = - d_- = \frac{\sinh \frac{\theta}{2}}{\cosh \theta}\ .
\end{eqnarray}
The entries satisfy the {\it free-fermion condition}:
\begin{eqnarray}
a_+ a_- + b_+ b_- = c_+ c_- + d_+ d_-\ .
\end{eqnarray}
The monodromy matrix $\mathcal{T}_0$ and the associated transfer matrix $\mathcal{T}$ are defined as
\begin{eqnarray}
\mathcal{T}_0 =  S_{01} (\theta - \theta_1) ... S_{0N} (\theta - \theta_N)\ , \qquad\qquad 
\mathcal{T} = \textrm{tr}_0 \mathcal{T}_0 \ , 
\end{eqnarray}
 
Then we define a new S-matrix $S^{(1)}$, which takes the same form as $S$ but with entries reshuffled as follows 
\begin{eqnarray}
&&a_\pm \to a^{(1)}_\pm \equiv - b_\pm\ , \qquad\, b_\pm \to b^{(1)}_\pm \equiv a_\pm\ , \nonumber\\
&&c_\pm \to c^{(1)}_\pm \equiv c_\pm\ , \qquad\quad d_\pm \to d^{(1)}_\pm \equiv - d_\pm\ .
\end{eqnarray}  
One can promptly notice that $S^{(1)}$ still satisfies the free-fermion condition. Next, we consider
\begin{eqnarray}
\label{traccia}
\mathcal{T} \, \mathcal{T}^{(1)} &=& \textrm{tr}_0 \Big[ S_{01} (\theta - \theta_1) ... S_{0N} (\theta - \theta_N) \big] \textrm{tr}_{0'} \Big[S^{(1)}_{0'1} (\theta - \theta_1) ... S^{(1)}_{0'N} (\theta - \theta_N)\Big] \nonumber \\
&=& \textrm{tr}_{0 \otimes 0'} \prod_{i=1}^N S_{0i} (\theta - \theta_i) \otimes S^{(1)}_{0'i}(\theta - \theta_i)\ ,  
\end{eqnarray} 
where the tensor product is between the two auxiliary spaces $0$ and $0'$ pertaining to $\mathcal{T}$ and $\mathcal{T}^{(1)}$, respectively.

The trick is now to find a similarity transformation $X$ acting on the tensor-product matrix $S_{0i} (\theta - \theta_i) \otimes S^{(1)}_{0'i}(\theta - \theta_i)$, which transforms it in the upper triangular form. Such a transformation is performed at each site, but it should not depend on the site-specific variables $\theta_i$ (i.e. no {\it inhomogeneities}). The dependence of $S$ only on the difference $\theta_0 - \theta_i$, together with the free-fermion condition, implies that the similarity matrix $X$ is constant\footnote{This was first empirically noted by Felderhof in \cite{Felderhof, Felderhof2, Felderhof3}. }. 
It is only in this fashion that the similarity matrices will all cancel out in the expression (\ref{traccia}), and the task of taking the trace will become straightforward.

In fact, one can prove that the following matrix:
\begin{eqnarray}
\label{x}
X = \frac{1}{\sqrt{2}} \begin{pmatrix}0&1&1&0\\1&0&0&1\\1&0&0&-1\\0&1&-1&0\end{pmatrix} = X^{-1} 
\end{eqnarray}
is such that
\begin{eqnarray}
\label{suchthatx}
\notag
X (S_{0i} \otimes S^{(1)}_{0'i}) X^{-1} &=& X \begin{pmatrix}A A^{(1)}&A B^{(1)}&B A^{(1)}&B B^{(1)}\\A C^{(1)}&A D^{(1)}&B C^{(1)}&B D^{(1)}\\C A^{(1)}&C B^{(1)}&D A^{(1)}&D B^{(1)}\\C C^{(1)}&C D^{(1)}&D C^{(1)}&D D^{(1)}\end{pmatrix} X^{-1} \\
&=& \begin{pmatrix} m_+ & * & * & *\\0 & n_+ & * & * \\ 0 & 0 & n_- & * \\ 0 & 0 & 0 & m_-\end{pmatrix},
\end{eqnarray}
where
\begin{eqnarray}
&&m_\pm \equiv \frac{1}{2 \cosh^2 (\theta - \theta_i)} \Big[\pm \cosh  (\theta - \theta_i) + \cosh 2 (\theta - \theta_i)\Big] \mathds{1}\ , \nonumber \\
&&n_\pm \equiv \frac{1}{2 \cosh^2 (\theta - \theta_i)} \Big[\pm \sinh  (\theta - \theta_i) + \sinh 2 (\theta - \theta_i)\Big] \sigma_3\ ,
\end{eqnarray}
and 
\begin{equation}
\mathds{1} = \mathsf{E}_{11} + \mathsf{E}_{22} \ , \qquad\qquad
\sigma_3 = \mathsf{E}_{11} - \mathsf{E}_{22} \ .
\end{equation} 
As $\rm tr_{0 \otimes 0'} = \rm tr_4$, it follows immediately that
\begin{eqnarray}
\label{traccia2}
\mathcal{T} \, \mathcal{T}^{(1)} &=& \prod_{i=1}^N m_+ (\theta - \theta_i) +  \prod_{i=1}^N m_- (\theta - \theta_i) + \prod_{i=1}^N n_+ (\theta - \theta_i) + \prod_{i=1}^N n_- (\theta - \theta_i) \ .  
\end{eqnarray}  
We now make use of a very particular relation between $S$ and $S^{(1)}$. One can check that
\begin{eqnarray}
\label{rela}
S_{0'i}^{(1)} = \zeta \, \sigma_1 \, S_{0'i}(\theta - \theta_i + i \pi) \sigma_1^{-1} \, \zeta^{-1}\ , 
\end{eqnarray}
where
\begin{eqnarray}
\sigma_1 = \mathsf{E}_{12} + \mathsf{E}_{21}\ , \qquad \qquad
\zeta = \mathsf{E}_{11} + i \mathsf{E}_{22}\ ,
\end{eqnarray}
and where the similarity transformation (\ref{rela}) is performed in the auxiliary space $0'$.

This implies that 
\begin{eqnarray}
\label{that}
\mathcal{T}^{(1)} (\theta) = \mathcal{T}(\theta + i \pi)\ , 
\end{eqnarray}
which turns (\ref{traccia2}) into a crossing-type equation, denoted as {\it inversion relation} \cite{Zamolodchikov:1991vh}. This is consistent with the property that $m_+$ maps into $m_-$ and $n_+$ into $n_-$, under $\theta \to \theta + i \pi$ (for small $N$, we checked explicitly that (\ref{traccia2}) is correct).

The eigenvalues of $\mathcal{T} \, \mathcal{T}^{(1)}$ are given by the same expression as (\ref{traccia2}), with $\sigma_3$ replaced by the fermionic degree of the particular eigenstate where $\mathcal{T} \, \mathcal{T}^{(1)}$ is acting on. The final task is then to factorise such expressions into a product of two functions, namely $f(\theta)f(\theta + i \pi)$. As it is familiar from solving crossing-symmetry relations for S-matrices, this is a difficult problem in general, whose study relies on analyticity assumptions. Here, we shall restrict ourselves to derive a condition which identifies potential zeroes of the transfer-matrix eigenvalues, which turns out to lead to the auxiliary Bethe-ansatz type equations. First, we split (\ref{traccia2}) as 
\begin{eqnarray}
\label{times}
\notag
&&\mathcal{T} \mathcal{T}^{(1)} = \bigg[\prod_{i=1}^N A(\theta - \theta_i) + (-1)^F \prod_{i=1}^NB(\theta - \theta_i)\bigg]\\
&&\times \bigg[\prod_{i=1}^N C(\theta - \theta_i) + (-1)^F \prod_{i=1}^ND(\theta - \theta_i)\bigg]\frac{1}{\prod_{i=1}^N 2\cosh^2(\theta - \theta_i)}\ ,
\end{eqnarray}
where $(-1)^F$ is the fermionic degree of the particular state where $\mathcal{T}\mathcal{T}^{(1)}$ acts on ($F=0, 1$ for boson, fermion respectively), and
\begin{eqnarray}
A(\theta - \theta_i) \equiv \frac{c_{0i}^+}{C_{0i}}\ , \qquad B(\theta - \theta_i) \equiv \frac{s_{0i}^-}{C_{0i}}\ , \qquad D(\theta - \theta_i) \equiv \frac{s_{0i}^+}{c_{0i}^+} C_{0i}\ ,
\end{eqnarray}
where $C_{0i}$ is a freedom of this rewriting, and we have defined
\begin{eqnarray}
&&c_{0i}^\pm = \pm \cosh (\theta - \theta_i)+ \cosh 2  (\theta - \theta_i)\ , \nonumber \\
&&s_{0i}^\pm = \pm \sinh  (\theta - \theta_i)+ \sinh 2  (\theta - \theta_i)\ ,
\end{eqnarray}
and used the fact that
\begin{eqnarray}
\label{thanks}
c_{0i}^+ \, c_{0i}^- \, = \, s_{0i}^+ \, s_{0i}^-\ .
\end{eqnarray}

The eigenvalue has $N$ potential poles, determined by the hyperbolic cosines at the denominator, and a certain number of potential zeroes, possibly coincident, depending on the particular state. The possible location of them will be determined below via a set of auxiliary Bethe-ansatz conditions. Moreover, the eigenvalues are periodic of period $2 \pi i$, as can be seen by the fact that, by shifting (\ref{traccia2}) of a further $+i \pi$ and using (\ref{that}), one gets
\begin{eqnarray}
\mathcal{T}(\theta + i \pi ) \mathcal{T}(\theta + 2 i \pi) = \mathcal{T}(\theta) \mathcal{T}(\theta + i \pi)\ , 
\end{eqnarray}
where we have also explicitly made use of the invariance of the r.h.s. of (\ref{traccia2}) under shift of $i \pi$.
Since the RTT relations implies (\ref{T_commutes}), we have
\begin{equation}
\mathcal{T}(\theta + 2 i \pi) = \mathcal{T}(\theta)\ . 
\end{equation}
Hence it suffices to study the location of poles and zeroes of $\mathcal{T}$ in the strip $\theta \in [-\pi, \pi)$. We recall that the dressing factor - say, $\Phi$ - does not affect these considerations, since we already know how we will have to decorate the eigenvalue of $\mathcal{T}$ obtained at the end, {\it i.e.} by a product of dressing factors $\prod_{i=1}^N \Phi(\theta-\theta_i)$. 

From (\ref{times}), one sees that potential zeroes of $\mathcal{T}$ can come from
\begin{eqnarray}
\prod_{i=1}^N \frac{A(z_k - \theta_i)}{B(z_k - \theta_i)} = (-1)^{F+1}\ , \qquad\mbox{or}\qquad \prod_{i=1}^N \frac{C(z_k - \theta_i)}{D(z_k - \theta_i)}= (-1)^{F+1}\ .
\end{eqnarray}
Plugging the explicit formulas, we see that the freedom of $C_{0i}$ is indeed irrelevant, and we obtain that the potential zeroes $\{ z^{\pm}_k \}$ can come from either of two conditions:
\begin{eqnarray}
\label{aux}
\prod_{i=1}^N \coth \frac{z^+_k - \theta_i}{2} = (-1)^{F+1}\ , \qquad\mbox{or}\qquad \prod_{i=1}^N \coth \frac{3(z^-_k - \theta_i)}{2} = (-1)^{F+1}\ ,
\end{eqnarray}
At this stage, we can identify (\ref{aux}) as a subset of the \emph{auxiliary Bethe equations}.
Here we use the important general fact that the zeroes of the transfer matrix eigenvalues and the auxiliary Bethe roots, i.e. solutions to the auxiliary Bethe equations, are the same.  The fact that this is a subset and not a one-to-one correspondence is because there is an ambiguity in identifying the zeroes of $\mathcal{T}$. As we shall see for the S-matrices considered later, this ambiguity will not show up, and the identification of zeroes of $\mathcal{T}$ with auxiliary Bethe roots will be exact. 
Notice that, thanks to (\ref{thanks}), each of two set of potential auxiliary Bethe equations maps into itself under $z_k^\pm \to z_k^\pm + i \pi$.

One needs at this point to identify the actual set of zeroes of $\mathcal{T}$ {\it vs.} those of $\mathcal{T}^{(1)}$ in order to extract the eigenvalues of $\mathcal{T}$. This can then be used to write down the ({\it momentum-carrying}) Bethe equation. 
As described in section \ref{Bethe}, one has to solve the following eigenvalue problem
\begin{eqnarray}
\label{eige}
 \, \mathcal{T}(p_0 | p_1,...,p_N) |\psi\rangle = e^{- i p_0 L}|\psi \rangle\ ,
\end{eqnarray} 
where $p_a = e^{\theta_a}$, $a=0,...N$, and the dressing factors are reintroduced in the transfer matrix. The eigenvalues of (\ref{eige}) are  parametrised by their potential poles and zeroes, which are solutions to the auxiliary Bethe equations, i.e. a subset of (\ref{aux}). 

Some experimenting with small $N$ seems to reveal that this final step is not straightforward, and would require a separate analysis. This holds also for the next cases which we discuss in this thesis. This observation  is clearly related to the fact that our eigenvalues may either tend to zero (as in this case) or have an essential singularity (as in the next section) at $\theta = \infty$, at odds with the situation in \cite{MC,Ahn:1993qa} - and \cite{Zamolodchikov:1991vh}, after going to a {\it reduced} transfer matrix with no essential singularities. This is easily evinced by studying the asymptotics of the corresponding S-matrices. Therefore, even the knowledge of the zeroes and poles of the meromorphic periodic function $\mathcal{T}(\theta)$ would not allow us to completely reconstruct it, as one cannot eliminate the ambiguity of factors which are entirely periodic functions of $\theta$ and depend on all $\theta_i$'s. 

A conjectured formula for the eigenvalues of (\ref{eige}) is given in \cite{Torrielli:2017nab}, where the eigenvalues have been found explicitly up to $N=4$, and then conjectured for a generic $N$. 

\subsection{Bethe-ansatz condition for Solution 3}
\label{sec:VIB}
In this section we study the Bethe ansatz for the Solution 3. Such solution still satisfies the free-fermion condition, and therefore the technique applied before to the Solution 2 is still valid here. The advantage of having performed the process on Solution 2 (Fendley's) is manifest from the fact that Solution 3 can formally be obtained from it in a large $\theta$ asymptotic expansion. This does not mean that one can indiscriminately expand at large $\theta$ all the previous formulas, but in several cases it implies that similar algebraic manipulations will apply.

We can be concise on the intermediate steps, and write now
\begin{eqnarray}
&&a_+ = a_- = 1\ , \qquad\quad b_- = - b_+ = 1\ , \nonumber\\
&&c_+ = c_- = e^{-\frac{\theta}{2}}\ , \qquad d_+ = - d_- = e^{-\frac{\theta}{2}}\ ,
\end{eqnarray}
such that
\begin{eqnarray}
a_+ a_- + b_+ b_- = c_+ c_- + d_+ d_-\ .
\end{eqnarray}
and
\begin{eqnarray}
&&a_\pm \to a^{(1)}_\pm = - b_\pm\ , \qquad b_\pm \to b^{(1)}_\pm = a_\pm\ , \nonumber\\
&&c_\pm \to c^{(1)}_\pm = c_\pm\ , \qquad\quad d_\pm \to d^{(1)}_\pm = - d_\pm\ .
\end{eqnarray}    
The very same transformation $X$ in (\ref{x}) and (\ref{suchthatx}) works for this case as well, and one obtains an upper triangular form like (\ref{suchthatx}) for $\mathcal{T} \mathcal{T}^{(1)}$, this time with diagonal entries
\begin{eqnarray}
&&m_\pm = \big(1\pm e^{-\theta}\big) \mathds{1}\ , \qquad n_\pm =  \big(1\pm e^{-\theta}\big) \sigma_3\ .
\end{eqnarray}
The great simplification with respect to the previous section is now that the product (\ref{traccia2}) of the two eigenvalues reduces to
\begin{eqnarray}
\label{fromm}
\mathcal{T} \mathcal{T}^{(1)} = (1+(-1)^F) \Bigg[\prod_{i=1}^N \Big(1 + e^{-(\theta - \theta_i)}\Big) +   \prod_{i=1}^N \Big(1 - e^{-(\theta - \theta_i)}\Big)\Bigg]\ ,
\end{eqnarray}
where $(-1)^F$ denotes again the fermionic number of the particular eigenstate under consideration, $F=0,1$ for a boson or fermion respectively. Formula (\ref{fromm}) shows that, in this case, we access only part of the spectrum, as $\mathcal{T} \mathcal{T}^{(1)}$ annihilates all fermionic eigenstates. A subset of the potential zeroes $\beta_k$ of the bosonic transfer-matrix eigenvalues provides the auxiliary Bethe equations
\begin{equation}
\label{sub}
\prod_{i=1}^N \tanh \frac{\beta_k - \theta_i}{2} = -1\ , \quad k=1,...,M\ .
\end{equation}
Furthermore, one can verify that the relations (\ref{rela}) and (\ref{that}) work exactly the same way, hence one can rely on the very same $2 \pi i$ periodicity property of the eigenvalue of $\mathcal{T}$. One would then write the momentum-carrying equation
\begin{eqnarray}
\label{fixo}
e^{i e^{\theta_0} L} \, \Bigg[\prod_{i=1}^N \Omega(\theta_0 - \theta_i)\Bigg] \Lambda(\theta_0|\theta_1,...,\theta_N| \beta_1,...,\beta_M)= 1\ ,
\end{eqnarray}
subject to (\ref{sub}), where $\Lambda$ is the transfer-matrix eigenvalue normalised to $a_+=1$, and we have inserted the dressing factor obtained in section \ref{sec:IV}. As we shall see next, the condition (\ref{sub}) on the potential zeroes of $\Lambda$ matches the naive massless relativistic limit of the auxiliary Bethe equations for $AdS_2$ \cite{Sorokin:2011rr}.

\subsection{Bethe-ansatz condition for non-relativistic massless $AdS_2$}
\label{sec:VIC}
Finally, we consider the {\it non-relativistic} massless S-matrix, which remarkably it also satisfies the free-fermion condition. 
One can straightforwardly repeat the entire line of argument developed in the previous sections, just by replacing (consistent with the relativistic limit)
\begin{eqnarray}
e^{\frac{\theta - \theta_i}{2}} \to \sqrt{\frac{\tan \frac{p_0}{4}}{\tan \frac{p_i}{4}}}\ , \qquad\quad 1 \pm e^{\theta - \theta_i}\to 1 \pm \frac{\tan \frac{p_0}{4}}{\tan \frac{p_i}{4}}\ ,
\end{eqnarray} 
where $\theta \to \theta + i \pi$ is replaced by $p_0\to -p_0$, and under the square-root this is prescribed to give $i$. This means that the auxiliary Bethe equations now read
\begin{equation}
\label{just}
\prod_{i=1}^N \frac{\sin \frac{q_k + p_i}{4}}{\sin \frac{q_k - p_i}{4}} = - 1\ , \qquad k=1,...,M\ .
\end{equation}
If we naively take the massless relativistic limit of half (corresponding to one wing of the $\mathfrak{psu}(1,1|2)$ Dynkin diagram) of the auxiliary Bethe equations conjectured in \cite{Sorokin:2011rr} (STWZ), we find that they seem to exactly match with the {\it square} of (\ref{just}), if we identify their auxiliary roots, say, $p_{k,3}$ with our $q_k$, switching {\it e.g.} the type $1$ roots off. This matching however is only indicative, since a matching performed directly in the massless sector is not possible: the results available in the literature pertain to the massive case, and we have simply taken a naive massless limit of those. A direct derivation in the massless sector was not performed in \cite{Sorokin:2011rr}.

The momentum-carrying equation can easily be obtained via the same naive limit from STWZ, and it can be simplified using momentum conservation $\sum_{j=1}^N p_{j,2}=0$ (in their conventions, translating into $\sum_{j=1}^N p_{j}=0$ in ours) to be expressed in terms of the same functions appearing in (\ref{just}), except for the dressing factors. However, once again a separate analysis would be required to obtain such an equation from the procedure we have described, in analogy to (\ref{fixo}). The relativistic limit should then ideally reproduce some variant of (\ref{fixo}), precisely like (\ref{just}) straightforwardly reduces to (\ref{sub}). The proposal for the dressing factor to appear in the Bethe ansatz made in \cite{Sorokin:2011rr} involves the inverse-square of the BES factor \cite{BES, BES2}, which should then be compared with $\Omega(\theta)$ of section \ref{sec:IV} in the appropriate massless relativistic limit along the lines of \cite{Bogdan}.

\chapter{\textbf{Geometry of $AdS_3$ massless scattering}}
\label{chapter: AdS3}

\section{The background algebra}
In this chapter we shall investigate properties of massless string excitations in the $AdS_3 \times S^3 \times T^4$ type IIB string sigma model. This background preserves 16 real supercharges, and the bosonic sector of the sigma model is described in terms of the coset space
\begin{equation}
\label{AdS3_coset}
\frac{SO(2,2) \times SO(4) \times U(1)^4}{SO(2,1) \times SO(3)} \ , 
\end{equation}
To include the fermionic contributions, the coset (\ref{AdS3_coset}) can be embedded into the supercoset space
\begin{equation}
\label{AdS3_super}
\frac{PSU(1,1 | 2)_L \times PSU(1,1 | 2)_R }{SU(1,1) \times SU(2)} \ . 
\end{equation} 
The isometry superalgebra of (\ref{AdS3_super}) is $\mathfrak{psu}(1,1|2)_L \oplus \mathfrak{psu}(1,1|2)_R$, however the choice of a vacuum state for the single particle representation reduces the isometry superalgebra to the little group $[\mathfrak{psu}(1|1)_L \oplus \mathfrak{psu}(1|1)_R]^2$, \cite{Ya3}. As explained in section \ref{sec:back_AdS2}, we shall consider just one copy of the isotropy algebra\footnote{We cannot further simplify $\mathfrak{psu}(1|1)_L \oplus \mathfrak{psu}(1|1)_R$, since the representations on the two branches are inequivalent.}, i.e. $\mathfrak{psu}(1|1)_L \oplus \mathfrak{psu}(1|1)_R$.

In order to describe physical properties of particles (e.g. mass, energy, momentum), we need to consider the centrally-extended $\mathfrak{psu}(1|1)_L \oplus \mathfrak{psu}(1|1)_R $ superalgebra, which we shall simply denote by $\mathfrak{su}(1|1)_L \oplus \mathfrak{su}(1|1)_R$. The non-vanishing graded commutation relations are
\begin{eqnarray}
\label{alge}
\{\mathcal{Q}_{L}, \mathcal{Q}_{R}\} = \mathcal{P} \ , \qquad  \{\mathcal{S}_{L}, \mathcal{S}_{R}\} = \mathcal{K} \ , \qquad 
\{\mathcal{Q}_{L}, \mathcal{S}_{L}\} = \mathcal{H}_{L} \ , \qquad \{\mathcal{Q}_{R}, \mathcal{S}_{R}\} = \mathcal{H}_{R}  \ ,
\end{eqnarray}
where $\mathcal{P}, \mathcal{K}, \mathcal{H}_L, \mathcal{H}_R$ are the central bosonic generators and $\mathcal{Q}_{L}, \mathcal{Q}_{R}, \mathcal{S}_{L}, \mathcal{S}_{R}$ are the fermionic generators. 

We represent the generators of $\mathfrak{su}(1|1)_L \oplus \mathfrak{su}(1|1)_R$ as $2\times2$ matrices acting on a boson-fermion doublet $( |\phi\rangle, |\psi\rangle )^T$. In this case, the algebra admits two inequivalent representations. 

The \emph{left} representation is 
\begin{equation}
\label{Q_S_2}
\mathcal{Q}_L = a \begin{pmatrix}
0 & 0 \\
1 & 0 
\end{pmatrix} \ , \qquad
\mathcal{S}_L = b \begin{pmatrix}
0 & 1 \\
0 & 0 
\end{pmatrix} \ , 
\end{equation}
and 
\begin{equation}
\label{Q_S_3}
\mathcal{Q}_R = c \begin{pmatrix}
0 &  1\\
0 & 0 
\end{pmatrix} \ , \qquad
\mathcal{S}_R = d \begin{pmatrix}
0 & 0 \\
1 & 0 
\end{pmatrix} \ , 
\end{equation}
and
\begin{equation}
\label{H_P_K_2}
\mathcal{H}_A = H_A \begin{pmatrix}
1 & 0 \\
0 & 1 
\end{pmatrix} \ , \qquad
\mathcal{P} = P \begin{pmatrix}
1 & 0 \\
0 & 1 
\end{pmatrix} \ , \qquad
\mathcal{K} = K \begin{pmatrix}
1 & 0 \\
0 & 1 
\end{pmatrix} \ ,
\end{equation}
where $A = L, R$ and $a, b, c, d, H_A, P, K \in \mathbb{C}$ are the \emph{representation parameters}, among which $a, b, c, d$ are the \emph{independent} ones. The only central generator of  $\mathfrak{su}(1|1)_L \oplus \mathfrak{su}(1|1)_R$ is the identity, therefore $\mathcal{P}, \mathcal{K}, \mathcal{H}_L, \mathcal{H}_R$ must be proportional to it. 

The \emph{right} representation is
\begin{equation}
\label{Q_S_3_right}
\mathcal{Q}_L = c \begin{pmatrix}
0 &  1\\
0 & 0 
\end{pmatrix} \ , \qquad
\mathcal{S}_L = d \begin{pmatrix}
0 & 0 \\
1 & 0 
\end{pmatrix} \ , 
\end{equation}
and
\begin{equation}
\label{Q_S_2_right}
\mathcal{Q}_R = a \begin{pmatrix}
0 & 0 \\
1 & 0 
\end{pmatrix} \ , \qquad
\mathcal{S}_R = b \begin{pmatrix}
0 & 1 \\
0 & 0 
\end{pmatrix} \ , 
\end{equation}
and
\begin{equation}
\label{H_P_K_2_right}
\mathcal{H}_A = H_A \begin{pmatrix}
1 & 0 \\
0 & 1 
\end{pmatrix} \ , \qquad
\mathcal{P} = P \begin{pmatrix}
1 & 0 \\
0 & 1 
\end{pmatrix} \ , \qquad
\mathcal{K} = K \begin{pmatrix}
1 & 0 \\
0 & 1 
\end{pmatrix} \ .
\end{equation}

\section{Representations of $\mathfrak{su}(1|1)_L \oplus \mathfrak{su}(1|1)_R$}
In this section we shall present two representations relevant for this chapter: the \emph{massive} and \emph{massless} representations. 

\subsubsection{Massive representation}
In the boson-fermion representation (\ref{Q_S_2}), (\ref{Q_S_3}) and (\ref{H_P_K_2}), we have 
\begin{align}
\label{AdS3_massive}
a =& \frac{1}{\sqrt2} \sqrt{\mathcal{E}+ m} \ ,  &\nonumber
b =& \frac{1}{\sqrt2} \frac{ h(1-e^{i p})}{\sqrt{\mathcal{E}+ m}} \ , \\
c =\, & \frac{1}{\sqrt2} \frac{ h(1-e^{-i p})}{\sqrt{\mathcal{E}+ m}} \ , &
d =\, & \frac{1}{\sqrt 2} \sqrt{\mathcal{E}+ m} \ ,
\end{align}
and 
\begin{eqnarray}
\label{AdS3_massive_2}
H_L = \frac{\mathcal{E}_L}{2} \ , \qquad 
H_R = \frac{\mathcal{E}_R}{2} \ , \qquad
 P = \frac{h}{2} (1 - e^{i  p} ) \ , \qquad 
 K = \frac{h}{2} (1 - e^{-i  p}) \ , 
\end{eqnarray}
where $\mathcal{E} \equiv \mathcal{E}_L + \mathcal{E}_R$ is the \emph{total energy}, $p$ is the \emph{momentum}, $m$ is the \emph{mass} and $h$ is the coupling constant.
The closure of the algebra implies the dispersion relation\footnote{For the same argument given in chapter \ref{Chapter: AdS2}, a quadratic Casimir cannot be constructed. However if one considers the secret symmetry, then a quadratic Casimir $\mathfrak{C}_2$ exists, and the dispersion relation (\ref{casimirAdS3}) is given by $\mathfrak{C}_2 = m^2$. } 
\begin{eqnarray}
\label{casimirAdS3}
\mathcal{E}^2  =  m^2 + 4 h^2 \sin^2 \frac{p}{2} \ . 
\end{eqnarray}

\subsubsection{Massless representation}
This representation is obtained by taking the following limit 
\begin{equation}
\label{massless_limit_2}
m \rightarrow 0 \ , \qquad\qquad h \quad \mbox{finite} \ , 
\end{equation}
of the massive representation. The representation parameters become
\begin{eqnarray}
\label{par1_massless}
&& a = \sqrt{h \sin \frac{p}{2} },  \qquad\qquad b = \pm  \sqrt{ h \sin \frac{p}{2}} \ , \nonumber \\
&& c = \pm  \sqrt{ h \sin \frac{p}{2} }, \qquad\ \  d = \sqrt{ h \sin \frac{p}{2}} \ , 
\end{eqnarray}
and
\begin{eqnarray}
\label{par2_massless}
H_L = \frac{\mathcal{E}_L}{2} \ , \qquad 
H_R = \frac{\mathcal{E}_R}{2} \ , \qquad
P = \frac{h}{2} (1 - e^{i  p} ) \ , \qquad 
K = \frac{h}{2} (1 - e^{-i  p}) \ , 
\end{eqnarray}
where $\mathfrak{Re} p \in [0 , \pi]$ for right movers, and $\mathfrak{Re} p \in [- \pi , 0]$ for left movers. 
By taking the massless limit of (\ref{casimirAdS3}), we obtain the massless dispersion relation\footnote{In terms of the quadratic Casimir, this corresponds to $\mathfrak{C}_2 = 0$. }
\begin{equation}
\label{massless_disp_AdS3}
\mathcal{E} = 2h \abs{\sin\frac{p}{2}} \ . 
\end{equation}

\section{Massless R-matrix}
In this section we consider the R-matrix in the massless scattering regime. For the purpose, we introduce the  
coproduct action of the $\mathfrak{su}(1|1)_L \oplus \mathfrak{su}(1|1)_R$ superalgebra on a 2-particle state in the massless representation as\footnote{We recall that in general the definition of a coproduct is not unique. The one defined above will be particularly useful for studying the boost symmetry of the $q$-deformed Poincar\'e superalgebra.}
\begin{equation}
\label{copro_1}
\Delta(\mathcal{Q}_A) =  \mathcal{Q}_A \otimes e^{i \frac{p}{4}} + e^{-i \frac{p}{4}} \otimes \mathcal{Q}_A\ , \qquad\qquad
 \Delta(\mathcal{S}_{A})= \, \mathcal{S}_{A} \otimes {e^{i \frac{p}{4}}} + {e^{-i \frac{p}{4}}} \otimes \mathcal{S}_{A}\ ,  
\end{equation}
and 
\begin{eqnarray}
\label{copro_2}
\notag
&&\Delta(\mathcal{H}_{A})= \, \mathcal{H}_{A} \otimes {e^{i \frac{p}{2}}} + {e^{-i \frac{p}{2}}} \otimes \mathcal{H}_{A}\ , \qquad\qquad
 \Delta(\mathcal{P})= \, \mathcal{P} \otimes e^{i \frac{p}{2}} + e^{-i \frac{p}{2}} \otimes \mathcal{P}\ , \\
&&\hspace{3cm} \Delta(\mathcal{K})= \, \mathcal{K} \otimes e^{i \frac{p}{2}} + e^{-i \frac{p}{2}} \otimes \mathcal{K}\ , 
\end{eqnarray}
where $A = L, R$. In \cite{Joakim, BogdanLatest} the authors found the following R-matrix for the left-left scattering
\begin{equation}
\label{Rchosen}
R = \Phi \begin{pmatrix}
1 & 0 & 0 & 0\\
0 & -\csc\frac{p_1 + p_2}{4}\sin \frac{p_1 - p_2}{4} & \csc\frac{p_1 + p_2}{4}\sqrt{\sin\frac{p_1}{2}\sin\frac{p_2}{2}} & 0 \\
0 & \csc\frac{p_1 + p_2}{4}\sqrt{\sin\frac{p_1}{2}\sin\frac{p_2}{2}}  & \csc\frac{p_1 + p_2}{4}\sin\frac{p_1 - p_2}{4} & 0 \\
0 & 0 & 0 & -1
\end{pmatrix} \ , 
\end{equation}
which is invariant under the $\mathfrak{su}(1|1)_L \oplus \mathfrak{su}(1|1)_R$ action
\begin{equation}
\label{DR=RD_AdS3}
  \Delta^{\text{op}} (\mathfrak{a})\,  R\ =\ R\, \Delta (\mathfrak{a}) \qquad\qquad \forall \  \mathfrak{a} \in \mathfrak{su}(1|1)_{L}\oplus \mathfrak{su}(1|1)_{R} \ ,
\end{equation}
and furthermore it satisfies the Yang-Baxter equation (\ref{YBE}) and braiding-unitarity (\ref{braid_unitarity}). It also satisfies crossing-symmetry (\ref{crossing}) with the dressing factor $\Phi$ found in \cite{Borsato:2016xns}.

\subsection{The $q$-deformed Poincar\'e superalgebra pseudo-invariance}

In this section we shall investigate whether the R-matrix (\ref{Rchosen}) admits a larger group of symmetries than just 
the $\mathfrak{su}(1|1)_L \oplus \mathfrak{su}(1|1)_R$ superalgebra. In particular we introduce the extra generators $\mathcal{J}_L, \mathcal{J}_R$, the so-called \emph{boost} generators, which they will act together with the old generators of $\mathfrak{su}(1|1)_L \oplus \mathfrak{su}(1|1)_R$ to generate two copies of the $q$-deformed Poincar\'e superalgebra in $1+1$ dimensions, denoted by $\mathfrak{E}_q(1,1)_{L} \oplus \mathfrak{E}_q(1,1)_{R}$. 

The non-vanishing graded commutation relations of $\mathfrak{E}_q(1,1)_{L} \oplus \mathfrak{E}_q(1,1)_{R}$ are \cite{Joakim}  
\begin{eqnarray}
\label{boostalgebra}
\notag
&&\{\mathcal{Q}_{L}, \mathcal{Q}_{R}\} = \mathcal{P} \ , \qquad  \{\mathcal{S}_{L}, \mathcal{S}_{R}\} = \mathcal{K} \ , \qquad 
\{\mathcal{Q}_{L}, \mathcal{S}_{L}\} = \mathcal{H}_{L} \ , \qquad \{\mathcal{Q}_{R}, \mathcal{S}_{R}\} = \mathcal{H}_{R}  \ , \\
\notag
&&[\mathcal{J}_A, \mathcal{Q}_B] \ = \frac{i}{2 \sqrt{\mu}} \frac{e^{i \frac{p}{2}} + e^{- i \frac{p}{2}}}{2} \mathcal{Q}_B\ , \qquad
 [\mathcal{J}_A, \mathcal{S}_B] = \frac{i}{2 \sqrt{\mu}} \frac{e^{i \frac{p}{2}} + e^{- i \frac{p}{2}}}{2} \, \mathcal{S}_B \ , \\
 && [\mathcal{J}_A, \mathcal{H}_B] \ = - [\mathcal{J}_A, \mathcal{P}] \ =  \ - [\mathcal{J}_A, \mathcal{K}] \ = \frac{e^{i p} - e^{-i p}}{2 \mu} \ , \qquad [\mathcal{J}_A, p] \ = \ i \mathcal{H}_A \ ,
\end{eqnarray}
where both $A$ and $B$ can be either $L$ or $R$, and $\mu \equiv \frac{4}{h^2}$.
The deformation parameter $q$ is related to the coupling constant $h$ via:
\begin{equation}
\log q = \frac{i}{h^2} \ .
\end{equation}
The massless dispersion relation (\ref{massless_disp_AdS3}) remains invariant after introducing the boost generators.
The boost operators act on a single particle state as
\begin{equation}
\label{boost_J}
\mathcal{J}_A = i \mathcal{H}_A \, \partial_p \ , \qquad A = L, R \ .   
\end{equation}
The coproduct action of $\mathfrak{E}_q(1,1)_{L} \oplus \mathfrak{E}_q(1,1)_{R}$ on a two-particle state is as follow. 
For the boost generators we have \cite{Gomez:2007zr, Charles}
\begin{eqnarray}
\label{deltaJ}
\notag
\Delta (\mathcal{J}_A) &=& \mathcal{J}_A \otimes (e^{i \frac{p}{2}}\mathds{1}) +  (e^{-i \frac{p}{2}}\mathds{1}) \otimes \mathcal{J}_A \\
&+& \frac{1}{2}\,  (\mathcal{Q}_A \, e^{- i \frac{p}{4}}) \otimes (\mathcal{S}_A \, e^{i \frac{p}{4}}) +  \frac{1}{2}\,  (\mathcal{S}_A \, e^{- i \frac{p}{4}}) \otimes (\mathcal{Q}_A \, e^{i \frac{p}{4}}) \ ,
\end{eqnarray}
while the coproduct of the remaining $\mathfrak{su}(1|1)_L \oplus \mathfrak{su}(1|1)_R$ generators is the same as in (\ref{copro_1}) and (\ref{copro_2}). 
The expression above for $\Delta(\mathcal{J}_A)$ is non-local, which is in agreement with a standard argument reviewed in \cite{rev}-VI.2.

In what follows, we shall understand if the R-matrix (\ref{Rchosen}) is invariant, in the sense of equation (\ref{DR=RD_AdS3}), under the enlarged algebra $\mathfrak{E}_q(1,1)_{L} \oplus \mathfrak{E}_q(1,1)_{R}$. 
Without loss of generality, we shall consider $\mathcal{J}_L$ only.  
We compute the action of $\mathcal{J}_L$ on (\ref{Rchosen}), and we find
\begin{equation}
\label{boost} 
\Delta(\mathcal{J}_L) R = 0 \ ,  \qquad \qquad\Delta^{op}(\mathcal{J}_L) R = 0  \ ,
\end{equation}
which is not the invariance condition of the R-matrix under the boost action, but this tells us that the R-matrix is annihilated by $\Delta(\mathcal{J}_L)$ and $\Delta^{op}(\mathcal{J}_L)$. 
Moreover, the boost generator (\ref{boost_J}) is not a symmetry of the Lagrangian.
We remark that to compute (\ref{boost}) we choose an analogous of the Heisemberg picture of quantum mechanics, where the states do not transform under boost, i.e. $\mathcal{J}_L |\phi\rangle =\mathcal{J}_L |\psi\rangle = 0$, while the algebra generators do transform\footnote{In analogy with the Heisenberg picture of quantum mechanics, the boost generator would play the role of time derivative, such that the states do not evolve in time, but the operators do evolve, and their evolution is dictated by their commutator with the Hamiltonian. From the world-sheet perspective \cite{Borsato:2017lpf}, it might be possible to write an expression analogous to the Heisenberg evolution equation 
\begin{equation}
J \mathfrak{a} = i [ \mathscr{H}, \mathfrak{a} ] \ ,  \qquad
\forall \ \mathfrak{a} \in \mathfrak{su}(1|1)_L \oplus \mathfrak{su}(1|1)_R \ . 
\end{equation}
where $\mathscr{H}$ is the world-sheet Hamiltonian. }.

\subsection{Geometric interpretation}
\label{sec:level2}
In this section we show that the annihilation of the R-matrix under the boost action (\ref{boost}) is instrumental for a geometric interpretation of the (massless) scattering. 

By taking linear combinations of the two equations in (\ref{boost}), we find that the R-matrix (\ref{Rchosen}) must satisfy the following equations
\begin{equation}
\label{DR=0}
D_M R \equiv \bigg[\frac{\partial}{\partial p_M} + \Gamma_M\bigg]R = 0, \qquad\qquad (M =1,2) \ ,
\end{equation}
where
\begin{eqnarray}
\label{connection}
&& \Gamma_1 \equiv -\frac{1}{4} \sqrt{\frac{\sin \frac{p_2}{2}}{\sin \frac{p_1}{2}}} \, \frac{\big[ \mathsf{E}_{12} \otimes \mathsf{E}_{21} + \mathsf{E}_{21} \otimes \mathsf{E}_{12}\big]}{\sin \frac{p_1 + p_2}{4}} \ ,\nonumber \\
&& \Gamma_2 \equiv \frac{1}{4} \sqrt{\frac{\sin \frac{p_1}{2}}{\sin \frac{p_2}{2}}} \, \frac{\big[ \mathsf{E}_{12} \otimes \mathsf{E}_{21} + \mathsf{E}_{21} \otimes \mathsf{E}_{12}\big]}{\sin \frac{p_1 + p_2}{4}} \ , 
\end{eqnarray} 
and
\begin{equation} 
\mathsf{E}_{12} \equiv \begin{pmatrix}0&1\\0&0\end{pmatrix}, \qquad \mathsf{E}_{21} \equiv \begin{pmatrix}0&0\\1&0\end{pmatrix} \ . 
\end{equation} 
We interpret $D_M$ in (\ref{DR=0}) as a \emph{covariant derivative} on a fibre bundle with a 2-dimensional manifold base space $\mathcal{B}$, equipped with real coordinates $(p_1,p_2)$. 
The entries of (\ref{Rchosen}) and the connection components (\ref{connection}) are periodic in $p_1$ and $p_2$, with period $8\pi$, which is four times the physical strip length of left- and right-moving massless modes \cite{Borsato:2016xns}. 
Therefore $\mathcal{B}$ is topologically equivalent to the 2-dimensional torus $T^2$. 

In this language, the R-matrix is a \emph{section} of the fibre bundle, which must be \emph{covariantly constant} under $D_M$.  
This motivates the choice of the fibre $\mathcal{F}$, as we shall explain. Since the (left part of the) R-matrix is an object defined in the universal enveloping algebra $\mathcal{U}[\mathfrak{su}(1|1)_L]\otimes\mathcal{U}[\mathfrak{su}(1|1)_L]$, then we must take $\mathcal{F} \subset \mathcal{U}[\mathfrak{su}(1|1)_L]\otimes\mathcal{U}[\mathfrak{su}(1|1)_L]$. Finally, the connection (\ref{connection}) takes values in $\mathfrak{su}(1|1)_L\oplus\mathfrak{su}(1|1)_L$, which is the algebra associated with the fibre $\mathcal{F}$. 

The connection (\ref{connection}) is characterised by infinitely many singularities on $\mathcal{B}$, given by the vanishing of the sine functions at the denominators. Moreover, the connection (\ref{connection}) is locally flat, i.e. away from singular points, the curvature $F_{MN}$ vanishes:
 \begin{equation}
\label{flatness}
F_{12} = \partial_1 \Gamma_2 - \partial_2 \Gamma_1 + [\Gamma_1,\Gamma_2] = 0 \ . 
\end{equation}
This implies that the connection $(\ref{connection})$ is a pure gauge connection, i.e. can be written as
\begin{equation}
\Gamma_M = g \partial_M  g^{-1} \ , 
\end{equation}
for some $g \in \mathcal{U}[\mathfrak{su}(1|1)_L]\otimes\mathcal{U}[\mathfrak{su}(1|1)_L]$. 

In what follows, we shall discuss how to restore the dressing factor $\Phi$. By multiplying the R-matrix (\ref{Rchosen}) with the dressing factor we obtain that the dressed R-matrix, denoted by $\tilde{R}$, must satisfy
\begin{equation}
\label{fase}
\bigg[\frac{\partial}{\partial p_M} + \Gamma_M - \frac{\partial}{\partial p_M} \log \Phi\bigg]\tilde{R} = 0 \, \qquad\qquad \tilde {R} \equiv \Phi R \ .
\end{equation}
Equation (\ref{fase}) can still be thought as a covariantly constant condition for the dressed R-matrix, but with respect to a new connection. The new connection is related to $\Gamma_M$ in (\ref{connection}) via a gauge transformation, as follows
\begin{equation}
\label{gammatilde}
\Gamma_M \rightarrow g\Gamma_M g^{-1} + g \partial_M g^{-1} \ , \qquad\qquad g = \Phi \mathds{1} \ .   
\end{equation}
A consequence of (\ref{fase}) and (\ref{gammatilde}) is that restoring the dressing factor does not alter the structure of the fibre bundle. The dressed R-matrix is again covariantly constant with respect to a pure gauge connection.

We shall remark that the R-matrix (\ref{Rchosen}) is symmetric, i.e. satisfies $\Pi(R)(p_2,p_1) = R(p_1,p_2)$. This implies that the two conditions in (\ref{DR=0}) are not independent. However we propose to consider (\ref{DR=0}) as a fundamental system of equations which must be satisfied by the most generic R-matrix solution. 
In the particular case of solution (\ref{Rchosen}), the two equations reduce just to one condition. Braiding unitarity of the $R$-matrix will then be considered as a constraint equation.

A consequence of the fact that (\ref{Rchosen}) is symmetric is also that $\Gamma_1$ and $\Gamma_2$ are proportional to the same matrix in $\mathfrak{su}(1|1)_L \oplus \mathfrak{su}(1|1)_L$. However we could interpret the connection (\ref{connection}) as a particular solution to the equations of motion plus the Bianchi identity. The most general solution to such system of equations would parametrise the proposed Universal R-matrix, as described in the next section.

\subsubsection{A comment on the relativistic limit}
We shall mention that if we take the relativistic limit of the $q$-deformed Poincar\'e superalgebra so far considered, then a choice of the boost coproduct, which is consistent with the algebra commutation relations, is the trivial one, i.e. 
\begin{equation}
\Delta(\mathcal{J}^{rel}_L) = \Delta^{op}(\mathcal{J}^{rel}_L) = \mathcal{J}^{rel}_L \otimes \mathds{1} +  \mathds{1} \otimes \mathcal{J}^{rel}_L \ ,
\end{equation}
where $\mathcal{J}^{rel}_L$ is the boost generator in the relativistic limit, which acts as follows
\begin{equation}
\mathcal{J}^{rel}_L = \frac{\partial}{\partial \theta} \ , 
\end{equation}
and $\theta$ is the rapidity of the particle where $\mathcal{J}^{rel}_L$ is acting on. Then equation (\ref{boost}) in the relativistic limit tells us that 
\begin{equation}
\label{singleo}
\Delta(\mathcal{J}^{rel}_L) R = \bigg[\frac{\partial}{\partial \theta_1} + \frac{\partial}{\partial \theta_2}\bigg] R = 0 \ ,
\end{equation}
which implies that the R-matrix depends only on the difference of the rapidities, i.e. $R = R(\theta_1 - \theta_2)$. This is consistent with the boost invariance of the R-matrix in relativistic integrable systems discussed in \cite{Charles}.

\section{Proposal for the Universal R-matrix}
\label{proposal}
In this section we derive an integral expression for the R-matrix (\ref{Rchosen}), which we propose as a candidate for the Universal R-matrix. 
Let us consider a curve $\gamma(\lambda)$ in the base space 
\begin{equation}
\gamma : [0, 1] \rightarrow \mathcal{B} \ , 
\end{equation}
and contract (\ref{DR=0}) with the velocity $\dot{p}^M= \frac{d p^M}{d \lambda}$ along $\gamma$, which gives us
\begin{equation}
\label{DR=0 contracted}
\bigg[\frac{d}{d\lambda} +  \dot{p}^M \Gamma_M \bigg] R = 0 \ . 
\end{equation}
For convenience, we apply the permutation operator $\Pi$ defined in (\ref{permutation_operator}) to (\ref{DR=0 contracted}), and we obtain
\begin{equation}
\label{DR=0 contracted2}
\bigg[\frac{d}{d\lambda} -  \dot{p}^M \Gamma_M \bigg]\Pi \circ R = 0 \ , 
\end{equation} 
where we used the following relation on states: 
\begin{equation}
\{\Pi, \Gamma_M\} = 0   \ .
\end{equation}
By integrating (\ref{DR=0 contracted2}) along the image of $\gamma$, we provide the following expression for the R-matrix
\begin{equation}
\label{int_R}
R \big[\gamma (\lambda)\big] = \Pi \circ \mathscr{P} \exp  \bigg(\int_{\gamma(0)}^{\gamma(\lambda)} dp^M \Gamma_M \bigg) \  , 
\end{equation}
where $\mathscr{P} \exp$ is the time-ordered exponential and we have used the fact that $\Pi^2=\mathds{1}\otimes \mathds{1}$. 

The normalisation in (\ref{int_R}) is chosen in the following way. We impose that the starting point of the curve $\gamma(0) = (\hat{p}_1, \hat{p}_2)$ is such that $\hat{p}_1 = \hat{p}_2$. Then for $\lambda = 0 $, i.e. particles with equal momenta, the R-matrix (\ref{Rchosen}) reduces to $\Pi$.
In Appendix \ref{consistency}, by choosing $\gamma$ to be a straight line, we check that the R-matrix in the integral formulation (\ref{int_R}) reproduces correctly the given R-matrix (\ref{Rchosen}).

\begin{figure}[h]
\begin{center}
 \includegraphics[scale=0.6]{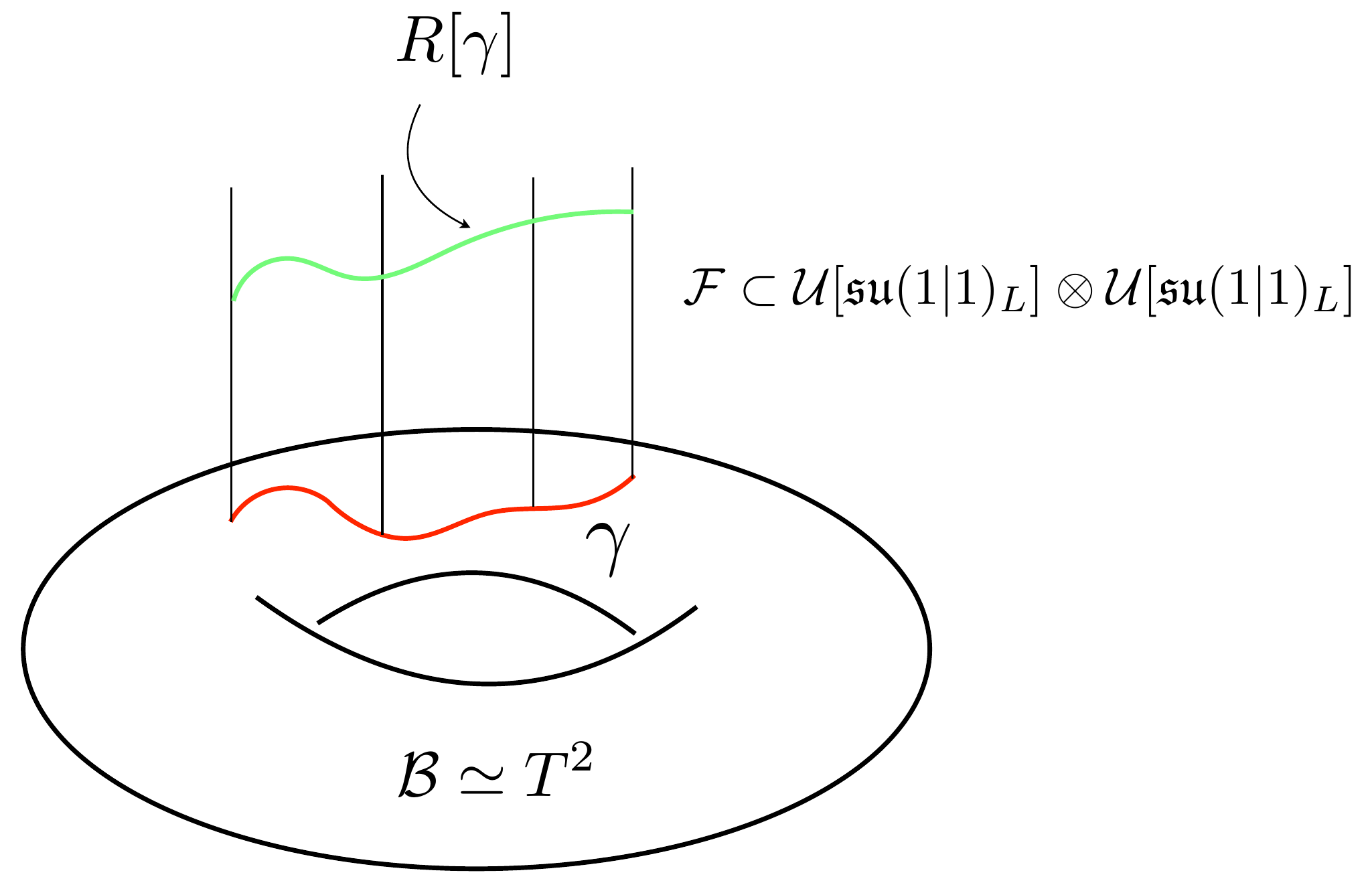}
\end{center}
\caption{Fibre bundle interpretation of the massless scattering.}
\end{figure}

Of course, what discussed above applies away from singularities and for paths entirely contained within regular regions. This is analogous to the Aharonov-Bohm type effects, and would obstruct the trivialisation of the underlying fibre bundle.

Equation (\ref{int_R}) has been derived by integrating the constraint (\ref{DR=0}), where the connection $\Gamma_M$ is the specific one given in (\ref{connection}). This leads us to the following
\begin{cnj}
Given a Hopf algebra, the Universal R-matrix is of the form (\ref{int_R}).
\end{cnj}

The idea is to deform the connection in (\ref{int_R}) and to still obtain a physical R-matrix, solution to the algebraic equations. This implies that the physical connections must also satisfy certain equations, which we expect to impose conditions on the singularities. The algebraic problem of determining the R-matrix translates into studying the moduli spaces of physical connections, where generalised Completeness Theorems of the R-matrix solutions could be formulated.

\subsubsection{Link to gauge theories}

In the case in which the fibre bundle so far constructed could be reduced to a principle bundle, the path-ordered expression (\ref{int_R}) in a gauge theory language would be the so-called {\it Wilson line} of the (almost everywhere flat) gauge connection $\Gamma$. 
Moreover if we choose $\gamma$ to be a closed curve and we trace (\ref{int_R}) over the superspin states, we obtain the \textit{Wilson loop} associated with $\Gamma$.
This is a consequence of the fact that the  R-matrix in the integral form (\ref{int_R}) contains the information about the \emph{holonomy} of $\Gamma$.

It is interesting to speculate whether there might exist a gauge theory rewriting our problem, where the gauge field $\Gamma_M$ lives on $\mathcal{B} \simeq T^2$, with gauge algebra given by two copies of $\mathfrak{su}(1|1)$. It would then be curious to investigate what a {\it gauge transformation} might correspond to in the original physical picture. This should turn into a {\it local} tranformation (i.e. momentum-dependent) of the basis of two-particle scattering states\footnote{This is a local redefinition of the Faddeev-Zamolodchikov operators \cite{rev2}}. Because of its local nature, it should tie in with the $\mathfrak{sl}(2)$ outer-automorphism of $\mathfrak{su}(1|1)_L \oplus \mathfrak{su}(1|1)_R$.
The link between integrable scatterings and gauge theories has also been noticed in \cite{Costello:2017dso, Costello:2018gyb}.

\section{Fundamental R-matrix equations revisited}
So far we have provided a geometric interpretation of the boost invariance in terms of a fibre bundle picture, where the boost generator defines a connection, and the R-matrix is a covariantly constant section of the bundle. In this section we shall comment on a geometric interpretation of the fundamental set of equations presented in section \ref{sec:fundamental_eq_R}, which every R-matrix describing an integrable scattering should satisfy.

\subsubsection{Braiding unitarity}
The braiding unitarity condition
\begin{eqnarray}
\label{braiding2}
\Pi(R)(p_2,p_1) \, R(p_1,p_2) = \mathds{1} \otimes \mathds{1} \ , 
\end{eqnarray}
could be potentially interpreted as a \emph{path inversion} condition on the base space $\mathcal{B}$. This is because there is a potential cancellation between the exponentials coming from $R$ and $\Pi (R)$ in (\ref{braiding2}) when the R-matrix is expressed in the integral form (\ref{int_R}), which might be due to inversion of the path of integration. Further investigation is needed thought. 

\subsubsection{Yang-Baxter equation}
We recall that defining a connection on a fibre bundle means to split uniquely the tangent space of the fibre bundle into the vertical and horizontal subspaces. 
Furthermore, the holonomy of a connection is trivial if given a \emph{closed} curve $\gamma$ on the base space, the horizontal lift $\tilde{\gamma}$ is also a \emph{closed} curve. 
Recall that the horizontal lift $\tilde{\gamma}$ of a given curve $\gamma$ in the base space is a curve in the whole fibre bundle such that if projected down to the base space one recovers $\gamma$, and in addition the tangent vectors to $\tilde{\gamma}$ always belong to the horizontal space. 
 
Let us consider the embedding of the base space $\mathcal{B}\simeq T^2$ into $\mathbb{R}^3$ equipped with coordinates $(p_1,p_2,p_3)$, where $p_3$ stands for the momentum of the third auxiliary particle. Let us choose three different points $X_1, X_2, X_3\in T^2$ and let us denote by $\gamma_{ij}$ a generic path from $X_i$ to $X_j$, which do not wrap around any singularity.
Let us consider the closed curve $\gamma \equiv \gamma_{12}\circ \gamma_{23} \circ \gamma_{31}$, which we shall assume to not contain any singularity, and let $\tilde{\gamma}$ be the horizontal lift of $\gamma$. 

If the holonomy of $\Gamma$ would be trivial, then the initial and final points of $\tilde{\gamma}$ must coincide. 
Naively, let us take the R-matrix as the fibre completion of $\gamma$ in the horizontal lift $\tilde{\gamma}$.
Then the closure of $\tilde{\gamma}$ reads 
\begin{equation}
R_{12}R_{23} = R_{13} \ .
\end{equation}
However this is \emph{not} the Yang-Baxter equation (\ref{YBE}), which instead points towards the non-triviality of the holonomy of $\Gamma$. 

The fact that the connection is flat allows us to provide extra information on the holonomy of $\Gamma$. 
A flat connection defines a homomorphism between the fundamental group $\pi_1$ of $\mathcal{B}$ and the holonomy group of $\Gamma$. Since the fundamental group of $\mathcal{B}\simeq T^2$ is $\mathbb{Z}\times \mathbb{Z}$, the holonomy of $\Gamma$ depends only on a pair of integers $(n_1, n_2)$, which are the winding numbers of the loop $\gamma$ on $T^2$, and do not depend on the particular shape of the loop $\gamma$. This analysis of course do not take into consideration any contribution coming from the singularities. 
It would be interesting to determine how this relates to the singularities of the connection appearing in (\ref{flatness}).

\subsubsection{Crossing symmetry}

A geometric interpretation of the crossing symmetry equation (\ref{crossing}) requires a suitable continuation to complex momenta, and therefore a  complexification of the fibre bundle, where naively the base space $\mathcal{B} \simeq T^2$ is promoted to $\bar{\mathcal{B}} \simeq \mathbb{C}^2$  . The path in the complexified base space $\bar{\gamma}$ entering in (\ref{int_R}) integrates the complexified connection $\bar{\Gamma}$ between {\it crossed} regions \cite{Borsato:2016xns} and again it should avoid the singularities of $\bar{\Gamma}$ on $\bar{\mathcal{B}}$.

\section{Quantum interpretation}
\label{sec:level22}
We shall show that equation (\ref{DR=0}) can also be interpreted as an auxiliary Schr\"odinger problem. In particular, the R-matrix can be identified with the time evolution operator of a one-particle state, which must satisfy the following Schr\"odinger equation:
\begin{equation}
\label{purp}
i \hbar \frac{d}{dt} U(t; t_0 ) = H U(t; t_0 ) \ .
\end{equation}
To show this, let us consider a curve on the momenta space $(p_1, p_2)$, which is topologically $T^2$, and let us parametrise the curve with the real parameter $t \in [t_0, +\infty]$, such that $p^M(t_0) = (p_0,p_0)$. 
Contracting (\ref{DR=0}) with the velocity $\dot{p}^M = dp^M / dt$, and applying the permutation operator $\Pi$, we obtain 
\begin{eqnarray}
\label{purp2}
\frac{d}{d t} \Pi \circ R = \dot{p}^M \Gamma_M \, \Pi \circ R \ . 
\end{eqnarray}
Comparing (\ref{purp2}) with (\ref{purp}), we have the following identification
\begin{equation}
\label{H_aux}
H \equiv i\hbar\, \dot{p}^M \Gamma_M \ , \qquad\qquad
U(t; t_0 ) \equiv \Pi \circ R \ . 
\end{equation}
The R-matrix is now identified with the propagator $U(t_0; t)$:
\begin{equation}
R = \Pi \circ \mathscr{P} \exp \bigg( - \frac{i}{\hbar} \int_{t_0}^t H \, dt\bigg) \ , 
\end{equation}
where $\mathscr{P} \exp$ is the time-ordered exponential. Any trajectory provides an alternative Schr\"odinger problem. 
The flatness of $\Gamma_M$, and the fact (proved by straightforward computation) that $\dot{p}^M \Gamma_M$ is identically zero along the line $p^1=p^2$, make all these quantum mechanical problems equivalent\footnote{This is true provided that the path chosen does not wrap around any singularity or any cycle of the torus.}.

It is not artificial to think of $\dot{p}^M \Gamma_M$ as a Hamiltonian (although possibly singular). In the next section, we show that the Hamiltonian-density emerging from a simplified algebraic Bethe ansatz and $\dot{p}^M \Gamma_M$ are proportional to each other. In (\ref{purp}), however, $H$ does not act on spin-chain sites, but on the superspin degrees of freedom of an auxiliary quantum-mechanical particle. 

Moreover, we notice that the auxiliary Hamiltonian (\ref{H_aux}) is  proportional to the 
symmetrised tensor-product of the two supercharges proportional to $\mathrm{E}_{12}$ and $\mathrm{E}_{21}$ entering in (7.28).
This is in strong analogy with ${\cal{N}}=1$ supersymmetric quantum mechanics, where the energy is fixed by the $\mathfrak{su}(1|1)$ supersymmetry algebra via the anti-commutator $\{Q,Q^\dagger\}$.

Possible alternative interpretations of equation (\ref{DR=0}) are given in Appendix \ref{alternative}, and in particular the Berry phase argument might link together the geometric and quantum interpretations. 

\subsection{Spin-chain Hamiltonian}
\label{sec:level3}

In this section, we shall formulate a simplified coordinate Bethe ansatz for the purpose of motivating the Hamiltonian approach presented in \ref{sec:level22}. Following \cite{Kulish}, we shall show that by using the R-matrix (\ref{Rchosen}) we can construct a gapless spin-chain Hamiltonian $\mathscr{H} $. To begin, it is useful to write the momenta $p_1, p_2$ as follows
\begin{eqnarray}
p_1 = i \nu + \lambda \ , \qquad p_2 = i \nu - \lambda \ ,
\end{eqnarray}
where $\nu, \lambda \in \mathbb{C}$. 
Next, we notice that
\begin{eqnarray}
R(\nu, \lambda=0) = \Pi \ .
\end{eqnarray}
We consider a spin-chain with $N$ sites, with periodicity $N+1 \equiv 1$. We set the momenta of the $n$-th site to be
\begin{equation}
p_n = i \nu_n + \lambda_n \ , \qquad p_{n+1} = i \nu_n - \lambda_n \ ,
\end{equation}
and we construct the local spin-chain Hamiltonian $h_{n, n+1}$, which acts on neighbour sites $n$ and $n+1$ as the logarithmic derivative of the R-matrix (\ref{Rchosen}), i.e.
\begin{equation}
h_{n,n+1} = \frac{\partial}{\partial \lambda_n} \log R(\nu, \lambda_n)\big|_{ \lambda_n=0} = \, \Pi \circ \frac{\partial}{\partial \lambda_n} R(\nu, 0) \ . 
\end{equation}
where we made the simplification $\nu_n = \nu \, ,\  \forall n$. 
The parameter $\nu_n$ plays the physical role
of an inhomogeneity along the spin-chain, which however we take to be
site-independent, by setting $\nu_n=\nu$ for simplicity.
This gives us 
\begin{eqnarray}
\label{ham_density}
h_{n,n+1} = \frac{i}{2 \sinh \frac{\nu}{2}} \, (\mathsf{E}_{12} \otimes \mathsf{E}_{21} + \mathsf{E}_{21} \otimes \mathsf{E}_{12}) \ ,
\end{eqnarray}
which is a Hermitian operator for real $\nu$ (taking into account fermionic signs).
The Hamiltonian density (\ref{ham_density}) coincides, up to an overall factor, with the algebraic part of the boost coproduct, or equivalently with the components of the connection $\Gamma_M$.

We construct the monodromy matrix
\begin{equation}
\mathcal{T}_0 (\lambda) \equiv R_{01} (\lambda) \cdots R_{0N} (\lambda) \ ,
\end{equation}
where $R_{0i}$ is the scattering matrix between the auxiliary particle and $i$-th particle. Here $R_{0i}$ plays also the role of Lax matrix. Then the Hamiltonian becomes
\begin{equation}
{\mathscr{H} } = \frac{d}{d\lambda} \log \text{tr}_0 \mathcal{T}_0 (0) = -\sum_{n=1}^N h_{n,n+1} \ .
\end{equation}
Consider now an infinite chain, with pseudo-vacuum \begin{eqnarray}
|\Omega\rangle = |... \phi \otimes \phi \otimes \phi ... \rangle \ ,
\end{eqnarray}
which is an eigenstate of ${\mathscr{H} }$ with zero energy. The spectrum of one-particle excitations above $|\Omega\rangle$ is {\it gapless}:
\begin{eqnarray}
\label{dis}
&&|\Psi_p\rangle = \sum_m e^{i p m} |... \phi_{m-1} \otimes \psi_m \otimes \phi_{m+1} ... \rangle \ ,\nonumber\\
&& {\mathscr{H} }|\Omega \rangle = 0 \ , \nonumber \\
&&{\mathscr{H} }|\Psi_p\rangle = \epsilon \sin p \, |\Psi_p\rangle, \qquad \epsilon = \frac{1}{\sinh \frac{\nu}{2}} \ .
\end{eqnarray}
There are intriguing similarities between the dispersion relation (\ref{dis}) and the one of massless {\it spinons}, which are massless excitations of the {\it antiferromagnetic} Heisenberg spin-chain \cite{Faddeev:1981ip}, with dispersion relation
\begin{equation}
\label{spinon}
E_{sp} = \frac{\pi}{2} \, |\sin p | \ .
\end{equation}
There is a similarity between the massless spinons dispersion relation (\ref{spinon}), and the massless (ferromagnetic) magnons dispersion relation (\ref{massless_disp_AdS3}). 
One may wonder of any possible spectral duality between the two. In analogy to the $AdS_5\times S^5$ case, where antiferromagnets appear in specific regimes with some peculiar parameters scaling \cite{H1, H11 ,H2, H21, H22}, it would be interesting to study the antiferromagnetic limit of the $AdS_3$ massless sector and test the spectral duality, particularly in view of \cite{PerLinus}.

\chapter{{\bf Conclusions}}

In the first part of this thesis, we focused on various aspects of black holes in higher dimensions in the context of string theory. 
In chapter \ref{chap:intro_BH} we introduced the notion of a near-horizon geometry with some illustrative examples (supersymmetric black ring and BMPV black hole). 
We showed that a near-horizon geometry always admits two obvious isometries, which are generated by the Killing vectors $V$ (\ref{iso_V}) and $D$ (\ref{iso_D}). The question is whether the near-horizon geometry admits even further symmetries. 
In the context of supersymmetric near-horizon geometries with non-trivial fluxes, there are evidences that the answer to this question is yes. This is contemplated in the horizon conjecture, which we enunciated in \ref{hor_conj}, and we showed the salient key points of the proof in the context of type IIA supergravity. 

In chapter \ref{ch:het_NHG}, we investigated the horizon conjecture when one includes the string corrections to the supergravity approximation in the context of heterotic theory. 
We decomposed all heterotic bosonic fields in Gaussian null coordinates and applied the near-horizon limit. We assumed supersymmetry, and we integrated the gravitino KSE along the light-cone directions. We decomposed all KSEs in terms of the bosonic near-horizon data and the spinors $\eta_{\pm}$, which are defined on the spatial cross section $\cS$. In this way we obtained the so-called reduced KSEs. We showed that the reduced KSEs impose some conditions on the bosonic near-horizon fields.

Then we investigated whether the near-horizon geometry admits supersymmetry enhancement. First, we showed that the reduced KSEs can be simplified to a set of necessary and sufficient conditions to establish supersymmetry, which consist of a gravitino and dilatino equations on $\cS$.  Then we reviewed the argument to establish the supersymmetry enhancement in uncorrected heterotic supergravity. The mechanism consists in assuming the existence of at least one Killing spinor $\eta_{\pm}$ from which one can generate a second linearly independent Killing spinor of opposite chirality $\eta'_{\mp}$ by using the $\slashed h$ map. This argument relies on various properties of $h$ which are derived from some global analysis of the Laplacian acting on $h^2$. In particular the fact that $h$ is parallel with respect to $\tilde{\nabla}^{(-)}$ is crucial. Remarkably, in the uncorrected heterotic case the generalised Lichenrowicz theorem is not needed to establish the supersymmetry enhancement. 
Then we investigated if the $\slashed h$ map can also be constructed when $\alpha'$ corrections are included. We showed that the global analysis of $h^2$ is insufficient to imply the useful properties of $h$ up to $\mathcal{O}(\alpha'^2)$, and therefore the $\slashed h$ map does not guarantee supersymmetry enhancement.  For instance, we obtained  that $h$ is parallel with respect to $\tilde{\nabla}^{(-)}$ at zeroth order in $\alpha'$, but not up to $\mathcal{O}(\alpha'^2)$. 

Therefore we investigated if a generalised Lichnerowicz theorem, including $\alpha'$ corrections, can be formulated. We showed that the zero modes, up to $\mathcal{O}(\alpha'^2)$ terms, of the twisted Dirac operator are solutions to the KSEs at zeroth order in $\alpha'$. The Lichnerowicz theorem in this case is insufficient to imply the KSEs up to $\mathcal{O}(\alpha'^2)$ terms. 
However we showed that if the near-horizon geometry admits at least one non-vanishing at zeroth order in $\alpha'$ negative light-cone chirality spinor, $\eta^{[0]}_- \neq 0$, then supersymmetry is enhanced. Such condition on the spinor implies that $h$ is parallel with respect to $\tilde{\nabla}^{(-)}$ up to $\mathcal{O}(\alpha'^2)$ terms, and it leaves all near-horizon heterotic fields invariant. This is sufficient to guarantee supersymmetry enhancement via the $\slashed h$ map. 

We proved the no-go theorem \ref{th:no_AdS2}, which states that there are no $AdS_2$ backgrounds in heterotic supergravity, for which all fields are smooth and the internal space is smooth, compact and without boundary. The result holds up to second order corrections in $\alpha'$, provided that there exists at least one $\eta_-^{[0]} \neq 0$. 
To prove this theorem, one needs the facts that $\Delta$ vanishes, and $h$ is an isometry.

In the case that supersymmetry is enhanced, we described the spacetime geometry of the near-horizon solutions. 
We recall that when supersymmetry is enhanced the description of the geometry is analogous to the uncorrected case, except of having to solve an anomaly corrected Bianchi identity for $H$. 
Therefore non-trivial heterotic near-horizon geometries, i.e. solutions with non vanishing fluxes, admit 2, 4, 6 and 8 preserved supersymmetries. The spacetime geometry for near-horizon geometries preserving 2 and 4 supersymmetries was described in section \ref{sec:geometry}. 

Finally, we considered non-supersymmetric near-horizon geometries for which supersymmetry is explicitly broken in a certain way. 
These type of horizons admit a spinor which is a solution to the gravitino KSE, but which do not satisfy the algebraic KSE. 
Therefore the algebraic KSE generates a second spinor, which we showed to also satisfy the gravitino KSE. Furthermore we showed that from these two $\tilde{\nabla}^{(-)}$-parallel spinors one can generate a vector which is a symmetry of the full solution. This fact allowed us to describe the geometry of the spatial cross section $\cS$. In this thesis we reported only the $G_2$ case, i.e. when there is exactly one spinor satisfying the initial assumptions on the KSEs. The cases where there are more of such spinors can be found in \cite{Fontanella:2016aok}.  

In chapter \ref{chap:bulk}, we considered the problem of extending a given near-horizon geometry into the bulk along the radial direction. This problem can be embedded into the more general question of determining all black hole solutions which share the same near-horizon geometry. 
We showed that the first order radial deformations of the horizon fields must satisfy an elliptic system of PDEs, which implies that the space of moduli is finite dimensional (theorem \ref{th:finiteness_moduli}). 
We showed that this result also holds when matter fields are coupled to the metric field. In particular, we considered the uncorrected heterotic supergravity and $D=11$ supergravity theories. In \cite{Fontanella:2016lzo}, the theorem \ref{th:finiteness_moduli} is also proved in Einstein-Maxwell-Dilaton theory in any dimension, with a topological term in four and five dimensions.  

We shall discuss the limitations which affect our approach to study the bulk extension of a near-horizon geometry. 
First we recall that the Gaussian null coordinates is a particular set of coordinates adapted to a neighbourhood of the Killing horizon $\cH$. This means that if we move sufficiently away from the horizon, we expect that the radial expansion breaks down. 
This happens because the metric diverges when $r$ is sufficiently large. 
Furthermore, computations at second order in $r$ becomes quite prohibitive. 
The general issue of obstructions for extending the metric of a submanifold into an ambient space has been considered by Fefferman and Graham \cite{Fefferman_Graham}. In particular the authors have studied at which order in $r$ the expansion breaks down. However a systematic classification of all possible obstructions is still missing. 

The fact that we are unable to extend to spatial infinity the near-horizon geometry written in Gaussian null coordinates implies that our approach is not sensitive to the asymptotic data (e.g. mass, charge and angular momentum) of the black hole, and neither to the asymptotic geometry of the spacetime. Therefore our question is different from the idea which is behind the uniqueness theorem of Hawking et al., which states that in $D=4$ Einstein theory the asymptotic data uniquely specify the (asymptotically flat) black hole. However if one would be able to extend up to infinity a near-horizon geometry, then we expect to recover the black hole uniqueness theorem in $D=4$.

We emphasize that a priori there is no guarantee that a given near-horizon geometry belongs to a genuine black hole, e.g. there exist near-horizon geometries with toroidal topology, which cannot belong to a black hole by the generalization of the Hawking's topology theorem. General criteria which allows to systematically exclude near-horizon geometries, like necessary and sufficient conditions for a horizon to belong to a black hole, are still missing, though a list of necessary conditions exist \cite{Booth:2005qc}.  

We comment that one of our key assumption to study near-horizon geometries is the analyticity of the horizon fields in the radial coordinate $r$. 
Such assumption is also behind the proof of the uniqueness theorem of Hawking et al., and in this case it has been attempted to relax the analyticity condition to the milder assumption that the horizon fields are smooth functions in $r$, see e.g. \cite{Alexakis:2013rya} for some progress and \cite{Ionescu:2015dna} for a review. However the problem of understanding whether the uniqueness theorem holds for a general class of smooth horizons, without making any further assumption, is still a very difficult problem.

\vspace{0.5cm}
In what follows we discuss two possible projects connected to the research developed in this thesis, which are part of my future planned investigation. 
\begin{enumerate}
\item Investigate the horizon conjecture \ref{hor_conj} for near-horizon geometries in $D=7$ (minimal) gauged supergravity. 
The novelty of this project is the presence of a non-abelian 1-form field which does not decouple from the dynamic of the rest of the fields, which happens in heterotic theory in $\alpha'$ perturbation expansion. There is very little work in the literature about near-horizon geometries with non-abelian gauging, and it would be interesting to determine if the geometric structure present in the abelian case also persists in the non-abelian case.

\item Infinitesimal deformations of near-horizon geometries with $G_2$ structure. In this project we shall consider near-horizon geometries which preserve $N = 2$ supersymmetries in heterotic supergravity, without $\alpha'$ corrections. The internal space of these geometries is a $G_2$ structure manifold. Instead of the radial deformations studied in this thesis, we shall consider generic \emph{supersymmetric deformations} of the near-horizon geometry. There are several motivations for this project. First, we are interested in finding global conditions which restrict the number of possible moduli, and the heterotic theory represents a manageable theory where to do such investigation. 
Ideally, we aim to construct scalar functions which depends on the moduli. By considering the action of the Laplacian on such functions and by applying the Hopf maximum principle, we expect to obtain some conditions on the moduli. 
Moreover, supersymmetric deformations will preserve the $G_2$ structure. This will be instrumental to find such global conditions on the moduli, which are important to further restrict the moduli space.  

As a second motivation, the presence of non-trivial moduli might lead to a mechanism for supersymmetry enhancement. This is also supported by the fact that there are no explicit heterotic near-horizon geometries which preserve $N = 2$ supersymmetries, and the known examples preserve at least $N = 4$. We believe that finding the supersymmetry enhancement for deformed $G_2$ near-horizon geometries would answer this question.

A third motivation is related to the work of X. de la Ossa et al. \cite{delaOssa:2017pqy} who considered infinitesimal deformations of heterotic geometries of the type $AdS_3 \times Y^7$, where $Y^7$ is a $G_2$ structure manifold. They deformed only the internal space $Y^7$, and kept the direct product $AdS_3 \times Y^7$ while deforming. They have shown that the deformations of the fields, must satisfy elliptic PDEs, which implies that the moduli space is finite dimensional. Furthermore, they proved that the elliptic system of PDEs implies the existence of a cohomology operator acting on the space of moduli. Recently this result opened new interesting connections with worldsheet BRST operators \cite{Fiset:2017auc}.
In this project we shall address the same issue considered by X. de la Ossa et al. also in the context of heterotic near-horizon geometries, which turns out to be more general, and it includes the $AdS_3 \times Y^7$ case as a particular limit. In our approach, we shall deform the whole spacetime near-horizon geometry, and not just the internal space $\cS$. Second, we shall not assume any factorization of the spacetime of the type $\mathcal{M}_3 \times Y^7$, with $\mathcal{M}_3$ a Lorentzian non-compact manifold, but we shall begin just with a generic near-horizon geometry which preserves $N = 2$ supersymmetries.
\end{enumerate}

\vspace{2cm}

In the second part of this thesis we have considered various aspects of integrable theories which arise in string theory. 
In chapter \ref{ch:Intro_Integra}, we introduced concepts, formalisms and techniques which are instrumental for chapters \ref{Chapter: AdS2} and \ref{chapter: AdS3}.  First, we presented our philosophy to study the gauge/gravity duality in wider contexts than the celebrated $AdS_5$/CFT$_4$ Maldacena's duality. This led us to study the duality on integrable supercoset spaces which are also string backgrounds. We provided the notion of classical integrability in $1+1$ dimensional field theories in terms of the existence of a Lax pair, and we showed some known examples of integrable theories (KdV and Sine-Gordon equations). We sketched the proof of the fact that a generic supercoset, in order to be integrable, requires the existence of a $\mathbb{Z}_4$ outer automorphism. 

Then we moved onto the topic of the scattering matrix. First, we introduced the notion of the single particle representation and the related concept of isometry algebra breaking due to the introduction of a vacuum state.  
We explained why for integrable scatterings one needs only to consider $2 \rightarrow 2$ processes, and we showed the set of fundamental equations which a scattering matrix must satisfy: braiding unitarity, crossing symmetry, Yang-Baxter and the algebra invariance.
Furthermore, we briefly discussed how the crossing symmetry equation becomes an equation for the dressing factor only.  

Then we came back again on the fundamental equations for the integrable scattering, but this time presenting them in an appropriate mathematical language: the Hopf algebras. 
Finally, we concluded the chapter with a discussion on the Bethe ansatz, and the standard procedure which one follows for its formulation, based on the existence of the pseudo-vacuum state.

In chapter \ref{Chapter: AdS2}, we investigated aspects of scatterings in the $AdS_2 \times S^2 \times T^6$ string background. We introduced the algebra relevant for this background, and the various representations (massive, massless, relativistic massless). Then we posed the problem of determining the R-matrix in the relativistic massless regime. In our approach we imposed only the algebra invariance condition and the Yang-Baxter equation, which were sufficient to completely fix the R-matrix entries, up to the dressing factor. Crossing symmetry and braiding unitarity are conditions which we checked a posteriori. Some of the solutions found do not satisfy both conditions. This is explained in terms of collinear scatterings, for which in the massless limit the associated R-matrix is a non-perturbative object, which describes CFTs at the fixed points of the RG flow rather than a physical scattering. This is in perfect agreement with Zamolodchikov's conjecture.

We showed that some of the solutions found can be reproduced by taking the relativistic limit of the parental non-relativistic massless R-matrices. Furthermore, such solutions can also be asymptotically reproduced by taking the large $\theta$ limit of the Fendley's solutions. This observation allowed us to provide an example of braiding unitarity breaking, in consonance with the Zamolodchikov's conjecture. 
The dressing factors for Solutions 2 and 3 were also found. 

In connection with chapter \ref{chapter: AdS3}, we showed that Solutions 2, 3, 4, 5 satisfy first order differential equations in the variable $\theta$, with a purely algebraic term $\Gamma^{(i)}_{\theta}$. Such algebraic term might be interpreted as a connection on a fibre bundle where the base space has been contracted to $S^1$ due to the relativistic limit. It is still unclear if this geometric structure exists for non-relativistic massless R-matrices in $AdS_2$ background. 

Finally, we worked towards a Bethe ansatz for the R-matrices Solutions 2 and 3, and for the non-relativistic massless solution. The technique used is based on the free-fermion condition, which is an algebraic condition for the S-matrix entries. The technique consists in considering a certain S-matrix and to define a second one, which is related to the first one via a reshuffle of the entries. 
One can show that the product of the two transfer matrices, which are constructed in terms of the two S-matrices just introduced, can be factorised. 
In the case of Solution 2, we found that $\mathcal{T} \mathcal{T}^{(1)}$ factorises into two block terms. This identifies the potential zeroes of $\mathcal{T}$, up to an ambiguity, and therefore the auxiliary Bethe equations. 
In the cases of Solution 3 and the non-relativistic S-matrix, the zeroes of $\mathcal{T} \mathcal{T}^{(1)}$ are identified with no ambiguity. In the non-relativistic case, the auxiliary Bethe equations found match the one conjectured by reading off the Dynkin diagram of the algebra in \cite{Sorokin:2011rr}. The fact that the eigenvalues of $\mathcal{T}$ go to zero or have an essential singularity at $\theta = \infty$, which is not what happens in \cite{Ahn:1993qa, MC}, implies that we are not able to reconstruct them with the knowledge of their zeroes and poles. However in \cite{Torrielli:2017nab} the transfer matrix has been diagonalized by brute force and a conjectured formula for the eigenvalues for all number of sites has been given.  

It is interesting to observe that the free-fermion condition is also satisfied by the S-matrices in $AdS_3$ backgrounds, in which case a pseudo-vacuum state can be defined. Therefore the technique based on the free-fermion condition seems to have a larger field of applicability that the standard one which relies on defining a reference state. Possible future research would be to study the Bethe ansatz for the \emph{massive} S-matrix via the free-fermion condition technique, and the thermodynamic Bethe ansatz, i.e. the large number of sites limit.

In Chapter \ref{chapter: AdS3}, we focussed on massless string excitations in the $AdS_3 \times S^3 \times T^4$ background. After introducing the massive and massless representations of the $\mathfrak{su}(1|1)_L \oplus \mathfrak{su}(1|1)_R$ superalgebra, we considered a specific R-matrix solution and asked whether this solution is invariant under a larger algebra. In particular, we considered the $q$-deformed super Poincar\'e algebra $\mathfrak{E}_q(1,1)_{L} \oplus \mathfrak{E}_q(1,1)_{R}$, which introduces the \emph{boost} generators $\mathcal{J}_L$ and $\mathcal{J}_R$ in addition to the generators of the $\mathfrak{su}(1|1)_L \oplus \mathfrak{su}(1|1)_R$ algebra. We showed that the boost generator annihilates the R-matrix instead of commuting with it. The boost action was reinterpreted as a parallel condition for the R-matrix with respect to a connection living on a certain fibre bundle. The base space of the fibre bundle is a 2-dimensional torus, which needs to be complexified to $\mathbb{C}^2$ in order to consider the analytic properties of the R-matrix, and the fibre is a subset of $\mathcal{U}[\mathfrak{su}(1|1)]_L \otimes \mathcal{U}[\mathfrak{su}(1|1)]_R$. 

We showed that restoring the R-matrix dressing factor can be interpreted as a pure gauge shift of the connection. By integrating the parallel condition, we provided an integral expression along a curve of the base space for the R-matrix, which we proposed as a candidate Universal R-matrix. Since the connection is characterized by infinitely many singularities, the curve considered is chosen in such a way it does not make any loop around any of them, which we expect to be a source of Aharonov-Bohm type effects. 

We revisited the fundamental R-matrix equations in this geometric language. Braiding unitarity may be interpreted as a path inversion in the base space, while the Yang-Baxter equation is a holonomy condition. Crossing symmetry has still to be explored, since it requires a complexification of the fibre bundle.    

We provided a second interpretation of the boost pseudo-invariance of the R-matrix as an auxiliary Sch\"odinger problem, where the algebraic part of the boost coproduct is interpreted as a Hamiltonian and the R-matrix as the time evolution operator of a one-particle state.
To support the quantum approach, we constructed a simplified coordinate Bethe ansatz by using the particular R-matrix chosen, and we constructed a gapless spin-chain Hamiltonian which is proportional to the algebraic part of the boost coproduct, whose spectrum is in a close analogy with antiferromagnetic excitation of the Heisenberg spin-chain.

The geometric and quantum interpretations could be related by a Berry phase argument  described in appendix \ref{alternative}. In \cite{Joakim}, it has been found an analogy between massless excitations in $AdS_3\times S^3 \times T^4$ and phonons, which identify vibrations of ions, in terms of sharing the same dispersion relation. 
In the regime where the underlying crystal is slowly vibrating, there might be an adiabatic variable for the scattering, which suggests for a Berry phase argument.

In connection to what discussed in this thesis, we propose three possible future projects in the context of integrability in lower dimensional AdS/CFT, and in particular in connection with the new emergent geometric interpretation of the scattering process. 

\begin{enumerate}
\item  Study the boost invariance of the R-matrix, and the associated fibre bundle, in $AdS_2\times S^2 \times T^6$ superstring. This project aims to investigate the boost invariance previously found in $AdS_3 \times S^3 \times T^4$ also in the $AdS_2$ background.  
In particular, we will investigate in the context of $AdS_2\times S^2 \times T^6$ background if the boost invariance of the R-matrix again provides a notion of a connection on a fibre bundle. This would allow us to find further evidence of the interesting geometric picture of the scattering process. 

As a hint, we have already shown in section \ref{connections} that the R-matrix must satisfy a differential equation with a non-trivial algebraic part, which potentially can be reproduced by a boost invariance condition.

\item Moduli spaces of connections and Universal R-matrix.
In the geometric picture described above, the R-matrix gains the meaning of holonomy of the connection, due to the fact that it can be written in terms of a path ordered exponential of the connection.
We conjecture that the R-matrix expressed in such a way in terms of the representation-free connection is a candidate Universal S-matrix. This implies that every physical R-matrix is specified by the holonomy of the connection, and different connections corresponds to different R-matrices. 
Instead of the R-matrix, in our approach we treat the connection as the fundamental object. 

In this project we will determine the system of equations which a \emph{physical connection} must satisfy. A physical connection is a connection which reproduces a physical R-matrix via exponentiation. 
Once the system of equations are identified, we will study infinitesimal deformations of a connection solution to this system of equations. We will investigate whether the moduli must satisfy elliptic PDEs, which would imply that the moduli space is finite dimensional.

This project is important for understanding the moduli space of the physical connection, which potentially encodes information about all possible R-matrix solutions. The aim is to work towards a formulation of a generalised Completeness Theorem for a R-matrix solutions classification.

\item Fibre bundle structure group and connection singularities. In this project we will investigate aspects of the structure of the fibre bundle. The structure group, which determines the transition functions on non-empty patch overlaps, is still undetermined. This also relates to the notion of gauge transformations, and the physical implementation on fields, which is of particular interest. 

Furthermore the fibre bundle should be complexified, because the
coordinates on the base space, which are the particle momenta, are analytically continued to complex values. 
After complexification, the fibre bundle base space would be a 4D manifold. This is a hint that there might be a connection between our geometric picture of the scattering and the duality found by Costello, Witten and Yamazaki \cite{Costello:2017dso, Costello:2018gyb} between integrable 2D scatterings and 4D gauge theories. 
There are of course significant differences, in that we have for instance a complexification of ${\mathbb{R}}^{1,1}$ instead of ${\mathbb{R}}^{1,3}$ as in \cite{Costello:2017dso, Costello:2018gyb}.

The connection has infinitely many singularities: it would be interesting to  investigate if these singularities are related to notions of bound states, resonances and/or Aharonov-Bohm type effects. 

\end{enumerate}

\clearpage{\pagestyle{empty}\cleardoublepage} 

\appendices

\chapter{Conventions and useful formulae}
\label{apx:A}
\section{Conventions}
Metric signature: $ \qquad (- + .... +)$

\noindent $p$-forms
\be
\omega = \frac{1}{p!} \omega_{\mu_1 ... \mu_p} \mathbf{e}^{\mu_1} \w \cdots \w  \mathbf{e}^{\mu_p} \ .
\ee
Antisymmetric products of gamma matricies
\be
\Gamma^{\mu_1 ... \mu_p} = \Gamma^{[\mu_1} \cdots \Gamma^{\mu_p]} \ , \qquad \qquad \Gamma^{[\mu}\Gamma^{\nu]} = \frac{1}{2} (\G^\mu \G^\nu - \G^\nu\G^\mu) \ .
\ee
Feynman slash notation of $p$-forms
\be
\slashed \omega = \G^{\mu_1 ... \mu_p} \omega_{\mu_1 ... \mu_p} \ . 
\ee
Curvature of a connection $\Gamma$
\begin{eqnarray}
R_{\mu\nu,}{}^{\rho}{}_{\sigma} = \partial_{\mu} \Gamma_{\nu}{}^{\rho}{}_{\sigma}-\partial_{\nu} \Gamma_{\mu}{}^{\rho}{}_{\sigma} + \Gamma_{\mu}{}^{\rho}{}_{\la} \Gamma_{\nu}{}^{\la}{}_{\sigma}-\Gamma_{\nu}{}^{\rho}{}_{\la}
\Gamma_{\mu}{}^{\la}{}_{\sigma}~,
\end{eqnarray}
Connections with torsion $\nabla^{(\pm)}$ of the spacetime
\begin{equation}
\label{connection_torsion}
\nabla^{(\pm)}_{\mu} Y^{\nu} = \nabla_{\mu} Y^{\nu} \pm \frac{1}{2} H_{\mu}{}^{\nu}{}_{\rho} Y^{\rho} \ .  
\end{equation}
Connections with torsion $\nabla^{(\pm)}$ of $\cS$
\begin{equation}
\tilde{\nabla}^{(\pm)}_{i} \xi^{j} = \tilde{\nabla}_{i} \xi^{j} \pm \frac{1}{2} W_{i}{}^{j}{}_{k} \xi^{k} \ .  
\end{equation}

\subsection{The $R^{(\pm)}$ curvature tensors}

The $R^{(+)}$ curvature tensor, which is the curvature of the connection $\nabla^{(+)}$, can also be written as
\begin{align}
R^{(+)}{}_{\mu\nu,\rho\sigma} = { R}_{\mu\nu\rho\sigma}
-{1 \over 2} \nabla_{\mu} H_{\rho\nu\sigma}+{1 \over 2} \nabla_{\nu} H_{\rho\mu\sigma}
+{1 \over 4} H_{\rho\mu\la} H^{\la}{}_{\nu\sigma} -{1 \over 4} H_{\rho\nu\la}H^{\la}{}_{\mu\sigma}~,
\end{align}
The relation between $R^{(+)}$ and $R^{(-)}$ is the following
\begin{eqnarray}
\label{curvcross}
R^{(+)}{}_{\mu\nu,\rho\sigma}-R^{(-)}{}_{\rho\sigma,\mu\nu}= {1 \over 2}(dH)_{\mu\nu\rho\sigma} \ .
\end{eqnarray}
The non-vanishing components of the $R^{(+)}$ curvature tensor in the basis ({\ref{nhbasis}}) are
\begin{eqnarray}
\label{rmin}
R^{(+)}{}_{-i,+j} &=& \tilde{\nabla}_j h_i +{1 \over 2} h^\ell W_{\ell ij}
~,~~~
R^{(+)}{}_{ij,+-} = dh_{ij}~,
\nonumber \\
R^{(+)}{}_{ij,+k} &=& r \bigg( \tilde{\nabla}_k dh_{ij} -h_k dh_{ij}
+{1 \over 2} (dh)_i{}^m W_{mjk} -{1 \over 2} (dh)_j{}^m W_{mik} \bigg)~,
\nonumber \\
R^{(+)}{}_{ij,k\ell} &=& {\tilde{R}}_{ijk\ell}- {1 \over 2}\tilde{\nabla}_i W_{kj\ell}
+{1 \over 2} \tilde{\nabla}_j W_{ki \ell} +{1 \over 4} W_{kim} W^m{}_{j \ell}
-{1 \over 4} W_{kjm} W^m{}_{i \ell}
\nonumber
\\
&=& \tilde{R}^{(+)}{}_{ij,k \ell}~,
\end{eqnarray}
where in the above expression, we have set $\Delta=0$, $N=h$ and $Y=dh$.
Note that the $R^{(+)}{}_{-i,+j}$ and $R^{(+)}{}_{ij,+k}$ terms give no
contribution to the Bianchi identity of $H$ or to the Einstein equations,
because $R^{(+)}{}_{\mu\nu,-i}=0$ for all $\mu,\nu$.

\subsection{Useful formulae}
For heterotic near-horizon geometries the non-vanishing components of the Hessian of $\Phi$, are given by
\begin{eqnarray}
\label{hessian1}
\nabla_+ \nabla_- \Phi &=& -{1 \over 2} h^i \tilde{\nabla}_i \Phi~,
\nonumber \\
\nabla_+ \nabla_i \Phi &=& -{1 \over 2} r (dh)_i{}^j \tilde{\nabla}_j \Phi~,
\nonumber \\
\nabla_i \nabla_j \Phi &=& \tilde{\nabla}_i \tilde{\nabla}_j \Phi~,
\end{eqnarray}
where in the above expression, we have set $\Delta=0$.

If $f$ is any function of spacetime, then frame derivatives are expressed in terms of co-ordinate derivatives  as
\begin{eqnarray}
\partial_+ f &=& \partial_u f +{1 \over 2} r^2 \Delta \partial_r f~,~~
\partial_- f = \partial_r f~,~~
\partial_i f = {\tilde{\partial}}_i f -r h_i \partial_r f  \ .
\end{eqnarray}

\section{Heterotic spinorial geometry}
Spinors can be described in terms of forms \cite{spin_geometry, harvey_spin}. In this section we shall summarize the information to describe $Spin (9,1)$ spinors in terms of forms. 

Let $(e_0, e_1,  ..., e_9)$ be a orthonormal basis for $T^*\mathbb{R}^{9,1}$, with respect to the Lorentzian inner product.
Consider the subspace $U \equiv \mathbb{C} \langle e_1, ... , e_5 \rangle$, which consists of all possible linear combinations of $e_1, .... , e_5$, with complex coefficients. 
Then Dirac spinors are in a one-to-one correspondence with poli-forms on $T^*\mathbb{R}^{9,1}$, whose space is here denoted as $\Lambda ( U )$.   
A Dirac spinor decomposes into two complex chiral representations, the so-called Weyl representations. The two inequivalent Weyl representations corresponds to poli-forms of even or odd degree, whose space is denoted as $\Lambda^{even}(U)$ and $\Lambda^{odd}(U)$ respectively. 
The gamma matrices act on a spinor $\eta$ in $\Lambda ( U )$ as 
\begin{eqnarray}
\notag
&&\G_0 \eta = - e_5 \w \eta + e_5 \lrcorner \eta \ , \qquad\qquad
\G_i \eta = e_i \w \eta + e_i \lrcorner \eta \ , \\
&&\G_5 \eta = e_5 \w \eta + e_5 \lrcorner \eta \ , \qquad\qquad
\G_{5+i} \eta = i e_i \w \eta - i e_i \lrcorner \eta \ ,
\end{eqnarray}
where $i = 1, ... , 4$. In our convention, $\G_A$ with $A= 1, ... , 9$ are Hermitian, while $\G_0$ is anti-Hermitian with respect to the auxiliary inner product
\begin{equation}
\label{aux_inner}
\langle z^a e_a , w^b e_b \rangle = \sum_{a=1}^5 (z^a)^* w^a \ , 
\end{equation} 
where $(z^a)^*$ is the standard complex conjugate of $z^a$. 
The auxiliary inner product (\ref{aux_inner}) is \emph{not} invariant under $Spin (9, 1)$ transformations on spinors, and therefore it cannot be used as a Dirac product. 
However one can define a $Spin(9, 1)$ invariant Dirac inner product between the two Dirac spinors $\epsilon$ and $\eta$ as follows
\begin{equation}
D ( \epsilon , \eta ) \equiv \langle \G_0 \epsilon , \eta \rangle \ , 
\end{equation}
where $\langle \cdot \, ,  \cdot \rangle$ is defined as in (\ref{aux_inner}). 

In addition, in even dimensions one can always define two Spin invariant Majorana inner products. So in our case we have two inequivalent $Spin (9,1)$ invariant Majorana inner products, defined as follows
\begin{equation}
M(\epsilon, \eta ) = \langle \G_{12345} \epsilon^* , \eta \rangle \ , 
\end{equation} 
and 
\begin{equation}
\tilde{M} (\epsilon , \eta ) = \langle \G_{06789} \epsilon^* , \eta \rangle \ . 
\end{equation}
The two inner products $M$ and $\tilde{M}$ are non-vanishing only if $\epsilon$ and $\eta$ are associated with forms of opposite degree (e.g. $\epsilon$ is even and $\eta$ is odd).  

It is known that $Spin(9, 1)$ admits Majorana-Weyl representations. These are obtained by imposing on complex Weyl representation the reality condition, also called Majorana conditions. There are two possible ways of imposing the Majorana condition, which reflects the two possible Majorana inner products $M$ and $\tilde{M}$, and they are
\begin{equation}
\label{real_1}
\eta = \G_{012345} \eta^* \ , 
\end{equation}
or 
\begin{equation}
\label{real_2}
\eta = \G_{6789} \eta^* \ . 
\end{equation}
This can be expressed in terms of charge conjugation matrix
\begin{equation}
C \equiv \G_{012345} \ , \qquad\qquad
\tilde{C} \equiv \G_{6789} \ , 
\end{equation}
and the reality conditions (\ref{real_1}) and (\ref{real_2}) becomes
\begin{equation}
\eta = C * \eta \ , \qquad\qquad
\eta = \tilde{C} * \eta  \ , 
\end{equation}
where the operator $C*$ is defined as $C* \eta \equiv C \eta^*$. 
The reality conditions map poli-forms of even (odd) degree to poli-forms of even (odd) degree, and select \emph{real} subspaces inside $\Lambda^{even}(U)$ and $\Lambda^{odd}(U)$, which we denote as $\Lambda^{even}_{Maj}(U)$ and $\Lambda^{odd}_{Maj}(U)$ respectively. 
For instance, $1$ and $e_{1234}$ are poli-forms in $\Lambda^{even}(U)$ associated with complex Weyl spinors of positive chirality. However one can consider the combination $a 1 + b \, e_{1234}$ and impose the reality condition (\ref{real_2}) and find the Majorana-Weyl spinor of positive chirality
\begin{equation}
\eta = a 1 + a^* e_{1234} \ .
\end{equation}
Therefore $1 + e_{1234}$ and $i (1 - e_{1234} )$ are two example of linearly independent Majorana-Weyl spinors of positive chirality, i.e. elements of $\Lambda^{even}_{Maj}(U)$. 
Spacetime forms can be constructed in terms of spinor bilinears as follows, by using the $\tilde{M}$ inner product
\begin{equation}
\zeta(\epsilon, \eta) = \frac{1}{k!} \langle \G_{06789} \epsilon^* , \G_{\mu_1... \mu_k} \eta \rangle e^{\mu_1} \w ... \w e^{\mu_k} \ , 
\end{equation}
which is equivalent by using the reality condition (\ref{real_2}) to 
\begin{equation}
\zeta(\epsilon, \eta) = \frac{1}{k!} \langle \G_{0} \epsilon , \G_{\mu_1... \mu_k} \eta \rangle e^{\mu_1} \w ... \w e^{\mu_k} \ .
\end{equation}
If both $\epsilon , \eta \in  \Lambda^{even}_{Maj}(U)$ (or equivalently, both in $\Lambda^{odd}_{Maj}(U)$), then the only non-vanishing spinor bilinears are the 1-, 3- and 5-forms. When the two spinors are of the same chirality, it suffices to compute forms up to degree five since the forms of higher degree are obtained by Hodge duality. The 5-forms are either self of anti-self dual.

One can introduce the following basis
\begin{equation}
\G_{\bar{\alpha}} = \frac{1}{\sqrt{2}} (\G_{\alpha} + i \G_{5+\alpha} ) \ , \qquad
\G_{\pm} = \frac{1}{\sqrt{2}} (\G_5 \pm \G_0 ) \ , \qquad
\G_{\alpha} = \frac{1}{\sqrt{2}} (\G_{\alpha} - i \G_{5 + \alpha} ) \ , 
\end{equation}
where the gamma matrices become ladder type operators, 
 \begin{equation}
 \G_+ = \sqrt{2} e_5 \lrcorner  \ , \qquad
 \G_- = \sqrt{2} e_5 \w \ , \qquad
 \G_{\alpha} = \sqrt{2} e_{\alpha} \w \ , \qquad
 \G_{\bar{\alpha}} = \sqrt{2} e_{\alpha} \lrcorner \ . 
  \end{equation}
The spinor $1$ is a \emph{Clifford vacuum}, which is annihilated by $\G_+$ and $\G_{\bar{\alpha}}$. Then every spinor can be created by acting on $1$ with the gamma matrices $\G_-$ and $\G_{\alpha}$.

\section{Manifolds with $G_2$ structure}
Let $(e^1, ... , e^7 )$ be a basis on $T^* \mathbb{R}^7$, and define a 3-form as 
\begin{equation}
\label{3_form}
\varphi = e^{123} + e^{145} + e^{167} + e^{246} - e^{257} - e^{347} - e^{356} \ . 
\end{equation} 
The subgroup of $GL(7, \mathbb{R})$ preserving $\varphi$ is the exceptional Lie group $G_2$, which is compact, connected, simply-connected, semisimple and 14-dimensional. 

Consider a 7-dimensional manifold $\mathcal{B}^7$.
The manifold $\mathcal{B}^7$ admits a $G_2$ \emph{structure} if it is orientable and spin, or equivalently its first and second Stiefel-Whitney classes vanishes. When this is the case, there exists a nowhere-vanishing Majorana spinor $\eta$, such that the 3-form (\ref{3_form}) can be constructed as spinor bilinear. 
Furthermore, $\mathcal{B}^7$ admits a $G_2$ \emph{holonomy} if 
\begin{equation}
\nabla \varphi = 0 \ , 
\end{equation}
or equivalently the 3-form $\varphi$ is closed and co-closed. Here $\nabla$ is the Levi-Civita connection. Manifolds which admits $G_2$ holonomy are Ricci-flat\footnote{We remark that this is not what happens in the heterotic case, where the holonomy of the connection with torsion is a subgroup of $G_2$. Indeed, heterotic near-horizon geometries with non trivial fluxes are not Ricci flat.}. 

The $G_2$ structure induces a splitting of the bundles of tensors on $\mathcal{B}^7$ into irreducible components. 
Let $\Lambda^p (T^* \mathcal{B}^7)$ be the space of $p$-forms on $\mathcal{B}^7$, and $\Lambda^p_{{\bf k}} (T^* \mathcal{B}^7)$ the subspace of $\Lambda^p (T^* \mathcal{B}^7)$ of $p$-forms which transform in the ${\bf k}$-dimensional irreducible representation of $G_2$. Then we have
\begin{eqnarray}
\notag
\Lambda^0 &=& \Lambda^0_{{\bf 1}} \\
\notag
\Lambda^1 &=& \Lambda^1_{{\bf 7}} \quad ( = T^* \mathcal{B}^7 )\\
\notag
\Lambda^2 &=& \Lambda^2_{{\bf 7}} \oplus \Lambda^2_{{\bf 14}} \\
\notag
\Lambda^3 &=& \Lambda^3_{{\bf 1}} \oplus \Lambda^3_{{\bf 7}} \oplus \Lambda^3_{{\bf 27}} \\
\notag
\Lambda^4 &=& \Lambda^4_{{\bf 1}} \oplus \Lambda^4_{{\bf 7}}\oplus \Lambda^4_{{\bf 27}} \\
\notag
\Lambda^5 &=& \Lambda^5_{{\bf 7}} 	\oplus
\Lambda^5_{{\bf 14}}  \\
\notag
\Lambda^6 &=& \Lambda^6_{{\bf 7}}\\
 \Lambda^7 &=& \Lambda^7_{{\bf 1}} \ .
\end{eqnarray}
The decomposition of $p = 4, 5, 6, 7$ forms follows from Hodge decomposition of $p= 0, 1, 2, 3$ forms. 

The projectors associated with the 2, 3, and 4-forms are given here explicitly. 
\begin{itemize}
\item 2-form $\alpha$  
\begin{eqnarray}
\notag
(P^{{\bf 7}} \alpha )_{i_1 i_2} &=& \frac{1}{3} \alpha_{i_1 i_2} + \frac{1}{6} (\star \varphi)_{i_1 i_2}{}^{\ell_1\ell_2} \alpha_{\ell_1\ell_2} \ , \\
(P^{{\bf 14}} \alpha )_{i_1 i_2} &=& \frac{2}{3} \alpha_{i_1 i_2} - \frac{1}{6} (\star \varphi)_{i_1 i_2}{}^{\ell_1\ell_2} \alpha_{\ell_1\ell_2}   \ . 
\end{eqnarray}
\item 3-form $\beta$ 
\begin{eqnarray}
\notag
(P^{{\bf 1}} \beta )_{i_1 i_2 i_3} &=& \frac{1}{42} \varphi^{\ell_1\ell_2\ell_3} \beta_{\ell_1\ell_2\ell_3} \varphi_{i_1 i_2 i_3} \ , \\
\notag
(P^{{\bf 7}} \beta )_{i_1 i_2 i_3} &=& \frac{1}{4} \beta_{i_1 i_2 i_3} - \frac{1}{24} \varphi^{\ell_1\ell_2\ell_3} \beta_{\ell_1\ell_2\ell_3} \varphi_{i_1 i_2 i_3} + \frac{3}{8} \beta_{\ell_1 \ell_2 [ i_1} \star \varphi_{i_2 i_3]}{}^{\ell_1\ell_2} \ ,  \\
(P^{{\bf 27}} \beta )_{i_1 i_2 i_3} &=& \frac{3}{4} \beta_{i_1 i_2 i_3} + \frac{1}{56} \varphi^{\ell_1\ell_2\ell_3} \beta_{\ell_1\ell_2\ell_3} \varphi_{i_1 i_2 i_3} - \frac{3}{8} \beta_{\ell_1 \ell_2 [ i_1} \star \varphi_{i_2 i_3]}{}^{\ell_1\ell_2} \ . 
\end{eqnarray}
\item 4-form $\gamma$ 
\begin{eqnarray}
\notag
(P^{{\bf 1}} \gamma )_{i_1 i_2 i_3 i_4} &=& \frac{1}{168} \gamma^{\ell_1\ell_2\ell_3\ell_4} (\star \varphi)_{\ell_1\ell_2\ell_3\ell_4} (\star \varphi)_{i_1 i_2i_3i_4} \ , \\
\notag
(P^{{\bf 7}} \gamma )_{i_1 i_2 i_3 i_4} &=& \frac{1}{4} \gamma_{i_1 i_2 i_3 i_4} - \frac{1}{96} \gamma^{\ell_1\ell_2\ell_3\ell_4} (\star \varphi)_{\ell_1\ell_2\ell_3\ell_4} (\star \varphi)_{i_1i_2i_3i_4} + \frac{3}{4} (\star \varphi)^{\ell_1\ell_2}{}_{[i_1i_2} \gamma_{i_3i_4] \ell_1\ell_2} \ , \\
\notag
(P^{{\bf 27}} \gamma )_{i_1 i_2 i_3 i_4} &=& \frac{3}{4} \gamma_{i_1 i_2 i_3 i_4} + \frac{1}{224} \gamma^{\ell_1\ell_2\ell_3\ell_4} (\star \varphi)_{\ell_1\ell_2\ell_3\ell_4} (\star \varphi)_{i_1i_2i_3i_4} - \frac{3}{4} (\star \varphi)^{\ell_1\ell_2}{}_{[i_1i_2} \gamma_{i_3i_4] \ell_1\ell_2} \ . \\
\end{eqnarray}
\end{itemize}

\clearpage{\pagestyle{empty}\cleardoublepage} 


\chapter{Spin Connection and Ricci Tensor}

In this appendix we list the components of the spin connection and Ricci tensor in the light-cone basis ({\ref{nhbasis}}) of the metric (\ref{NHG}), where the metric components $\{ \D, h, \gamma \}$ are allowed to depend on $r$, or equivalently of the metric (\ref{Gauss_metric}) in the extremal case. 

From the following general expressions, one can obtain the spin connection and Ricci tensor of the near-horizon metric by suppressing the $r$-dependence of $\{ \D, h, \gamma \}$.

\section{Spin Connection}
The non-vanishing components of the spin connection in
the frame basis ({\ref{nhbasis}}) of the horizon metric (\ref{Gauss_metric}) are
\be
\notag
\O_{-,+i} &=& -\frac{1}{2} h_i - \frac{1}{2} r  \dot{h}_i \ , \\
\notag
\O_{-, ij} &=& \dot{e}^k{}_I e^I{}_{[i} \d_{j] k} \ , \\ 
\notag
\O_{+,+-} &=& -r \D - \frac{1}{2}r^2 \dot{\D} \ ,  \\
\notag
\O_{+,+i} &=& \frac{1}{2} r^2(  \Delta h_i - \tilde{\partial}_i \Delta) + \frac{1}{2} r^3 \dot{\D} h_i  -\frac{1}{2} r^3 \D \dot{h}_i  \ , \\
\notag
\O_{+,-i} &=& -\frac{1}{2} h_i - \frac{1}{2}r \dot{h}_i \ , \\
\notag
\O_{+,ij} &=& -\frac{1}{2} r \tilde{d} h_{ij} + r^2  h_{[i} \dot{h}_{j]} + \frac{1}{2} r^2 \D \dot{e}^k{}_I e^I{}_{[i} \d_{j]k} \ , \\
\notag
\O_{i,+-} &=& \frac{1}{2} h_i + \frac{1}{2}r \dot{h}_i \ , \\ 
\notag
\O_{i,+j} &=& - \frac{1}{2} r \tilde{d} h_{ij} + r^2  h_{[i} \dot{h}_{j]} - \frac{1}{2} r^2 \D \dot{e}^k{}_I e^I{}_{(i} \d_{j)k} \ , \\
\notag
\O_{i,j-} &=& \dot{e}^k{}_I e^I{}_{(i} \d_{j) k} \ , \\
\O_{i,jk} &=& \tilde\O_{i,jk} + r \d_{j\ell}\dot{e}^{\ell}{}_I  e^I{}_{[k} h_{i]} - r \d_{i\ell}\dot{e}^{\ell}{}_I  e^I{}_{[j} h_{k]} + r\d_{k\ell} \dot{e}^{\ell}{}_I  e^I{}_{[i} h_{j]} \ . 
\ee
where $\tilde\Omega$ denotes the spin-connection of the spatial horizon section  ${{\cal{S}}}$ in the ${\bf{e}}^i$ basis.

\section{Ricci Tensor}
The Ricci tensor components in the frame basis ({\ref{nhbasis}}) are 
\begin{eqnarray}
\nonumber
R_{++} &=& r^2 \bigg( \frac{1}{2}{{\tilde{\nabla}}}_i {{\tilde{\nabla}}}^i \Delta  -\frac{3}{2} h^i {{\tilde{\nabla}}}_i \Delta - \frac{1}{2} \Delta {{\tilde{\nabla}}}_i h^i + \Delta h^i h_i + \frac{1}{4} \tilde{d}h_{ij} \tilde{d}h^{ij} \bigg) \\
\nonumber
&+& r^3 \bigg( -  h^i {{\tilde{\nabla}}}_i \dot{\Delta} - \frac{1}{2} \dot{\Delta} {{\tilde{\nabla}}}_i h^i + \frac{1}{2} \dot{h}^i {{\tilde{\nabla}}}_i \Delta + \frac{1}{2} \Delta {{\tilde{\nabla}}}_i \dot{h}^i + 2 \dot{\Delta} h^i h_i - \Delta h^i \dot{h}_i \\
&+&  \frac{1}{4}\Delta \dot{g}_k{}^k h_i h^i - \frac{1}{2} \Delta \dot{g}_{ij} h^i h^j  - \frac{1}{4} \dot{g}_k{}^k h_i {{\tilde{\nabla}}}^i \Delta  + \frac{1}{2} \dot{g}_{ij} h^i {{\tilde{\nabla}}}^j\Delta- h^i \dot{h}^j \tilde{d}h_{ij} \bigg) + {\cal{O}}(2) \ ,
\nonumber \\
\end{eqnarray}
\begin{eqnarray}
\nonumber
R_{+-} &=& \frac{1}{2}{{\tilde{\nabla}}}_i h^i  - \Delta - \frac{1}{2} h^ih_i \\ 
&+& r \bigg( \frac{1}{2}{{\tilde{\nabla}}}_i \dot{h}^i - \frac{1}{2}\Delta \dot{g}_k{}^k  - \frac{1}{4} \dot{g}_k{}^k h_i h^i + \frac{1}{2} \dot{g}_{ij} h^i h^j -2\dot{\Delta} - 2 h^i \dot{h}_i  \bigg) + {\cal{O}}(2) \ ,
\nonumber \\
\end{eqnarray}
\begin{eqnarray}
R_{--} ={\cal{O}}(2)  \ ,
\end{eqnarray}
\begin{eqnarray}
\nonumber
R_{+i} &=&  r \bigg( \frac{1}{2}{{\tilde{\nabla}}}^k \tilde{d}h_{ik} +  h^j \tilde{d}h_{ji} + \Delta h_i - {{\tilde{\nabla}}}_i \Delta   \bigg) \\
\nonumber
&+& r^2 \bigg(-\frac{1}{2} \Delta \dot{h}_i  + \frac{1}{2} \dot{h}_i {{\tilde{\nabla}}}_j h^j + h^j {{\tilde{\nabla}}}_j \dot{h}_i  - \frac{1}{2} h_i {{\tilde{\nabla}}}_j \dot{h}^j - \frac{1}{2} {{\tilde{\nabla}}}_i (h_j \dot{h}^j )\\
\nonumber
&+& 2\dot{\Delta} h_i - \frac{1}{2} {{\tilde{\nabla}}}_i \dot{\Delta} + 3 h^j h_{[i} \dot{h}_{j]} - \frac{3}{4} \Delta \dot{g}_{ij} h^j  +  \frac{1}{2} \dot{g}_{ij} {{\tilde{\nabla}}}^j \Delta + \frac{1}{4} \Delta {{\tilde{\nabla}}}^j\dot{g}_{ij}   \\
&-&  \frac{1}{4} {{\tilde{\nabla}}}_i (\Delta  \dot{g}_k{}^k) + \frac{3}{8}\Delta h_i \dot{g}_k{}^k  + \frac{1}{2} \dot{g}_i{}^j {\tilde{d}} h_{jk} h^k + \frac{1}{4} \dot{g}_k{}^k h^j {\tilde{d}} h_{ji} + \frac{1}{2} {\tilde{d}} h_i{}^j \dot{g}_{jk} h^k   \bigg) + {\cal{O}}(2)  \ , 
\nonumber \\
\end{eqnarray}
\begin{eqnarray}
R_{-i} &=& \dot{h}_i + \frac{1}{2} {{\tilde{\nabla}}}^j \dot{g}_{ji} - \frac{1}{2} {{\tilde{\nabla}}}_i \dot{g}_k{}^k + \frac{1}{4} h_i \dot{g}_k{}^k - \frac{1}{2} \dot{g}_{ij} h^j + {\cal{O}}(2) \ ,
\end{eqnarray}
\begin{eqnarray}
\nonumber
R_{ij} &=& {\tilde{{\mathcal R}}}_{ij}
+{{\tilde{\nabla}}}_{(i} h_{j)} -{1 \over 2} h_i h_j
\nonumber \\
&+& r \bigg({{\tilde{\nabla}}}_{(i} {{\dot{h}}}_{j)}-3h_{(i} {{\dot{h}}}_{j)}
+\big(-\Delta+{1 \over 2}{{\tilde{\nabla}}}_k h^k -h_k h^k\big)
{\dot{g}}_{ij}
-{\dot{g}}_{(i}{}^k {{\tilde{\nabla}}}_{|k|} h_{j)}
\nonumber \\
&-& h^k {{\tilde{\nabla}}}_{(i} {\dot{g}}_{j) k}
+h^k {{\tilde{\nabla}}}_k {\dot{g}}_{ij}
-h_{(i} {{\tilde{\nabla}}}^k {\dot{g}}_{j) k} +2 h_k h_{(i} {\dot{g}}_{j)}{}^k+ h_{(i} {{\tilde{\nabla}}}_{j)} {\dot{g}}_k{}^k
\nonumber \\
&+&{1 \over 2}{\dot{g}}_k{}^k \big({{\tilde{\nabla}}}_{(i} h_{j)}-h_i h_j \big)   \bigg) + {\cal{O}}(2) \ .
\end{eqnarray}

Here ${\cal{O}}(2)$ consists of terms linear in ${\ddot{h}}, {\ddot{\Delta}}, {\ddot{g}}$,
and terms quadratic in ${\dot{h}}, {\dot{\Delta}}, {\dot{g}}$, which play no role in the moduli
space calculations.

\clearpage{\pagestyle{empty}\cleardoublepage} 


\chapter{Field Equations}
\label{appx:field_eqns}
In this appendix we list the various bosonic field equations of anomaly corrected and uncorrected heterotic supergravity and $D=11$ supergravity. In each theory, we decompose the field equations in terms of the horizon data, which is not taken in the near-horizon limit. One can obtain the equations of motion for the fields in the near-horizon limit by suppressing the $r$-dependence of all fields. 

We denote by ${\tilde{\nabla}}$ the 
Levi-Civita connection on ${\cal{S}}$, restricted to $r=const.$.

\section{Anomaly corrected heterotic bosonic field equations}
The Bianchi identity associated with the 3-form is
\begin{eqnarray}
\label{bian}
dH = - {\alpha' \over 4} \bigg( {\rm tr}(R^{(+)} \wedge R^{(+)}) - {\rm tr}(F \wedge F) \bigg) + \mathcal{O}(\alpha'^2) \ ,
\end{eqnarray}
where ${\rm tr} (F \wedge F) = F^a{}_b \wedge F^b{}_a$ ($a, b$ are gauge indices on $F$).

The Einstein equation is
\begin{eqnarray}
\label{ein}
&&R_{\mu\nu} -{1 \over 4} H_{\mu \la_1 \la_2} H_{\nu}{}^{\la_1 \la_2}
+2 \nabla_{\mu} \nabla_{\nu} \Phi
\cr &&~~~~~~~
+ {\alpha' \over 4} \bigg(R^{(+)}{}_{\mu \la_1, \la_2 \la_3}
R^{(+)}{}_{\nu}{}^{\la_1, \la_2 \la_3}-F_{\mu \la ab}F_{\nu}{}^{\la ab} \bigg)=\mathcal{O}(\alpha'^2)~.
\nonumber \\
\end{eqnarray}
The gauge field equations are
\begin{eqnarray}
\label{geq1}
\nabla^{	\mu} \bigg(e^{-2 \Phi} H_{\mu \nu_1 \nu_2}\bigg)= \mathcal{O}(\alpha'^2)~,
\end{eqnarray}
and
\begin{eqnarray}
\label{geq2}
\nabla^{\mu} \bigg(e^{-2 \Phi}F_{\mu\nu} \bigg)+{1 \over 2} e^{-2 \Phi} H_{\nu\la_1 \la_2} F^{\la_1 \la_2}= \mathcal{O}(\alpha')~.
\end{eqnarray}
The dilaton field equation is
\begin{eqnarray}
\label{deq}
\nabla_{\mu} \nabla^{\mu} \Phi &=& 2 \nabla_{\mu} \Phi \nabla^{\mu} \Phi -{1 \over 12} H_{\la_1 \la_2 \la_3} H^{\la_1 \la_2 \la_3}
\nonumber \\
&+& {\alpha' \over 16} \bigg(R^{(+)}{}_{\la_1 \la_2, \la_3 \la_4}
R^{(+)}{}^{\ \la_1 \la_2, \la_3 \la_4}-F_{\la_1 \la_2 ab}F^{\la_1 \la_2 ab} \bigg) + \mathcal{O}(\alpha'^2) \ .
\end{eqnarray}
This completes the list of field equations. 

\section{Uncorrected heterotic bosonic field equations}
\label{appx:unc_het}
In this section we list the uncorrected heterotic bosonic field equations in the Einstein frame, which turns out to be convenient for the moduli space computation. Since we neglect $\alpha'$ corrections, the dynamic of non-abelian 2-form field strength $F$ decouples from the rest of the heterotic fields, and here we set $F = 0$. 

After the Weyl rescaling $g \rightarrow e^{-\frac{1}{2}\Phi} g $, the gauge field equation is:
\be
\label{gaugeH}
\nabla^{\mu} \bigg( e^{- \Phi} H_{\mu\lambda_1 \lambda_2} \bigg) = 0 \ ,  
\ee
The Einstein equation is:
\be
\label{eins}
R_{\mu\nu} - \frac{1}{2} \nabla_{\mu} \Phi \nabla_{\nu} \Phi + \frac{1}{48} e^{-\Phi} g_{\mu\nu} H_{\lambda_1\lambda_2\lambda_3} H^{\lambda_1\lambda_2\lambda_3} - \frac{1}{4} e^{- \Phi} H_{\mu\lambda_1\lambda_2} H_{\nu}{}^{\lambda_1\lambda_2} = 0 \ . 
\ee
The dilaton field equation is:
\be
\label{dilaton}
\nabla_{\mu}\nabla^{\mu} \Phi + \frac{1}{12} e^{-\Phi} H_{\lambda_1\lambda_2\lambda_3} H^{\lambda_1\lambda_2\lambda_3} = 0 \ . 
\ee

\subsection{Decomposition in Gaussian null coordinates}
The components of the Einstein equation (\ref{eins}) are:
{\flushleft{The $++$ component: } }
\be 
\notag
&&R_{++} + r^2 e^{-\Phi} \bigg( \frac{1}{2} (h\w N)_{ij} Y^{ij} - \frac{1}{4} Y_{ij} Y^{ij} - \frac{1}{4} (h\w N)_{ij} (h\w N)^{ij} \bigg) \\
\notag
&&+ r^3 e^{-\Phi} \bigg( - \frac{1}{4} \D Z_{ij} Y^{ij} + \frac{1}{4} \D Z_{ij} (h\w N)^{ij} \bigg) \\
&&+ r^4 \bigg( - \frac{1}{16} e^{-\Phi} \D^2 Z_{ij} Z^{ij} - \frac{1}{8} \D^2 \dot{\Phi}^2  \bigg)  = 0 \ . 
\ee
The $--$ component:
\be
R_{--} - \frac{1}{4} e^{-\Phi} Z_{ij} Z^{ij} - \frac{1}{2} \dot{\Phi} \dot{\Phi} = 0 \ . 
\ee
The $+-$ component: 
\be
\label{+- ein}
\notag
&&R_{+-} + \frac{3}{8} e^{-\Phi} N_i N^i  + \frac{1}{48} e^{-\Phi} W_{ijk} W^{ijk}  \\
\notag
&&+ r e^{-\Phi} \bigg( \frac{1}{8} (h\w N)_{ij} Z^{ij} - \frac{1}{8} Y_{ij} Z^{ij} - \frac{1}{24} (h\w Z)_{ijk} W^{ijk}  \bigg) \\
&&+ r^2 \bigg(  - \frac{1}{16}  e^{-\Phi} \D Z_{ij} Z^{ij} + \frac{1}{48} e^{-\Phi} (h\w Z)_{ijk} (h\w Z)^{ijk} - \frac{1}{4} \D \dot{\Phi}^2  \bigg) = 0 \ . 
\ee
The $+i$ component: 
\be
\notag
&&R_{+i} + r e^{-\Phi} \bigg( \frac{1}{2} Y_{ij} N^j - \frac{1}{2} N^j (h\w N)_{ij} - \frac{1}{4} Y^{jk} W_{ijk} + \frac{1}{4} (h\w N)^{jk} W_{ijk} \bigg)  \\
\notag
&&+ r^2 e^{-\Phi} \bigg( \frac{1}{4} \D N^j Z_{ij} - \frac{1}{8} \D Z^{jk} W_{ijk} + \frac{1}{4} Y^{jk} (h\w Z)_{ijk} - \frac{1}{4} (h\w N)^{jk} (h \w Z)_{ijk} \bigg) \\
&&- \frac{1}{4} r^2 \D \dot{\Phi} \hn_i \Phi + r^3 \bigg( - \frac{1}{8} e^{-\Phi} \D Z^{jk} (h\w Z)_{ijk} + \frac{1}{4} \D h_i \dot{\Phi}^2 \bigg) = 0 \ . 
\ee
The $-i$ component: 
\be
\label{-i ein}
\notag
&&R_{-i} - \frac{1}{2} e^{-\Phi} N^j Z_{ij} - \frac{1}{4} e^{-\Phi} Z^{jk} W_{ijk} - \frac{1}{2} \dot{\Phi} \hn_i \Phi  \\
&&+ r \bigg( \frac{1}{4} e^{-\Phi} Z^{jk} (h\w Z)_{ijk} + \frac{1}{2} \dot{\Phi}^2 h_i \bigg) = 0 \ . 
\ee
The $ij$ component:
\be 
\notag
&&R_{ij} + \frac{1}{2}e^{-\Phi} N_i N_j - \frac{1}{4} e^{-\Phi} W_{i\ell_1\ell_2} W_j{}^{\ell_1\ell_2}  -\frac{1}{8}e^{-\Phi} \gamma_{ij} N_k N^k \\
\notag
&&+ \frac{1}{48} e^{-\Phi} \gamma_{ij} W_{\ell_1\ell_2\ell_3} W^{\ell_1\ell_2\ell_3} - \frac{1}{2} \hn_i \Phi \hn_j \Phi \\
\notag
&& + r e^{-\Phi} \bigg( Y^k{}_{(i} Z_{j)k} - (h\w N)^k{}_{(i} Z_{j) k} + \frac{1}{2} W^{\ell_1\ell_2}{}_{(i} (h\w Z)_{j) \ell_1\ell_2} + \frac{1}{8}
\gamma_{ij} Y_{\ell_1\ell_2} Z^{\ell_1\ell_2} \\
\notag
&& - \frac{1}{8} \gamma_{ij} (h\w N)_{\ell_1\ell_2} Z^{\ell_1\ell_2} - \frac{1}{24} \gamma_{ij} (h\w Z)_{\ell_1\ell_2\ell_3} W^{\ell_1\ell_2\ell_3} \bigg) + r \dot{\Phi} h_{(i} \hn_{j)} \Phi  \\
\notag 
&& + r^2 e^{-\Phi} \bigg( \frac{1}{2} \D Z^k{}_{(i} Z_{j)k}  - \frac{1}{4} (h\w Z)_{i \ell_1\ell_2} (h\w Z)_j{}^{\ell_1\ell_2} + \frac{1}{16} \gamma_{ij} \D Z_{\ell_1\ell_2} Z^{\ell_1\ell_2} \\
&& + \frac{1}{48} \gamma_{ij} (h\w Z)_{\ell_1\ell_2\ell_3} (h\w Z)^{\ell_1\ell_2\ell_3}  \bigg) - \frac{1}{2} r^2 h_i h_j \dot{\Phi}^2 = 0 \ . 
\ee
The components of the gauge field equation (\ref{gaugeH}) are:

{\flushleft{The $+-$ component: }}
\be
\notag
- N_i \hn^i \Phi + \hn^i N_i &+& r \bigg(  \dot{\Phi} h^i N_i - h^i N_i + h^i N^j \dot{\gamma}_{ij} \\
 &-& \frac{1}{2} \dot{\gamma}_k{}^k N^i h_i - \frac{1}{2} \tilde{d} h_{ij} Z^{ij} \bigg) + r^2 h_i \dot{h}_j Z^{ij} = 0 \ .  
\ee
The $+i$ component: 
\be
\notag
&&\hn^j ( Y_{ij} - (h\w N)_{ij} + \frac{1}{2} r \D Z_{ij} ) - \tilde{d}h_{ij} N^j - \hn^j \Phi Y_{ij} \\
\notag
&& +  \hn^j \Phi (h\w N)_{ij} - Y_{ij} h^j + (h\w N)_{ij}h^j - \frac{1}{2} \tilde{d}h_{\ell_1\ell_2} W^{\ell_1\ell_2}{}_i \\
\notag
&& + r \bigg( \frac{1}{2}\D \dot{\Phi} N_i - \frac{1}{2} \D \dot{N}_i + \frac{1}{2}\D \dot{\gamma}_{ij} N^j - \frac{1}{2} \D h^j Z_{ij} + \frac{1}{2} \hn^j \D Z_{ij} - h_i \dot{h}_j N^j + h_j \dot{h}_i N^j  \\
\notag
&& - \frac{1}{2} \D \hn^j \Phi  Z_{ij} +  \dot{\Phi} Y_{ij} h^j -  \dot{\Phi} h^j (h\w N)_{ij} - h^j \dot{Y}_{ij} + h^j (\dot{h} \w N )_{ij} + h^j (h\w \dot{N})_{ij} \\
\notag
&&+ \frac{1}{2} h^j \dot{\gamma}_i{}^k ( Y_{kj} - (h\w N)_{kj} ) + h_j \dot{\gamma}^{kj} ( Y_{ik} - (h\w N)_{ik} ) - \frac{1}{2} \dot{\gamma}_k{}^k h^j(Y_{ij} - (h\w N)_{ij}) \\
\notag
&& - \frac{1}{2} \dot{\gamma}_{ij} h_k (Y^{kj} - (h\w N)^{kj} ) + \frac{1}{4} \D \dot{\gamma}_k{}^k N_i + \frac{1}{2} \tilde{d}h_{\ell_1\ell_2} (h\w W)^{\ell_1\ell_2}{}_i + h_{\ell_1} \dot{h}_{\ell_2} W^{\ell_1\ell_2}{}_i \bigg) \\
\notag
&&+ r^2 \bigg( \frac{1}{2} (\D \dot{h}^j-  \dot{\D} h^j ) Z_{ij} + \frac{1}{2} h^j \dot{\Phi} \D Z_{ij} - \frac{1}{2} h^j ( \dot{\D}Z_{ij} + \D \dot{Z}_{ij} ) + \frac{1}{4} h_j \D \dot{\gamma}_{ki} Z^{kj} \\
&& + \frac{1}{2} h_j \D \dot{\gamma}^{kj} Z_{ik} - \frac{1}{4} \D \dot{\gamma}_k{}^k h^j Z_{ij} - \frac{1}{4} \D h_k \dot{\gamma}_{ij} Z^{kj} - h_{\ell_1} \dot{h}_{\ell_2} (h\w W)^{\ell_1\ell_2}{}_i \bigg) = 0 \ .
\ee
The $-i$ component: 
\be
\notag
&&\hn^j Z_{ij} + \dot{N}_i -  \dot{\Phi} N_i  - \dot{\gamma}_{ik} N^k + \frac{1}{2} \dot{\gamma}_k{}^k N_i - h^j Z_{ij} -  \hn^j \Phi Z_{ij} \\
&& + r \bigg( \dot{\Phi} h^j Z_{ij} - \dot{h}^j Z_{ij} - h^j \dot{Z}_{ij} + h^j Z_{kj} \dot{\gamma}^k{}_i + h^j Z_{ik} \dot{\gamma}^k{}_j - \frac{1}{2} \dot{\gamma}_k{}^k Z_{ij} h^j \bigg) = 0 \ . 
\ee
The $ij$ component: 
\be 
\label{ijgauge}
\notag
&& Y_{ij} - (h\w N)_{ij} - h^k W_{ijk} -  \hn^k \Phi W_{kij} + \hn^k W_{kij} \\
\notag
&& + r \bigg( \dot{Y}_{ij} -  \dot{\Phi} Y_{ij} +  \dot{\Phi} (h\w N)_{ij} - (\dot{h}\w N)_{ij} - (h\w \dot{N})_{ij} + 2\D Z_{ij}  \\
\notag
&&- \dot{\gamma}_i{}^k ( Y - h\w N)_{kj} - \dot{\gamma}_j{}^k (Y - h\w N)_{ik} + \frac{1}{2} \dot{\gamma}_k{}^k ( Y - h \w N)_{ij}  \\
\notag
&& + 2h^k (h\w Z)_{ijk} - \dot{h}^k W_{ijk} +  \hn^k \Phi (h\w Z)_{kij} + \dot{\Phi} h^k W_{kij} - \hn^k (h\w Z)_{kij} \\
\notag 
&& + h^k \dot{W}_{kij} + \dot{\gamma}^{\ell}{}_k h^k W_{\ell ij} - \frac{1}{2} \dot{\gamma}_k{}^k h^{\ell} W_{\ell ij} + \dot{\gamma}_i{}^{\ell} h^k W_{k\ell j}  + \dot{\gamma}^{\ell}{}_j h^k W_{ki\ell} \bigg) \\
\notag
&& + r^2 \bigg( - \dot{\Phi} \D Z_{ij} + \D \dot{Z}_{ij} -  \dot{\gamma}_i{}^k \D Z_{kj} -  \dot{\gamma}_j{}^k \D Z_{ik} + \frac{1}{2} \dot{\gamma}_k{}^k \D Z_{ij} + \dot{h}^k (h\w Z)_{ijk}  \\
\notag 
&& -  \dot{\Phi} h^k (h\w Z)_{kij} + h^k (\dot{h}\w Z)_{kij} + h^k (h\w \dot{Z})_{kij}  + \frac{1}{2} \dot{\gamma}_k{}^k h^{\ell} (h \w Z)_{\ell ij}\\
&& - \dot{\gamma}^{\ell}{}_k h^k (h\w Z)_{\ell ij} - \dot{\gamma}_i{}^{\ell} h^k (h\w Z)_{k\ell j} - \dot{\gamma}^{\ell}{}_j h^k (h\w Z)_{ki\ell} \bigg) = 0 \ . 
\ee
The dilaton field equation (\ref{dilaton}) decomposes as:
\be
\label{lin_dilaton}
\notag
&&\hn_i \hn^i \Phi - h_i \hn^i \Phi - \frac{1}{2}e^{-\Phi}  N_i N^i + \frac{1}{12}e^{-\Phi}  W_{ijk} W^{ijk} \\
\notag
&&+ r \bigg( 2 \D \dot{\Phi} - 2 h^i \hn_i \dot{\Phi} - \dot{\Phi} \hn_i h^i + 2 h_i h^i \dot{\Phi} + \dot{\gamma}_{ij} h^i \hn^j \Phi - \dot{h}_i \hn^i \Phi  \\
\notag
&& - \frac{1}{2} \dot{\gamma}_k{}^k h_i \hn^i \Phi + \frac{1}{2} e^{-\Phi}  Y_{ij} Z^{ij} - \frac{1}{2} e^{-\Phi}  (h\w N)_{ij} Z^{ij} - \frac{1}{6} e^{-\Phi}  W^{ijk} (h\w Z)_{ijk} \bigg) \\
\notag
&& + r^2 \bigg( \D \ddot{\Phi} + \dot{\D} \dot{\Phi} + 2h^i \dot{h}_i \dot{\Phi} + h_i h^i \ddot{\Phi} - h^i h^j \dot{\gamma}_{ij} \dot{\Phi} + \frac{1}{2} \dot{\gamma}_k{}^k ( h_i h^i + \D )  \dot{\Phi} \\
&&  + \frac{1}{4} e^{-\Phi} \D Z_{ij} Z^{ij}  + \frac{1}{12} e^{-\Phi} (h\w Z )_{ijk} (h\w Z)^{ijk} \bigg) = 0 \ . 
\ee

\section{$D=11$ supergravity bosonic field equations}
\label{appx:D=11}
The Einstein equations:
\begin{eqnarray}
\label{d11ein}
R_{\mu \nu}={1 \over 12} F_{\mu \lambda_1 \lambda_2 \lambda_3} F_\nu{}^{\lambda_1 \lambda_2 \lambda_3}
-{1 \over 144} g_{\mu \nu} F_{\lambda_1 \lambda_2 \lambda_3 \lambda_4} F^{\lambda_1 \lambda_2 \lambda_3 \lambda_4} \ .
\end{eqnarray}
The gauge field equations are given by
\begin{eqnarray}
\label{g4eq}
\nabla^\nu F_{\nu \lambda_1 \lambda_2 \lambda_3}={q \over (4!)^2} \epsilon_{\lambda_1 \lambda_2 \lambda_3}{}^{\mu_1 \mu_2 \mu_3 \mu_4 \mu_5 \mu_6 \mu_7 \mu_8} F_{\mu_1 \mu_2 \mu_3 \mu_4} F_{\mu_5 \mu_6 \mu_7 \mu_8}
\end{eqnarray}
where $q$ is a constant. Here we have included a topological term in the action proportional to
$q F \wedge F \wedge C$. On imposing supersymmetry, the value of $q$ is fixed by
requiring consistency of the gauge field equations with the integrability conditions
of the gravitino Killing spinor equations. However, here our analysis is purely
in the bosonic sector, so the value of $q$ is kept arbitrary.

\subsection{Decomposition in Gaussian null coordinates}
The gauge field equation ({\ref{g4eq}}) decomposes into the following components. 
{\flushleft{The $+-k$ component:}}
\begin{eqnarray}
\label{g4eq1}
{{\tilde{\nabla}}}^\ell \Psi_{\ell k} -r h^\ell {\dot{\Psi}}_{\ell k}
-{1 \over 2}r ({\tilde{d}} h)^{mn} Z_{mnk} +r h^\ell {\dot \gamma}_\ell{}^q \Psi_{qk}
-{1 \over 2}r h^q \Psi_{qk} {\dot{\gamma}}_m{}^m +r h^\ell {\dot \gamma}_k{}^q \Psi_{\ell q}
\nonumber \\
= {q \over 576} \epsilon_k{}^{\ell_1 \ell_2 \ell_3 \ell_4 \ell_5 \ell_6 \ell_7 \ell_8} X_{\ell_1 \ell_2 \ell_3 \ell_4}
X_{\ell_5 \ell_6 \ell_7 \ell_8}-{qr \over 72}
\epsilon_k{}^{\ell_1 \ell_2 \ell_3 \ell_4 \ell_5 \ell_6 \ell_7 \ell_8} h_{\ell_1} Z_{\ell_2 \ell_3 \ell_4} X_{\ell_5 \ell_6 \ell_7 \ell_8} \ .
\nonumber \\
\end{eqnarray}

{\flushleft{The $+k_1 k_2$ component:}}
\begin{eqnarray}
\label{g4eq2}
{{\tilde{\nabla}}}^\ell (W-h \wedge \Psi)_{\ell k_1 k_2}+
2 r \Delta h^\ell Z_{\ell k_1 k_2}
+h^\ell (W-h \wedge \Psi)_{\ell k_1 k_2}
\nonumber \\
+r h^\ell({\dot{W}}-{\dot{h}}\wedge \Psi - h \wedge {\dot{\Psi}})_{\ell k_1 k_2}
+{1 \over 2}(-{\tilde{d}} h^{mn}+r h^m {\dot{h}}^n) X_{mn k_1 k_2}
\nonumber \\
+{1 \over 2} r{\tilde{d}} h^{mn}(h \wedge Z)_{mn k_1 k_2}
+{1 \over 2} r (-2 h^\ell {\dot{\gamma}}_\ell{}^q +h^q {\dot{\gamma}}_m{}^m ) (W - h \wedge \Psi)_{q k_1 k_2}
\nonumber \\
-r h^\ell {\dot{\gamma}}_{k_1}{}^q (W-h \wedge \Psi)_{\ell q k_2}
+r h^\ell {\dot{\gamma}}_{k_2}{}^q (W-h \wedge \Psi)_{\ell q k_1}
\nonumber \\
={q \over 72} \epsilon_{k_1 k_2}{}^{\ell_1 \ell_2 \ell_3 \ell_4 \ell_5 \ell_6 \ell_7} (W-h \wedge \Psi)_{\ell_1 \ell_2 \ell_3}
X_{\ell_4 \ell_5 \ell_6 \ell_7}
\nonumber \\
-{qr \over 18} \epsilon_{k_1 k_2}{}^{\ell_1 \ell_2 \ell_3 \ell_4 \ell_5 \ell_6 \ell_7} h_{\ell_1} Z_{\ell_2 \ell_3 \ell_4} W_{\ell_5 \ell_6 \ell_7} \ .
\nonumber \\
\end{eqnarray}

{\flushleft{The $-k_1 k_2$ component:}}
\begin{eqnarray}
\label{g4eq3}
{{\tilde{\nabla}}}^\ell Z_{\ell k_1 k_2} &=& {\dot{\Psi}}_{k_1 k_2}
+h^\ell Z_{\ell k_1 k_2}
-{\dot{\gamma}}_{k_1}{}^\ell \Psi_{\ell k_2}
+{\dot{\gamma}}_{k_2}{}^\ell \Psi_{\ell k_1}
+{1 \over 2}{\dot{\gamma}}_m{}^m \Psi_{k_1 k_2}
\nonumber \\
&+&{q \over 72} \epsilon_{k_1 k_2}{}^{\ell_1 \ell_2 \ell_3 \ell_4 \ell_5 \ell_6 \ell_7} Z_{\ell_1 \ell_2 \ell_3}
X_{\ell_4 \ell_5 \ell_6 \ell_7}
\end{eqnarray}
and the $k_1 k_2 k_3$ component:
\begin{eqnarray}
\label{g4eq4}
{{\tilde{\nabla}}}^\ell (X-r h \wedge Z)_{\ell k_1 k_2 k_3}
+2r \Delta Z_{k_1 k_2 k_3}
-(h^\ell+r {\dot{h}}^\ell) X_{\ell k_1 k_2 k_3}
+(W-h \wedge \Psi)_{k_1 k_2 k_3}
\nonumber \\
+r (\dot{W}-{\dot{h}} \wedge \Psi -h \wedge {\dot \Psi})_{k_1 k_2 k_3}
-3r {\dot{\gamma}}^\ell{}_{[k_1} (W-h \wedge \Psi)_{k_2 k_3] \ell}
-r h^\ell {\dot{X}}_{\ell k_1 k_2 k_3}
\nonumber \\
+2r h^\ell(h \wedge Z)_{\ell k_1 k_2 k_3}
+{1 \over 2} r {\dot{\gamma}}_m{}^m (W-h \wedge \Psi)_{k_1 k_2 k_3}
\nonumber \\
-{1 \over 2} r(-2 h^\ell {\dot{\gamma}}_\ell{}^q + h^q {\dot{\gamma}}_m{}^m)
X_{q k_1 k_2 k_3} -3r h^q {\dot{\gamma}}_{[k_1}{}^\ell X_{k_2 k_3] \ell q}
\nonumber \\
=-{q \over 24} \epsilon_{k_1 k_2 k_3}{}^{\ell_1 \ell_2 \ell_3 \ell_4 \ell_5 \ell_6} \Psi_{\ell_1 \ell_2} (X-r h \wedge Z)_{\ell_3 \ell_4 \ell_5 \ell_6}
\nonumber \\
+{qr \over 18} \epsilon_{k_1 k_2 k_3}{}^{\ell_1 \ell_2 \ell_3 \ell_4 \ell_5 \ell_6} (W-h \wedge \Psi)_{\ell_1 \ell_2 \ell_3}
Z_{\ell_4 \ell_5 \ell_6} \ .
\nonumber \\
\end{eqnarray}

It should be noted that in ({\ref{g4eq1}})-({\ref{g4eq4}}), we have supppressed
the appearance of terms of the form $\dot{Z}$, and also
$\dot{\Delta} Z, {\dot{h}} Z, {\dot{\gamma}} Z$, because as we explained in section \ref{sec:mod_space_11D}, $Z$ is linear in the moduli, and hence these
terms are suppressed in the moduli space calculation.
Furthermore, ({\ref{g4eq2}}) has been simplified by making
use of ({\ref{g4eq3}}) to eliminate the ${{\tilde{\nabla}}}^\ell Z_{\ell k_1 k_2}$ term from ({\ref{g4eq2}}).

The Einstein equation (\ref{d11ein}) decomposes into the following components:

{\flushleft{The $++$ component:}}
\begin{eqnarray}
R_{++} = \frac{1}{12} r^2 (W- h\wedge \Psi)_{\ell_1\ell_2\ell_3} (W - h\wedge \Psi + r \Delta Z)^{\ell_1\ell_2\ell_3}   \ .
\end{eqnarray}
The $+-$ component:
\begin{eqnarray}
\label{+-ein}
\nonumber
R_{+-} &=& - \frac{1}{6} \Psi_{\ell_1\ell_2} \Psi^{\ell_1\ell_2} + \frac{1}{36} r (W - h\wedge \Psi )_{\ell_1\ell_2\ell_3} Z^{\ell_1\ell_2\ell_3} \\
&-& \frac{1}{144}  X_{\ell_1\ell_2\ell_3\ell_4} X^{\ell_1\ell_2\ell_3\ell_4} + \frac{1}{72} r (h\wedge Z)_{\ell_1\ell_2\ell_3\ell_4} X^{\ell_1\ell_2\ell_3\ell_4} \ .
\end{eqnarray}
The $+i$ component:
\begin{eqnarray}
\nonumber
R_{+i} &=& - \frac{1}{4} r \Psi_{\ell_1\ell_2} (W- h\wedge \Psi + \frac{1}{2} r \Delta Z)_i{}^{\ell_1\ell_2} 
\nonumber \\
&-& \frac{1}{12} r^2 (W - h\wedge \Psi)_{\ell_1\ell_2\ell_3} (h\wedge Z)_i{}^{\ell_1\ell_2\ell_3} \nonumber \\
&+& \frac{1}{12} r X_i{}^{\ell_1\ell_2\ell_3} ( W -h\wedge \Psi + \frac{1}{2} r \Delta Z )_{\ell_1\ell_2\ell_3} \ .
\end{eqnarray}
The $-i$ component:
\begin{eqnarray}
\label{-iein}
R_{-i} = \frac{1}{4} \Psi_{\ell_1\ell_2} Z_i{}^{\ell_1\ell_2} + \frac{1}{12} Z_{\ell_1\ell_2\ell_3} X_i{}^{\ell_1\ell_2\ell_3} \ .
\end{eqnarray}
The $ij$ component:
\begin{eqnarray}
\nonumber
R_{ij} &=& - \frac{1}{2} \Psi_{i\ell} \Psi_j{}^{\ell} + \frac{1}{2} r (W - h\wedge\Psi + \frac{1}{2} r \Delta Z)_{\ell_1\ell_2 (i} Z_{j)}{}^{\ell_1\ell_2} + \frac{1}{12} X_{i\ell_1\ell_2\ell_3} X_j{}^{\ell_1\ell_2\ell_3} \\
\nonumber
&+& \frac{1}{6} r (h\wedge Z)_{\ell_1\ell_2\ell_3 (i} X_{j)}{}^{\ell_1\ell_2\ell_3} + \frac{1}{12} \gamma_{ij} \Psi_{\ell_1\ell_2} \Psi^{\ell_1\ell_2} - \frac{1}{144} \gamma_{ij} X_{\ell_1\ell_2\ell_3\ell_4} X^{\ell_1\ell_2\ell_3\ell_4} \\
&-& \frac{1}{18} r \gamma_{ij}(W - h\wedge \Psi )_{\ell_1\ell_2\ell_3} Z^{\ell_1\ell_2\ell_3}  + \frac{1}{72} r \gamma_{ij} (h\wedge Z)_{\ell_1\ell_2\ell_3\ell_4} X^{\ell_1\ell_2\ell_3\ell_4} \ .
\end{eqnarray}

\clearpage{\pagestyle{empty}\cleardoublepage} 


\chapter{Simplification of the reduced KSEs}
\label{apx:simp_KSEs}

In this appendix, we show the detail of the simplification of the reduced KSEs obtained in section (\ref{sec:KSE}).  We shall first consider separately the cases whether $\phi^{[0]}_+ \equiv 0$ and $\phi^{[0]}_+ \neq 0$. In these two cases, the analysis makes use of global arguments (e.g. the Hopf maximum principle). The conditions on the bosonic fields obtained will be the same. Finally, we show that by using local arguments we can simplify the reduced KSEs to the pair (\ref{gravsimp}) and (\ref{algsimpmax}), which are the naive restriction of the gravitino and dilatino KSEs to $\cS$.

\section{Solutions with vanishing positive chirality spinors}
\label{apx:simp_KSEs_1}

Suppose that there exists a Killing spinor $\epsilon$ with
$\epsilon^{[0]} \not \equiv 0$, but $\phi_+^{[0]} \equiv 0$ and $\eta_+^{[0]} \equiv 0$. 
Such a spinor must therefore have $\eta_-^{[0]} \not \equiv 0$, and hence from (\ref{mu-2}) it follows that
\begin{eqnarray}
h^{[0]}+N^{[0]}=0 \ .
\end{eqnarray}
Then ({\ref{par2}}) implies that
\begin{eqnarray}
\label{partrans}
d \parallel \eta_-^{[0]} \parallel^2 = - \parallel \eta_-^{[0]} \parallel^2
h^{[0]} \ .
\end{eqnarray}
In particular, this condition implies that if
$\eta_-^{[0]}$ vanishes at any point on the horizon
section, then $\eta_-^{[0]}=0$ everywhere.
So, $\eta_-^{[0]}$ must be everywhere non-vanishing.

On taking the divergence of ({\ref{partrans}}), and
making use of the $N_1=+, N_2=-$ component of the 2-form gauge potential field equation ({\ref{geq1}}), one obtains the following condition

\begin{eqnarray}
\tilde{\nabla}^{[0] i} \tilde{\nabla}^{[0]}_i \parallel \eta_-^{[0]} \parallel^2 - \big(2 \tilde{\nabla}^i \Phi^{[0]} + \parallel \eta_-^{[0]} \parallel^{-2} \tilde{\nabla}^{[0] i}  \parallel \eta_-^{[0]} \parallel^2 \big) \tilde{\nabla}^{[0]}_i  \parallel \eta_-^{[0]} \parallel^2 =0 \ .
\end{eqnarray}
As $ \parallel \eta_-^{[0]} \parallel^2$ is nowhere vanishing, an application of the maximum principle
implies that $ \parallel \eta_-^{[0]} \parallel^2=const.$, and hence ({\ref{partrans}})
gives that
\begin{eqnarray}
h^{[0]}=0 \ , \qquad N^{[0]}=0 \ .
\end{eqnarray}
These conditions, together with  ({\ref{alg4a}}), imply that
\begin{eqnarray}
\Delta=\mathcal{O}(\alpha'^2) \ .
\end{eqnarray}
Then the dilaton field equation ({\ref{deq}}) implies that
\begin{eqnarray}
\tilde{\nabla}^i \tilde{\nabla}_i (e^{-2 \Phi}) ={1 \over 6} e^{-2 \Phi} W_{ijk} W^{ijk} + \mathcal{O}(\alpha')~,
\end{eqnarray}
and hence it follows that
\begin{eqnarray}
\Phi^{[0]}=const, \qquad W^{[0]}=0 \ .
\end{eqnarray}
Furthermore, this then implies that
\begin{eqnarray}
H=du \wedge dr \wedge N +r du \wedge Y + W + \mathcal{O}(\alpha'^2)~,
\end{eqnarray}
and hence
\begin{eqnarray}
dH = du \wedge dr \wedge (dN-Y)-r du \wedge dY +dW + \mathcal{O}(\alpha'^2)~.
\end{eqnarray}
As the $ruij$ component on the RHS of the Bianchi identity is $\mathcal{O}(\alpha'^2)$
this implies that
\begin{eqnarray}
Y=dN+\mathcal{O}(\alpha'^2) \ ,
\end{eqnarray}
and in particular, $Y^{[0]}=0$.

Next consider the gauge equations. The $+-$ component of the 2-form gauge potential field equations ({\ref{geq1}}) is
\begin{eqnarray}
\label{dfree1}
\tilde{\nabla}^i N_i = \mathcal{O}(\alpha'^2)~.
\end{eqnarray}
Also, the $u$-dependent part of ({\ref{par3}}) implies that
\begin{eqnarray}
\tilde{\nabla}_i (h+N)_j \Gamma^j \eta_- = \mathcal{O}(\alpha'^2)~,
\end{eqnarray}
which gives that
\begin{eqnarray}
\label{udepsimp1}
\tilde{\nabla}_i(h+N)_j = \mathcal{O}(\alpha'^2)~.
\end{eqnarray}
Taking the trace of this expression, and using ({\ref{dfree2}}) yields
\begin{eqnarray}
\label{dfree2}
\tilde{\nabla}^i h_i = \mathcal{O}(\alpha'^2)~.
\end{eqnarray}
Next, recall that the gravitino KSE ({\ref{par4}}) implies
\begin{eqnarray}
\label{dnsq1}
\tilde{\nabla}_i \parallel \eta_- \parallel^2 = -{1 \over 2}(h-N)_i  \parallel \eta_- \parallel^2 + \mathcal{O}(\alpha'^2) \ .
\end{eqnarray}
Taking the divergence yields, together with ({\ref{dfree1}}) and ({\ref{dfree2}}) the condition
\begin{eqnarray}
\tilde{\nabla}^i \tilde{\nabla}_i  \parallel \eta_- \parallel^2 = \mathcal{O}(\alpha'^2)\ ,
\end{eqnarray}
which implies that $ \parallel \eta_- \parallel^2= const + \mathcal{O}(\alpha'^2)$.
Substituting back into ({\ref{dnsq1}}) gives the condition
$N=h+\mathcal{O}(\alpha'^2)$, and hence ({\ref{udepsimp1}}) implies
that
\begin{eqnarray}
\tilde{\nabla}_i h_j = \mathcal{O}(\alpha'^2) \ .
\end{eqnarray}

So, to summarize, for this class of solutions, we have obtained the following
 conditions on the fields
\begin{eqnarray}
\label{bossimp1}
N=h+\mathcal{O}(\alpha'^2), && \quad h^{[0]}=0, \quad Y=\mathcal{O}(\alpha'^2), \quad \tilde{\nabla}_i h_j= \mathcal{O}(\alpha'^2),
\nonumber \\
\Delta = \mathcal{O}(\alpha'^2), && \quad H^{[0]}=0, \quad \Phi^{[0]}=const~,
\end{eqnarray}
and it is straightforward to check that the generic conditions on
$\phi_+$ then simplify to
\begin{eqnarray}
\label{par3bb}
\tilde{\nabla}_i \phi_+ -{1 \over 8}W_{ijk} \Gamma^{jk} \phi_+= \mathcal{O}(\alpha'^2)~,
\end{eqnarray}
and
\begin{eqnarray}
\label{auxalg1bbb}
\bigg(\Gamma^i \tilde{\nabla}_i \Phi +{1 \over 2} h_i \Gamma^i -{1 \over 12} W_{ijk} \Gamma^{ijk} \bigg) \phi_+= \mathcal{O}(\alpha'^2)\ ,
\end{eqnarray}
and
\begin{eqnarray}
\label{auxalg1cbb}
{\tilde{F}}_{ij} \Gamma^{ij} \phi_+ = \mathcal{O}(\alpha') \ .
\end{eqnarray}

The generic conditions on $\eta_-$ also simplify to
\begin{eqnarray}
\label{par4bb}
\tilde{\nabla}_i \eta_- -{1 \over 8}W_{ijk} \Gamma^{jk} \eta_-= \mathcal{O}(\alpha'^2) \ ,
\end{eqnarray}
and
\begin{eqnarray}
\label{auxalg2bbb}
\bigg(\Gamma^i \tilde{\nabla}_i \Phi -{1 \over 2} h_i \Gamma^i -{1 \over 12} W_{ijk} \Gamma^{ijk} \bigg) \eta_-= \mathcal{O}(\alpha'^2)\ ,
\end{eqnarray}
and
\begin{eqnarray}
\label{auxalg2cbb}
{\tilde{F}}_{ij} \Gamma^{ij} \eta_- = \mathcal{O}(\alpha') \ .
\end{eqnarray}

In the next section, we shall  consider the case for which there exists a Killing spinor with
$\phi_+^{[0]} \not \equiv 0$.
It will be shown that the  conditions ({\ref{bossimp1}}) on the bosonic fields
and the simplified KSEs listed above correspond to special cases
of the corresponding  conditions on the fields and simplified KSEs
of  $\phi_+^{[0]} \not \equiv 0$. In particular,
this will allow the KSEs for $\phi_+^{[0]} \equiv 0$
and $\phi_+^{[0]} \not \equiv 0$ to be written in a unified way.

\section{Solutions with non-vanishing positive chirality spinors}
\label{apx:simp_KSEs_2}

Suppose that there exists a Killing
spinor $\epsilon$, with $\epsilon^{[0]} \not \equiv 0$ and
$\phi_+^{[0]} \not \equiv 0$. Then consider ({\ref{par1}}); this implies that
\begin{eqnarray}
\label{pt1}
\tilde{\nabla}_i \parallel \phi_+ \parallel^2 = {1 \over 2}(h_i-N_i)\parallel \phi_+ \parallel^2 + \mathcal{O}(\alpha'^2)~,
\end{eqnarray}
and ({\ref{alg7}}) gives that
\begin{eqnarray}
\label{alg7b}
\tilde{\nabla}_i (h-N)_j + {1 \over 2}(h_i N_j - h_j N_i)
-{1 \over 2}(h_i h_j -N_i N_j)
\nonumber \\
-(dh-Y)_{ij} -{1 \over 2} W_{ijk}(h-N)^k = \mathcal{O}(\alpha'^2)~.
\end{eqnarray}
Taking the divergence of ({\ref{pt1}}), and using ({\ref{par1}})
together with the trace of ({\ref{alg7b}}), we find that
\begin{eqnarray}
\label{lapsq1}
\tilde{\nabla}^i \tilde{\nabla}_i \parallel \phi_+ \parallel^2 - h^i \tilde{\nabla}_i \parallel \phi_+ \parallel^2 = \mathcal{O}(\alpha'^2)~.
\end{eqnarray}
An application of the maximum principle (see e.g. \cite{maxp})
then yields the condition
\begin{eqnarray}
\tilde{\nabla}_i \parallel \phi_+ \parallel^2= \mathcal{O}(\alpha'^2)~.
\end{eqnarray}

To see this, note that to zeroth order in $\alpha'$,
({\ref{lapsq1}}) implies that $\tilde{\nabla}^{[0]}_i \parallel \phi_+^{[0]}\parallel^2=0$, on applying the maximum principle.
Then ({\ref{pt1}}) and ({\ref{alg7b}}) imply that $N^{[0]}=h^{[0]}$ and $Y^{[0]}=dh^{[0]}$; and from ({\ref{alg4a}}) we also have $\Delta^{[0]}=0$.
Then it is useful to consider the  field equations of the 2-form gauge potential
({\ref{geq1}}), which imply that
\begin{eqnarray}
\label{bcx1}
\tilde{\nabla}^i \bigg( e^{-2 \Phi} h_i \bigg)= \mathcal{O}(\alpha')~,
\end{eqnarray}
and
\begin{eqnarray}
\label{bcx2}
e^{2 \Phi} \tilde{\nabla}^j \big(e^{-2 \Phi} dh_{ji}\big)
+{1 \over 2} W_{ijk} dh^{jk} + h^j dh_{ji}= \mathcal{O}(\alpha')~,
\end{eqnarray}
and the Einstein equations imply that
\begin{eqnarray}
\label{bcx3}
{\tilde{R}}_{ij} + \tilde{\nabla}_{(i} h_{j)} -{1 \over 4} W_{imn} W_j{}^{mn}
+2 \tilde{\nabla}_i \tilde{\nabla}_j \Phi  = \mathcal{O}(\alpha')~.
\end{eqnarray}
Using ({\ref{bcx1}}), ({\ref{bcx2}}) and ({\ref{bcx3}})
it follows that{\footnote{We remark that
the condition ({\ref{bcx4}}) was also obtained
in \cite{hethor}. In that case, a bilinear matching condition
was imposed in order to find $N^{[0]}=h^{[0]}, Y^{[0]}=dh^{[0]}$.
Here we do not assume such a bilinear matching condition, but nevertheless
we find the same condition.}}
\begin{eqnarray}
\label{bcx4}
&&\tilde{\nabla}^i \tilde{\nabla}_i h^2 + (h-2 d \Phi)^j \tilde{\nabla}_j h^2
= 2 \tilde{\nabla}^{(i} h^{j)} \tilde{\nabla}_{(i} h_{j)}
\cr
&&~~~~~~~~~~~~~~~~
+{1 \over 2}(dh - i_h W)_{ij} (dh-i_h W)^{ij} + \mathcal{O}(\alpha')~.
\end{eqnarray}
In particular, ({\ref{bcx4}}) implies that $\tilde{\nabla}^{[0] i} h^{[0]}_i =0$
on applying the maximum principle.
It follows from ({\ref{lapsq1}}) that
\begin{eqnarray}
\tilde{\nabla}^{[0]i} \tilde{\nabla}^{[0]}_i \langle \phi_+^{[0]}, \phi_+^{[1]} \rangle
- h^{[0]i}  \tilde{\nabla}^{[0]}_i \langle \phi_+^{[0]}, \phi_+^{[1]} \rangle =0~.
\end{eqnarray}
On multiplying this condition by $\langle \phi_+^{[0]}, \phi_+^{[1]} \rangle$
and integrating by parts, using $\tilde{\nabla}^{[0] i} h^{[0]}_i =0$, one finds that
$\tilde{\nabla}^{[0]}_i \langle \phi_+^{[0]}, \phi_+^{[1]} \rangle =0$ as well.
So, it follows that $\tilde{\nabla}_i \parallel \phi_+ \parallel^2 = \mathcal{O}(\alpha'^2)$.

Then, ({\ref{pt1}}) also implies that $N=h+\mathcal{O}(\alpha'^2)$.
Substituting these conditions back into ({\ref{alg4a}}),
we find that $\Delta^{[1]}=0$ as well, so $\Delta=\mathcal{O}(\alpha'^2)$.
Also, ({\ref{alg7b}}) implies that
\begin{eqnarray}
Y -dh= {\cal{O}}(\alpha'^2)~.
\end{eqnarray}

To summarize the conditions on the bosonic fields;
we have shown that for solutions with $\phi_+^{[0]} \neq 0$, we must have
\begin{eqnarray}
\label{bossimp2}
\Delta=\mathcal{O}(\alpha'^2), \qquad N=h+\mathcal{O}(\alpha'^2), \qquad Y=dh+\mathcal{O}(\alpha'^2)
\end{eqnarray}
which implies that
\begin{eqnarray}
H = d (\mathbf{e}^- \wedge \mathbf{e}^+) + W + \mathcal{O}(\alpha'^2)~.
\end{eqnarray}
The  field equation ({\ref{geq1}}) of the 2-form gauge potential
can then be rewritten in terms of the near-horizon data
as
\begin{eqnarray}
\label{geq1a}
\tilde{\nabla}^i \big( e^{-2 \Phi} h_i \big)= \mathcal{O}(\alpha'^2)~,
\end{eqnarray}
\begin{eqnarray}
\label{geq1b}
e^{2 \Phi} \tilde{\nabla}^j \big(e^{-2 \Phi} dh_{ji}\big)
+{1 \over 2} W_{ijk} dh^{jk} + h^j dh_{ji}= \mathcal{O}(\alpha'^2)~,
\end{eqnarray}
and
\begin{eqnarray}
\label{geq1c}
e^{2 \Phi} \tilde{\nabla}^k \big(e^{-2 \Phi} W_{kij}\big)
+ dh_{ij} - h^k W_{kij} = \mathcal{O}(\alpha'^2)~.
\end{eqnarray}
In addition, ${\cal{P}}=\mathcal{O}(\alpha') $ and so $F= {\tilde{F}} + \mathcal{O}(\alpha')$.
The $i,j$ component of the Einstein equation then simplifies to
\begin{eqnarray}
\label{einsp}
{\tilde{R}}_{ij} + \tilde{\nabla}_{(i} h_{j)} -{1 \over 4} W_{imn} W_j{}^{mn}
+2 \tilde{\nabla}_i \tilde{\nabla}_j \Phi
\nonumber \\
+{\alpha' \over 4} \bigg(-2 dh_{i \ell}
dh_j{}^\ell + \tilde{R}^{(+)}{}_{i \ell_1, \ell_2 \ell_3}
\tilde{R}^{(+)}{}_j{}^{\ell_1, \ell_2 \ell_3}
- {\tilde{F}}_{i\ell}{}^{ab} {\tilde{F}}_j{}^\ell{}_{ab} \bigg) =\mathcal{O}(\alpha'^2)~.
\end{eqnarray}
Furthermore, the dilaton field equation can be written as
\begin{eqnarray}
\label{deqsimp1}
\tilde{\nabla}^i \tilde{\nabla}_i \Phi - h^i \tilde{\nabla}_i \Phi -2 \tilde{\nabla}^i \Phi \tilde{\nabla}_i \Phi -{1 \over 2} h_i h^i
+{1 \over 12} W_{ijk} W^{ijk}
\nonumber \\
+{\alpha' \over 16} \big(2 dh_{ij} dh^{ij}
+ {\tilde{F}}_{ij}{}^{ab} {\tilde{F}}^{ij}{}_{ab}
- \tilde{R}^{(+)}{}_{\ell_1 \ell_2, \ell_3 \ell_4}
\tilde{R}^{(+)}{}^{\ell_1 \ell_2, \ell_3 \ell_4} \big) = \mathcal{O}(\alpha'^2) \ .
\end{eqnarray}

On making use of the conditions (\ref{bossimp2})  on the bosonic fields, the KSEs on
$\phi_+$ then simplify further to
\begin{eqnarray}
\label{par3}
\tilde{\nabla}_i \phi_+ -{1 \over 8}W_{ijk} \Gamma^{jk} \phi_+= \mathcal{O}(\alpha'^2)~,
\end{eqnarray}
\begin{eqnarray}
\label{auxalg1}
dh_{ij} \Gamma^{ij} \phi_+= \mathcal{O}(\alpha'^2)~,
\end{eqnarray}
\begin{eqnarray}
\label{auxalg1b}
\bigg(\Gamma^i \tilde{\nabla}_i \Phi +{1 \over 2} h_i \Gamma^i -{1 \over 12} W_{ijk} \Gamma^{ijk} \bigg) \phi_+= \mathcal{O}(\alpha'^2)~,
\end{eqnarray}
and
\begin{eqnarray}
\label{auxalg1c}
{\tilde{F}}_{ij} \Gamma^{ij} \phi_+ = \mathcal{O}(\alpha') \ .
\end{eqnarray}

Furthermore, KSEs on $\eta_-$ also simplify to
\begin{eqnarray}
\label{par4}
\tilde{\nabla}_i \eta_- -{1 \over 8}W_{ijk} \Gamma^{jk} \eta_-= \mathcal{O}(\alpha'^2)~,
\end{eqnarray}
\begin{eqnarray}
\label{auxalg2}
dh_{ij} \Gamma^{ij} \eta_-= \mathcal{O}(\alpha'^2)~,
\end{eqnarray}
\begin{eqnarray}
\label{auxalg2b}
\bigg(\Gamma^i \tilde{\nabla}_i \Phi -{1 \over 2} h_i \Gamma^i -{1 \over 12} W_{ijk} \Gamma^{ijk} \bigg) \eta_-= \mathcal{O}(\alpha'^2)~,
\end{eqnarray}
and
\begin{eqnarray}
\label{auxalg2c}
{\tilde{F}}_{ij} \Gamma^{ij} \eta_- = \mathcal{O}(\alpha') \ .
\end{eqnarray}
In both cases above, (\ref{par3}) and (\ref{par4}) are a consequence of the gravitino KSE, (\ref{auxalg1b}) and (\ref{auxalg2b}) are associated to the dilatino KSE,
while (\ref{auxalg1c}) and (\ref{auxalg2c}) are derived from the gaugino KSE. The two additional conditions (\ref{auxalg1}) and (\ref{auxalg2}) can be thought of
as integrability conditions.

\section{Further Simplification of the KSEs}

\label{ind}

Here we shall show that the independent  KSEs are given in (\ref{gravsimp}) and (\ref{algsimpmax}).
We first note that the  conditions on the bosonic fields  ({\ref{bossimp1}})
(obtained from the case when $\phi_+^{[0]} \equiv 0$)
actually imply those of  ({\ref{bossimp2}})
(corresponding to the $\phi_+^{[0]} \not \equiv 0$ case). Furthermore, the KSEs ({\ref{par3bb}}), ({\ref{auxalg1bbb}}),
({\ref{auxalg1cbb}}), ({\ref{par4bb}}), ({\ref{auxalg2bbb}})
and ({\ref{auxalg2cbb}}) are identical to
the KSE ({\ref{par3}}), ({\ref{auxalg1b}}), ({\ref{auxalg1c}}), ({\ref{par4bb}}), ({\ref{auxalg2b}})
and ({\ref{auxalg2c}}).
Hence, we shall concentrate on the simplification of the KSEs associated
with the case $\phi_+^{[0]} \not \equiv 0$, as the simplification of
the KSEs in the case $\phi_+^{[0]} \equiv 0$ follows in exactly the same way.

\subsection{Elimination of conditions ({\ref{auxalg1}}),
({\ref{auxalg1c}}), ({\ref{auxalg2}}), ({\ref{auxalg2c}})}

Let us assume ({\ref{par3}}), ({\ref{auxalg1b}}),
({\ref{par4}}) and ({\ref{auxalg2b}}). Then acting on the algebraic conditions
({\ref{auxalg1b}}) and ({\ref{auxalg2b}}) with the Dirac operator
$\Gamma^\ell \tilde{\nabla}_\ell$, one obtains
\begin{eqnarray}
\bigg(\tilde{\nabla}_i \tilde{\nabla}^i \Phi \mp h^i \tilde{\nabla}_i \Phi -2 \tilde{\nabla}^i \Phi \tilde{\nabla}_i \Phi
-{1 \over 2} h_i h^i +{1 \over 12} W_{ijk} W^{ijk}+{1 \over 4} (1\pm1) dh_{ij} \Gamma^{ij}
\nonumber \\
 +{1 \over 4} (-1\pm 1)h^k W_{kij}
\Gamma^{ij} -{1 \over 48} dW_{\ell_1 \ell_2 \ell_3 \ell_4}
\Gamma^{\ell_1 \ell_2 \ell_3 \ell_4} \bigg) \phi_\pm = \mathcal{O}(\alpha'^2)~,
\end{eqnarray}
where we have made use of the field equations ({\ref{geq1a}}) and
({\ref{geq1c}}), together with the algebraic conditions
({\ref{auxalg1b}}) and ({\ref{auxalg2b}}).
Next, on substituting the dilaton equation and the Bianchi identity
into the above expression, one finds
\begin{eqnarray}
\label{reduc1}
\bigg({1 \over 2}(1 \mp 1) \tilde{\nabla}^i h_i
+{1 \over 4} (1 \pm 1) dh_{ij} \Gamma^{ij}
+{1 \over 4}(-1 \pm 1) (i_h W)_{ij} \Gamma^{ij}
\nonumber \\
\hspace{-5mm}+{\alpha' \over 32} \big( 2dh_{ij} \Gamma^{ij} dh_{pq} \Gamma^{pq}
+{\tilde{F}}_{ij}{}^{ab} \Gamma^{ij}
{\tilde{F}}_{pq ab}\Gamma^{pq}
- \tilde{R}^{(+)}{}_{ij,}{}^{mn}
\Gamma^{ij} \tilde{R}^{(+)}{}_{pq,mn} \Gamma^{pq}\big) \bigg) \phi_\pm
=\mathcal{O}(\alpha'^2)~.
\end{eqnarray}
Further simplification can be obtained by noting that
the integrability conditions of the KSE ({\ref{par3}}) and ({\ref{par4}})
are
\begin{eqnarray}
\tilde{R}^{(-)}{}_{ij,pq} \Gamma^{pq} \phi_\pm = \mathcal{O}(\alpha'^2)~,
\end{eqnarray}
and hence
\begin{eqnarray}
\tilde{R}^{(+)}{}_{pq,ij} \Gamma^{pq} \phi_\pm = \mathcal{O}(\alpha')~,
\end{eqnarray}
from which it follows that the final term on the RHS of
({\ref{reduc1}}) is $\mathcal{O}(\alpha'^2)$ and hence can be neglected.
So, ({\ref{reduc1}}) is equivalent to
\begin{eqnarray}
\label{reduc1b}
\bigg({1 \over 2}(1 \mp 1) \tilde{\nabla}^i h_i
+{1 \over 4} (1 \pm 1) dh_{ij} \Gamma^{ij}
+{1 \over 4}(-1 \pm 1) (i_h W)_{ij} \Gamma^{ij}
\nonumber \\
+{\alpha' \over 16} dh_{ij} \Gamma^{ij} dh_{pq} \Gamma^{pq}
+{\alpha' \over 32} {\tilde{F}}_{ij}{}^{ab} \Gamma^{ij}
{\tilde{F}}_{pqab}\Gamma^{pq} \bigg) \phi_\pm
=\mathcal{O}(\alpha'^2)~.
\end{eqnarray}

We begin by considering the condition which ({\ref{reduc1b}}) imposes
on $\phi_+$:
\begin{eqnarray}
\label{reduc1c}
\bigg({1 \over 2} dh_{ij} \Gamma^{ij}
+{\alpha' \over 16} dh_{ij} \Gamma^{ij} dh_{pq} \Gamma^{pq}
+{\alpha' \over 32} {\tilde{F}}_{ij}{}^{ab} \Gamma^{ij}
{\tilde{F}}_{pqab}\Gamma^{pq} \bigg) \phi_+
=\mathcal{O}(\alpha'^2) \ .
\end{eqnarray}
To zeroth order this gives
\begin{eqnarray}
dh_{ij} \Gamma^{ij} \phi_+ = \mathcal{O}(\alpha')~,
\end{eqnarray}
which implies that the second term on the LHS of ({\ref{reduc1c}}) is
of $\mathcal{O}(\alpha'^2)$, and hence can be neglected. Using this,
({\ref{reduc1c}}) gives that
\begin{eqnarray}
\alpha' \langle {\tilde{F}}_{ij}{}^{ab} \Gamma^{ij} \phi_+,
{\tilde{F}}_{pqab} \Gamma^{pq} \phi_+ \rangle = \mathcal{O}(\alpha'^2)~,
\end{eqnarray}
which implies that
\begin{eqnarray}
{\tilde{F}}_{ij}{}^{ab} \Gamma^{ij} \phi_+ = \mathcal{O}(\alpha') \ .
\end{eqnarray}
Using this the third term on the LHS of ({\ref{reduc1c}})
is also of $\mathcal{O}(\alpha'^2)$. So, the remaining content
of ({\ref{reduc1c}}) is
\begin{eqnarray}
dh_{ij} \Gamma^{ij} \phi_+ = \mathcal{O}(\alpha'^2) \ .
\end{eqnarray}
Hence, we have proven that the KSE ({\ref{par3}}) and ({\ref{auxalg1b}})
imply the algebraic KSE ({\ref{auxalg1}}) and  ({\ref{auxalg1c}}).

Next, we consider the condition which  ({\ref{reduc1b}}) imposes
on $\phi_-$, which is
\begin{eqnarray}
\label{reduc1d}
\notag
\bigg(\tilde{\nabla}^i h_i &-&{1 \over 2} (i_h W)_{ij} \Gamma^{ij} \\
+{\alpha' \over 32}\big( 2dh_{ij} \Gamma^{ij} dh_{pq} \Gamma^{pq}
&+& {\tilde{F}}_{ij}{}^{ab} \Gamma^{ij}
{\tilde{F}}_{pq ab}\Gamma^{pq}\big) \bigg) \phi_-
=\mathcal{O}(\alpha'^2)~.
\end{eqnarray}
However, note also that the $u$-dependent part of
({\ref{par3}}), with ({\ref{par4}}), implies that
\begin{eqnarray}
\label{udep1}
\bigg(\tilde{\nabla}_i h_j - {1 \over 2} W_{ijk} h^k \bigg) \Gamma^j \phi_-= \mathcal{O}(\alpha'^2)~.
\end{eqnarray}
On contracting this expression with $\Gamma^i$,
we find
\begin{eqnarray}
\label{udep1b}
\bigg(\tilde{\nabla}^i h_i +{1 \over 2} dh_{ij}\Gamma^{ij} -{1 \over 2}
(i_h W)_{ij} \Gamma^{ij} \bigg) \phi_- = \mathcal{O}(\alpha'^2)~,
\end{eqnarray}
and on substituting this expression into ({\ref{reduc1d}}) we get
\begin{eqnarray}
\label{reduc1e}
\bigg(-{1 \over 2} dh_{ij} \Gamma^{ij}
+{\alpha' \over 16} dh_{ij} \Gamma^{ij} dh_{pq} \Gamma^{pq}
+{\alpha' \over 32} {\tilde{F}}_{ij}{}^{ab} \Gamma^{ij}
{\tilde{F}}_{pqab}\Gamma^{pq} \bigg) \phi_-
=\mathcal{O}(\alpha'^2)~.
\end{eqnarray}
Hence, we find from exactly the
same reasoning which was used to analyse the conditions on $
\phi_+$, that ({\ref{par4}}) and ({\ref{auxalg2b}}) imply
({\ref{auxalg2}}) and ({\ref{auxalg2c}}).

So, on making use of the  field equations, it follows that
the necessary and sufficient conditions for supersymmetry simplify
to the conditions ({\ref{par3}}) and ({\ref{auxalg1b}}) on
$\phi_+$, and to ({\ref{par4}}) and ({\ref{auxalg2b}}) on $\eta_-$.
We remark that the $u$-dependent parts of
the conditions ({\ref{par3}}) and ({\ref{auxalg1b}}) also impose
conditions on $\eta_-$. We shall examine the conditions on
$\eta_-$ further in the next section, and show how these may be
simplified.

\subsection{Elimination of $u$-dependent parts of ({\ref{par3}}) and
({\ref{auxalg1b}})}

We begin by considering the $u$-dependent parts of ({\ref{par3}}) and
({\ref{auxalg1b}}), assuming that ({\ref{par4}}) and ({\ref{auxalg2b}})
hold. The $u$-dependent part of the condition on $\phi_+$
obtained from ({\ref{par3}}) is
\begin{eqnarray}
\label{udepa}
\bigg(\tilde{\nabla}_i h_j - {1 \over 2} W_{ijk} h^k \bigg) \Gamma^j \eta_-= \mathcal{O}(\alpha'^2)~,
\end{eqnarray}
and the $u$-dependent part of the algebraic condition
({\ref{auxalg1b}}) is given by
\begin{eqnarray}
\label{udepb}
\bigg(\Gamma^i \tilde{\nabla}_i \Phi +{1 \over 2} h_i \Gamma^i
-{1 \over 12} W_{ijk}\Gamma^{ijk} \bigg) h_\ell \Gamma^\ell \eta_- = \mathcal{O}(\alpha'^2)~.
\end{eqnarray}
On adding $h_\ell \Gamma^\ell$ acting on ({\ref{auxalg2b}}) to the
above expression, we find that ({\ref{udepb}}) is equivalent to
the condition
\begin{eqnarray}
\label{udepc}
\bigg(\tilde{\nabla}^i h_i -{1 \over 2} h^i W_{ijk} \Gamma^{jk} \bigg) \eta_- = \mathcal{O}(\alpha'^2) \ ,
\end{eqnarray}
where we have also made use of the field equation ({\ref{geq1a}}).
On contracting ({\ref{udepa}}) with $\Gamma^i$, it then follows
that ({\ref{udepc}}) is equivalent
to
\begin{eqnarray}
dh_{ij} \Gamma^{ij} \eta_- = \mathcal{O}(\alpha'^2) \ .
\end{eqnarray}
However, as shown in the previous section, this condition
is implied by ({\ref{par4}}) and ({\ref{auxalg2b}})
on making use of the  field equations.

So, it remains to consider the condition ({\ref{udepa}}).
First, recall that the integrability conditions
of the gravitino equation of ({\ref{par4}}) is given
by
\begin{eqnarray}
\tilde{R}^{(-)}{}_{ij,k\ell} \Gamma^{k \ell} \eta_- = \mathcal{O}(\alpha'^2)~.
\end{eqnarray}
On contracting with $\Gamma^j$, one then obtains
\begin{eqnarray}
\label{minint1}
\bigg( \big(-2 {\tilde{R}}_{ij} +{1 \over 2} W_{imn}W_j{}^{mn}
-2 \tilde{\nabla}^k \Phi W_{kij} + dh_{ij} - h^k W_{kij} \big) \Gamma^j
\nonumber \\
+ \big(-{1 \over 6} (dW)_{ijk\ell}-{1 \over 3} \tilde{\nabla}_i W_{jk\ell}
+{1 \over 2} W_{ij}{}^m W_{k \ell m} \big) \Gamma^{jk\ell} \bigg)
\eta_- = \mathcal{O}(\alpha'^2)~,
\end{eqnarray}
where we have used the gauge equation ({\ref{geq1c}}).
Also, on taking the covariant derivative of the algebraic condition
({\ref{auxalg2b}}), and using ({\ref{par4}}), one also finds the
following mixed integrability condition
\begin{eqnarray}
\label{minint2}
\bigg( \big(\tilde{\nabla}_i \tilde{\nabla}_j \Phi -{1 \over 2} \tilde{\nabla}_i h_j +{1 \over 2} W_{ikj}
\tilde{\nabla}^k \Phi -{1 \over 4} W_{ikj} h^k \big) \Gamma^j
\nonumber \\
+\Gamma^{jk\ell} \big(-{1 \over 12} \tilde{\nabla}_i W_{jk\ell}
+{1 \over 8} W_{jkm}W_{i\ell}{}^m \big) \bigg) \eta_-=\mathcal{O}(\alpha'^2)~.
\end{eqnarray}
 On eliminating the $\tilde{\nabla}_i W_{jk\ell} \Gamma^{jk\ell}$ terms
between ({\ref{minint1}}) and ({\ref{minint2}}), one obtains
the condition
\begin{eqnarray}
\bigg(\big(-2 {\tilde{R}}_{ij} +{1 \over 2} W_{imn}W_j{}^{mn}
+dh_{ij} -2 h^k W_{kij} -4 \tilde{\nabla}_i \tilde{\nabla}_j \Phi
+2 \tilde{\nabla}_i h_j \big) \Gamma^j
\nonumber \\
-{1 \over 6} dW_{ijk\ell} \Gamma^{jk\ell}
\bigg) \eta_-=\mathcal{O}(\alpha'^2)~.
\end{eqnarray}
Next, we substitute the Einstein equation ({\ref{einsp}}) in order
to eliminate the Ricci tensor, and also use the Bianchi identity for $dW$.
One then obtains, after some rearrangement of terms, the following
condition
\begin{eqnarray}
\label{minint3}
\bigg( \big(4 \tilde{\nabla}_i h_j -2 h^k W_{kij} \big) \Gamma^j
+ \alpha' \big( {1 \over 2} dh_{ij} \Gamma^j dh_{k \ell} \Gamma^{k \ell}
+{1 \over 4} {\tilde{F}}_{ijab} \Gamma^j {\tilde{F}}_{k \ell}{}^{ab}
\Gamma^{k \ell}
\nonumber \\
 -{1 \over 4} \tilde{R}^{(+)}{}_{ij,}{}^{mn}
\Gamma^j \tilde{R}^{(+)}{}_{k \ell,mn} \Gamma^{k\ell} \big) \bigg)
\eta_- = \mathcal{O}(\alpha'^2)~.
\end{eqnarray}
The $\alpha'$ terms in the above expression can be neglected, as they
all give rise to terms which are in fact $\mathcal{O}(\alpha'^2)$.
This is because of the conditions ({\ref{auxalg2}}) and ({\ref{auxalg2c}}),
which we have already shown follow from ({\ref{par4}}) and ({\ref{auxalg2b}}), together with the bosonic conditions,
as well as the fact that
\begin{eqnarray}
\tilde{R}^{(+)}{}_{k\ell, mn} \Gamma^{k \ell}\eta_-=\mathcal{O}(\alpha')~,
\end{eqnarray}
which follows from the integrability condition of ({\ref{par4}}).
It follows that ({\ref{minint3}}) implies ({\ref{udepa}}).

\clearpage{\pagestyle{empty}\cleardoublepage} 


\chapter{A consistency condition in heterotic theory}

Suppose that we consider
the Bianchi identity associated with the 3-form as
\begin{eqnarray}
\label{bianx}
dH = - {\alpha' \over 4} \bigg( {\rm tr}({\cal R}  \wedge {\cal R}) - {\rm tr}(F \wedge F) \bigg) + \mathcal{O}(\alpha'^2)~,
\end{eqnarray}
where $ {\cal R}$ is a spacetime curvature which will be specified later.
Also observe that the 2-form gauge potential and the Einstein equation can be written together as
\begin{eqnarray}
\label{einx}
R^{(-)}{}_{MN}
+2\nabla^{(-)}_M \nabla_N \Phi
+ {\alpha' \over 4} \bigg({\cal R}_{M L_1, L_2 L_3}
{\cal R}_N{}^{L_1, L_2 L_3}-F_{M Lab}F_N{}^{Lab} \bigg)=\mathcal{O}(\alpha'^2)~.
\end{eqnarray}
Then one can establish by direct computation that
\begin{eqnarray}
R^{(-)}{}_{M[N,PQ]}=-{1\over3} \nabla^{(-)}_M H_{NPQ}-{1\over6} dH_{MNPQ}~.
\label{bianx}
\end{eqnarray}
Using this and the field equations of the theory, one can derive the relation
\begin{eqnarray}
&&R^{(-)}{}_{MN,PQ}\Gamma^N \Gamma^{PQ}\epsilon=-{1\over3} \nabla^{(-)}_M\big(H_{LPQ} \Gamma^{LPQ}-12 \partial_L\Phi \Gamma^L\big)\epsilon
\cr
&&~~~~~~-{\alpha'\over4} [{\cal R}_{MN,EF} {\cal R}_{PQ,}{}^{EF}- F_{MN ab} F_{PQ}{}^{ab}] \Gamma^N \Gamma^{PQ}\epsilon + \mathcal{O}(\alpha'^2)~.
\label{xxxid}
\end{eqnarray}
If $\epsilon$ satisfies the gravitino KSE, the left hand side of this relation vanishes.  Furthermore the right-hand-side vanishes
as well provided that the dilatino and gaugino KSEs are satisfied, and in addition
\begin{eqnarray}
 {\cal R}_{PQ,}{}^{EF}\Gamma^{PQ}\epsilon=\mathcal{O}(\alpha')~.
\label{instcon}
\end{eqnarray}
Of course in heterotic string perturbation theory
\begin{eqnarray}
R^{(+)}{}_{PQ,}{}^{EF}\Gamma^{PQ}\epsilon={\cal O}(\alpha')~,
 \end{eqnarray}
 as a consequence of the gravitino KSE and the closure of $H$ at that order. Thus one can set $ {\cal R}=R^{(+)}$ and the identity
 (\ref{xxxid}) will hold up to order $\alpha'^2$.

One consequence of the identity (\ref{xxxid}) is that if the gravitino KSE and gaugino KSEs are satisfied as well as (\ref{instcon}) but the dilatino is not, then the gravitino KSE admits an additional parallel spinor of the opposite chirality.  Such kind of identities have been established before for special cases in \cite{howegp}. Here we have shown that this result is generic in the context of heterotic theory.

\clearpage{\pagestyle{empty}\cleardoublepage} 


\chapter{Lichnerowicz Theorem Computation}
\label{calcId}

In this appendix, we present the details for the calculation of
the functional ${\cal{I}}$ defined in ({\ref{I functional}}),
and show how the constants $q$ and $c$ are fixed by requiring
that certain types of terms which arise in the calculation should
vanish. We begin by considering the calculation at zeroth order
in $\alpha'$, and then include the corrections at first order in
$\alpha'$. We remark that we shall retain terms of the type
$h^i \tilde{\nabla}_i \Phi$ throughout. This is because although
these terms vanish at zeroth order in $\alpha'$ as a consequence
of the analysis in Section 8, it does not follow from this analysis
that ${\cal{L}}_h \Phi = \mathcal{O}(\alpha'^2)$. However, as we shall see, it turns out that the
coefficient multiplying the terms $h^i \tilde{\nabla}_i \Phi$, which depends
on the  constants $q$ and $c$, vanishes when one requires that
several other terms in ${\cal{I}}$ vanish as well. So these
terms do not give any contribution to ${\cal{I}}$ at either zeroth or
first order in $\alpha'$.

\setcounter{subsection}{0}

\subsection{Computations at zeroth order in $\alpha'$}
\label{I zeroth order}
Throughout the following analysis, we assume Einstein equations, dilaton field equation and Bianchi identity at zeroth order in $\alpha'$.
To proceed, we expand out the definition of ${\nabla}^{(\kappa)}_i$ and $\mathcal{D}$ in $\mathcal{I}$, obtaining the following expression
\begin{align}
\label{expansion}
\notag
\mathcal{I} &= \int_{\mathcal{S}} e^{c\Phi} 2(\kappa - q) \langle \Gamma^i \mathcal{A} \eta_{\pm} , \tilde{\nabla}^{(-)}_i \eta_{\pm} \rangle + e^{c\Phi} (8\kappa^2 - q^2) \langle \eta_{\pm} , \mathcal{A}^{\dagger} \mathcal{A} \eta_{\pm} \rangle  \\
& \qquad\quad - e^{c\Phi} \langle \tilde{\nabla}^{(-)}_i \eta_{\pm} , \Gamma^{ij} \tilde{\nabla}^{(-)}_j \eta_{\pm} \rangle \ .
\end{align}

Now, after writing $\tilde{\nabla}^{(-)}$ in terms of the Levi-Civita connection $\tilde{\nabla}$ and after integrating by parts, the expression (\ref{expansion}) decomposes into
\begin{eqnarray}
\mathcal{I} = \mathcal{I}_1 +\mathcal{I}_2 + \mathcal{I}_3 \ ,
\end{eqnarray}
where
\begin{align}
\mathcal{I}_1 = \int_{\mathcal{S}}  &e^{c\Phi} 2(\kappa- q) \langle \eta_{\pm} , \mathcal{A}^{\dagger} \mathcal{D} \eta_{\pm} \rangle + e^{c\Phi} (8\kappa^2 - 2 \kappa q + q^2) \langle \eta_{\pm} , \mathcal{A}^{\dagger}\mathcal{A} \eta_{\pm} \rangle   \\
&- \frac{1}{64} e^{c\Phi}\langle \eta_{\pm} , \Gamma^{\ell_1\ell_2} \Gamma^{ij} \Gamma^{\ell_3\ell_4} W_{i\ell_1\ell_2}W_{j\ell_3\ell_4} \eta_{\pm} \rangle  \ ,
\end{align}
and
\begin{align}
\notag
\mathcal{I}_2 = \int_{\mathcal{S}} &c e^{c\Phi} \langle \eta_{\pm} , \Gamma^{ij} \tilde{\nabla}_j \eta_{\pm} \rangle  + \frac{1}{8} e^{c\Phi} \langle \tilde{\nabla}_i \eta_{\pm}, \Gamma^{ij}\Gamma^{\ell_1\ell_2} W_{j\ell_1\ell_2} \eta_{\pm} \rangle  \\
&- \frac{1}{8} e^{c\Phi} \langle \eta_{\pm}, \Gamma^{\ell_1\ell_2} \Gamma^{ij} W_{j\ell_1\ell_2} \tilde{\nabla}\eta_{\pm} \rangle  \ ,
\end{align}
and
\begin{eqnarray}
\mathcal{I}_3 = \int_{\mathcal{S}} - e^{c\Phi} \langle \tilde{\nabla}_i \eta_{\pm} , \Gamma^{ij} \tilde{\nabla}_j \eta_{\pm} \rangle \ .
\end{eqnarray}
In particular, we note the identity
\begin{eqnarray}
\Gamma^{\ell_1\ell_2} \Gamma^{ij} \Gamma^{\ell_3\ell_4} W_{i\ell_1\ell_2}W_{j\ell_3\ell_4} = 8 W^i{}_{\ell_1\ell_2}W_{i\ell_3\ell_4} \Gamma^{\ell_1\ell_2\ell_3\ell_4} - 4 W_{ijk}W^{ijk} \ ,
\end{eqnarray}
which simplifies $\mathcal{I}_1$. After integrating by parts the second term in $\mathcal{I}_2$, we have
\begin{align}
\label{I_2 parts}
\notag
\mathcal{I}_2 = \int_{\mathcal{S}} &c e^{c\Phi} \langle \eta_{\pm} , \Gamma^{ij} \tilde{\nabla}_j \eta_{\pm} \rangle  - \frac{1}{8} e^{c\Phi} \langle \eta_{\pm} , \left( \Gamma^{ij}\Gamma_{mn} - \Gamma_{mn}\Gamma^{ij} \right) W_j{}^{mn} \tilde{\nabla}_i \eta_{\pm} \rangle \\
& - \frac{c}{8} e^{c\Phi} \langle \eta_{\pm} , \Gamma^{i\ell_1\ell_2\ell_3} \tilde{\nabla}_i \Phi W_{\ell_1\ell_2\ell_3} \eta_{\pm} \rangle    - \frac{1}{8}e^{c\Phi} \langle \eta_{\pm} , \Gamma^{\ell_1\ell_2\ell_3\ell_4} \tilde{\nabla}_{\ell_1}W_{\ell_2\ell_3\ell_4} \eta_{\pm} \rangle  \ ,
\end{align}
where the last term is order $\alpha'$, so we shall neglect it. Now we shall focus on the second term of (\ref{I_2 parts}). First note that
\begin{eqnarray}
\left( \Gamma^{ij}\Gamma_{mn} - \Gamma_{mn}\Gamma^{ij} \right) W_j{}^{mn} = - 4 \Gamma^{mn} W^i{}_{mn} =  \frac{4}{3} W_{\ell_1\ell_2\ell_3} \left( \Gamma^{\ell_1\ell_2\ell_3}\Gamma^i + \Gamma^{i\ell_1\ell_2\ell_3} \right) \ .
\end{eqnarray}
Then, after an integration by parts and after writing $\tilde{\nabla}$ in terms of $\mathcal{D}$, we have
\begin{align}
\label{GGexp}
\notag
\int_{\mathcal{S}}- \frac{1}{8} e^{c\Phi} \langle \eta_{\pm} , &\left( \Gamma^{ij}\Gamma_{mn} -  \Gamma_{mn}\Gamma^{ij} \right) W_j{}^{mn} \tilde{\nabla}_i \eta_{\pm} \rangle = \int_{\mathcal{S}}  - \frac{1}{6}e^{c\Phi} \langle \eta_{\pm} , W_{\ell_1\ell_2\ell_3}\Gamma^{\ell_1\ell_2\ell_3} \mathcal{D} \eta_{\pm} \rangle  \\
\notag
&+ \frac{q}{6} e^{c\Phi} \langle \eta_{\pm} , W_{\ell_1\ell_2\ell_3}\Gamma^{\ell_1\ell_2\ell_3}\mathcal{A} \eta_{\pm} \rangle  - \frac{1}{48} e^{c\Phi} \langle \eta_{\pm} , W_{\ell_1\ell_2\ell_3}\Gamma^{\ell_1\ell_2\ell_3} W_{ijk}\Gamma^{ijk} \eta_{\pm} \rangle  \\
&+ \frac{c}{12} e^{c\Phi} \langle \eta_{\pm} , \Gamma^{i\ell_1\ell_2\ell_3} \tilde{\nabla}_i \Phi W_{\ell_1\ell_2\ell_3} \eta_{\pm} \rangle  + \frac{1}{12} e^{c\Phi} \langle \eta_{\pm} , \Gamma^{\ell_1\ell_2\ell_3\ell_4}\tilde{\nabla}_{\ell_1} W_{\ell_2\ell_3\ell_4}  \eta_{\pm} \rangle \ .
\end{align}
The last term of (\ref{GGexp}) is order $\alpha'$, so we shall neglect it. To proceed further, we shall substitute $W_{ijk}\Gamma^{ijk}$ in terms of $\mathcal{A}$, using its definition. This produces terms proportional to the norm squared of $\mathcal{A}\, \eta_{\pm}$, together with a number of counterterms. In detail, one obtains
\begin{align}
\notag
\int_{\mathcal{S}}- \frac{1}{8} e^{c\Phi} \langle \eta_{\pm} , &\left( \Gamma^{ij}\Gamma_{mn} -  \Gamma_{mn}\Gamma^{ij} \right) W_j{}^{mn} \tilde{\nabla}_i \eta_{\pm} \rangle = \int_{\mathcal{S}}  - \frac{1}{6}e^{c\Phi} \langle \eta_{\pm} , W_{\ell_1\ell_2\ell_3}\Gamma^{\ell_1\ell_2\ell_3} \mathcal{D} \eta_{\pm} \rangle  \\
\notag
& + e^{c\Phi} \left(\frac{1}{48} - \frac{q}{6}\right) \langle \eta_{\pm} , \mathcal{A}^{\dagger}\mathcal{A} \eta_{\pm} \rangle + e^{c\Phi}\left(\frac{1}{2} - 2q\right) \langle \eta_{\pm} , \Gamma^i \tilde{\nabla}_i\Phi \mathcal{A} \eta_{\pm} \rangle \\
\notag
& \pm e^{c\Phi} \left(\frac{1}{4}-q \right) \langle \eta_{\pm} , \Gamma^i h_i \mathcal{A} \eta_{\pm} \rangle + 3 e^{c\Phi} \langle \eta_{\pm} , \tilde{\nabla}_i \Phi \tilde{\nabla}^i \Phi \eta_{\pm} \rangle \pm 3e^{c\Phi}\langle\eta_{\pm} , h^i\tilde{\nabla}_i\Phi \eta_{\pm} \rangle \\
& + \frac{3}{4} e^{c\Phi} \langle \eta_{\pm} , h_i h^i \eta_{\pm} \rangle + \frac{c}{12} e^{c\Phi} \langle \eta_{\pm} , \Gamma^{i\ell_1\ell_2\ell_3} \tilde{\nabla}_i \Phi W_{\ell_1\ell_2\ell_3} \eta_{\pm} \rangle + \mathcal{O}(\alpha')\ .
\end{align}
Let us focus now on the first term of (\ref{I_2 parts}). After writing $\Gamma^{ij}$ as $\Gamma^i\Gamma^j - \delta^{ij}$ and after integrating by parts, we have
\begin{align}
\label{I_2first}
\notag
\int_{\mathcal{S}} c e^{c\Phi} \langle \eta_{\pm} , \Gamma^{ij} \tilde{\nabla}_j \eta_{\pm} \rangle = \int_{\mathcal{S}} &c e^{c\Phi} \langle \eta_{\pm} , \Gamma^\ell \tilde{\nabla}_\ell\Phi \Gamma^i \tilde{\nabla}_i \eta_{\pm} \rangle \\
&+ \frac{c}{2} e^{c\Phi} \langle \eta_{\pm} , \tilde{\nabla}_i\tilde{\nabla}^i \Phi  \eta_{\pm} \rangle +  \frac{c^2}{2} e^{c\Phi}\langle \eta_{\pm} , \tilde{\nabla}_i\Phi\tilde{\nabla}^i\Phi  \eta_{\pm} \rangle \ .
\end{align}
The first term in the RHS of (\ref{I_2first}) can be rewritten in terms of the modified Dirac operator $\mathcal{D}$ after subtracting suitable terms. The second term on the RHS can be further simplified using the dilaton field equation at zeroth order in $\alpha'$. On performing these calculations, we have
\begin{align}
\notag
\int_{\mathcal{S}} c e^{c\Phi} \langle \eta_{\pm} , \Gamma^{ij} \tilde{\nabla}_j \eta_{\pm} \rangle = \int_{\mathcal{S}} &c e^{c\Phi} \langle \eta_{\pm} , \Gamma^\ell \tilde{\nabla}_\ell\Phi \mathcal{D} \eta_{\pm} \rangle -\frac{c}{24} e^{c\Phi} \langle \eta_{\pm} , W_{ijk}W^{ijk} \eta_{\pm} \rangle \\
\notag
& + c\left( \frac{1}{8} - q \right) e^{c\Phi} \langle \eta_{\pm} , \Gamma^{i\ell_1\ell_2\ell_3} \tilde{\nabla}_i\Phi W_{\ell_1\ell_2\ell_3} \eta_{\pm} \rangle + \frac{c}{4} e^{c\Phi} \langle \eta_{\pm} , h_i h^i \eta_{\pm} \rangle \\
\notag
&+ 12c \left(\frac{1}{12}+ \frac{c}{24} + q \right) e^{c\Phi} \langle \eta_{\pm} , \tilde{\nabla}_i \Phi \tilde{\nabla}^i \Phi \eta_{\pm} \rangle \\
&+ 6c\left(\frac{1}{12}\pm q\right)e^{c\Phi}\langle\eta_{\pm} , h^i\tilde{\nabla}_i\Phi \eta_{\pm} \rangle + \mathcal{O}(\alpha')~.
\end{align}
Let us now focus on $\mathcal{I}_3$. Recall that
\begin{eqnarray}
\label{Ricci scalar}
\Gamma^{ij}\tilde{\nabla}_i\tilde{\nabla}_j \eta_{\pm} = - \frac{1}{4} \tilde{R}\, \eta_{\pm} \ .
\end{eqnarray}
Therefore after integrating by parts and using (\ref{Ricci scalar}) neglecting $\alpha'$ corrections from Einstein equations, $\mathcal{I}_3$ becomes
\begin{align}
\notag
\mathcal{I}_3 = \int_{\mathcal{S}} - \frac{5}{48} e^{c\Phi} \langle \eta_{\pm} , W_{ijk}W^{ijk} \eta_{\pm} \rangle +& e^{c\Phi} \langle \eta_{\pm} , \tilde{\nabla}_i\Phi \tilde{\nabla}^i \Phi \eta_{\pm} \rangle + \frac{1}{4} e^{c\Phi} \langle \eta_{\pm} , h_i h^i \eta_{\pm} \rangle \\
& + e^{c\Phi}\langle\eta_{\pm} , h^i\tilde{\nabla}_i\Phi \eta_{\pm} \rangle + \mathcal{O}(\alpha') \ .
\end{align}
Collecting together all terms and substituting $h_ih^i$ by inverting the zeroth order in $\alpha'$  dilaton filed equation, one finally gets
\begin{align}
\label{I final}
\notag
\mathcal{I} = \int_{\mathcal{S}} & e^{c\Phi} \langle \eta_{\pm} , \left(c\Gamma^\ell\tilde{\nabla}_\ell\Phi - \frac{1}{6} W_{\ell_1\ell_2\ell_3}\Gamma^{\ell_1\ell_2\ell_3} + 2(\kappa - q) \mathcal{A}^{\dagger} \right) \mathcal{D} \eta_{\pm} \rangle \\
\notag
& + (8\kappa^2 -2\kappa q- \frac{q}{12} + q^2) e^{c\Phi} \langle \eta_{\pm} , \mathcal{A}^{\dagger} \mathcal{A} \eta_{\pm} \rangle \\
\notag
& + \frac{3}{4}\left( q - \frac{1}{12}\right) e^{c\Phi} \langle \eta_{\pm} , W^i{}_{\ell_1\ell_2} W^{i\ell_3\ell_4} \Gamma^{\ell_1\ell_2\ell_3\ell_4} \eta_{\pm} \rangle \\
\notag
&-  c \left(q -\frac{1}{12}\right) e^{c\Phi} \langle \eta_{\pm} , \Gamma^{i\ell_1\ell_2\ell_3}\tilde{\nabla}_i\Phi W_{\ell_1\ell_2\ell_3} \eta_{\pm} \rangle \\
\notag
& + 6 \left(\frac{1}{12} + q + \frac{c}{12}\right) e^{c\Phi} \langle \eta_{\pm} , \tilde{\nabla}_i\tilde{\nabla}^i \Phi \eta_{\pm} \rangle \\
\notag
&+ 12c \left( q + \frac{c}{24} \right) e^{c\Phi} \langle \eta_{\pm}, \tilde{\nabla}_i\Phi\tilde{\nabla}^i\Phi \eta_{\pm} \rangle \\
& + \left( \frac{1}{2}-6q \pm 6q(c+2)\right) e^{c\Phi}\langle\eta_{\pm} , h^i\tilde{\nabla}_i\Phi \eta_{\pm} \rangle + \mathcal{O}(\alpha')  \ .
\end{align}
In order to eliminate the term $\langle \eta_\pm, W^i{}_{\ell_1 \ell_2}
W_{i \ell_3 \ell_4} \Gamma^{\ell_1 \ell_2 \ell_3 \ell_4} \eta_\pm \rangle$,
which has no sign and cannot be rewritten in terms of ${\cal{D}}$
or $\mathcal{A}^\dagger \mathcal{A}$,
we must set
\begin{eqnarray}
q= {1 \over 12} + \mathcal{O}(\alpha') \ .
\end{eqnarray}
and then in order to eliminate the $\langle \eta_\pm, \tilde{\nabla}^i \tilde{\nabla}_i \Phi \eta_\pm \rangle$
term we must further set
\begin{eqnarray}
c=-2+\mathcal{O}(\alpha') \ .
\end{eqnarray}
Then (\ref{I final}) simplifies to
\begin{eqnarray}
\mathcal{I} = \int_{\mathcal{S}}  e^{-2\Phi} \langle \eta_{\pm} , \Psi \mathcal{D} \eta_{\pm} \rangle + \left(8\kappa^2 -\frac{\kappa}{6}\right) \int_{\mathcal{S}} e^{-2\Phi} \parallel \mathcal{A} \, \eta_{\pm} \parallel^2 +\, \mathcal{O}(\alpha') \ ,
\end{eqnarray}
where
\begin{eqnarray}
\label{Psi}
\Psi \equiv -2\Gamma^\ell\tilde{\nabla}_\ell\Phi - \frac{1}{6} W_{\ell_1\ell_2\ell_3}\Gamma^{\ell_1\ell_2\ell_3} + 2\left(\kappa - \frac{1}{12}\right) \mathcal{A}^{\dagger} \ .
\end{eqnarray}

\subsection{Computations at first order in $\alpha'$}
In this section we shall consider corrections at first order in $\alpha'$. $\mathcal{I}_2$ and $\mathcal{I}_3$ gain $\alpha'$ corrections from bosonic field equations and Bianchi identity, while $\mathcal{I}_1$ does not. Therefore we have
\begin{align}
\notag
\mathcal{I}_1 = &\int_{\mathcal{S}} e^{c\Phi} 2(\kappa- q) \langle \eta_{\pm} , \mathcal{A}^{\dagger} \mathcal{D} \eta_{\pm} \rangle + e^{c\Phi} (8\kappa^2 - 2 \kappa q + q^2) \langle \eta_{\pm} , \mathcal{A} ^{\dagger}\mathcal{A} \eta_{\pm} \rangle   \\
&- \frac{1}{8} e^{c\Phi}\langle \eta_{\pm} ,  W^i{}_{\ell_1\ell_2}W_{i\ell_3\ell_4}\Gamma^{\ell_1\ell_2\ell_3\ell_4}\eta_{\pm} \rangle  + \frac{1}{16}e^{c\Phi} \langle\eta_{\pm} , W_{ijk}W^{ijk}\eta_{\pm} \rangle + \mathcal{O}(\alpha'^2)\ ,
\end{align}
and
\begin{align}
\notag
\mathcal{I}_2 = \int_{\mathcal{S}}  &\ c e^{c\Phi} \langle \eta_{\pm} , \left(\Gamma^\ell \tilde{\nabla}_\ell\Phi  -\frac{1}{6}W_{\ell_1\ell_2\ell_3}\Gamma^{\ell_1\ell_2\ell_3}\right)\mathcal{D} \eta_{\pm} \rangle -\frac{c}{24} e^{c\Phi} \langle \eta_{\pm} , W_{ijk}W^{ijk} \eta_{\pm} \rangle \\
\notag
& + c\left( \frac{5}{24} - q \right) e^{c\Phi} \langle \eta_{\pm} , \Gamma^{i\ell_1\ell_2\ell_3} \tilde{\nabla}_i\Phi W_{\ell_1\ell_2\ell_3} \eta_{\pm} \rangle + e^{c\Phi} \left(\frac{1}{48} - \frac{q}{6}\right) \langle \eta_{\pm} , \mathcal{A}^{\dagger}\mathcal{A} \eta_{\pm} \rangle \\
\notag
&  + e^{c\Phi}\left(\frac{1}{2} - 2q\right) \langle \eta_{\pm} , \Gamma^i \tilde{\nabla}_i\Phi \mathcal{A} \eta_{\pm} \rangle \pm e^{c\Phi} \left(\frac{1}{4}-q \right) \langle \eta_{\pm} , \Gamma^i h_i \mathcal{A} \eta_{\pm} \rangle \\
\notag
& + \left(\frac{3}{4} + \frac{c}{4}\right) e^{c\Phi} \langle \eta_{\pm} , h_i h^i \eta_{\pm} \rangle + \left(c + \frac{c^2}{2} + 12cq + 3 \right) e^{c\Phi} \langle \eta_{\pm} , \tilde{\nabla}_i \Phi \tilde{\nabla}^i \Phi \eta_{\pm} \rangle  \\
\notag
& + \left(\frac{c}{2}\pm 6cq\pm 3\right) e^{c\Phi}\langle\eta_{\pm} , h^i\tilde{\nabla}_i\Phi \eta_{\pm} \rangle \\
\notag
&-\frac{1}{24}e^{c\Phi} \langle \eta_{\pm} , \Gamma^{\ell_1\ell_2\ell_3\ell_4} \tilde{\nabla}_{\ell_1}W_{\ell_2\ell_3\ell_4} \eta_{\pm} \rangle + \alpha' \frac{c}{32} e^{c\Phi} \bigg( -2 \langle \eta_{\pm},  dh_{ij}dh^{ij} \eta_{\pm} \rangle  \\
&+ \langle \eta_{\pm} , \tilde{R}^{(+)}{}_{\ell_1\ell_2,\ell_3\ell_4}\tilde{R}^{(+)}{}^{\ell_1\ell_2,\ell_3\ell_4} \eta_{\pm} \rangle - \langle \eta_{\pm} , \tilde{F}_{ij}{}^{ab}\tilde{F}^{ij}{}_{ab} \eta_{\pm} \rangle \bigg) + \mathcal{O}(\alpha'^2) \ ,
\end{align}
and
\begin{align}
\notag
\mathcal{I}_3 = \int_{\mathcal{S}} & - \frac{5}{48} e^{c\Phi} \langle \eta_{\pm} , W_{ijk}W^{ijk} \eta_{\pm} \rangle + e^{c\Phi} \langle \eta_{\pm} , \tilde{\nabla}_i\Phi \tilde{\nabla}^i \Phi \eta_{\pm} \rangle + \frac{1}{4} e^{c\Phi} \langle \eta_{\pm} , h_i h^i \eta_{\pm} \rangle  \\
\notag
&+ e^{c\Phi}\langle\eta_{\pm} , h^i\tilde{\nabla}_i\Phi \eta_{\pm} \rangle +\alpha' \frac{3}{32} e^{c\Phi} \bigg( -2 \langle \eta_{\pm},  dh_{ij}dh^{ij} \eta_{\pm} \rangle \\
&+ \langle \eta_{\pm} , \tilde{R}^{(+)}{}_{\ell_1\ell_2,\ell_3\ell_4}\tilde{R}^{(+)}{}^{\ell_1\ell_2,\ell_3\ell_4} \eta_{\pm} \rangle - \langle \eta_{\pm} , \tilde{F}_{ij}{}^{ab}\tilde{F}^{ij}{}_{ab} \eta_{\pm} \rangle \bigg) + \mathcal{O}(\alpha'^2) \ .
\end{align}
Combining all together and considering $\alpha'$ corrections from substituting $h_ih^i$ by inverting the dilaton field equations, we have
\begin{align}
\label{I alpha' 1}
\notag
\mathcal{I} = \int_{\mathcal{S}} & e^{c\Phi} \langle \eta_{\pm} , \left(c\Gamma^\ell\tilde{\nabla}_\ell\Phi - \frac{1}{6} W_{\ell_1\ell_2\ell_3}\Gamma^{\ell_1\ell_2\ell_3} + 2(\kappa - q) \mathcal{A}^{\dagger} \right) \mathcal{D} \eta_{\pm} \rangle \\
\notag
& + (8\kappa^2 -2\kappa q- \frac{q}{12} + q^2) e^{c\Phi} \langle \eta_{\pm} , \mathcal{A}^{\dagger} \mathcal{A} \eta_{\pm} \rangle \\
\notag
& + \frac{3}{4}\left( q - \frac{1}{12}\right) e^{c\Phi} \langle \eta_{\pm} , W^i{}_{\ell_1\ell_2} W^{i\ell_3\ell_4} \Gamma^{\ell_1\ell_2\ell_3\ell_4} \eta_{\pm} \rangle \\
\notag
&-  c \left(q -\frac{1}{12}\right) e^{c\Phi} \langle \eta_{\pm} , \Gamma^{i\ell_1\ell_2\ell_3}\tilde{\nabla}_i\Phi W_{\ell_1\ell_2\ell_3} \eta_{\pm} \rangle + 12c \left( q + \frac{c}{24} \right) e^{c\Phi} \langle \eta_{\pm}, \tilde{\nabla}_i\Phi\tilde{\nabla}^i\Phi \eta_{\pm} \rangle \\
\notag
& + 6 \left(\frac{1}{12} + q + \frac{c}{12}\right) e^{c\Phi} \langle \eta_{\pm} , \tilde{\nabla}_i\tilde{\nabla}^i \Phi \eta_{\pm} \rangle  + \left( \frac{1}{2}-6q \pm 6q(c+2)\right)e^{c\Phi}\langle\eta_{\pm} , h^i\tilde{\nabla}_i\Phi \eta_{\pm} \rangle \\
\notag
&+ \frac{\alpha'}{64}e^{c\Phi} \bigg( 2 \langle\eta_{\pm}, \Gamma^{\ell_1\ell_2\ell_3\ell_4}dh_{\ell_1\ell_2}dh_{\ell_3\ell_4}\rangle - \langle\eta_{\pm}, \Gamma^{\ell_1\ell_2\ell_3\ell_4}\tilde{R}^{(+)}{}_{\ell_1\ell_2, ij}\tilde{R}^{(+)}{}_{\ell_3\ell_4,}{}^{ij}\eta_{\pm}\rangle  \\
\notag
& \hspace{7cm}+ \langle\eta_{\pm}, \Gamma^{\ell_1\ell_2\ell_3\ell_4}\tilde{F}_{\ell_1\ell_2,\, ab}\tilde{F}_{\ell_3\ell_4}{}^{ab}\eta_{\pm}\rangle \bigg)  \\
\notag
&+ \alpha' \frac{3}{8}\left(\frac{1}{6}-q\right) e^{c\Phi}\bigg( -2 \langle \eta_{\pm},  dh_{ij}dh^{ij} \eta_{\pm} \rangle \\
&+ \langle \eta_{\pm} , \tilde{R}^{(+)}{}_{\ell_1\ell_2,\ell_3\ell_4}\tilde{R}^{(+)}{}^{\ell_1\ell_2,\ell_3\ell_4} \eta_{\pm} \rangle - \langle \eta_{\pm} , \tilde{F}_{ij}{}^{ab}\tilde{F}^{ij}{}_{ab} \eta_{\pm} \rangle \bigg) + \mathcal{O}(\alpha'^2) \ .
\end{align}

To further simplify (\ref{I alpha' 1}), we note the following identity
\begin{eqnarray}
\label{dh identity}
\notag
\langle \eta_{\pm} , \Gamma^{\ell_1\ell_2\ell_3\ell_4} dh_{\ell_1\ell_2} dh_{\ell_3\ell_4} \eta_{\pm} \rangle &=& 
\langle \eta_{\pm} , \Gamma^{\ell_1\ell_2}dh_{\ell_1\ell_2} \Gamma^{\ell_3\ell_4}dh_{\ell_3\ell_4} \eta_{\pm} \rangle  \\
&+& 2 \langle \eta_{\pm} , dh_{ij}dh^{ij} \eta_{\pm} \rangle  \ .
\end{eqnarray}
Identities analogous to (\ref{dh identity}) hold also for the terms which involve $\tilde{R}^{(+)}{}_{ij,k\ell}$ and $\tilde{F}_{ij}{}^{ab}$.
This leads to
\begin{align}
\label{I alpha' 2}
\notag
\mathcal{I} = \int_{\mathcal{S}} & e^{c\Phi} \langle \eta_{\pm} , \left(c\Gamma^\ell\tilde{\nabla}_\ell\Phi - \frac{1}{6} W_{\ell_1\ell_2\ell_3}\Gamma^{\ell_1\ell_2\ell_3} + 2(\kappa - q) \mathcal{A}^{\dagger} \right) \mathcal{D} \eta_{\pm} \rangle \\
\notag
& + (8\kappa^2 -2\kappa q- \frac{q}{12} + q^2) e^{c\Phi} \langle \eta_{\pm} , \mathcal{A}^{\dagger} \mathcal{A} \eta_{\pm} \rangle \\
\notag
& + \frac{3}{4}\left( q - \frac{1}{12}\right) e^{c\Phi} \langle \eta_{\pm} , W^i{}_{\ell_1\ell_2} W^{i\ell_3\ell_4} \Gamma^{\ell_1\ell_2\ell_3\ell_4} \eta_{\pm} \rangle \\
\notag
&-  c \left(q -\frac{1}{12}\right) e^{c\Phi} \langle \eta_{\pm} , \Gamma^{i\ell_1\ell_2\ell_3}\tilde{\nabla}_i\Phi W_{\ell_1\ell_2\ell_3} \eta_{\pm} \rangle + 12c \left( q + \frac{c}{24} \right) e^{c\Phi} \langle \eta_{\pm}, \tilde{\nabla}_i\Phi\tilde{\nabla}^i\Phi \eta_{\pm} \rangle \\
\notag
& + 6 \left(\frac{1}{12} + q + \frac{c}{12}\right) e^{c\Phi} \langle \eta_{\pm} , \tilde{\nabla}_i\tilde{\nabla}^i \Phi \eta_{\pm} \rangle  + \left( \frac{1}{2}-6q \pm 6q(c+2)\right)e^{c\Phi}\langle\eta_{\pm} , h^i\tilde{\nabla}_i\Phi \eta_{\pm} \rangle\\
\notag
&+{3 \over 8}\alpha' (q-{1 \over 12})e^{c \Phi}
\bigg(2 dh_{ij} dh^{ij} + {\tilde{F}}_{ij}{}^{ab} {\tilde{F}}^{ij}{}_{ab}
- \tilde{R}^{(+)}{}_{\ell_1 \ell_2,\ell_3 \ell_4}
\tilde{R}^{(+)}{}^{ \ell_1 \ell_2, \ell_3 \ell_4} \bigg) \parallel \eta_{\pm} \parallel^2 \\
\notag
& - \frac{\alpha'}{32} e^{c\Phi} \parallel \slashed{dh}\, \eta_{\pm}\parallel ^2
- \frac{\alpha'}{64} e^{c\Phi} \parallel \slashed{\tilde{F}}\, \eta_{\pm} \parallel ^2 + \frac{\alpha'}{64} e^{c\Phi}\langle \tilde{R}^{(+)}{}_{\ell_1\ell_2,\, ij}\Gamma^{\ell_1\ell_2}\eta_{\pm}, \tilde{R}^{(+)}{}^{ ij}_{\ell_3\ell_4,}\Gamma^{\ell_3\ell_4}\eta_{\pm}\rangle
\\ \notag
 &+ \, \mathcal{O}(\alpha'^2) \ .
\end{align}
In order to eliminate the term $\langle \eta_\pm, W^i{}_{\ell_1 \ell_2}
W_{i \ell_3 \ell_4} \Gamma^{\ell_1 \ell_2 \ell_3 \ell_4} \eta_\pm \rangle$,
which has no sign and cannot be rewritten in terms of ${\cal{D}}$
or $\mathcal{A}^\dagger \mathcal{A}$,
we must set
\begin{eqnarray}
q= {1 \over 12} + \mathcal{O}(\alpha'^2) \ .
\end{eqnarray}
and then in order to eliminate the $\langle \eta_\pm, \tilde{\nabla}^i \tilde{\nabla}_i \Phi \eta_\pm \rangle$
term we must further set
\begin{eqnarray}
c=-2+\mathcal{O}(\alpha'^2) \ .
\end{eqnarray}
Then (\ref{I alpha' 2}) is significantly simplified to
\begin{align}
\notag
\mathcal{I}  & = \left(8\kappa^2 - \frac{1}{6} \kappa \right) \int_{\mathcal{S}} e^{-2 \Phi} \parallel \mathcal{A}\,  \eta_{\pm} \parallel^2
+ \int_{\mathcal{S}} e^{-2\Phi} \langle \eta_{\pm}, \Psi \mathcal{D} \eta_{\pm} \rangle \\
&- \frac{\alpha'}{64} \int_{\mathcal{S}} e^{-2\Phi} \left( 2 \parallel \slashed{dh}\, \eta_{\pm} \parallel^2 + \parallel \slashed{\tilde{F}} \eta_{\pm} \parallel^2 - \langle \tilde{R}^{(+)}{}_{\ell_1\ell_2,\, ij}\Gamma^{\ell_1\ell_2}\eta_{\pm}, \tilde{R}^{(+)}{}^{ij}_{\ell_3\ell_4,}\Gamma^{\ell_3\ell_4}\eta_{\pm}\rangle \right) + \mathcal{O}(\alpha'^2)\ ,
\end{align}
where $\Psi$ is defined in (\ref{Psi}).

\clearpage{\pagestyle{empty}\cleardoublepage} 


\chapter{$AdS_{n+1}$ as warped product over $AdS_n$}
\label{apx:AdS_n}
The $AdS_{n+1}$ space can be written  as a warped product over $AdS_n$. This has been observed before in \cite{strominger} for $AdS_3$ and elsewhere, e.g. \cite{slicex}. For this, we
label all geometrical objects defined on $AdS_{n+1}$ and $AdS_n$ by $n+1$ and $n$ respectively, e.g. $ds^2_{n+1}$ is the metric on $AdS_{n+1}$ and $ds^2_n$ is the metric on $AdS_n$. In principle $AdS_{n+1}$ and $AdS_n$ can have different radii, which are indicated by $\ell_{n+1}$ and $\ell_n$ respectively.
Coordinates on $AdS_{n+1}$ are taken to be as follows
\begin{eqnarray}
x^{I} = (x^0, x^i) \ , \qquad\qquad x^0 \equiv y \ , \qquad i= 1, ... , n \ .
\end{eqnarray}

We shall begin with an ansatz for the metric on $AdS_{n+1}$ as a warped product over $AdS_n$, i.e.
\begin{eqnarray}
\label{metric ansatz n+1}
ds_{n+1}^2 = dy^2 + f(y)^2 ds_n^2 \ .
\end{eqnarray}
We want to determine the necessary and sufficient conditions to impose on $f(y)$ in order for $ds^2_{n+1}$ to be the metric on $AdS_{n+1}$. To succeed, we have to impose the fact $AdS_{n+1}$ is a maximally symmetric space. Locally, the necessary and sufficient condition is that the Riemann tensor must assume the following form
\begin{eqnarray}
\label{maximally_symm}
R^{(n+1)}_{IJKL} = -\frac{1}{\ell^2_{n+1}} \left(g^{(n+1)}_{IK}g^{(n+1)}_{JL} - g^{(n+1)}_{JK}g^{(n+1)}_{IL} \right) \ ,
\end{eqnarray}
Equation (\ref{maximally_symm}) implies also that the metric (\ref{metric ansatz n+1}) is Einstein and the curvature scalar is constant and negative, i.e.
\begin{eqnarray}
\label{Einstein}
R^{(n+1)}_{IJ} = - \frac{n}{\ell^2_{n+1}} g^{(n+1)}_{IJ} \ , \qquad\qquad R^{(n+1)}= -\frac{1}{\ell^2_{n+1}}n(n+1) \ .
\end{eqnarray}

The non-vanishing Christoffel symbols of (\ref{metric ansatz n+1}) are:
\begin{eqnarray}
\Gamma^{(n+1)\, k}_{\qquad i \  0} = \frac{f'(y)}{f(y)} \delta^k{}_i \ , \quad \Gamma^{(n+1)\, 0}_{\qquad i \  j} = - f(y)f'(y)g^{(n+1)}_{ij} \ , \quad
\Gamma^{(n+1)\, k}_{\qquad i \  j} = \Gamma^{(n)\, k}_{\quad\, i \  j}  \ .
\end{eqnarray}
The non-vanishing Riemann tensor components are:
\begin{align}
\notag
R^{(n+1)}{}_{i0,}{}^{k}{}_0 &= - \frac{f''(y)}{f(y)} \delta^k{}_i \ , \\
\notag
R^{(n+1)}{}_{i0,}{}^0{}_{\ell} &= f(y)f''(y) g^{(n)}_{i\ell} \ , \\
R^{(n+1)}{}_{ij,}{}^k{}_{\ell} & =  R^{(n)}{}_{ij,}{}^k{}_{\ell} + f'(y)^2\left(\delta^k{}_j g^{(n)}_{i\ell} - \delta^k{}_i g^{(n)}_{j\ell} \right) \ ,
\end{align}
and
\begin{align}
\notag
R^{(n+1)}_{i0,k0} &= - f(y)f''(y) g_{ik}^{(n)} \ , \\
R^{(n+1)}_{ij, kl} &= f(y)^2 R^{(n)}_{ij,kl} - f(y)^2f'(y)^2\left(g^{(n)}_{ik}g^{(n)}_{jl} - g^{(n)}_{jk}g^{(n)}_{il} \right) \ .
\end{align}
The non-vanishing Ricci tensor components are:
\begin{align}
\label{Ricci n+1}
\notag
R^{(n+1)}_{00} &= - n \frac{f''(y)}{f(y)} \ , \\
R^{(n+1)}_{ij} &= R^{(n)}_{ij} + \left[f'(y)^2(1-n) - f(y)f''(y)\right] g^{(n)}_{ij} \ .
\end{align}
The Riemann tensor on $AdS_n$ must assume the following form
\begin{eqnarray}
R^{(n)}_{ijk\ell} = - \frac{1}{\ell^2_{n}}\left(g_{ik}^{(n)}g_{j\ell}^{(n)} - g_{jk}^{(n)}g_{i\ell}^{(n)} \right) \ .
\end{eqnarray}
Now we impose (\ref{maximally_symm}). The $(i0,k0)$-components provide the first ordinary differential equation for $f$
\begin{eqnarray}
\label{1_ODE}
f''(y) = \frac{1}{\ell^2_{n+1}} f(y) \ .
\end{eqnarray}
The $(ij,kl)$-components provide the second ordinary differential equation for $f$
\begin{eqnarray}
\label{2_ODE}
f'(y)^2 - \frac{1}{\ell^2_{n+1}} f(y)^2 + \frac{1}{\ell^2_n} = 0 \ .
\end{eqnarray}
Since equations in (\ref{Einstein}) are derived from (\ref{maximally_symm}), they would imply again (\ref{1_ODE}) and (\ref{2_ODE}), so there is nothing further to be learned from those conditions.
The general solution of (\ref{1_ODE}) and (\ref{2_ODE}) is
\begin{eqnarray}
\label{warped}
f(y) = \alpha \cosh\left(\frac{y}{\ell_{n+1}}\right) + \beta \sinh\left(\frac{y}{\ell_{n+1}}\right)\ ,
\end{eqnarray}
where $\alpha$ and $\beta$ are constants which satisfy
\begin{eqnarray}
\alpha^2 - \beta^2 = \frac{\ell^2_{n+1}}{\ell^2_{n}} \ .
\end{eqnarray}
The solution (\ref{warped}) leads us to the following conclusions
\begin{enumerate}
\item if $y\in (-\infty, +\infty)$, then locally the $AdS_{n+1}$ metric can be written as $AdS_n\times_w \mathbb{R}$.
\item if $y\in [0, 1]$, then locally the $AdS_{n+1}$ metric can be written as  $AdS_n\times_w [0, 1]$ as the warp factor is not periodic.
\item if $y\in [0, 1]$ and force periodicity on $y$, then the metric of $AdS_n\times_w S^1$ is discontinuous as the warp factor is not periodic.
\end{enumerate}
From the perspective of near horizon geometries where $n=2$, the first case violates the compactness condition of the partial horizon section. The second case implies that the spatial horizon
 section has a  boundary. The third case violates  smoothness condition since (\ref{warped}) is not periodic.
Hence  all cases violate one or more  of the  assumptions required to prove that there are no  $AdS_2$ horizons in the heterotic theory.

\clearpage{\pagestyle{empty}\cleardoublepage} 


\chapter{A consistency check for the integral R-matrix}
\label{consistency}
In this appendix, we compute the path-ordered formula (\ref{int_R}) for the R-matrix on a specific contour, which is most convenient for the calculation, and check that it reproduces (\ref{Rchosen}). We choose the path of integration $\gamma$ to be a straight line with initial point $(p_2, p_2)$ and final point $(p_1, p_2)$. Therefore we only need to integrate along the $p_1$-axis as follows
\begin{equation}
\label{Wilsone}
R (p_1, p_2) = \Pi \circ \mathscr{P} \exp{\int_{p_2}^{p_1} dx \Gamma_1 (x,p_2)} \ ,
\end{equation} 
where $\mathscr{P} \exp$ is the path-ordered exponential. 
We begin by expanding out (\ref{Wilsone}) as follows
\begin{eqnarray}
\notag
\mathscr{P} \exp{\int_{p_2}^{p_1} dx \, \Gamma_1 (x,p_2)} = \mathds{1} + \int_{p_2}^{p_1} dx \, \Gamma_1 (x,p_2) + \int_{p_2}^{p_1} dx \int_{p_2}^x dy \, \Gamma_1 (x,p_2) \, \Gamma_1(y,p_2) + \mathcal{O}(\Gamma^3) \ . \\
\end{eqnarray}
By using the expression for $\Gamma_1$ in (\ref{connection}), we obtain
\begin{eqnarray}
\label{retu}
R &=& \mathsf{E}_{11} \otimes \mathsf{E}_{11} - \mathsf{E}_{22} \otimes \mathsf{E}_{22} + \big(\mathsf{E}_{11} \otimes \mathsf{E}_{22} - \mathsf{E}_{22} \otimes \mathsf{E}_{11}\big) \, \sin \int_{p_2}^{p_1} g(x, p_2) \, dx \nonumber \\
&+& \big(\mathsf{E}_{21} \otimes \mathsf{E}_{12} - \mathsf{E}_{12} \otimes \mathsf{E}_{21}\big) \, \cos \int_{p_2}^{p_1} g(x,p_2) \, dx \ ,
\end{eqnarray}
where 
\begin{equation}
\mathsf{E}_{11} \equiv \begin{pmatrix}1&0\\0&0\end{pmatrix}\ ,
\qquad 
\mathsf{E}_{22} \equiv \begin{pmatrix}0&0\\0&1\end{pmatrix} \ , \qquad
\mathsf{E}_{12} \equiv \begin{pmatrix}0&1\\0&0\end{pmatrix}\ ,
\qquad
\mathsf{E}_{21} \equiv \begin{pmatrix}0&0\\1&0\end{pmatrix}\ ,
\end{equation}
and
\begin{equation}
\label{g}
g(p_1,p_2) \equiv -\frac{1}{4} \sqrt{\frac{\sin \frac{p_2}{2}}{\sin \frac{p_1}{2}}} \, \frac{1}{\sin \frac{p_1 + p_2}{4}} \ . 
\end{equation}
In terms of the matrix representation, (\ref{retu}) can be written as
\begin{equation}
\label{R_check}
R = \begin{pmatrix}1&0&0&0\\0&\sin \sigma &\cos \sigma&0\\0&\cos \sigma&-\sin \sigma&0\\0&0&0&-1\end{pmatrix}, \qquad \sigma \equiv \int _{p_2}^{p_1} g(x,p_2) \, dx \ ,
\end{equation}
where we used the fact that $\mathfrak{a} \otimes \mathfrak{b} \, |v\rangle \otimes |w\rangle = (-)^{|\mathfrak{b}||v|} \mathfrak{a}|v\rangle \otimes \mathfrak{b}|w\rangle$. 
After integrating $\sigma$, one can explicitly verify that (\ref{R_check}) reproduces (\ref{Rchosen}), up to the dressing factor. 

This proves that the integral expression for the R-matrix, which we found by using the boost symmetry, exactly reproduces the massless left-left R-matrix which we chose at the beginning.

\clearpage{\pagestyle{empty}\cleardoublepage} 


\chapter{Alternative interpretations of the boost action}
\label{alternative}

In this appendix, we collect some speculative ideas which might provide alternative descriptions of the $q$-deformed Poincar\'e boost generators $\mathcal{J}_L, \mathcal{J}_R$.

\vspace{5mm}
\begin{center}
{\bf Berry phase}
\end{center}

One might conceive regimes where the geometric and quantum mechanical interpretations we have outlined in the main text converge into a single picture, inspired by the notion of {\it Berry phase}. This might tie in with the link drawn in \cite{Joakim} with the physics of phonons, excitations created by particles moving in the potential of slowly-vibrating ions in a crystal. From the viewpoint of the $q$-Poincar\'e algebra, the momenta $(p_1, p_2)$ cohere as a single phonon \cite{Joakim}, which could be described by a single-particle quantum mechanics. A Berry-phase picture could link to our fibre bundle, with the momenta $p^M(\tau)$ as adiabatically-changing variables\footnote{This might be described by the so-called {\it vacuum bundles} \cite{Hori}, pointed out to us by J. McOrist.}.

\vspace{5mm}
\begin{center}
{\bf Lax pairs}
\end{center}

If we read the flatness of $\Gamma$ in terms of a Lax pair, then this could define a {\it classically integrable} system (although it is not a mathematical implication). If this were the case, this could be yet another subsidiary interpretation. $R$ would then be the solution of the auxiliary linear problem, therefore  connected to the {\it Gel'fand-Levitan-Marchenko} equation (reviewed {\it e.g.} in \cite{revDurham}) giving soliton solutions {\it via} the classical inverse scattering method. One issue is that $[\Gamma_1,\Gamma_2]$ vanishes on its own. Such Lax pairs are sometimes called {\it weak} \cite{colorado} - as the momenta became {\it coordinates}, we have no spectral parameter, and conservation laws trivialise. One could envisage introducing a spectral parameter ({\it baxterisation}). This might affect (or perhaps resolve) some of the singularities of $\Gamma$. It could also provide a link with the recent results of \cite{Klose:2016uur}.

\vspace{5mm}
\begin{center}
{\bf Universal $R$-matrix}
\end{center}

We observe that (\ref{int_R}) could be rewritten in terms of the supercharges $\mathcal{Q}$ and $\mathcal{S}$, by recombining suitable factors of $\sqrt{\sin \frac{p}{2}}$ in the exponent. When so expressed, we believe this should provide equivalent rewritings of the {\it universal $R$-matrix} of the $q$-deformed Poincar\'e Hopf-superalgebra. Not surprisingly, universal $R$-matrices are traditionally given by exponential formulas. It would be interesting to verify this claim from first principles in view of \cite{Beisert:2016qei}, and get an algebraic expression for the scalar factor \cite{Borsato:2016xns}.  

When regarded in this perspective, our approach is very reminiscent of the one developed in \cite{Maillet}. Here, the standpoint is slightly different, as in our particular case the ordinary classical $r$-matrix cannot be defined \cite{Joakim}. Nevertheless, the two procedures become very close in appearance when considering $\check{r}$ \cite{Joakim}, with the crucial distinction that the latter is not a solution of the classical Yang-Baxter equation. We feel however that there should be a strong relationship, given the striking resemblance.

\vspace{5mm}
\begin{center}
{\bf Similarities}
\end{center}

It is interesting to mention that $\Delta(\mathcal{J})$ resembles a (deformed) super-Poincar\'e generator in two dimensions, whose typical undeformed version reads in superspace
$
J^{\alpha \beta} = x^\alpha P^\beta - x^\beta P^\alpha - \frac{1}{2} P_\gamma \bar{\theta} \gamma^{\alpha \beta \gamma} \theta.
$
This would present $R$ as a $q$-super-translation invariant. 

\smallskip

Additionally, (\ref{DR=0}) reminds of Knizhnik-Zamolodchikov (KZ) equations and their quantisation \cite{FR}. The analogy with the KZ equation becomes stronger if we  consider that the algebraic part of (\ref{DR=0}) is proportional to $\check{r}$ of \cite{Joakim}. It would be fascinating to connect this to $q$-CFTs \cite{Sara1, Sara2}, or form factors in integrable models \cite{Smirnov:1993gp}.

\clearpage{\pagestyle{empty}\cleardoublepage} 




\fancyhead[RO,LE]{\thepage}
\fancyhead[CO,CE]{}
\fancyhead[LO,RE]{References}
\fancyfoot{}




\fancyhead[RO,LE]{\thepage}
\fancyhead[CO,CE]{}
\fancyhead[LO,RE]{Curriculum vitae}
\fancyfoot{}





\end{document}